\documentclass[%
 reprint,
 amsmath,amssymb,
 aps,
]{revtex4}

\usepackage{graphicx}% Include figure files
\usepackage{dcolumn}% Align table columns on decimal point
\usepackage{bm}% bold math
\usepackage{longtable}
\usepackage{float}
\usepackage{subfigure}
\usepackage{overpic}
\usepackage{changepage}
\usepackage[version=4]{mhchem}
\usepackage[dvipsnames]{xcolor}
\def\simge{\mathrel{\rlap{\raise 0.511ex
       \hbox{$>$}}{\lower 0.511ex \hbox{$\sim$}}}}
\def\simle{\mathrel{\rlap{\raise 0.511ex
        \hbox{$<$}}{\lower 0.511ex \hbox{$\sim$}}}}

\begin{document}

\preprint{APS/123-QED}

\title{Compiled Properties of Nucleonic Matter and Nuclear and Neutron Star Models from Non-Relativistic and Relativistic Interactions}% 

\author{Boyang Sun}

\author{Saketh Bhattiprolu}
 
 \author{James M. Lattimer}

\affiliation{%
 Department of Physics \& Astronomy, Stony Brook University, Stony Brook, NY 11794 USA}%

\date{\today}

\begin{abstract}
This paper compiles the model parameters and zero-temperature properties of an extensive collection of published theoretical nuclear interactions, including 255 non-relativistic (Skyrme-like) forces, 270 relativistic mean field (RMF) and point-coupling (RMF-PC) forces, and 13 Gogny-like forces.  This forms the most exhaustive tabulation of model parameters to date. The properties of uniform symmetric matter and pure neutron matter at the saturation density are determined.  Symmetry properties found from the second-order term of a Taylor expansion in neutron excess are compared to the energy difference of pure neutron and symmetric nuclear matter at the saturation density.  Selected liquid-droplet model parameters, including the surface tension and surface symmetry energy, are determined for semi-infinite surfaces. Theoretical liquid droplet model neutron skin thicknesses of the neutron-rich closed-shell nuclei $^{48}$Ca and $^{208}$Pb are compared to published theoretical Hartree-Fock and experimental results.  A similar comparison is made between theoretical liquid-droplet and experimental values of dipole polarizabilities. In addition, radii, binding energies, moments of inertia and tidal deformabilities of 1.2, 1.4 and 1.6 M$_\odot$ neutron stars are computed.  An extensive correlation analysis of bulk matter, nuclear structure, and low-mass neutron star properties is performed and compared to nuclear experiments and astrophysical observations.  

\end{abstract}

\maketitle

\section{\label{sec:level1}Introduction}
The properties of uniform matter consisting of neutrons and protons at zero temperature are functions of the total baryon density $n$ and the proton fraction $x$.  The equation of state of uniform matter between about half the nuclear saturation density $n_0\approx0.16$ fm$^{-3}$ and $2n_0$ is an essential ingredient in models of neutron stars and in calculations of gravitational collapse supernovae and in neutron star mergers.  For example, the radii of typical neutron stars between $1.3M_\odot$ and $1.6M_\odot$ depends strongly on the symmetry energy near $n_0$, and the amount of mass ejected in a binary neutron star merger, in which both components are likely in this mass range, is very sensitive to the neutron star radius.   

Many microscopic models of dense matter begin with a two-body nucleon-nucleon interaction with parameters fitted to nucleon-nucleon phase shifts and to properties of the deuteron. But many-body formalisms based on these inputs have historically had difficulty in matching the saturation properties established, for example, from experimental binding energies using the empirical liquid droplet or Hartree-Fock nuclear models.  
As a result, the vast majority of published interactions have instead followed more empirical approaches.

The simplest ones employ non-relativistic effective density-dependent zero-range two-and three-nucleon interactions, whose parameters are fit to the properties of nuclei and/or symmetric matter at the saturation density.  These zero-range non-relativistic interactions are collectively known as Skyrme-type interactions, and lead to an energy density that is, effectively, a power-law series expressed in terms of the wave number or baryon density.  A more complicated non-relativistic approach (Gogny-type) utilizes  short finite-range interactions.

A distinctly different theoretical relativistic mean field (RMF) approach is based on quantum hadrodynamics and involves forces mediated by two to four meson fields, which leads to an energy density that depends not only on the baryon density and the proton fraction, but also on the meson field strengths. The meson fields are determined at each baryon density and proton fraction by energy minimization, which leads to a more complex density-dependence of the energy density than for Skyrme-type interactions.  Several variations of the RMF approach, depending on whether the couplings are linear or non-linear in the fields, have been proposed.

The uniform matter properties for  hundreds of published models using these two main approaches were compiled by Dutra et al. in Ref. \cite{dutra2012skyrme} for 240 Skyrme-like interactions and
in Ref. \cite{dutra2014relativistic} for 263 RMF interactions (see also Ref.\cite{PhysRevC.107.035805}).  The main purposes of this paper are to:
\begin{itemize}
   \item Create a database of interaction parameters for Skyrme-, RMF- and Gogny-type forces, which were not tabulated in Refs. \cite{dutra2012skyrme,dutra2014relativistic};
   \item  remove duplicated forces and include additional forces, as well as correcting typos in previous works;
\item compare symmetry properties of these forces found by both differencing the energies of pure neutron and symmetric matter and computing the quadratic term of an expansion of the energy in powers of the neutron excess;
    \item compute  properties of a semi-infinite surface, including the surface tension and surface symmetry energy; 
    \item tabulate the liquid droplet model (DM) neutron skin thicknesses and dipole polarizabilities of $^{48}$Ca and $^{208}$Pb and compare the former to Hartree-Fock computations and experiments;
    \item compute low-mass neutron star properties, such as radii, binding energies, moments of inertia and tidal deformabilities; and
    \item produce extensive correlation analyses among bulk matter, nuclear structure, and low-mass neutron star properties. 
    \end{itemize}
The creation of this parameter database will greatly ease future studies, alleviating the need to peruse hundreds of individual publications for this information.

The bulk of the force models we consider were included in Refs. \cite{dutra2012skyrme,dutra2014relativistic}.  We include additional force models and correct a few published parameter sets.  At the same time, in the case of RMF models, we selectively omit a few models because of ambiguities in their published parameters or the models' complexity.  We did not include some recent RMF forces because they do not fit into the parameterization schemes of Ref. \cite{dutra2014relativistic}, such as the 18 forces [V,S,M][P,Z,N][E,R] in Ref. \cite{typel2018relativistic}.  These will be included in future work.

%\begin{widetext}
\section{Skyrme-Type Interactions}

\subsection{Energy Density}
The energy density of bulk matter in Skyrme-type models contains multiple parameters $t_i,x_i$ and $\sigma_i$:
\begin{eqnarray}
{\cal E} &=&\frac{3\hbar^2}{10M}\left(\frac{3\pi^2}{2}\right)^{2/3} n^{5/3}H_{5/3}+\frac{t_0}{4} n^2\left[x_0+2-\left(x_0+\frac{1}{2}\right)\textit{H}_2\right]+\frac{1}{24}\sum\limits_{i=1}^3t_{3i} n^{\sigma_i+2}\left[x_{3i}+2-\left(x_{3i}+\frac{1}{2}\right)\textit{H}_2\right]\nonumber\\
&+&\frac{3}{40}\left(\frac{3\pi^2}{2}\right)^{2/3} n^{8/3}(a\textit{H}_{5/3}+b\textit{H}_{8/3})+\frac{3}{40}\left(\frac{3\pi^2}{2}\right)^{2/3} n^{8/3+\delta}\left[t_4(x_4+2)\textit{H}_{5/3}-t_4\left(x_4+\frac{1}{2}\right)\textit{H}_{8/3}\right],\nonumber\\
&+&\frac{3}{40}\left(\frac{3\pi^2}{2}\right)^{2/3} n^{8/3+\gamma}\left[t_5(x_5+2)\textit{H}_{5/3}+t_5\left(x_5+\frac{1}{2}\right)\textit{H}_{8/3}\right]\\\nonumber
&+& \frac{1}{2}\left[Q_{nn}(\vec{\nabla} n_n)^2+2Q_{np}\vec{\nabla} n_n \cdot\vec{\nabla} n_p+Q_{pp}(\vec\nabla n_p)^2\right], 
\label{eq:skyrme}\end{eqnarray}
with $a$ and $b$ being the parameter combinations
\begin{equation}
a = t_1(x_1+2)+t_2(x_2+2), \qquad b = t_2\left(x_2+{\frac{1}{2}}\right)-t_1\left(x_1+{\frac{1}{2}}\right) \label{eq:ab},
\end{equation}
and $H_n(x)= 2^{n-1}[x^n+(1-x)^n]$, such that $H_n(x=0) = 2^{n-1}$ for pure neutron matter (PNM) and $H_n(x=1/2) = 1$ for symmetric nuclear matter (SNM). The coefficients $Q_{ij}$ in the spatially varying part of the Skyrme Hamiltonian density are 
\begin{eqnarray}
Q_{nn}=Q_{pp} &=& {3\over16} \left[ t_1 \left(1 - x_1 \right) - t_2\left( 1 + x_2 \right) +t_4n^\delta\left(1-x_4+{4\delta\over3}\left[1+{x_4\over2}-x_t\left({1\over2}+x_4\right)\right]\right)-t_5n^\gamma(1+x_5)\right]\,,\nonumber\\
Q_{np}=Q_{pn} &=& {1\over8}\left[ 3 t_1 \left(1 + \frac{x_1}{2} \right) - t_2 \left( 1 + \frac{x_2}{2} \right)+{3t_4n^\delta\over2}(2+x_4+\delta)-t_5n^\gamma\left(1+{x_5\over2}\right) \right]\,. \label{eq:qnn}
\end{eqnarray}
The first term in Eq. (1) is the kinetic energy density for non-interacting neutrons and protons.  Together with the terms involving $a,b$ and $t_4$, it forms the total kinetic energy density.  Equivalently, the bare nucleon mass $M$ is replaced by the effective neutron ($t=n$) and proton ($t=p$) masses $M^*_t$:
\begin{equation}
    {\hbar^2\over2M_t^*}={\hbar^2\over2M}+{n\over4}\left({a\over2}+bx_t+t_4n^\delta\left[1+{x_4\over2}-\left({1\over2}+x_4\right)x_t\right]+t_5n^\gamma\left[1+{x_5\over2}+\left({1\over2}+x_5\right)x_t\right]\right).
\label{eq:eff}\end{equation} 
where $x_n=1-x$ and $x_p=x$.  We ignored contributions from Coulomb and spin-orbit interactions which do not contribute to the energy density of uniform bulk matter.

Modifications to the conventional, original form of ${\cal E}$, as described by Vautherin and Brink in Ref. \cite{vautherin1972hartree}, include the summation over index i in the 
$t_{3i}$ term introduced by Agrawal \emph{et al} in Ref. \cite{agrawal2006exploring} and additional terms involving $t_4, x_4, t_5$ and $x_5$ used by Chamel \emph{et al} in Refs. \cite{chamel2008further,chamel2009further}.
 
\subsection{Uniform Nuclear Matter}
In uniform matter, the density gradients $\vec{\nabla} n_n=\vec{\nabla} n_p = 0$.  The energy per particle $E={\cal E}_B/n$ is then obtained as
\begin{eqnarray}
E(n,x)&=& \frac{3\hbar^2}{10M}\left(\frac{3\pi^2}{2}\right)^{2/3} n^{2/3}\textit{H}_{5/3}+\frac{t_0}{4} n\left[x_0+2-\left(x_0+{1\over2}\right)\textit{H}_2\right]+\frac{1}{24}\sum\limits_{i=1}^3t_{3i} n^{\sigma_i+1}\left[x_{3i}+2-\left(x_{3i}+{1\over2}\right)\textit{H}_2\right]\nonumber\\
&+&\frac{3}{40}\left(\frac{3\pi^2}{2}\right)^{2/3} n^{5/3}(a\textit{H}_{5/3}+b\textit{H}_{8/3})+\frac{3t_4}{40}\left(\frac{3\pi^2}{2}\right)^{2/3} n^{5/3+\delta}\left[(x_4+2)\textit{H}_{5/3}-\left(x_4+\frac{1}{2}\right)\textit{H}_{8/3}\right]\nonumber\\
&+&\frac{3t_5}{40}\left(\frac{3\pi^2}{2}\right)^{2/3} n^{5/3+\gamma}\left[(x_5+2)\textit{H}_{5/3}+\left(x_5+\frac{1}{2}\right)\textit{H}_{8/3}\right], 
\label{eq:eskyrme}
\end{eqnarray}

The pressure is given as
\begin{eqnarray}
P( n, x) &=& n^2\frac{\partial E}{\partial n}=\frac{\hbar^2}{5M}\left(\frac{3\pi^2}{2}\right)^{2/3} n^{5/3}\textit{H}_{5/3}+\frac{1}{48}\sum\limits_{i=1}^3t_{3i}(\sigma_i+1) n^{\sigma_i+2}\left[2(x_{3i}+2)-(2x_{3i}+1)\textit{H}_2\right]  \nonumber\\
&+&\frac{1}{8}\left(\frac{3\pi^2}{2}\right)^{2/3} n^{8/3}(a\textit{H}_{5/3}+b\textit{H}_{8/3})+\frac{t_4}{40}\left(\frac{3\pi^2}{2}\right)^{2/3}(5+3\delta) n^{\frac{8}{3}+\delta}\left[(x_4+2)\textit{H}_{5/3}-\left(x_4+\frac{1}{2}\right)\textit{H}_{8/3}\right]  \nonumber\\
&+&\frac{t_0}{8} n^2\left[2(x_0+2)-(2x_0+1)\textit{H}_2\right]+\frac{t_5}{40}\left(\frac{3\pi^2}{2}\right)^{2/3}(5+3\gamma) n^{\frac{8}{3}+\gamma}\left[(x_5+2)\textit{H}_{5/3}+\left(x_5+\frac{1}{2}\right)\textit{H}_{8/3}\right].
\end{eqnarray}
The saturation density $n_0$ is defined by $P(n_0, 1/2)=0$.  The volume incompressibility is defined as
\begin{eqnarray}
\mathcal{K}( n,x)&=&9 n^2\frac{\partial^2E}{\partial n^2}=-\frac{3\hbar^2}{5M}\left(\frac{3\pi^2}{2}\right)^{2/3} n^{2/3}\textit{H}_{5/3} \nonumber\\
&+&\frac{3}{16}\sum\limits_{i=1}^3t_{3i}\sigma_i(\sigma_i+1) n^{\sigma_i+1}\left[2(x_{3i}+2)-(2x_{3i}+1)\textit{H}_2\right]+3\left(\frac{3\pi^2}{2}\right)^{2/3} n^{5/3}(a\textit{H}_{5/3}+b\textit{H}_{8/3}) \nonumber\\
&+&\frac{3t_4}{40}\left(\frac{3\pi^2}{2}\right)^{2/3}(2+3\delta)(5+3\delta) n^{\frac{5}{3}+\delta}\left[(x_4+2)\textit{H}_{5/3}-\left(x_4+\frac{1}{2}\right)\textit{H}_{8/3}\right] \nonumber\\
&+&\frac{3t_5}{40}\left(\frac{3\pi^2}{2}\right)^{2/3}(2+3\gamma)(5+3\gamma) n^{\frac{5}{3}+\gamma}\left[(x_5+2)\textit{H}_{5/3}+\left(x_5+\frac{1}{2}\right)\textit{H}_{8/3}\right].
\end{eqnarray}
Finally, the skewness is defined as
\begin{eqnarray}
\mathcal{Q}(n,x)&=&27 n^3\frac{\partial^3E}{\partial n^3}=\frac{12\hbar^2}{5M}\left(\frac{3\pi^2}{2}\right)^{2/3} n^{2/3}\textit{H}_{5/3}-\frac{3}{4}\left(\frac{3\pi^2}{2}\right)^{2/3} n^{5/3}(a\textit{H}_{5/3}+b\textit{H}_{8/3}) \nonumber\\
&+&\frac{9}{16}\sum\limits_{i=1}^3t_{3i}\sigma_i(\sigma_i^2-1) n^{\sigma_i+1}\left[2(x_{3i}+2)-(2x_{3i}+1)\textit{H}_2\right] \nonumber\\
&+&\frac{3t_4}{40}\left(\frac{3\pi^2}{2}\right)^{2/3}(2+3\delta)(5+3\delta)(3\delta-1) n^{\frac{5}{3}+\delta}\left[(x_4+2)\textit{H}_{5/3}-\left(x_4+\frac{1}{2}\right)\textit{H}_{8/3}\right] \nonumber\\
&+&\frac{3t_5}{40}\left(\frac{3\pi^2}{2}\right)^{2/3}(2+3\gamma)(5+3\gamma)(3\gamma-1) n^{\frac{5}{3}+\gamma}\left[(x_5+2)\textit{H}_{5/3}+\left(x_5+\frac{1}{2}\right)\textit{H}_{8/3}\right].
\end{eqnarray}
Relevant properties of SNM include $E_0=E(n_0,1/2)$, $P_0=P(n_0,1/2)=0$, $K_0 = \mathcal{K}( n_0, 1/2)$ and $Q_0 = \mathcal{Q}( n_0, 1/2)$ and those of PNM include $E_{N0}=E(n_0,0)$, $P_{N0}=P(n_0,0)$, $K_{N0}=\mathcal{K}(n_0,0)$ and $Q_{N0}=\mathcal{Q}(n_0,0)$.

\subsection{Symmetry Energy}
The symmetry energy can be defined in two ways.  First it can be expressed as the difference of the PNM energy and the SNM energy
\begin{equation}
\mathcal{S}_1( n) = E( n, 0)-E( n, 1/2),\label{eq:s1}
\end{equation}
and second, as the second derivative of $E$ with respect to the neutron excess $1-2x$ at $x=1/2$,
\begin{equation}
    \mathcal{S}_2(n)={1\over8}\left({\partial^2E\over\partial x^2}\right)_{x=1/2}.\label{eq:s2}
\end{equation}
If higher than quadratic terms in an expansion of $S$ in $1-2x$ are negligible, then $S_1$ and $S_2$ would be identical.
The symmetry energy defined in either way can be expanded in a Taylor series in $y=( n- n_0)/(3 n_0)$ around the saturation density $n_0$:
\begin{equation}
\mathcal{S}_{1,2}( n) = J_{1,2} + L_{1,2}y + \frac{1}{2!}K_{sym1,2}y^2+\frac{1}{3!}Q_{sym1,2}y^3+O(y^4),
\end{equation}
where $J_{1,2}, L_{1,2}, K_{sym1,2}$ and $Q_{sym1,2}$ are the so-called symmetry energy parameters.

In the first case, one finds
\begin{eqnarray}
J_1 &=& E_{N0}-E_0= \frac{3\hbar^2}{10M}\left(\frac{3\pi^2n_0}{2}\right)^{2/3}(2^{2/3}-1)-\frac{t_0}{8} n_0(2x_0+1)-\frac{1}{48}\sum\limits_{i=1}^3t_{3i} n_0^{\sigma_i+1}(2x_{3i}+1)  \nonumber\\
&+&\frac{3}{40}\left(\frac{3\pi^2}{2}\right)^{2/3} n_0^{5/3}\left[(2^{2/3}-1)a+(2^{5/3}-1)b\right]+\frac{3t_4}{40}\left(\frac{3\pi^2}{2}\right)^{2/3} n_0^{5/3+\delta}\left[2^{2/3}(1-x_4)-\frac{3}{2}\right]  \nonumber\\
&+&\frac{3t_5}{40}\left(\frac{3\pi^2}{2}\right)^{2/3} n_0^{5/3+\gamma}\left[3\cdot2^{2/3}(1+x_5)-\frac{5}{2}\left(1+\frac{4}{5}x_5\right)\right],\nonumber\\
L_1 &=&{3\over n_0}(P_{N0}-P_0)=\frac{3\hbar^2}{5M}\left({3\pi^2 n_0\over2}\right)^{2/3}\left(2^{2/3}-1\right)-{3t_0\over8}n_0(2x_0+1)-{1\over16}\sum_{i=1}^3t_{3i}n_0^{\sigma_i+1}(\sigma_i+1)(2x_{3i}+1)\nonumber\\
&+&{3\over8}\left({3\pi^2\over2}\right)^{2/3}n_0^{5/3}\left[\left(2^{2/3}-1\right)a+\left(2^{5/3}-1\right)b\right]+{3t_4\over40}\left({3\pi^2\over2}\right)^{2/3}n_0^{5/3+\delta}(5+3\delta)\left[2^{2/3}(1-x_4)-{3\over2}\right]\nonumber\\
&+&{3t_5\over40}\left({3\pi^2\over2}\right)^{2/3}n_0^{5/3+\gamma}(5+3\gamma)\left[3\cdot2^{2/3}(1+x_5)-{5\over2}\left(1+{4x_5\over5}\right)\right],\nonumber\\
%L_1 &=&{3P_{N0}\over n_0}=\frac{3\hbar^2}{5M}\left(3\pi^2 n_0\right)^{2/3}+\frac{3}{16}\sum\limits_{i=1}^3t_{3i}(\sigma_i+1) n_0^{\sigma_i+1}+\frac{3}{8}\left(3\pi^2\right)^{2/3} n_0^{5/3}(a+2b)  \nonumber\\
%&+&\frac{9}{8}t_0 n_0+\frac{3t_4}{40}\left(3\pi^2\right)^{2/3}(5+3\delta) n_0^{\frac{5}{3}+\delta}(1-x_4)+\frac{9t_5}{40}\left(3\pi^2\right)^{2/3}(5+3\gamma) n_0^{\frac{5}{3}+\gamma}(1+x_5),\nonumber\\
K_{sym1} &=&K_{N0}-K_0=-\frac{3\hbar^2}{5M}\left(\frac{3\pi^2}{2} n_0\right)^{2/3}(2^{2/3}-1)+3\left(\frac{3\pi^2}{2}\right)^{2/3} n_0^{5/3}\left[(2^{2/3}-1)a+(2^{5/3}-1)b\right] \nonumber\\
&-&\frac{3}{16}\sum\limits_{i=1}^3t_{3i}\sigma_i(\sigma_i+1) n_0^{\sigma_i+1}(2x_{3i}+1)+\frac{3t_4}{40}\left(\frac{3\pi^2}{2}\right)^{2/3}(2+3\delta)(5+3\delta) n_0^{\frac{5}{3}+\delta}\left[2^{2/3}(1-x_4)-\frac{3}{2}\right] \nonumber\\
&+&\frac{3t_5}{40}\left(\frac{3\pi^2}{2}\right)^{2/3}(2+3\gamma)(5+3\gamma) n_0^{\frac{5}{3}+\gamma}\left[3\cdot2^{2/3}(1+x_5)-\frac{5}{2}\left(1+\frac{4}{5}x_5\right)\right],\nonumber\\
Q_{sym1} &=& Q_{N0}-Q_0 = \frac{12\hbar^2}{5M}\left(\frac{3\pi^2}{2}\right)^{2/3} n_0^{2/3}(2^{2/3}-1) \nonumber\\
&-&\frac{3}{4}\left(\frac{3\pi^2}{2}\right)^{2/3} n_0^{5/3}\left[(2^{2/3}-1)a+(2^{5/3}-1)b\right]-\frac{9}{16}\sum\limits_{i=1}^3t_{3i}\sigma_i(\sigma_i+1)(\sigma_i-1) n_0^{\sigma_i+1}(2x_{3i}+1)\nonumber\\
&+&\frac{3t_4}{40}\left(\frac{3\pi^2}{2}\right)^{2/3}(2+3\delta)(5+3\delta)(3\delta-1) n_0^{\frac{5}{3}+\delta}\left[2^{2/3}(1-x_4)-\frac{3}{2}\right] \nonumber\\
&+&\frac{3t_5}{40}\left(\frac{3\pi^2}{2}\right)^{2/3}(2+3\gamma)(5+3\gamma)(3\gamma-1) n_0^{\frac{5}{3}+\gamma}\left[3\cdot2^{2/3}(1+x_5)-\frac{5}{2}\left(1+\frac{4}{5}x_5\right)\right].
\end{eqnarray}

In the second case,
\begin{eqnarray}
\mathcal{S}_2(n) &=&\frac{\hbar^2}{6M}\left(\frac{3\pi^2}{2}\right)^{2/3} n^{2/3}-\frac{t_0}{8}(2x_0+1) n-\frac{1}{48}\sum\limits_{i=1}^3t_{3i}(2x_{3i}+1) n^{\sigma_i+1}  \nonumber\\
&+&\frac{1}{24}\left(\frac{3\pi^2}{2}\right)^{2/3}(a+4b) n^{5/3}-\frac{t_4}{8}\left(\frac{3\pi^2}{2}\right)^{2/3}x_4 n^{\frac{5}{3}+\delta}+\frac{t_5}{24}\left(\frac{3\pi^2}{2}\right)^{2/3}(5x_5+4) n^{\frac{5}{3}+\gamma}.
\end{eqnarray}
At the saturation density, we have $J_2=\mathcal{S}_2(n_0)$. Other saturation properties are consequently given as
\begin{eqnarray}
L_2 &=& 3 n_0\left(\frac{\partial\mathcal{S}_2}{\partial  n}\right)_{ n= n_0} =\frac{\hbar^2}{3M}\left(\frac{3\pi^2}{2}\right)^{2/3} n_0^{2/3}-\frac{3t_0}{8}(2x_0+1) n_0-\frac{1}{16}\sum\limits_{i=1}^3t_{3i}(2x_{3i}+1)(\sigma_i+1) n_0^{\sigma_i+1} \nonumber\\
&+&\frac{5}{24}\left(\frac{3\pi^2}{2}\right)^{2/3}(a+4b) n_0^{5/3}-\frac{t_4}{8}\left(\frac{3\pi^2}{2}\right)^{2/3}(5+3\delta)x_4 n_0^{\frac{5}{3}+\delta}
+\frac{t_5}{24}\left(\frac{3\pi^2}{2}\right)^{2/3}(5+3\gamma)(5x_5+4) n_0^{\frac{5}{3}+\gamma},\nonumber \\
K_{sym2} &=&9n_0^2\left({\partial^2\mathcal{S}_2\over\partial n^2}\right)_{n=n_0}= -\frac{\hbar^2}{3M}\left(\frac{3\pi^2}{2}\right)^{2/3} n_0^{2/3}-\frac{3}{16}\sum\limits_{i=1}^3t_{3i}(2x_{3i}+1)(\sigma_i+1)\sigma_i n_0^{\sigma_i+1}+\nonumber\\
&+&\frac{5}{12}\left(\frac{3\pi^2}{2}\right)^{2/3}(a+4b) n_0^{5/3} -\frac{t_4}{8}\left(\frac{3\pi^2}{2}\right)^{2/3}(5+3\delta)(2+3\delta)x_4 n_0^{\frac{5}{3}+\delta}\nonumber\\
&+&\frac{t_5}{24}\left(\frac{3\pi^2}{2}\right)^{2/3}(5+3\gamma)(2+3\gamma)(5x_5+4) n_0^{\frac{5}{3}+\gamma}, \nonumber\\
Q_{sym2} &=&27n_0^3\left({\partial^3\mathcal{S}_2\over\partial n^3}\right)_{n=n_0}=
\frac{4\hbar^2}{3M}\left(\frac{3\pi^2}{2}\right)^{2/3} n_0^{2/3}-\frac{9}{16}\sum\limits_{i=1}^3t_{3i}(2x_{3i}+1)(\sigma_i+1)\sigma_i(\sigma_i-1) n_0^{\sigma_i+1}\nonumber\\ 
&-&\frac{5}{12}\left(\frac{3\pi^2}{2}\right)^{2/3}(a+4b) n_0^{5/3} 
-\frac{t_4}{8}\left(\frac{3\pi^2}{2}\right)^{2/3}(5+3\delta)(2+3\delta)(3\delta-1)x_4 n_0^{\frac{5}{3}+\delta} \nonumber\\
&+&\frac{t_5}{24}\left(\frac{3\pi^2}{2}\right)^{2/3}(5+3\gamma)(2+3\gamma)(3\gamma-1)(5x_5+4) n_0^{\frac{5}{3}+\gamma}.
\end{eqnarray}
The list of saturation properties of Skyrme forces is given in Table II of Appendix A.

\section{Relativistic Mean-Field Models}
In contrast to Skyrme-like models, RMF models describe the strong interaction by an exchange of mesons which are usually assumed to couple minimally to nucleons.  These include the isospin scalar-scalar $\sigma$, the isospin scalar-vector  $\omega$, and, in some cases, the isospin vector-vector $\rho$ and, in additional cases, the isospin vector-scalar  $\delta$ mesons.  The meson fields are approximated by their mean fields.  An interesting property of RMF models is the occurrence of two different particle number densities: vector densities  $n_n$ and $n_p$, which are equivalent to the baryon densities occurring in Skyrme-type interactions, and scalar densities $n_{sn}$ and $n_{sp}$. Their interplay is essential to describe the saturation of nuclear matter.  Most models considered assume the meson-nucleon couplings to be constant, but we also consider models in which either the couplings depend on baryon-density or the nucleons interact with each other only through effective point-like couplings not involving mesons.

\subsection{Models with Constant Couplings}

\subsubsection{Energy Density}
The total energy density of infinite bulk nuclear matter is assumed to be dependent both on the baryon densities $n_i$ ($i=n,p$) and the meson fields $\sigma, \omega$, and $\rho$ and $\delta$ when the model contains them:
\begin{eqnarray}
\mathcal{E}_B&=&\sum_{i=n,p}\mathcal{E}_{kin,i}+g_\omega\omega n+{1\over2}g_\rho\rho(n_n-n_p)+V(\sigma)-F(\omega,\sigma)-G(\rho,\omega,\sigma)+I(\delta),\label{eq:eps}\end{eqnarray}

with potential energy densities
\begin{eqnarray}
V(\sigma)&=&{\sigma^2\over(\hbar c)^3}\left[{1\over2}m_\sigma^2+\sigma\left({A\over3}+{B\over4}\sigma\right)\right],\quad
F(\omega,\sigma)={\omega^2\over(\hbar c)^3}\left[{1\over2}m_\omega^2+g_\omega^2\left(g_\sigma\sigma\left[\alpha_1+{\alpha_1^\prime\over2}g_\sigma\sigma\right]+{C\over4}g_\omega^2\omega^2\right)\right],\cr
G(\rho,\omega,\sigma)&=&{\rho^2\over(\hbar c)^3}\left[{1\over2}m_\rho^2+g_\rho^2\left(g_\sigma\sigma\left[\alpha_2+{\alpha_2^\prime\over2}g_\sigma\sigma\right]+{1\over2}\alpha_3^\prime g_\omega^2\omega^2+{1\over4}Dg_\rho^2\rho^2\right)\right],\quad
    I(\delta) = \frac{\delta^2}{(\hbar c)^3}{m_\delta^2\over2}.
\label{eq:potentials}
\end{eqnarray}
The kinetic contributions to the energy density are
\begin{equation}
\mathcal{E}_{kin,i}= \frac{1}{\pi^2}\int_0^{k_{Fi}}k^2\sqrt{(\hbar ck)^2+M^{*2}_i}dk={3\over4}E_{Fi}n_i+{1\over4}M^*_in_{si},\label{eq:ekin}
\end{equation}
with Fermi energies for nucleons $E_{Fi}=\sqrt{M^{*2}_i+\left(\hbar ck_{Fi}\right)^2}$, wave numbers $k_{Fi}=(3\pi^2n_i)^{1/3}$, and effective masses 
\begin{equation}
    M^*_n=M-g_\sigma\sigma-g_\delta\delta,\qquad M^*_p=M-g_\sigma\sigma+g_\delta\delta.
\end{equation}
As opposed to the case with Skyrme interactions, the neutron and proton effective masses differ only when $\delta$ mesons are included. The total vector and scalar densities are defined as $n = n_p +  n_n$ and $n_s =  n_{sp}+ n_{sn}$, where
\begin{equation}
n_{si}=\frac{M^*_i}{\pi^2}\int_0^{k_{Fi}}\frac{k^2dk}{\sqrt{(\hbar ck)^2+M_i^{*2}}}={1\over2\pi^2}\left({M_i^*\over\hbar c}\right)^3\left[{\hbar ck_{Fi}E_{Fi}\over M_i^{*2}}-\ln\left({\hbar ck_{F,i}+E_{F,i}\over M_i^*}\right)\right].
\label{eq:ns}
\end{equation}
    Our notation follows that of \cite{dutra2014relativistic} with the exception that the opposite sign is used for the $\rho$ and $\delta$ meson fields, such that here they are both positive for neutron-rich matter.

The values for the fields are determined from energy minimization at fixed baryon densities.  As discussed below in \S \ref{sec:surface}, in spatially-varying matter the total energy density for RMF forces also receives contributions from gradient terms:
\begin{eqnarray}
\mathcal{E}&=&{\cal E}_B+{1\over2\hbar c}\left[\left({d\sigma\over dr}\right)^2-\left({d\omega\over dr}\right)^2-\left({d\rho\over dr}\right)^2+\left({d\delta\over dr}\right)^2\right],
\label{eq:eps0}\end{eqnarray}
The Euler-Lagrange equations expressing minimization of the total energy density [Eq. (\ref{eq:eps0})] with respect to the meson fields, in  spatially non-uniform semi-infinite systems where curvature corrections can be ignored, and for fixed $k_{Fn}$ and $k_{Fp}$, are~\cite{steiner2005isospin}
\begin{eqnarray}
{d^2\sigma\over dr^2}&=&\hbar c\left[-g_\sigma n_s+{\partial V(\sigma)\over\partial\sigma}-{\partial F(\omega,\sigma)\over\partial\sigma}-{\partial G(\rho,\omega,\sigma)\over\partial\sigma}\right],\nonumber\\
{d^2\omega\over dr^2}&=&\hbar c\left[-g_\omega n+{\partial F(\omega,\sigma)\over\partial\omega}+{\partial G(\rho,\omega,\sigma)\over\partial\omega}\right],\nonumber\\
{d^2\rho\over dr^2}&=&\hbar c\left[-{g_\rho\over2}(n_n-n_p)+{\partial G(\rho,\omega,\sigma)\over\partial\rho}\right],
\nonumber\\
    \frac{d^2\delta}{dr^2} &=& \hbar c\left[-g_\delta (n_{sn}-n_{sp})+\frac{\partial I(\delta)}{\partial \delta}\right]. \label{eq:EL}
\end{eqnarray}
 We used the facts that 
 \begin{equation}({\partial\mathcal{E}_{kin,i}/\partial M_i^*})_{k_{Fi}}=n_{si},\qquad (\partial\mathcal{E}_{kin,i}/\partial\sigma)_{k_{Fi},\delta}=-g_\sigma n_{si},\qquad
 (\partial {\cal E}_{kin,i}/\partial\delta)_{k_{Fi},\sigma}=-\tau_i g_\delta n_{si},
 \end{equation}
where $\tau_i=\pm 1$ for neutrons and protons, respectively. %The difference between the neutron and proton number densities can also be written in terms of the proton fraction $x$, $n_n-n_p=(1-2x)n$. 
We can write Eq. (\ref{eq:EL}) as a vector equation:
\begin{equation}
{d^2{\bf X}\over dr^2}=\hbar c({\bf W}-{\bf H}),\qquad{\bf X}=\begin{pmatrix}\sigma\\\omega\\\rho\\\delta\end{pmatrix},\qquad {\bf H}={\partial (F+G-V-I)\over\partial{\bf X}},\qquad {\bf W}=\begin{pmatrix}-g_\sigma n_s \\
g_\omega n \\
\mbox{\Large${g_\rho\over2}$}(1-2x)n \\
-g_\delta(n_{sn}-n_{sp})\end{pmatrix},
\end{equation}
where the proton fraction is defined as $x=n_p/n$.

\subsubsection{Uniform Nuclear Matter}
In uniform matter, the gradients are zero, and the meson fields can be solved implicitly in terms of the densities:
\begin{eqnarray}
g_\sigma n_s&=&{\sigma\over(\hbar c)^3}\left[m_\sigma^2+\sigma(A+B\sigma)-g_\omega^2g_\sigma\omega^2(\alpha_1+\alpha_1^\prime g_\sigma\sigma)-g_\rho^2g_\sigma\rho^2(\alpha_2+\alpha_2^\prime g_\sigma\sigma)\right],\cr 
g_\omega n&=&{\omega\over(\hbar c)^3}\left[m_\omega^2+g_\omega^2\left( g_\sigma\sigma\left[\alpha_1+{\alpha_1^\prime\over2}g_\sigma\sigma\right]+Cg_\omega^2\omega^2+g_\rho^2\rho^2\alpha_3^\prime\right)\right],\cr 
{g_\rho\over2}(n_n-n_p)&=&{\rho\over(\hbar c)^3}\left[m_\rho^2+g_\rho^2\left(g_\sigma\sigma\left[\alpha_2+{\alpha_2^\prime\over2}g_\sigma\sigma\right]+{\alpha_3^\prime\over2}g_\omega^2\omega^2+Dg_\rho^2\rho^2\right)\right],\cr
g_\delta(n_{s,n}-n_{s,p})&=&{\delta\over(\hbar c)^3}m_\delta^2.
\label{eq:field}
\end{eqnarray}
Then the energy density of bulk uniform matter can be expressed as
\begin{eqnarray}
\mathcal{E}_B&=&\sum_{i=n,p}\mathcal{E}_{kin,i}+V-F-G+I+g_\omega\omega n+{g_\rho\over2}\rho(n_n-n_p).
\label{eq:eps1}\end{eqnarray}

We note the fact that, operating on Eq. (\ref{eq:eps0}),
\begin{eqnarray}
{d\mathcal{E}_B\over dn}\equiv\mathcal{E}_B^\prime&=&\left({\partial\mathcal{E}_B\over\partial n}\right)_{\sigma,\omega,\rho,\delta}\!\!\!+\left({\partial\mathcal{E}_B\over\partial\sigma}\right)_{n,\omega,\rho,\delta}\left({\partial\sigma\over\partial n}\right)_{\omega,\rho,\delta}\!\!\!+\left({\partial\mathcal{E}_B\over\partial\omega}\right)_{n,\sigma,\rho,\delta}\left({\partial\omega\over\partial n}\right)_{\sigma,\rho,\delta}\!\!\! \nonumber\\
&\,&+\left({\partial\mathcal{E}_B\over\partial\rho}\right)_{n,\sigma,\omega,\delta}\left({\partial\rho\over\partial n}\right)_{\sigma,\omega,\delta}\!\!\!+\left({\partial\mathcal{E}_B\over\partial\delta}\right)_{n,\sigma,\omega,\rho}\left({\partial\delta\over\partial n}\right)_{\sigma,\omega,\rho}\!\!\!=\left({\partial\mathcal{E}_B\over\partial n}\right)_{\sigma,\omega,\rho,\delta},
\end{eqnarray}
where we denote total derivatives with respect to the density with primes hereafter. The final four terms vanish due to the fact that the fields are determined from the Euler-Lagrange equations expressing energy minimization.  Since $M^*$ and $n_s$ are functions of $\sigma$ and $\delta$, they are treated as constants in determining the first derivative of $\mathcal{E}_B$ (but not for higher-order derivatives).  Thus one obtains, operating on Eq. (\ref{eq:eps}), and using $\mu_i$ for the nucleon chemical potentials,
\begin{equation}
{\partial\mathcal{E}_B\over\partial n_i}=\mu_i=E_{Fi}+g_\omega\omega+{g_\rho\rho\over2} \tau_i. \label{eq:der}
\end{equation}
  We then also have $\mathcal{E}_B+P_B=\sum_{i=n,p}\mu_in_i$ and $\mathcal{E}_{kin,i}+P_{kin,i}=\sum_{i=n,p}E_{Fi}n_i$. The kinetic contribution to the pressure is
\begin{equation}
P_{kin,i}= \frac{\hbar c}{3\pi^2}\int_0^{k_{Fi}}\frac{k^4dk}{\sqrt{(\hbar ck)^2+M_i^{*2}}}=(E_{Fi}n_i-M_i^*n_{si})/4.
\end{equation}
The pressure and total chemical potential are
\begin{eqnarray}
P_B&=&n^2{d({\cal E}_B/n)\over d n}=\sum_{i=n,p} P_{kin,i}-V+F+G-I,\cr
\mu&=&{d\mathcal{E}_B\over dn}=(1-x)\mu_n+x\mu_p=(1-x)E_{Fn}+xE_{Fp}+g_\omega\omega+{g_\rho\rho\over2}(1-2x).
\end{eqnarray}

For general fixed proton fraction $x$, we need the solution of the simultaneous equations for $\sigma$ and $\omega$, and in the case of asymmetric matter, additionally $\rho$ and, if $g_\delta\ne0$, $\delta$, i.e., Eq. (\ref{eq:field}), which we express as ${\bf H}={\bf W}$. 
For nearly all the RMF forces described by Eq. (\ref{eq:eps}), however, this set of equations can be dimensionally reduced by eliminating $\rho$ and $\delta$.  First, almost all the force models have $D=0$, for which 
\begin{eqnarray}
\rho&=&g_\rho(\hbar c)^3 n\left[m_\rho^2+g_\rho^2\left({\alpha_3^\prime\over2}g_\omega^2\omega^2+g_\sigma\sigma\left[\alpha_2+{\alpha_2^\prime\over2}g_\sigma\sigma\right]\right)\right]^{-1},\cr
\rho^\prime&=&{\rho\over n}\left[1-{g_\rho\rho\over(\hbar c)^3}\left(\alpha_3^\prime g_\omega^2\omega\omega^\prime+g_\sigma\sigma^\prime\left[\alpha_2+\alpha_2^\prime g_\sigma\sigma\right] \right) \right],\cr
\rho^{\prime\prime}&=&2\left({\rho^{\prime2}\over\rho}-{\rho^\prime\over n}\right)-{g_\rho\rho^2\over n(\hbar c)^3}\left[\alpha_3^\prime g_\omega^2(\omega^{\prime2}+\omega\omega^{\prime\prime})+g_\sigma(\alpha_2\sigma^{\prime\prime}+\alpha_2^\prime g_\sigma(\sigma^{\prime\prime}\sigma+\sigma^{\prime2})\right],\cr
\rho^{\prime\prime\prime}&=&2\left(3{\rho^\prime\rho^{\prime\prime}\over\rho}-3{\rho^{\prime3}\over\rho^2}-{\rho^{\prime\prime}\over n}+{\rho^\prime\over n^2}+{\rho^{\prime2}\over\rho n}\right)\cr
&\,&-{g_\rho\rho^2\over n(\hbar c)^3}\{\alpha_3^\prime g_\omega^2\left(3\omega^\prime\omega^{\prime\prime}+\omega\omega^{\prime\prime\prime}\right)+g_\sigma\left[\alpha_2\sigma^{\prime\prime\prime}+g_\sigma\alpha_2^\prime\left(3\sigma^{\prime\prime}\sigma^\prime+\sigma^{\prime\prime\prime}\sigma\right)\right]\}.
\label{eq:rhoder}\end{eqnarray}
Second, in these models with $\delta$ mesons, $\delta$ can be expressed solely in terms of the scalar densities,
\begin{eqnarray}
\delta&=&{d\over g_\delta}\sum_i\tau_in_{si},\cr
\delta^\prime&=&{d\over g_\delta}\sum_i\tau_i\left({\partial n_{si}\over\partial n}-g_\sigma\mathcal{A}_i\sigma^\prime\right)\left(1+d\sum_i\mathcal{A}_i\right)^{-1}\!\!\!,\cr
\delta^{\prime\prime}&=&{d\over g_\delta}\sum_i\tau_i\left({\partial^2n_{si}\over\partial n^2}+2M_i^{*\prime}{\partial^2n_{si}\over\partial n\partial M_i^*}+M_i^{*\prime2}{\partial n_{si}^2\over\partial M_i^{*2}}-g_\sigma\mathcal{A}_i\sigma^{\prime\prime}\right)\left(1+d\sum_i\mathcal{A}_i\right)^{-1}\!\!\!,\cr\delta^{\prime\prime\prime}&=&{d\over g_\delta}\sum_i\tau_i\left({\partial^3n_{si}\over\partial n^3}+3{\partial^3n_{si}\over\partial n^2\partial M_i^*}M_i^{*\prime}+3{\partial^3n_{si}\over\partial n\partial M_i^{*2}}M_i^{*\prime2}+{\partial^3 n_{si}\over\partial M_i^{*3}}M_i^{*\prime3}+{\partial^2n_{s,i}\over\partial n\partial M_i^*}M_i^{*\prime}\right.\cr
&\,&\left.+{\partial^2n_{si}\over\partial M_i^{*2}}M_i^{*\prime}M_i^{*\prime\prime}-g_\sigma\mathcal{A}_i\sigma^{\prime\prime\prime}\right)
\left(1+d\sum_i\mathcal{A}_i\right)^{-1}\!\!\!,\label{eq:deltader}\end{eqnarray}
where $d=(\hbar c)^3g_\delta^2/m_\delta^2$.

The needed derivatives of the scalar densities are
\begin{equation}
{\partial n_{si}\over\partial n_i}={M_i^*\over E_{F,i}}, \qquad{\partial n_{si}\over\partial M_i^*}=-{1\over g_\sigma}{\partial n_{si}\over\partial\sigma}=-{\tau_i\over g_\delta}{\partial n_{si}\over\partial\delta}=3\left({n_{si}\over M_i^*}-{n_i\over E_{Fi}}\right)\equiv\mathcal{A}_i, \label{eq:dnsr}
\end{equation}
\begin{eqnarray}
{\partial^2n_{si}\over\partial n_i^2}&=&-{M_i^*(\hbar c k_{Fi})^2\over3n_i E_{Fi}^3},\qquad
{\partial^2n_{si}\over\partial n_i\partial M_i^*}={(\hbar ck_{Fi})^2\over E_{Fi}^3},\cr
{\partial^2n_{si}\over\partial M_i^{*2}}&=&{3\over M_i^*}\left[{2n_{si}\over M_i^*}-{n_i\over E_{Fi}^3}\left(3E_{Fi}^2-M_i^{*2}\right)\right],
\label{eq:d2nsr}
\end{eqnarray}
and
\begin{eqnarray}
{\partial^3n_{si}\over\partial n_i^3}&=&{M_i^*(\hbar ck_{Fi})^2\over9n_i^2 E_{Fi}^5}\left(4E_{Fi}^2-3M_i^{*2}\right),\qquad {\partial^3n_{si}\over\partial n_i^2\partial M_i^*}={(\hbar ck_{Fi})^2\over 3n_iE_{Fi}^5}\left(3M_i^{*2}-E_{Fi}^2\right),\cr
{\partial^3n_s\over\partial n_i\partial M_i^{*2}}&=&-{3M_i^*(\hbar ck_{Fi})^2\over E_{Fi}^5},\qquad {\partial^3n_{s}\over\partial M_i^{*3}}={9\over M_i^{*2}}\left[{2n_{si}\over3M_i^*}-{n_i\over E_{Fi}}\left(1-{4M_i^{*2}\over3E_{Fi}^2}+{M_i^{*4}\over E_{Fi}^4}\right)\right].
\label{eq:d3nsr}
\end{eqnarray}
It follows that 
\begin{equation}
    {\partial n_s\over\partial n}=\sum_ix_i{\partial n_{si}\over\partial n_i}=\sum_ix_i{M_i^*\over E_{F,i}},\qquad{\partial^2n_s\over\partial n^2}=\sum_ix_i{\partial^2n_{si}\over\partial n_i^2},
\end{equation}
where $x_n=1-x$ and $x_p=x$.  We note the derivatives of the effective masses are
\begin{equation}
    M_i^{*\prime}=-g_\sigma\sigma^\prime-g_\delta\tau_i\delta^\prime,\qquad
M_i^{*\prime\prime}=-g_\sigma\sigma^{\prime\prime}-g_\delta\tau_i\delta^{\prime\prime},\qquad M_i^{*\prime\prime\prime}=-g_\sigma\sigma^{\prime\prime\prime}-g_\delta\tau_i\delta^{\prime\prime\prime}.
\label{eq:massder}\end{equation}

In order to compute the standard properties of matter at the saturation density, we are also required to determine ${\bf X^\prime}, {\bf X^{\prime\prime}}$ and ${\bf X^{\prime\prime\prime}}$.
The first-order field derivatives are found from ${\bf H}^\prime={\bf W}^\prime$, which we write as ${\bf X^\prime}={\bf A}^{-1}{\bf B}$ where
\begin{eqnarray}
{\bf A}&=&{\partial({\bf H}-{\bf W})\over\partial{\bf X}}=\begin{pmatrix}
\mbox{\Large${\partial^2(F+G-V)\over\partial\sigma^2}$}+g_\sigma\mbox{\Large${\partial n_s\over\partial\sigma}$}~~ & \mbox{\Large${\partial^2F\over\partial\sigma\partial\omega}$} & \mbox{\Large${\partial^2 G\over\partial\sigma\partial\rho}$}&~~g_\sigma\mbox{\Large${\frac{\partial n_{s}}{\partial \delta}}$} \\
\mbox{\Large${\partial^2F\over\partial\sigma\partial\omega}$} & \mbox{\Large${\partial^2(F+G)\over\partial\omega^2}$}~~ & \mbox{\Large${\partial^2 G\over\partial\omega\partial\rho}$} &0\\
\mbox{\Large${\partial^2 G\over\partial\sigma\partial\rho}$} & \mbox{\Large${\partial^2 G\over\partial\omega\partial\rho}$} & \mbox{\Large${\partial^2G\over\partial\rho^2}$} &0\\
g_\delta\sum_i\mbox{\Large${\frac{\partial n_{si}}{\partial \sigma}}$}\tau_i &0&0&-\mbox{\Large${\frac{\partial^2I}{\partial\delta^2}}$}+g_\delta\sum_i\mbox{\Large${\frac{\partial n_{si}}{\partial\delta}}$}\tau_i\end{pmatrix},\cr
{\bf B}&=&{\partial{\bf W}\over\partial n}=\begin{pmatrix}
-\mbox{$g_\sigma$}\mbox{\Large${\partial n_s\over\partial n}$} \\
\mbox{$g_\omega$} \\
\mbox{$g_\rho$}\mbox{\large${1-2x\over2}$}\\-g_\delta\sum_i\mbox{\Large${\frac{\partial n_{si}}{\partial n}}$}\tau_i\end{pmatrix}.%,\qquad
\label{eq:matrix}\end{eqnarray}

The second-order field derivatives then follow from ${\bf H}^{\prime\prime}={\bf W}^{\prime\prime}$, or
\begin{equation}
{\bf X^{\prime\prime}}={\bf A}^{-1}({\bf B^\prime}-{\bf A^\prime X^\prime})\equiv{\bf A}^{-1}{\bf C}
\label{eq:xderiv1}\end{equation}
where the vector ${\bf C}$ has components
\begin{equation}
C_i={\partial B_i\over\partial n}+2{\partial B_i\over\partial X_j}X_j^\prime-{\partial A_{ij}\over\partial X_k}X_k^\prime X_j^\prime;
\end{equation}
summation over repeated indices is assumed.  The fact that $\partial A_{ij}/\partial n=-\partial B_i/\partial X_j$ was used; note that only its $[1,1]$ component, $-g_\sigma^2\sum_i\partial^2n_{si}/(\partial n\partial M_i^*)$, and when $g_\delta$ is non-zero,  its $[1,4]$ and $[4,1]$ components, both $-g_\sigma g_\delta\sum_i\tau_i\partial n_{si}^2/(\partial n\partial M_i^*)$, and its $[4,4]$ component,  $-g_\delta^2\sum_i\partial^2n_{si}/(\partial n\partial M_i^*)$, are non-zero.  

The third-order derivatives, similarly, follow from ${\bf H}^{\prime\prime\prime}={\bf W}^{\prime\prime\prime}$, or
\begin{equation}
{\bf X^{\prime\prime\prime}}={\bf A}^{-1}({\bf B^{\prime\prime}}-{\bf A^{\prime\prime}X^\prime}-2{\bf A^\prime X^{\prime\prime}})\equiv {\bf A}^{-1}{\bf D},
\label{eq:xderiv2}\end{equation}
where the vector ${\bf D}$ has components
\begin{equation}
D_i={\partial^2B_i\over\partial n^2}+3\left({\partial^2B_i\over\partial n\partial X_j}X^\prime_j+{\partial^2B_i\over\partial X_k\partial X_j}X_k^\prime X_j^\prime+{\partial B_i\over\partial X_j}X^{\prime\prime}_j-{\partial A_{ij}\over\partial X_k}X_k^\prime X_j^{\prime\prime}\right)-{\partial^2A_{ij}\over\partial X_\ell\partial X_k}X_\ell^\prime X_k^\prime X_j^\prime.
\end{equation}

    The general expressions for the incompressiblity and skewness of uniform matter are given by
    \begin{eqnarray}
        K &=& 9 n^2\frac{d^2(\mathcal{E}_B/ n)}{d n^2} = 9\left( n\mathcal{E}_B^{\prime\prime}-2\mathcal{E}_B^\prime+\frac{2\mathcal{E}_B}{n}\right) = 9n\mathcal{E}_B^{\prime\prime}-18\frac{P_B}{n}, \nonumber\\
        Q &=& 27 n^3\frac{d^3(\mathcal{E}_B/ n)}{d n^3} = 27\left( n^2\mathcal{E}_B^{\prime\prime\prime}-3 n\mathcal{E}_B^{\prime\prime}+6\mathcal{E}_B^\prime-\frac{6\mathcal{E}_B}{ n}\right) =27n^2\mathcal{E}_B^{\prime\prime\prime}-9K,
    \label{eq:incskew}\end{eqnarray}
 where the second and third derivatives of energy density are
    \begin{eqnarray}
        \mathcal{E}_B^{\prime\prime}&=&x E_{Fp}^\prime+(1-x)E_{Fn}^\prime+g_\omega\omega^\prime+\frac{g_\rho}{2}(1-2x)\rho^\prime, \nonumber\\
        \mathcal{E}_B^{\prime\prime\prime}&=&xE_{Fp}^{\prime\prime}+(1-x)E_{Fn}^{\prime\prime}+g_\omega\omega^{\prime\prime}+\frac{g_\rho}{2}(1-2x)\rho^{\prime\prime}, \label{eq:d23er}
    \end{eqnarray}
    with 
    \begin{eqnarray}
        E_{Fi}^{\prime} &=& E_{Fi}^{-1}\left[\frac{(\hbar ck_{Fi})^2}{3 n}+M_i^*M_i^{*\prime}\right], \nonumber\\
        E_{Fi}^{\prime\prime} &=& E_{Fi}^{-1}\left[-\frac{(\hbar ck_{Fi})^2}{9 n^2}-E_{Fi}^{\prime\,2}+M_i^{*\prime2}+M_i^*M_i^{*\prime\prime}\right].
        \label{eq:efp}
    \end{eqnarray}

\subsubsection{Uniform Symmetric Matter}
In this case, $\rho=\delta=G=I=0$.  We define $k_F=k_{Fn}=k_{Fp}=(3\pi^2n/2)^{1/3}$, $M^*=M_n^*=M_p^*$, and $E_F=E_{Fn}=E_{Fp}=\sqrt{M^{*2}+(\hbar ck_F)^2}$, and therefore find
\begin{equation}
n_s={1\over\pi^2}\left({M^*\over\hbar c}\right)^3\left[{\hbar ck_FE_F\over M^{*2}}-\ln\left({\hbar ck_F+E_F\over M^*}\right)\right]. \label{eq:usmb}
\end{equation}
Note that
\begin{equation}
{\partial n_s\over\partial n}={M^*\over E_F},\qquad{\partial n_s\over\partial M^*}=3\left({n_s\over M^*}-{n\over E_F}\right)\equiv\mathcal{A},
\end{equation}
which defines ${\cal A}$. 

The internal energy per particle and the pressure are
\begin{equation}
E_{1/2}={3E_F+M^*n_s/n\over4}+g_\omega\omega+{V-F\over n}-M, \qquad
P_{1/2}={E_Fn-M^*n_s\over4}-V+F.
\end{equation}
The needed derivatives of the energy density of symmetric matter are
\begin{equation}
\mathcal{E}_{B,1/2}^\prime=E_F+g_\omega\omega, \qquad \mathcal{E}_{B,1/2}^{\prime\prime}=E_F^\prime+g_\omega\omega^\prime, \qquad
\mathcal{E}_{B,1/2}^{\prime\prime\prime}=E_F^{\prime\prime}+g_\omega\omega^{\prime\prime}.
\end{equation}
Derivatives of $E_{F}$ and $M^*$ are given by the same expressions as in Eqs. (\ref{eq:efp}) and (\ref{eq:massder}), respectively, but ignoring the subscript $i$.
The $\sigma$ and $\omega$ fields and derivatives are determined by the two-dimensional vector ${\bf H}={\bf W}$ and the $2\times2$ matrix ${\bf AX^\prime}={\bf B}$ equations.  The incompressibility and skewness are found from Eq. (\ref{eq:incskew}).

\subsubsection{Uniform Neutron Matter}
Now $k_{Fn}=(3\pi^2n)^{1/3}$ and $E_{Fn}=\sqrt{M_n^{*2}+(
\hbar ck_{Fn})^2}$.
The internal energy per particle and pressure of neutron matter are
\begin{eqnarray}
E_N&=&{3E_{Fn}+M^*n_{sn}/n\over4}+g_\omega\omega+{g_\rho\over2}\rho+{V-F-G+I\over n}%{(\hbar c)^3\over2}\left({g_\omega^2\over m_\omega^2}n+{g_\sigma^2\over m_\sigma^2}{n_{sN}^2\over n}+{g_\rho^2\over4m_\rho^2}n\right)
-M,\cr
P_N&=&{E_{Fn}n-M_n^*n_{sn}\over4}-V+F+G-I%+{(\hbar c)^3\over2}\left({g_\omega^2\over m_\omega^2}n^2-{g_\sigma^2\over m_\sigma^2}n_{sN}^2+{g_\rho^2\over4m_\rho^2}n^2\right)
.
\end{eqnarray}
where
\begin{equation}
n_{sn}={1\over2\pi^2}\left({M_n^*\over\hbar c}\right)^3\left[{\hbar ck_{Fn}E_{Fn}\over M_n^{*2}}-\ln\left({\hbar ck_{Fn}+E_{Fn}\over M_n^*}\right)\right].
\end{equation}

The values of the fields can be determined using three (or four, in the case $g_\delta\ne0$) simultaneous equations ${\bf H}={\bf W}$, although these can be reduced for the forces considered here as noted in Eqs. (\ref{eq:rhoder}) and (\ref{eq:deltader}).
Their $n^{\rm th}$ derivatives are obtained from ${\bf H}^{n\prime}={\bf W}^{n\prime}$.  The components of the matrices ${\bf A}$ and ${\bf B}$ are given by Eq. (\ref{eq:matrix}), noting that
\begin{equation}
{\partial n_{sn}\over\partial n}={M_n^*\over E_{Fn}},\qquad{\partial n_{sn}\over\partial M_n^*}=3\left({n_{sn}\over M_n^*}-{n\over E_{Fn}}\right)=\mathcal{A}_n.
\end{equation}.

The derivatives of the uniform neutron matter energy density become
\begin{equation}
\mathcal{E}_{N}^\prime=E_{Fn}+g_\omega\omega+{g_\rho\rho\over2}%\left({g_\omega^2\over m_\omega^2}+{g_\rho^2\over4m_\rho^2}\right)(\hbar c)^3n
,\qquad
\mathcal{E}_{N}^{\prime\prime}=E_{Fn}^\prime+g_\omega\omega^\prime+{g_\rho\rho^\prime\over2}%\left({g_\omega^2\over m_\omega^2}+{g_\rho^2\over4m_\rho^2}\right)(\hbar c)^3
,\qquad
\mathcal{E}_N^{\prime\prime\prime}=E_{Fn}^{\prime\prime}+g_\omega\omega^{\prime\prime}+{g_\rho\rho^{\prime\prime}\over2}.
\end{equation}
The derivatives of $E_{Fn}$ and $M^*_n$ are given in Eqs. (\ref{eq:efp}) and (\ref{eq:massder}), respectively.
The incompressibility and skewness are given by Eq. (\ref{eq:incskew}).

\subsubsection{Symmetry Energy}
 Two approaches for the calculation of symmetry energy have been shown in Eq.\,(\ref{eq:s1}) and Eq.\,(\ref{eq:s2}). For the constant coupling models, the second approach gives the formula for symmetry energy as
\begin{eqnarray}
{\mathcal{S}}_2 = {1\over8}{d^2(\mathcal{E}/ n)\over dx^2}\bigg|_{x=1/2}&=&{(\hbar ck_F)^2\over6E_F}+{g_\rho^2\over8}\left({\partial^2G\over\partial\rho^2}\right)^{-1} n+{g_\delta\over8}\left(4{\partial n_{s}\over\partial n}+{g_\delta\over n}{\partial n_{s}\over\partial M^*}{\partial\delta\over\partial x}\right){\partial\delta\over\partial x}+{1\over 8n}{\partial^2I\over\partial\delta^2}\left({\partial\delta\over\partial x}\right)^2,\cr
&=&\mathcal{S}_{kin,0}+{g_\rho^2\over8m_\rho^*}(\hbar c)^3n+\mathcal{S}_\delta%= {(\hbar ck_F)^2\over6E_F}+{g_\rho^2(\hbar c)^3\over8m_\rho^*} n,
\label{eq:s2rmf}\end{eqnarray}
with
\begin{equation}
    {\cal S}_{kin,0}={(\hbar ck_F)^2\over6E_F}
    \end{equation}
being the kinetic contribution in the absence of $\delta$ mesons, \begin{equation}
    m_\rho^* =(\hbar c)^3{\partial^2G\over\partial\rho^2}=m_\rho^2+g_\sigma g_\rho^2\sigma(2\alpha_2+\alpha_2^{\prime}g_\sigma\sigma)+\alpha_3^\prime g_\omega^2g_\rho^2\omega^2,
\end{equation}
and 
\begin{equation}
    {\partial\delta\over\partial x}=-4g_\delta{\partial n_s\over\partial n}\left({\partial^2I\over\partial\delta^2}+2g_\delta^2{\partial n_s\over\partial M^*}\right)^{-1}.
\end{equation}
$\mathcal{S}_\delta$ is the contribution of $\delta$ mesons to both the kinetic and potential symmetry energies,
\begin{equation}
    \mathcal{S}_\delta=-\frac{dM^{*2}n}{2E_F^2\left(1+d\mathcal{A}\right)}\equiv{d{\cal B}\over1+d{\cal A}},
\end{equation}
which defines ${\cal B}$. Note that the quantity $d=(\hbar c)^3g_\delta^2/m_\delta^2$ has been defined immediately following Eq. (\ref{eq:deltader}).
The slope, curvature and skewness of the symmetry energy are, respectively,
\begin{eqnarray}
\mathcal{L}_2 &=& 3 n{\cal S}_2^\prime=\mathcal{L}_{kin,0}-{3g_\rho^2(\hbar c)^3\over8m_\rho^*}\left(\frac{n}{m_\rho^*}m_\rho^{*\prime}-1\right)n+\mathcal{L}_\delta \nonumber\\
\mathcal{K}_{sym,2}&=&9 n^2{\cal S}_2^{\prime\prime}= \mathcal{K}_{kin,0}-\frac{9g_\rho^2(\hbar c)^3 }{4m_\rho^{*2}}\left(m_\rho^{*\prime}-\frac{ n}{m_\rho^*}m_\rho^{*\prime2}+\frac{ n}{2}m_\rho^{*\prime\prime}\right)n^2+\mathcal{K}_\delta\label{eq:sym2}\\
\hspace*{-1cm}\mathcal{Q}_{sym,2}&=&27 n^3{\cal S}_2^{\prime\prime\prime}= \mathcal{Q}_{kin,0}-\frac{27g_\rho^2 (\hbar c)^3}{4m_\rho^{*3}}\left(\frac{3 n}{m_\rho^*}m_\rho^{*\prime3}-3m_\rho^{*\prime2}+\frac{3}{2}m_\rho^* m_\rho^{*\prime\prime}-3 n m_\rho^{*\prime}m_\rho^{*\prime\prime}+\frac{1}{2} n m_\rho^*m_\rho^{*\prime\prime\prime}\right)n^3+\mathcal{Q}_\delta\nonumber
\end{eqnarray}
where the kinetic energy contributions in the absence of $\delta$ mesons are
\begin{eqnarray}
    \mathcal{L}_{kin,0} &=& 3n\mathcal{S}_{kin,0}^\prime=\mathcal{S}_{kin,0}\left(2-3n{E_F^\prime\over E_F}\right), \nonumber\\
%    {\mathcal{K}}_{sym,2}^{\rm \delta M}, \nonumber\\
    \mathcal{K}_{kin,0} &=& 9n^2\mathcal{S}^{\prime\prime}_{kin,0} = -2\mathcal{S}_{kin,0}-{\frac{3}{E_F}}\left(3n^2\mathcal{S}_{kin,0}E_F^{\prime\prime}+2n\mathcal{L}_{kin,0}E_F^\prime\right), \nonumber\\
    \mathcal{Q}_{kin,0} &=& 27n^3{\mathcal{S}}^{\prime\prime\prime}_{kin,0} = 8\mathcal{S}_{kin},0-\frac{9}{E_F}\left(3n^3S_{kin,0}E_F^{\prime\prime\prime}+3n^2\mathcal{L}_{kin,0}E_F^{\prime\prime}+n \mathcal{K}_{kin,0}E_F^\prime\right), \label{eq:spkin}
\end{eqnarray}
with
\begin{equation}
E_F^{\prime\prime\prime}=E_F^{-1}\left(-3E_F^{\prime\prime}E_F^\prime+{4(\hbar ck_F)^2\over27 n^3}+3M^{*\prime}M^{*\prime\prime}+M^*M^{*\prime\prime\prime}\right),
\end{equation}
and
\begin{eqnarray}
m_\rho^{*\prime}&=&2g_\rho^2\left[g_\sigma\sigma^\prime(\alpha_2+\alpha_2^\prime g_\sigma\sigma)+\alpha_3^\prime g_\omega^2\omega^\prime\omega+3Dg_\rho^2\rho\rho^\prime\right],\cr
m_\rho^{*\prime\prime}&=&2g_\rho^2\left[g_\sigma\alpha_2\sigma^{\prime\prime}+\alpha_2^\prime g_\sigma^2(\sigma\sigma^{\prime\prime}+\sigma^{\prime2})+\alpha_3^\prime g_\omega^2(\omega^{\prime2}+\omega\omega^{\prime\prime})+3Dg_\rho^2(\rho^{\prime2}+\rho\rho^{\prime\prime})\right],\cr
m_\rho^{*\prime\prime\prime}&=&2g_\rho^2\left[g_\sigma\alpha_2\sigma^{\prime\prime\prime}+\alpha_2^\prime g_\sigma^2(3\sigma^\prime\sigma^{\prime\prime}+\sigma\sigma^{\prime\prime\prime})+\alpha_3^\prime g_\omega^2(3\omega^\prime\omega^{\prime\prime}+\omega\omega^{\prime\prime\prime})+3Dg_\rho^2(3\rho^\prime\rho^{\prime\prime}+\rho\rho^{\prime\prime\prime})\right].
\end{eqnarray}
The contributions from the $\delta$ mesons are
\begin{eqnarray}
    \mathcal{L}_\delta&=&3n\mathcal{S}_\delta^\prime=3dn\left({\cal B}^\prime-{\cal S}_\delta{\cal A}^\prime\right)\left(1+d\mathcal{A}\right)^{-1}\nonumber\\
\mathcal{K}_\delta&=&9n^2{\cal S}_\delta^{\prime\prime}=9dn^2\left({\cal B}^{\prime\prime}-2{\cal S}_\delta^\prime{\cal A}^\prime-{\cal S}_\delta{\cal A}^{\prime\prime}\right)\left(1+d\mathcal{A}\right)^{-1},\nonumber\\
\mathcal{Q}_\delta&=&27n^3{\cal S}_\delta^{\prime\prime\prime}=27dn^3\left({\cal B}^{\prime\prime\prime}-3\left[{\cal S}_\delta^{\prime\prime}{\cal A}^\prime+{\cal S}_\delta^\prime{\cal A}^{\prime\prime}\right]-{\cal S}_\delta{\cal A}^{\prime\prime\prime}\right)\left(1+d\mathcal{A}\right)^{-1}
\end{eqnarray}
where, for example,
\begin{equation}
    \mathcal{A}^\prime={3M^{*\prime}\over M^*}\left(\mathcal{A}+{nM^{*2}\over E_F^3}\right)+{\left(\hbar ck_F\right)^2\over E_F^3},\qquad{\cal B}^\prime={\cal B}\left({1\over n}+{2M^{*\prime}\over M^*}-{2E_F^\prime\over E_F}\right).
\end{equation}
The higher-order derivatives of ${\cal A}$ and ${\cal B}$ are more complicated and are not explicitly indicated.  

To calculate $\mathcal{Q}_{sym,2}$, the following derivatives are also needed
\begin{eqnarray}
    {\partial^4n_s\over\partial n\partial M^{*3}}&=&-\frac{3(\hbar ck_{F})^2}{E_F^7}\left(E_F^2-5M^{*2}\right), \qquad 
    {\partial^4n_s\over\partial n^2\partial M^{*2}}=-{M^*\over 3n}{(\hbar ck_{F})^2\over E_{F}^7}\left(E_{F}^2+5M^{*2}\right),\nonumber\\
    {\partial^4n_s\over\partial n^3\partial M^*}&=&{1\over9n^2}{(\hbar ck_{F})^2\over E_{F}^5}\left(4E_{F}^2+\frac{15M^{*4}}{E_F^2}-21M^{*2}\right), \label{eq:d4nsr}
\end{eqnarray}
where all the symbols are for symmetric nuclear matter.

At the saturation density $n_0$, we have
\begin{eqnarray}
J_1&=&E_N(n_0)-E_0,\qquad L_1={3P_N(n_0)\over n_0},\cr
\quad K_{sym,1}&=&{\cal K}_N(n_0)-{\cal K}_{1/2}(n_0)\equiv K_N-K_{1/2},\qquad Q_{sym,1}={\cal Q}_N(n_0)-{\cal Q}_{1/2}(n_0)\equiv Q_N-Q_{1/2},\cr
J_2&=&{\cal S}_2(n_0),\qquad L_2={\cal L}_2(n_0), \qquad K_{sym,2}={\cal K}_{sym,2}(n_0),\qquad Q_{sym,2}={\cal Q}_{sym,2}(n_0).
\end{eqnarray}

These meson-exchange models with constant couplings can be divided into five types following Dutra et al's classification~\cite{dutra2014relativistic}:
\begin{itemize}
 \item {type 1} (linear finite range models): models in which
$g_\delta=A=B=C=D=\alpha_1=\alpha_2=\alpha_1^\prime=\alpha_2^\prime=\alpha_3^\prime=0$. 
This is the case of the linear Walecka model, and there are up to three effective parameters, including $g_\sigma/m_\sigma$, $g_\omega/m_\omega$, and $g_\rho/m_\rho$, which could correspond to unique values of $n_0, E_0$ and, in the case that $g_\rho\ne0$, $J_1$ (or $J_2$).

\item {type 2} ($\sigma^3+\sigma^4$ models): models in which
$g_\delta=C=D=\alpha_1=\alpha_2=\alpha_1^\prime=\alpha_2^\prime=\alpha_3^\prime=0$. 
This type corresponds to parameterizations related to the Boguta-Bodmer model, and has up to five effective parameters, including $g_\sigma/m_\sigma$, $g_\omega/m_\omega$, $g_\rho/m_\rho$, $A/m_\sigma^3$ and $B/m_\sigma^4$, which could correspond to unique values of $n_0, E_0$, $J_1$, $K_{1/2}$ and $L_1$.

\item {type 3} ($\sigma^3+\sigma^4+\omega^4$ models): models in which
$g_\delta=D=\alpha_1=\alpha_2=\alpha_1^\prime=\alpha_2^\prime=\alpha_3^\prime=0$. 
These parameterizations include a quartic self-interaction in the $\omega$ field, and have up to six effective parameters, including $g_\sigma/m_\sigma$, $g_\omega/m_\omega$, $g_\rho/m_\rho$, $A/m_\sigma^3$, $B/m_\sigma^4$ and $C$, which could correspond to unique values of $n_0, E_0$, $J_1$, $K_{1/2}$, $L_1$ and $K_N$.

\item {type 4} ($\sigma^3+\sigma^4+\omega^4$ + cross terms models): models in which $g_\delta=D=0$ and at least one of the coupling constants $\alpha_1$, $\alpha_2$,
$\alpha_1^\prime$, $\alpha_2^\prime$, or $\alpha_3^\prime$ is different from zero.
These models could have up to 11 effective parameters.

\item {type 7}: models which have $\delta$ mesons.  Since the models included here all have $D=\alpha_1=\alpha_2=\alpha_1^\prime=\alpha_2^\prime=0$, there are up to seven effective parameters, including $g_\sigma/m_\sigma$, $g_\omega/m_\omega$, $g_\rho/m_\rho$, $g_\delta/m_\delta$, $A/m_\sigma^3$, $B/m_\sigma^4,C$ and $\alpha_3^\prime$, which could correspond to unique values of $n_0, E_0$, $J_1$, $K_{1/2}$, $L_1$, $K_N$ and $Q_{1/2}$.
\end{itemize}

\subsection{Relativistic Models with Density-Dependent Couplings}
 In the Density-Dependent (DD) coupling model, denoted type 5 by Ref. \cite{dutra2014relativistic}, the nucleon-meson couplings between nucleons and mesons are taken to be functions of baryon density, and the parameters of the models we consider have $g_\delta=A=B=C=D=\alpha_1=\alpha_2=\alpha_1^\prime=\alpha_2^\prime=\alpha_3^\prime=0$. These models represent extensions of type 1 models with the substitutions
\begin{equation}
    g_\sigma \rightarrow \Gamma_\sigma(n), \quad g_\omega \rightarrow \Gamma_\omega(n), \quad g_\rho \rightarrow \Gamma_\rho(n), \label{eq:replace}
\end{equation}
where the density dependences are parameterized with 
\begin{eqnarray}
    \Gamma_{\alpha}(u) &=& \Gamma_\alpha(n_0)f_{\alpha}(u)= a_\alpha\frac{1+b_\alpha(u+d_\alpha)^2}{1+c_\alpha(u+d_\alpha)^2}\qquad\qquad \alpha\in[\sigma,\omega]\cr
    \Gamma_\alpha(u) &=& \Gamma_\alpha(n_0)e^{-a_\alpha(u-1)}\,\qquad\qquad\qquad\qquad\qquad\,\,\, \alpha\in[\rho],
    \end{eqnarray}
    with $u=n/n_0$.
The magnitudes of the meson fields follow from minimization of the energy density with respect to the meson fields, giving the equivalent of Eq.\,(\ref{eq:field}):
\begin{equation}
    \sigma = (\hbar c)^3\frac{\Gamma_\sigma}{m_\sigma^2}n_s, \qquad \omega = (\hbar c)^3\frac{\Gamma_\omega}{m_\omega^2} n, \qquad \rho = (\hbar c)^3\frac{\Gamma_\rho}{2m_\rho^2}(1-2x)n.
    \label{eq:mft5}
\end{equation}
The definition for scalar and vector densities, proton and neutron effective masses and Fermi energies are also the same as previously with the substitution Eq. (\ref{eq:replace}). The energy density and pressure, respectively, are therefore
\begin{equation}
    \mathcal{E}_{\rm DD} = \sum_{i=n,p}\mathcal{E}_{kin,i}+\frac{1}{2(\hbar c)^3}\left[m_\sigma^2\sigma^2-m_\omega^2\omega^2-m_\rho^2\rho^2\right]+\Gamma_\omega\omega n+\frac{1-2x}{2}\Gamma_\rho\rho n,
\end{equation}
and
\begin{equation}
    P_{\rm DD} = \sum_{i=n,p}P_{kin,i}-\frac{1}{2(\hbar c)^3}\left[m_\sigma^2\sigma^2-m_\omega^2\omega^2-m_\rho^2\rho^2\right]+n\Sigma_R(n),
\end{equation}
where $\Sigma_R(n)$ is an additional term involving the density derivatives of the density-dependent couplings
%\begin{equation}
%    \Sigma_R(n) = -\Gamma_\sigma^\prime\sigma n_s+\Gamma_\omega^\prime\omega n +\frac{1-2x}{2}\Gamma_\rho^\prime\rho n.
%\end{equation}
\begin{equation}
      \Sigma_R(n) = -\Gamma_\sigma^\prime\sigma n_s+\Gamma_\omega^\prime\omega n +\frac{1-2x}{2}\Gamma_\rho^\prime\rho n=-(\hbar c)^3{\Gamma_\sigma\Gamma_\sigma^\prime\over m_\sigma^2} n_s^2+\sum_{\alpha=\omega,\rho}C_\alpha\Gamma_\alpha\Gamma_\alpha^\prime n^2,
\end{equation}
where $C_\omega = (\hbar c)^3/m_\omega^2$ and $C_\rho=(1-2x)^2(\hbar c)^3/(4m_\rho^2)$, for the models considered here.  The chemical potentials are then analogs of Eq. (\ref{eq:der}), and the expressions for vectors $\bf{X},\; \bf{W},$ and $ \bf{H}$ are the same as previously, with the substitution of Eq.\,(\ref{eq:replace}).  For uniform matter, ${\bf H}={\bf W}$ is given by Eq. (\ref{eq:mft5}).

From Eq.\,(\ref{eq:mft5}), the matrix $\bf{A}$ becomes diagonal and the vector $\bf{B}$ now has additional terms containing vector field coupling derivatives,
\begin{eqnarray}
    \bf{A} &=& {\rm diag}\left(-\frac{m_\sigma}{{(\hbar c)^3}}+\Gamma_\sigma\frac{\partial n_s}{\partial \sigma},\quad \frac{m_\omega^2}{(\hbar c)^3},\quad \frac{m_\rho^2}{(\hbar c)^3}\right), \\
{\bf B}&=&\begin{pmatrix}
-\mbox{$\Gamma_\sigma$}\mbox{\Large${\partial n_s\over\partial n}$}-n_s\Gamma_\sigma^\prime, \\
\mbox{$\Gamma_\omega$}+n\Gamma_\omega^\prime \\
\mbox{$(\Gamma_\rho+n\Gamma_\rho^\prime)$}\mbox{\large${1-2x\over2}$}\end{pmatrix}.
\end{eqnarray}
The required derivatives of $\bf{X}$ and $n_s$ are identical to Eqs.\;(\ref{eq:dnsr})-(\ref{eq:d3nsr}), and the computation of ${\bf B}$ and the derivatives of ${\bf A}$ and ${\bf B}$ now require the coupling derivatives
\begin{eqnarray}
    \Gamma_\alpha^\prime &=& \frac{2a_\alpha\Gamma_\alpha(n_0)}{n_0}\frac{(b_\alpha-c_\alpha)(x+d_\alpha)}{\left[1+c_\alpha(x+d_\alpha)^2\right]^2}, \quad
    \Gamma_\alpha^{\prime\prime} = \frac{2a_\alpha\Gamma_\alpha(n_0)}{n_0^2}\frac{(b_\alpha-c_\alpha)\left[1-3c_\alpha(x+d_\alpha)^2\right]}{\left[1+c_\alpha(x+d_\alpha)^2\right]^3}, \nonumber\\
    \Gamma_\alpha^{\prime\prime\prime} &=& \frac{-24a_\alpha\Gamma_\alpha(n_0)}{n_0^3}\frac{(b_\alpha-c_\alpha)c_\alpha(x+d_\alpha)\left[1-c_\alpha(x+d_\alpha)^2\right]}{\left[1+c_\alpha(x+d_\alpha)^2\right]^4}
\end{eqnarray}
for $\alpha=\sigma,\omega$, and
\begin{equation}
    \Gamma_\rho^\prime = -a_\rho\Gamma_\rho/n_0, \quad \Gamma_\rho^{\prime\prime} = a_\rho^2\Gamma_\rho/n_0^2, \quad \Gamma_\rho^{\prime\prime\prime} = -a_\rho^3\Gamma_\rho/n_0^3.
\end{equation}
The effective masses for neutrons and protons are equal and their derivatives become
\begin{equation}
    M^{*\prime} = -(\Gamma_\sigma\sigma^\prime+\sigma\Gamma_\sigma^\prime), \quad
    M^{*\prime\prime} = -(\Gamma_\sigma\sigma^{\prime\prime}+2\sigma^\prime\Gamma_\sigma^\prime+\sigma\Gamma_\sigma^{\prime\prime}),\quad
    M^{*\prime\prime\prime} = -(\Gamma_\sigma\sigma^{\prime\prime\prime}+3\sigma^{\prime\prime}\Gamma_\sigma^\prime+3\sigma^\prime\Gamma_\sigma^{\prime\prime}+\sigma\Gamma_\sigma^{\prime\prime\prime}).
\end{equation}
The general formulae for the incompressibility and skewness of uniform ANM, Eq. (\ref{eq:incskew}), require different expressions for the derivatives of $\mathcal{E}_{\rm DD}$,

\begin{eqnarray}
    \mathcal{E}_{\rm DD}^{\prime\prime} &=& x E_{Fp}^\prime+(1-x)E_{Fn}^\prime+\sum_{\alpha=\omega,\rho}C_\alpha\left(\Gamma_\alpha^2+2\Gamma_\alpha\Gamma_\alpha^\prime n\right)+\Sigma_R^\prime, \\
    \mathcal{E}_{\rm DD}^{\prime\prime\prime} &=& x E_{Fp}^{\prime\prime}+(1-x)E_{Fn}^{\prime\prime}+\sum_{\alpha=\omega,\rho}C_\alpha\left[2n\left(\Gamma_\alpha\Gamma_\alpha^{\prime\prime}+\Gamma_\alpha^{\prime2}\right)+4\Gamma_\alpha\Gamma_\alpha^{\prime}\right]+\Sigma_R^{\prime\prime},
\end{eqnarray}
with
\begin{eqnarray}
    \Sigma_R^\prime &=& -\frac{(\hbar c)^3}{m_\sigma^2}\left[(\Gamma_\sigma^{\prime2}+\Gamma_\sigma\Gamma_\sigma^{\prime\prime})n_s^2+2\Gamma_\sigma\Gamma_\sigma^\prime n_sn_s^\prime\right]+\sum_{\alpha=\omega,\rho}C_\alpha\left[(\Gamma_\alpha^{\prime2}+\Gamma_\alpha\Gamma_\alpha^{\prime\prime})n^2+2n\Gamma_\alpha\Gamma_\alpha^\prime\right],\\
    \Sigma_R^{\prime\prime} &=& -\frac{(\hbar c)^3}{m_\sigma^2}\left[(\Gamma_\sigma\Gamma_\sigma^{\prime\prime\prime}+3\Gamma_\sigma^\prime\Gamma_\sigma^{\prime\prime})n_s^2+2\Gamma_\sigma\Gamma_\sigma^\prime n_sn_s^{\prime\prime}+2\Gamma_\sigma\Gamma_\sigma^\prime n_s^{\prime2}+4(\Gamma_\sigma^{\prime2}+\Gamma_\sigma\Gamma_\sigma^{\prime\prime})n_sn_s^\prime\right] \nonumber \\
    &\,& +\sum_{\alpha=\omega,\rho}C_\alpha\left[(\Gamma_\alpha\Gamma_\alpha^{\prime\prime\prime}+3\Gamma_\alpha^\prime\Gamma_\alpha^{\prime\prime})n^2+4n(\Gamma_\alpha^{\prime2}+\Gamma_\alpha\Gamma_\alpha^{\prime\prime})+2\Gamma_\alpha\Gamma_\alpha^\prime)\right].
\end{eqnarray}
Derivatives of the scalar density are needed, and can be found from
\begin{equation}
    n_{si}^\prime={\partial n_{si}\over\partial n}+{\partial n_{si}\over\partial M^*}M^{*\prime},\qquad n_{si}^{\prime\prime}={\partial^2 n_{si}\over\partial n^2}+2{\partial^2 n_{si}\over\partial n\partial  M^*}M^{*\prime}_i+{\partial n_{si}\over\partial M^*}M^{*\prime\prime},
    \end{equation}
where the partial derivatives are given in Eqs. (\ref{eq:dnsr}) and (\ref{eq:d2nsr}).  

Similarly, the symmetry energy functions are given as follows:
\begin{eqnarray}
    {\mathcal{S}}_2^{\rm DD} &=& {\mathcal{S}}_{kin,0}+{(\hbar c)^3\over8m_\rho^{2}}\Gamma_\rho^2 n, \nonumber\\
    {\mathcal{L}}_2^{\rm DD} &=& {\mathcal{L}}_{kin,0}-{3n(\hbar c)^3\over8m_\rho^{2}}\left(2\Gamma_\rho\Gamma_\rho^\prime n+\Gamma_\rho^2\right), \nonumber\\
    {\mathcal{K}}_{sym,2}^{\rm DD}&=& {\mathcal{K}}_{kin,0}-\frac{9n^2(\hbar c)^3 }{4m_\rho^{2}}\left[(\Gamma_\rho\Gamma_\rho^{\prime\prime}+\Gamma_\rho^{\prime2})n+2\Gamma_\rho\Gamma_\rho^\prime\right],\nonumber\\
    {\mathcal{Q}}_{sym,2}^{\rm DD}&=& {\mathcal{Q}}_{kin,0}-\frac{27n^3(\hbar c)^3}{4m_\rho^{2}}\left[(\Gamma_\rho\Gamma_\rho^{\prime\prime\prime}+3\Gamma_\rho^\prime\Gamma_\rho^{\prime\prime})n+3(\Gamma_\rho\Gamma_\rho^{\prime\prime}+\Gamma_\rho^{\prime2})\right],
\end{eqnarray}
where the kinetic parts are given in Eq.\;(\ref{eq:spkin}).

\subsection{Relativistic Nonlinear Point-Coupling Models}
In the nonlinear point-coupling (PC) model~\cite{nikolaus1992nuclear}, nucleons interact with each other only through effective point-like interactions, without exchanging mesons. Although this model is classified as a type 6 RMF model by Ref. \cite{dutra2014relativistic}, technically it is closer to a Skyrme-like model, albeit with a more complicated density dependence. 

\subsubsection{Energy Density}
The energy density and the pressure for uniform nuclear matter are
\begin{eqnarray}
    \mathcal{E}_{\rm PC}&=&\sum_{i=n,p}{\mathcal E}_{kin,i}+{\alpha^\prime_{\rm V}\over2}n^2+{\alpha^\prime_{\rm TV}\over2}(n_n-n_p)^2+{\gamma^\prime_{\rm V}\over4}n^4+{\gamma^\prime_{\rm TV}\over4}(n_n-n_p)^4\nonumber\\
    &\,&-\eta^\prime_2n_s^2n^2-{\alpha^\prime_s\over2}n_s^2-{2\beta^\prime_s\over3}n_s^3-{3\gamma^\prime_s\over4}n_s^4-{\alpha^\prime_{\rm TS}\over2}(n_{sn}-n_{sp})^2,\\
    P_{\rm PC}&=&\sum_{i=n,p}P_{kin,i}+{\alpha^\prime_{\rm V}\over2}n^2+{\alpha^\prime_{\rm TV}\over2}(n_n-n_p)^2+{3\gamma^\prime_{\rm TV}\over4}(n_n-n_p)^4\nonumber\\
    &\,&+2\eta^\prime_1n_sn^2+3\eta^\prime_2n_s^2n^2+2\eta^\prime_3n_s(n_n-n_p)^2+{\alpha^\prime_s\over2}n_s^2+{2\beta^\prime_s\over3}n_s^3+{3\gamma^\prime_s\over4}n_s^4+{\alpha^\prime_{\rm TS}\over2}(n_{sn}-n_{sp})^2,
\end{eqnarray}
where $\mathcal{E}_{kin}$ is the RMF kinetic energy density.  We simplified the expressions by using
\begin{equation}
    \alpha_{s,\rm V,TV,TS}^\prime=(\hbar c)^3\alpha_{s,\rm V,TV,TS},\quad \beta^\prime_s=(\hbar c)^6\beta_s,\quad \gamma^\prime_{s,\rm V,TV}=(\hbar c)^9\gamma_{s,\rm V,TV},\quad \eta^\prime_{1,3}=(\hbar c)^6\eta_{1,3},\quad \eta^\prime_2=(\hbar c)^9\eta_2. 
\end{equation}
The effective masses for protons and neutrons are defined as
\begin{equation}
M_t^*=M+\alpha^\prime_sn_s+\beta^\prime_sn_s^2+\gamma^\prime_sn_s^3+\eta^\prime_1n^2+2\eta^\prime_2n_sn^2
+\eta^\prime_ 3(n_n-n_p)^2-\tau_i\alpha^\prime_{TS}(n_{sn}-n_{sp}). 
%M_n^*=M+(\hbar c)^3\left[\alpha_s\mathfrak{n}_s+\beta_s\mathfrak{n}_s^2+\gamma_s\mathfrak{n}_s^3+\eta_1\mathfrak{n}^2+2\eta_2\mathfrak{n}_s\mathfrak{n}^2
%+\eta_ 3(\mathfrak{n}_n-\mathfrak{n}_p)^2-\alpha_{\mbox{\tiny TS}}(\mathfrak{n}_{s,n}-\mathfrak{n}_{s,p})\right].
\end{equation}
Note that the $\alpha$ parameters are related to constant coupling RMF parameters by 
\begin{eqnarray}
\alpha_s = -\frac{g_\sigma^2}{m_\sigma^2}, \qquad \alpha_{\rm V}=\frac{g_\omega^2}{m_\omega^2},\qquad
\alpha_{\rm TV} = \frac{g_\rho^2}{m_\rho^2}, \qquad \alpha_{\rm TS}=\frac{g_\delta^2}{m_\delta^2}
\end{eqnarray}
The incompressibility and skewness are given by analogous expressions to Eq. (\ref{eq:incskew}), which require derivatives of the energy density:
\begin{eqnarray}
\mathcal{E}_{\rm PC}^\prime &=&x E_{Fp}+(1-x)E_{Fn} +n\left[\alpha^\prime_{\rm V}+ \alpha^\prime_{\rm TV}(1-2x)^2\right] + n^3\left[\gamma^\prime_{\rm V}
+ \gamma^\prime_{\rm TV}(1-2x)^4\right] \nonumber\\
&\,&+ 2nn_s\left[\eta^\prime_1 + \eta^\prime_3(1-2x)^2+ \eta^\prime_2n_s\right],\nonumber\\
\mathcal{E}_{\rm PC}^{\prime\prime} &=&x E_{Fp}^\prime+(1-x)E_{Fn}^\prime + \alpha^\prime_{\rm V} + \alpha^\prime_{\rm TV}(1-2x)^2+ 3n^2\left[\gamma^\prime_{\rm V}
 + \gamma^\prime_{\rm TV}(1-2x)^4\right] \nonumber\\
&\,&+ 2(n_s^\prime n+n_s)\left[\eta_1^\prime+ \eta^\prime_3(1-2x)^2\right]+2\eta^\prime_2n_s(2n_s^\prime n+n_s) \nonumber\\
\mathcal{E}_{\rm PC}^{\prime\prime\prime} &=&x E_{Fp}^{\prime\prime}+(1-x)E_{Fn}^{\prime\prime} + 6n\left[\gamma^\prime_{\rm V}+ \gamma^\prime_{\rm TV}(1-2x)^4\right]\nonumber\\
&\,&+ 2(n_s^{\prime\prime}n+2n_s^\prime)\left[\eta_1^\prime+\eta^\prime_3(1-2x)^2\right]+ 4\eta^\prime_2(n_s^{\prime\prime}n_s n+n_s^{\prime2}n+2n_sn_s^\prime) .
\end{eqnarray}

\subsubsection{Uniform Symmetric Matter}

In the symmetric matter case, we can take $\alpha_{\rm TV}=\alpha_{\rm TS}=\gamma_{\rm TV}=0$, so we have $M^*_n=M^*_p=M^*$ as well as $k_F=k_{Fn}=k_{Fp}=(3\pi^2n/2)^{1/3}$ and $E_F=E_{Fn}=E_{Fp}=\sqrt{M^{*2}+(\hbar ck_F)^2}$. Formulae for the relevant properties at the saturation density are easily obtained in analogy with the RMF constant coupling case. 
The derivative of the scalar density is
\begin{equation}
    n_s^\prime={\partial n_s\over\partial n}+{\partial n_s\over\partial M^*}M^{*\prime}\equiv{\cal C}+{\cal A}M^{*\prime}
\end{equation}
and the derivatives of the effective mass are
\begin{eqnarray}
M^{*\prime}&=&n_s^\prime\left(\alpha^\prime_s+2\beta^\prime_sn_s+3\gamma^\prime_sn_s^2+2\eta^\prime_2 n^2\right)+2n\left(\eta_1^\prime+\eta^\prime_2n_s\right)\equiv n^\prime_sM_1+M_2,\nonumber\\
M^{*\prime\prime}&=&n_s^{\prime\prime}M_1+n_s^\prime M_1^\prime+M_2^\prime,\nonumber\\
M^{*\prime\prime\prime}&=&n_s^{\prime\prime\prime}M_1+2n_s^{\prime\prime} M_1^\prime+n_s^\prime M_1^{\prime\prime}+M_2^{\prime\prime}.
\label{eq:msPC}\end{eqnarray}
We therefore find
\begin{eqnarray}
    n_s^\prime&=&[{\cal C}+{\cal A}M_2](1-{\cal A}M_1)^{-1},\nonumber\\
    n_s^{\prime\prime}&=&\left[{\cal C}^\prime+{\cal A}^\prime(M_2+n_s^\prime M_1)+{\cal A}(M_2^\prime+n_s^\prime M_1^\prime)\right](1-{\cal A}M_1)^{-1},\nonumber\\
    n_s^{\prime\prime\prime}&=&\left[{\cal C}^{\prime\prime}+{\cal A}^{\prime\prime}(M_2+n_s^\prime M_1)+2{\cal A}^{\prime}(M_2^\prime+n_s^{\prime\prime}M_1^\prime+n_s^\prime M_1^\prime)+{\cal A}(M_2^{\prime\prime}+2n_s^{\prime\prime}M_1^\prime+n_s^\prime M_1^{\prime\prime})\right](1-{\cal A}M_1)^{-1},
\label{eq:nsPC}\end{eqnarray}

\subsubsection{Uniform Neutron Matter}

In the case of uniform neutron matter, the definitions for $k_{Fn}$ and $E_{Fn}$ are the same as previously. The derivative of the scalar density is
\begin{equation}
    n_{sn}^\prime={\partial n_{sn}\over\partial n}+{\partial n_{sn}\over\partial M^*_n}M^{*\prime}_n\equiv{\cal C}_n+{\cal A}_nM^{*\prime}_n
\end{equation} and the derivative of the effective mass is
\begin{equation}
    M^{*\prime}_n=n_s^\prime\left(\alpha_s^\prime-\alpha_{TS}^\prime+2\beta_s^\prime n_{sn}+3\gamma_s^\prime n_{sn}^2+2\eta^\prime_2n^2\right)+2n\left(\eta_1^\prime+\eta_3^\prime+\eta_2^\prime n_{sn}\right)\equiv n_{sn}^\prime M_{1n}+M_{2n},
\end{equation}
thus $M_{1n}=M_1-\alpha_{TS}^\prime$ and $M_{2n}=M_2+2n\eta_3^\prime$.  Higher order derivatives of $n_{sn}$ and $M^*_n$ follow in analogy to the symmetric matter case, Eqs. (\ref{eq:msPC}) and (\ref{eq:nsPC}).

\subsubsection{Symmetry Energy}

The symmetry energy functions for the second approach are
\begin{eqnarray}
     {\cal S}_2^{\rm PC} &=& \mathcal{S}_{kin,0}+\frac{\alpha^\prime_{\rm TV}}{2}n+\eta^\prime_3nn_s+{\cal S}_{\rm TS}, \nonumber\\
    {\mathcal{L}}_2^{\rm PC} &=& \mathcal{L}_{kin,0}+\frac{3\alpha^\prime_{\rm TV}}{2}n+3\eta^\prime_3n(nn_s^\prime+n_s)+{\cal L}_{\rm TS}, \nonumber\\
    {\mathcal{K}}_{sym,2}^{\rm PC} &=& \mathcal{K}_{kin,0}+9\eta^\prime_3n^2(2n_s^\prime+nn_s^{\prime\prime})+{\cal K}_{\rm TS}, \nonumber\\
   {\mathcal{Q}}_{sym,2}^{\rm PC} &=& \mathcal{Q}_{kin,0}+27\eta^\prime_3n^3(3n_s^{\prime\prime}+nn_s^{\prime\prime\prime})+{\cal Q}_{\rm TS},
\end{eqnarray}
where
\begin{eqnarray}
    {\cal S}_{\rm TS}&=&-\frac{\alpha^\prime_{\rm TS}}{2}\frac{{\cal B}}{1-\alpha_{TS}^\prime{\cal A}},\nonumber\\
    {\cal L}_{\rm TS}&=&3n{\cal S}_{\rm TS}^\prime=-\frac{3\alpha_{TS}^\prime}{2}n\frac{{\cal B}^\prime-2{\cal S}_{\rm TS}{\cal A}^\prime}{1-\alpha_{TS}^\prime{\cal A}},\nonumber\\
   {\cal K}_{\rm TS}&=&9n^2{\cal S}_{\rm TS}^{\prime\prime}=-{9\alpha^\prime_{\rm TS}\over2}n^2\frac{{\cal B}^{\prime\prime}-4{\cal S}_{\rm TS}^\prime{\cal A}^\prime-2{\cal S}_{\rm TS}{\cal A}^{\prime\prime}}{1-\alpha_{TS}^\prime{\cal A}}, \nonumber\\ 
{\cal Q}_{\rm TS}&=&27n^3{\cal S}_{\rm TS}^{\prime\prime\prime}=-\frac{27\alpha^\prime_{\rm TS}}{2}n^3\frac{{\cal B}^{\prime\prime\prime}-6{\cal S}_{\rm TS}^{\prime\prime}{\cal A}^\prime-6{\cal S}_{\rm TS}^\prime{\cal A}^{\prime\prime}-2{\cal S}_{\rm TS}{\cal A}^{\prime\prime\prime}}{1-\alpha_{TS}^\prime{\cal A}}.
\end{eqnarray}
All the needed quantities are given in Eqs. (\ref{eq:dnsr}) - (\ref{eq:d3nsr}) and (\ref{eq:d4nsr}). Table VIII of Appendix B lists the saturation properties for all types of RMF forces.

\section{Gogny Interactions}
%\subsubsection{Uniform Nuclear Matter}

The Gogny two-body effective nuclear model is characterized by potential energy contributions from a finite-range interaction and a spin-orbit interaction.  It bears a closer resemblance to a Skyrme-like than an RMF-like model. 

\subsubsection{Energy Density}
The total energy per particle can be expressed as an explicit function of number density and proton fraction, using the notation of Ref. \cite{gonzalez2017higher},
\begin{eqnarray}
E( n,x) &=& \frac{3\hbar^2k_F^2}{10M}H_{5/3}+\sum\limits_{i=1}^2{t_{3i}\over4} n^{\alpha_i+1}\left[x_{3i}+2-\left(x_{3i}+\frac{1}{2}\right)H_2\right]
+\frac{\pi^{3/2}}{2}n\sum\limits_{i=1}^2\mu_i^3\left[\mathcal{A}_i+\mathcal{B}_i(1-2x)^2\right]\nonumber\\
&\,&-\sum\limits_{i=1}^2\frac{\mathcal{C}_i\left[\mathbf{e}(k_{Fn}\mu_i)+\mathbf{e}(k_{Fp}\mu_i)\right]-\mathcal{D}_i\bar{\mathbf{e}}(k_{Fn}\mu_i,k_{Fp}\mu_i)}{2(k_F\mu_i)^3},
\end{eqnarray}
where the Fermi momentum of SNM is defined as $k_F=(3\pi^2 n/2)^{1/3}$ and $k_{Fn}$ and $k_{Fp}$ stand for the Fermi momenta of neutrons and protons, respectively.  The function $H_n(x)$ is defined in Eq.\,(\ref{eq:skyrme}). Note that the kinetic contribution, which is non-relativistic, and the zero-range many-body contribution correspond to the first and third terms of the Skyrme force in Eq.\,(\ref{eq:eskyrme}). The parameters $\mu_i$ are interaction lengths, and the coefficients $\mathcal{A}_i$, $\mathcal{B}_i$, $\mathcal{C}_i$ and $\mathcal{D}_i$ are combinations of the spin-isospin exchange parameters $B_i,{\bf H}_i, M_i$ and $W_i$:
\begin{eqnarray}
\mathcal{A}_i &=& \frac{1}{4}(4W_i+2B_i-2\mathbf{H}_i-M_i), \qquad
\mathcal{B}_i = -\frac{1}{4}(2{\bf H}_i+M_i),\qquad\nonumber\\
\mathcal{C}_i &=& \frac{1}{\sqrt{\pi}}(W_i+2B_i-\mathbf{H}_i-2M_i), \qquad
\mathcal{D}_i = \frac{1}{\sqrt{\pi}}(\mathbf{H}_i+2M_i).
\end{eqnarray}
$\mathbf{e}(\eta)$ and $\bar{\mathbf{e}}(\eta_1,\eta_2)$ are finite-range functions defined as
\begin{eqnarray}
\mathbf{e}(\eta) &=& \frac{\sqrt{\pi}}{2}\eta^3{\rm erf}(\eta)+\left(\frac{\eta^2}{2}-1\right)\exp\left(-\eta^2\right)-\frac{3\eta^2}{2}+1,\nonumber\\
\bar{\mathbf{e}}(\eta_1,\eta_2) &=& \sum\limits_{s=\pm 1}s\left[\frac{\sqrt{\pi}}{2}(\eta_1^3+s\eta_2^3){\rm erf}\left(\frac{\eta_1+s\eta_2}{2}\right)
+(\eta_1^2+\eta_2^2-s\eta_1\eta_2-2)\exp\left(-\frac{1}{4}[\eta_1+s\eta_2]^2\right)\right],
%\bar{\mathbf{e}}(\eta_1,\eta_2) &=& \sum\limits_{s=\pm 1}s\left[\frac{\sqrt{\pi}}{2}(\eta_1+s\eta_2)(\eta_1^2+\eta_2^2-s\eta_1\eta_2)\rm{erf}\left(\frac{\eta_1+s\eta_2}{2}\right)\right.\nonumber\\
%&\,&\left.+(\eta_1^2+\eta_2^2-s\eta_1\eta_2-2)\exp\left(-\frac{1}{4}[\eta_1+s\eta_2]^2\right)\right],
\end{eqnarray}
   where ${\rm erf} (x) = (2/\sqrt{\pi})\int_0^x \exp(-t^2)\,dt$ is the error function. Note that ${\bf e}(0)=\bar{\mathbf{e}}(\eta,0)=0$ and $\bar{\mathbf{e}}(\eta,\eta)=2\mathbf{e}(\eta)$.

The pressure in isospin asymmetric matter is
\begin{eqnarray}
P( n,x) &=& n^2E^\prime(n,x)=n\left[\frac{\hbar^2k_F^2}{5M}H_{5/3}+\sum\limits_{i=1}^2(\alpha_i+1){t_{3i}\over4} n^{\alpha_i+1}\left[x_{3i}+2-\left(x_{3i}+\frac{1}{2}\right)H_2\right]\right.\nonumber\\
+\frac{\pi^{3/2}}{2}\!\!\!\!&n&\!\!\!\!\left.\sum\limits_{i=1}^2\mu_i^3\left[\mathcal{A}_i+\mathcal{B}_i(1-2x)^2\right]-\sum\limits_{i=1}^2\left\{\mathcal{C}_i\left[(1-x)\mathbf{p}(k_{Fn}\mu_i)+x\mathbf{p}(k_{Fp}\mu_i)\right]-\frac{\mathcal{D}_i}{2}\bar{\mathbf{p}}(k_{Fn}\mu_i,k_{Fp}\mu_i)\right\}\right]
\end{eqnarray}
with
\begin{eqnarray}
    \mathbf{p}(\eta)&=&n\left[{{\bf e(\eta)}\over\eta^3}\right]^\prime=\frac{1}{2\eta}-\frac{1}{\eta^3}+\left(\frac{1}{\eta^3}+\frac{1}{2\eta}\right)\exp\left(-\eta^2\right),\nonumber\\
\bar{\mathbf{p}}(\eta_1,\eta_2) &=& n\left[{\bar{\bf e}(\eta_1,\eta_2)\over\eta_1^3+\eta_2^3}\right]^\prime=\frac{2}{\eta_1^3+\eta_2^3}\sum\limits_{s=\pm 1}(\eta_1\eta_2+2s)\exp\left(-\frac{1}{4}[\eta_1+s\eta_2]^2\right),
\end{eqnarray}
which, like $\bar{\bf e}$, also satisfies $\bar{\mathbf{p}}(\eta,\eta)=2\mathbf{p}(\eta)$ and $\bar{\mathbf{p}}(\eta,0)=0$.  Note that $\eta^\prime=\eta/(3n)$.
\subsubsection{Uniform Symmetric Matter}
For uniform symmetric matter, one finds
\begin{eqnarray}
E_{1/2}(n) &=& \frac{3\hbar^2k_F^2}{10M}+{3\over8}\sum\limits_{i=1}^2t_{3i} n^{\alpha_i+1}+\frac{\pi^{3/2}}{2} n\sum\limits_{i=1}^2\mu_i^3\mathcal{A}_i  -\sum\limits_{i=1}^2\frac{{\cal C}_i-{\cal D}_i}{(k_F\mu_i)^3}\mathbf{e}(k_{F}\mu_i), \nonumber\\
P_{1/2}(n) &=&n\left[\frac{\hbar^2k_F^2}{5M}+{3\over8}\sum\limits_{i=1}^2t_{3i}(\alpha_i+1) n^{\alpha_i+1}+\frac{\pi^{3/2}}{2}n\sum\limits_{i=1}^2\mu_i^3\mathcal{A}_i -\sum\limits_{i=1}^2\left(\mathcal{C}_i-{\cal D}_i\right)\mathbf{p}(k_{F}\mu_i)\right].
\end{eqnarray}
The incompressibility and skewness can be derived from the pressure
\begin{eqnarray}
{\cal K}_{1/2}(n) &=& 9P^\prime_{1/2}(n)-\frac{18P_{1/2}(n)}{n} = -\frac{3\hbar^2k_F^2}{5M}+\frac{27}{8}\sum\limits_{i=1}^2\alpha_i(\alpha_i+1)t_{3i} n^{\alpha_i+1}
\nonumber\\
&+&9\sum\limits_{i=1}^2(\mathcal{C}_i-\mathcal{D}_i)\left[\mathbf{p}(k_F\mu_i)-n\mathbf{p}^\prime(k_F\mu_i)\right], \nonumber\\
{\cal Q}_{1/2}(n) &=& 27 nP^{\prime\prime}_{1/2}(n)-\frac{54P_{1/2}( n)}{ n}-12{\cal K}_{1/2}( n) = \frac{12\hbar^2k_F^2}{5M}+\frac{81}{8}\sum\limits_{i=1}^2\alpha_i(\alpha_i^2-1)t_{3i} n^{\alpha_i+1}  \nonumber \\
&-& 27\sum\limits_{i=1}^2(\mathcal{C}_i-\mathcal{D}_i)\left[2\mathbf{p}(k_F\mu_i)-2n\mathbf{p}^\prime(k_F\mu_i)+n^2\mathbf{p}^{\prime\prime}(k_F\mu_i)\right],
\end{eqnarray}
where the logarithmic derivatives of ${\bf p}$ 
%$\mathbf{p}^{\prime}$, $\mathbf{p}^{\prime\prime}$ 
are
\begin{eqnarray}
    n\bf{p}^\prime(\eta) &=& n\frac{d{\bf p}}{dn}=\frac{\eta}{3}\frac{d{\bf p}}{d\eta} = \frac{1}{\eta^3}-\frac{1}{6\eta}-\left(\frac{1}{\eta^3}+\frac{5}{6\eta}+\frac{\eta}{3}\right)\exp\left({-\eta^2}\right), \nonumber\\
    n^2\mathbf{p}^{\prime\prime}(\eta) &=& \frac{\eta^2}{9}\frac{d^2{\bf p}}{d\eta^2}-\frac{2\eta}{9}\frac{d{\bf p}}{d\eta} =\frac{2}{9\eta}-\frac{2}{\eta^3}+\left(\frac{2}{\eta^3}+\frac{16}{9\eta}+\frac{7}{9}\eta+\frac{2}{9}\eta^3\right)\exp\left({-\eta^2}\right).
\end{eqnarray}
%When calculating the symmetric properties at saturation density, we have $K_0={\cal K}_{1/2}(n_0)$ and $Q_0={\cal Q}_{1/2}(n_0)$.
\subsubsection{Uniform Neutron Matter}
In the case of uniform neutron matter, $k_{Fn}=(3\pi^2 n)^{1/3}$, and 
\begin{eqnarray}
E_N(n) &=& \frac{3\hbar^2k_{Fn}^2}{10M}+\sum\limits_{i=1}^2{t_{3i}\over4} n^{\alpha_i+1}\left(1-x_{3i}\right)+\frac{\pi^{3/2}}{2} n\sum\limits_{i=1}^2\mu_i^3\left(\mathcal{A}_i+\mathcal{B}_i\right)  -\sum\limits_{i=1}^2\frac{{\cal C}_i\mathbf{e}(k_{Fn}\mu_i)}{2(k_{Fn}\mu_i)^3},\nonumber\\
P_N(n) &=&\!\! n\left[\frac{\hbar^2k_{Fn}^2}{5M}+\sum\limits_{i=1}^2{t_{3i}\over4}(\alpha_i+1) n^{\alpha_i+1}\left(1-x_{3i}\right)+\frac{\pi^{3/2}}{2} n\sum\limits_{i=1}^2\mu_i^3\left(\mathcal{A}_i+\mathcal{B}_i\right)-\sum\limits_{i=1}^2\mathcal{C}_i\mathbf{p}(k_{Fn}\mu_i)\right],\nonumber\\
{\cal K}_{N}(n) &=& -\frac{3\hbar^2k_{Fn}^2}{5M}+\frac{9}{4}\sum\limits_{i=1}^2\alpha_i(\alpha_i+1)t_{3i} n^{\alpha_i+1}(1-x_{3i})+9\sum\limits_{i=1}^2\mathcal{C}_i\left[\mathbf{p}(k_{Fn}\mu_i)-n\mathbf{p}^\prime(k_{Fn}\mu_i\right], \\
{\cal Q}_{N}(n) &=& \frac{12\hbar^2k_{Fn}^2}{5M}+\frac{27}{4}\sum\limits_{i=1}^2\alpha_i(\alpha_i^2-1)t_{3i} n^{\alpha_i+1}(1-x_{3i})-27\sum\limits_{i=1}^2\mathcal{C}_i\left[2\mathbf{p}(k_{Fn}\mu_i)-2n\mathbf{p}^\prime(k_{Fn}\mu_i)+n^2\mathbf{p}^{\prime\prime}(k_{Fn}\mu_i)\right],\nonumber
\end{eqnarray}

\subsubsection{Symmetry Properties}
The symmetry energy and its derivatives at saturation density can be expressed in the first approach as
\begin{eqnarray}
{\cal S}_1 = E_N(n)-E_{1/2}(n) &=& \frac{3\hbar^2\left(k_{Fn}^2-k_{F}^2\right)}{10M}-\sum\limits_{i=1}^2{t_{3i}\over8} n^{\alpha_i+1}(1+2x_{3i})+\frac{\pi^{3/2}}{2} n\sum\limits_{i=1}^2\mu_i^3\mathcal{B}_i  \nonumber\\
&-& \sum\limits_{i=1}^2\frac{\mathcal{C}_i\mathbf{e}(k_{Fn}\mu_i)-2(\mathcal{C}_i-\mathcal{D}_i)\mathbf{e}(k_{F}\mu_i)}{2(k_{F}\mu_i)^3},
\end{eqnarray}
The slope of symmetric energy can be calculated from $3\left[P_N(n)-P_{1/2}(n)\right]/n$:
\begin{eqnarray}
{\cal L}_1 &=& \frac{3\hbar^2(k_{Fn}^2-k_F^2)}{5M}-{3\over8}\sum\limits_{i=1}^2(\alpha_i+1)t_{3i} n^{\alpha_i+1}(1+2x_{3i})+\frac{3\pi^{3/2}}{2} n\sum\limits_{i=1}^2\mu_i^3\mathcal{B}_i
\nonumber\\&-&3\sum\limits_{i=1}^2\left[\mathcal{C}_i\mathbf{p}(k_{Fn}\mu_i)-({\cal C}_i-{\cal D}_i)\mathbf{p}(k_{F}\mu_i)\right].
\end{eqnarray}
The symmetry incompressibility and skewness are further obtained as
\begin{eqnarray}
{\cal K}_{sym1} &=& -\frac{3\hbar^2\left(k_{Fn}^2-k_{F}^2\right)}{5M}-\frac{9}{8}\sum\limits_{i=1}^2\alpha_i(\alpha_i+1)t_{3i} n^{\alpha_i+1}(1+2x_{3i})  \nonumber \\
&+& 9\sum\limits_{i=1}^2\left\{\mathcal{C}_i\left[\mathbf{p}(k_{Fn}\mu_i)-n\mathbf{p}^\prime(k_{Fn}\mu_i)\right]-(\mathcal{C}_i-\mathcal{D}_i)\left[\mathbf{p}(k_{F}\mu_i)-n{\bf p}^\prime(k_F\mu_i)\right]\right\}  \nonumber \\
{\cal Q}_{sym1} &=& \frac{12\hbar^2\left(k_{Fn}^2-k_{F}^2\right)}{5M}-\frac{27}{8}\sum\limits_{i=1}^2\alpha_i(\alpha_i^2-1)t_{3i} n^{\alpha_i+1}(1+2x_{3i}) \\
&-& 27\sum\limits_{i=1}^2\left\{\mathcal{C}_i\left[2\mathbf{p}(k_{Fn}\mu_i)-2n\mathbf{p}^\prime(k_{Fn}\mu_i)+n^2{\bf p}^{\prime\prime}(k_{Fn}\mu_i)\right]
-(\mathcal{C}_i-\mathcal{D}_i)\left[2\mathbf{p}(k_{F}\mu_i)-2n\mathbf{p}^\prime(k_{F}\mu_i)+n^2{\bf p}^{\prime\prime}(k_F\mu_i)\right]\right\}.\nonumber
\end{eqnarray}
The second approach gives the symmetry energy as
\begin{eqnarray}
{\cal S}_2 = \frac{1}{8}\frac{\partial^2 E( n, x)}{\partial x^2}\bigg|_{x=1/2} &=& \frac{\hbar^2}{6M}\left(\frac{3\pi^2}{2}\right)^{2/3} n^{2/3}-\frac{1}{8}\sum\limits_{i=1}^2t_{3i} n^{\alpha_i+1}(2x_{3i}+1)+\frac{\pi^{3/2}}{2} n\sum\limits_{i=1}^2\mu_i^3\mathcal{B}_i  \nonumber\\
&-&\frac{1}{6}\sum\limits_{i=1}^2\left[\mathcal{C}_i\mathbf{s}_1(k_{F}\mu_i)-\mathcal{D}_i\mathbf{s}_2(k_{F}\mu_i)\right],
\end{eqnarray}
with
\begin{equation}
    \mathbf{s}_1(\eta) = \frac{1}{\eta}-\left(\eta+\frac{1}{\eta}\right)\exp\left({-\eta^2}\right), \quad \mathbf{s}_2(\eta) = \frac{1}{\eta}-\eta-\frac{1}{\eta}\exp\left({-\eta^2}\right).
\end{equation}
Then  its derivatives can be obtained as
\begin{eqnarray}
{\cal L}_2 = 3 nS_2^\prime &=& \frac{\hbar^2}{3M}(3\pi^2 n/2)^{2/3}-\frac{3}{8}\sum\limits_{i=1}^2(\alpha_i+1)t_{3i} n^{\alpha_i+1}(2x_{3i}+1)+\frac{3\pi^{3/2}}{2} n\sum\limits_{i=1}^2\mu_i^3\mathcal{B}_i  \nonumber\\
&\,& -\frac{n}{2}\sum\limits_{i=1}^2\left[\mathcal{C}_i\mathbf{s}^\prime_1(k_{F}\mu_i)-\mathcal{D}_i\mathbf{s}^\prime_2(k_{F}\mu_i)\right];\nonumber\\
{\cal K}_{sym2} = 9 n^2S_2^{\prime\prime} &=& -\frac{\hbar^2}{3M}(3\pi^2 n/2)^{2/3}-\frac{9}{8}\sum\limits_{i=1}^2\alpha(\alpha_i+1)t_{3i} n^{\alpha_i+1}(2x_{3i}+1) \nonumber \\
&\,& -\frac{3n^2}{2}\sum\limits_{i=1}^2\left[{\cal C}_i{\bf s}^{\prime\prime}_1(k_{F}\mu_i)-{\cal D}_i{\bf s}^{\prime\prime}_2(k_F\mu_i)\right]
%\left[-\mathcal{C}_i\mathbf{l}_1(k_{F}\mu_i)+\mathcal{D}_i\mathbf{l}_2(k_{F}\mu_i)\right]+\frac{1}{6}\sum\limits_{i=1}^2(k_{F}\mu_i)^2\left[-\mathcal{C}_i\mathbf{k}_1(k_{F}\mu_i)+\mathcal{D}_i\mathbf{k}_2(k_{F}\mu_i)\right]
;\nonumber\\
{\cal Q}_{sym2} = 27 n^3S_3^{\prime\prime\prime} &=& -\frac{4\hbar^2}{3M}(3\pi^2 n/2)^{2/3}-\frac{27}{8}\sum\limits_{i=1}^2\alpha(\alpha_i^2-1)t_{3i} n^{\alpha_i+1}(2x_{3i}+1)  \\
&\,& -\frac{9n^3}{2}\sum\limits_{i=1}^2\left[{\cal C}_i{\bf s}^{\prime\prime\prime}_1(k_F\mu_i)-{\cal D}_i{\bf s}^{\prime\prime\prime}_2(k_F\mu_i)\right],\nonumber
%\left[-\mathcal{C}_i\mathbf{l}_1(k_{F}\mu_i)+\mathcal{D}_i\mathbf{l}_2(k_{F}\mu_i)\right]-\sum\limits_{i=1}^2(k_{F}\mu_i)^2\left[-\mathcal{C}_i\mathbf{k}_1(k_{F}\mu_i)+\mathcal{D}_i\mathbf{k}_2(k_{F}\mu_i)\right] \nonumber \\
%&\;& +\frac{1}{6}\sum\limits_{i=1}^2(k_{F}\mu_i)^3\left[-\mathcal{C}_i\mathbf{q}_1(k_{F}\mu_i)+\mathcal{D}_i\mathbf{q}_2(k_{F}\mu_i)\right],
\end{eqnarray}
where the logarithmic derivatives of $\mathbf{s}_{1,2}$ are:
\begin{eqnarray}
n\mathbf{s}^\prime_1 &=&\frac{\eta}{3}\frac{d{\bf s}_1}{d\eta}= -\frac{1}{3\eta}+\left(\frac{1}{3\eta}+\frac{\eta}{3}+\frac{2\eta^3}{3}\right)e^{-\eta^2}, \qquad
n\mathbf{s}^\prime_2 =\frac{\eta}{3}\frac{d{\bf s}_2}{d\eta}=  -\frac{1}{3\eta}-\frac{\eta}{3}+\left(\frac{1}{3\eta}+\frac{2\eta}{3}\right)e^{-\eta^2},\nonumber\\
n^2\mathbf{s}^{\prime\prime}_1 &=& \frac{4}{9\eta}-\left(\frac{4}{9\eta}+\frac{4\eta}{9}+\frac{2\eta^3}{9}+\frac{4\eta^5}{9}\right)e^{-\eta^2}, \quad
n^2\mathbf{s}^{\prime\prime}_2 =\frac{4}{9\eta}+\frac{2\eta}{9}-\left(\frac{4}{9\eta}+\frac{2\eta}{3}+\frac{4\eta^3}{9}\right)e^{-\eta^2},\\
%n^3\mathbf{s}^{\prime\prime\prime}_1 &=&-\frac{28}{27\eta}+\left(\frac{28}{27\eta}+\frac{28\eta}{27}+\frac{14\eta^3}{27}+\frac{8\eta^5}{27}+\frac{8\eta^7}{27}\right)e^{-\eta^2}, \quad
%n^3\mathbf{s}^{\prime\prime\prime}_2 =-\frac{28}{27\eta}-\frac{10\eta}{27}+\left(\frac{28}{27\eta}+\frac{38\eta}{27}+\frac{8\eta^3}{9}+\frac{8\eta^5}{27}\right)e^{-\eta^2}.\nonumber \\
n^3\mathbf{s}_1^{\prime\prime\prime}&=&-{7\over9\eta}+\left({7\over9\eta}+{7\eta\over9}+{26\eta^3\over27}-{26\eta^5\over27}+{8\eta^7\over27}\right)e^{-\eta^2},\quad
n^3\mathbf{s}_2^{\prime\prime\prime}=-{7\over9\eta}-{5\eta\over27}+\left({7\over9\eta}+{26\eta\over27}+{20\eta^3\over27}+{8\eta^5\over27}\right)e^{-\eta^2}.\nonumber
\end{eqnarray}
All the saturation properties of Gogny interactions are included in Table XII of Appendix C.
\section{Comparisons of Properties at the Saturation Density}

\begin{figure}[H]
\centering
\vspace*{0cm}
\begin{adjustwidth}{0cm}{0cm}
    \includegraphics[width=8.5cm,angle=0]{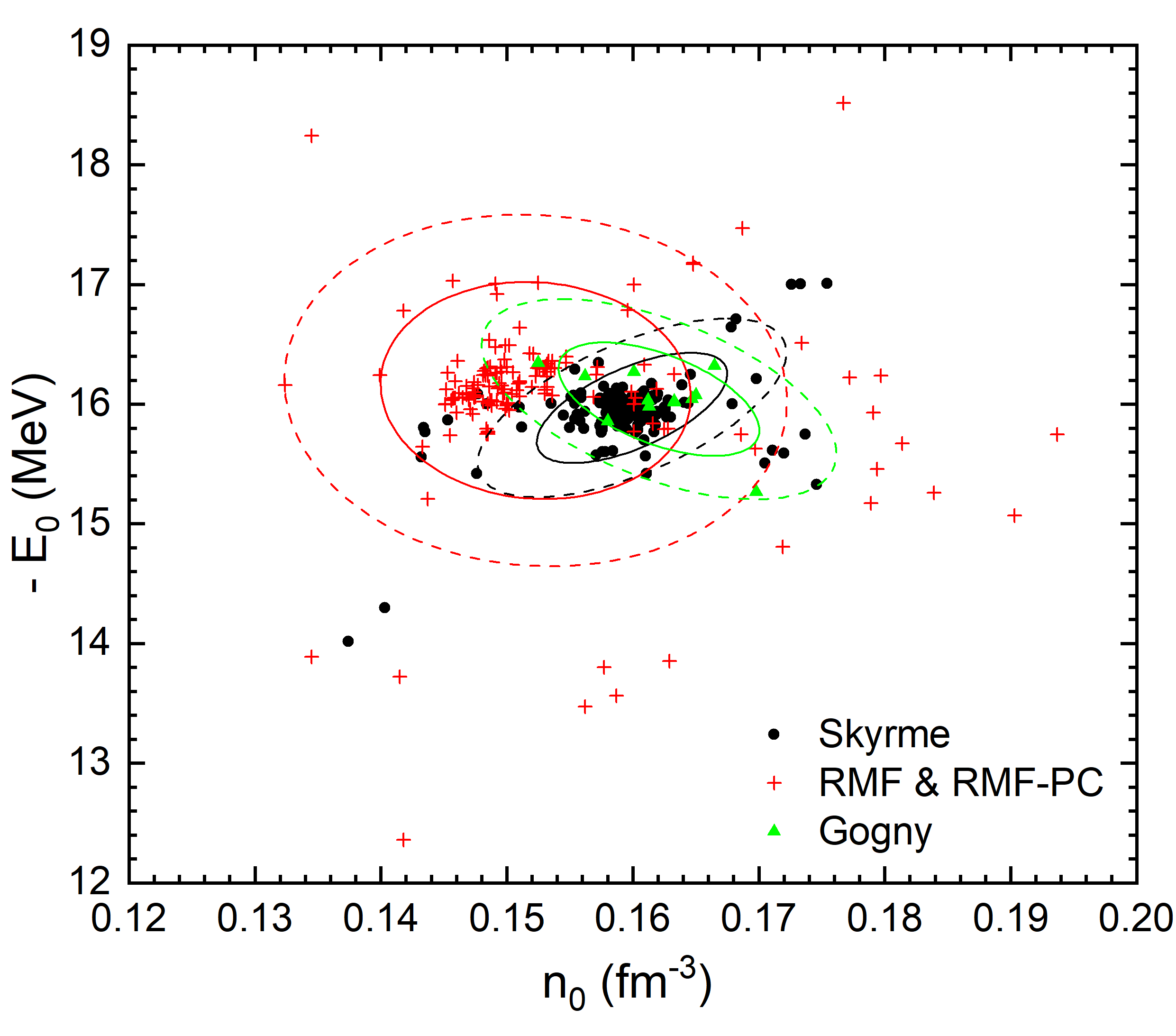}
     \includegraphics[width=8.5 cm,angle=0]{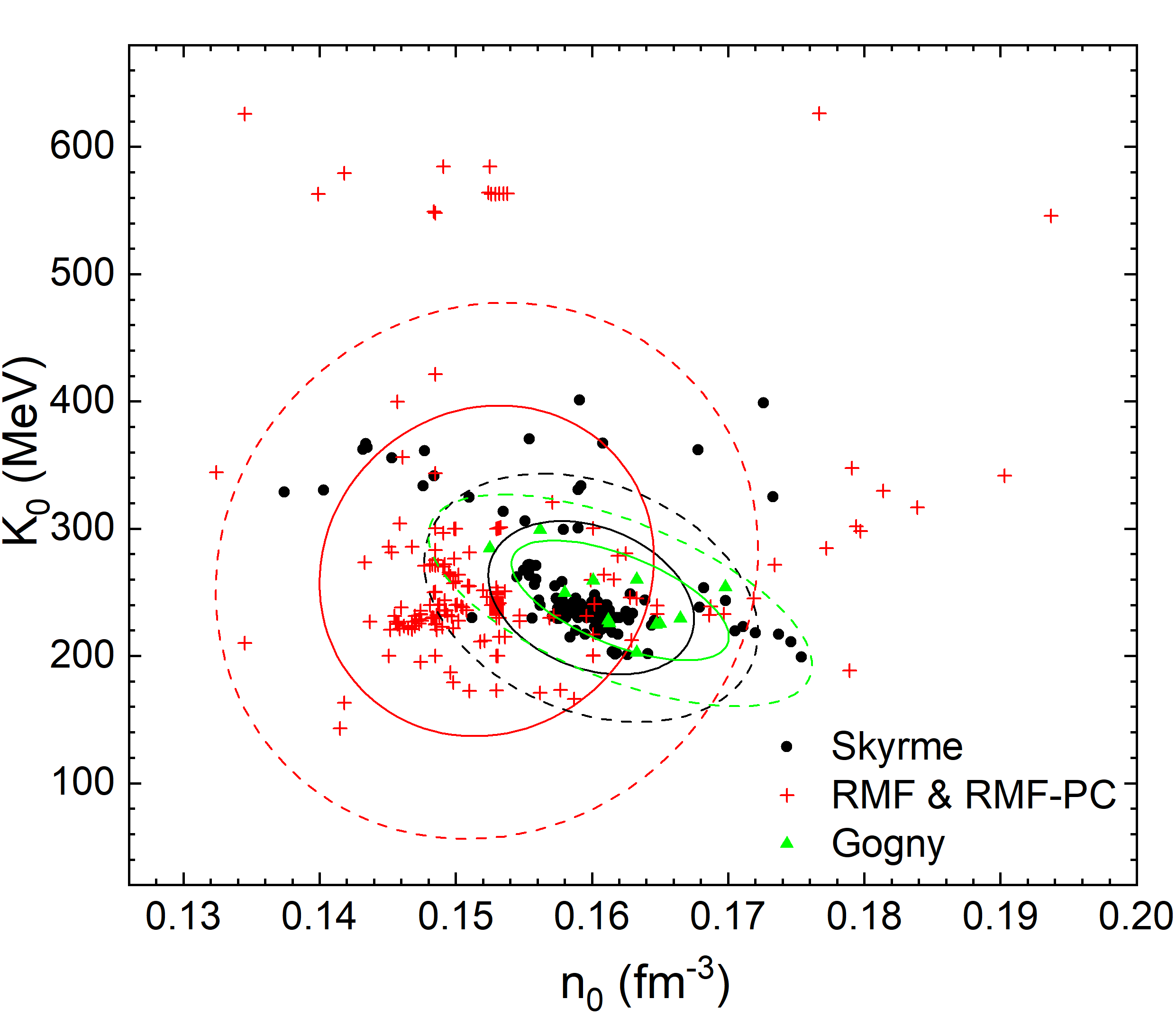}
\end{adjustwidth}
  \vspace*{-0.5cm}  \caption{The correlations between the saturation density $n_0$ and the bulk energy per particle $E_0$ (left panel) and the incompressibility $K_0$ (right panel) of symmetric matter at $n_0$ for Skyrme (black filled circles), Gogny (green triangles) and RMF (red pluses) models.  Solid and dashed lines indicate $1\sigma$ and $2\sigma$ confidence ellipses for each force type.}
  \label{fig:Bn0}
\end{figure}

\begin{figure}[H]
\centering
\begin{adjustwidth}{0cm}{0cm}
     \includegraphics[width=8.5 cm,angle=0]{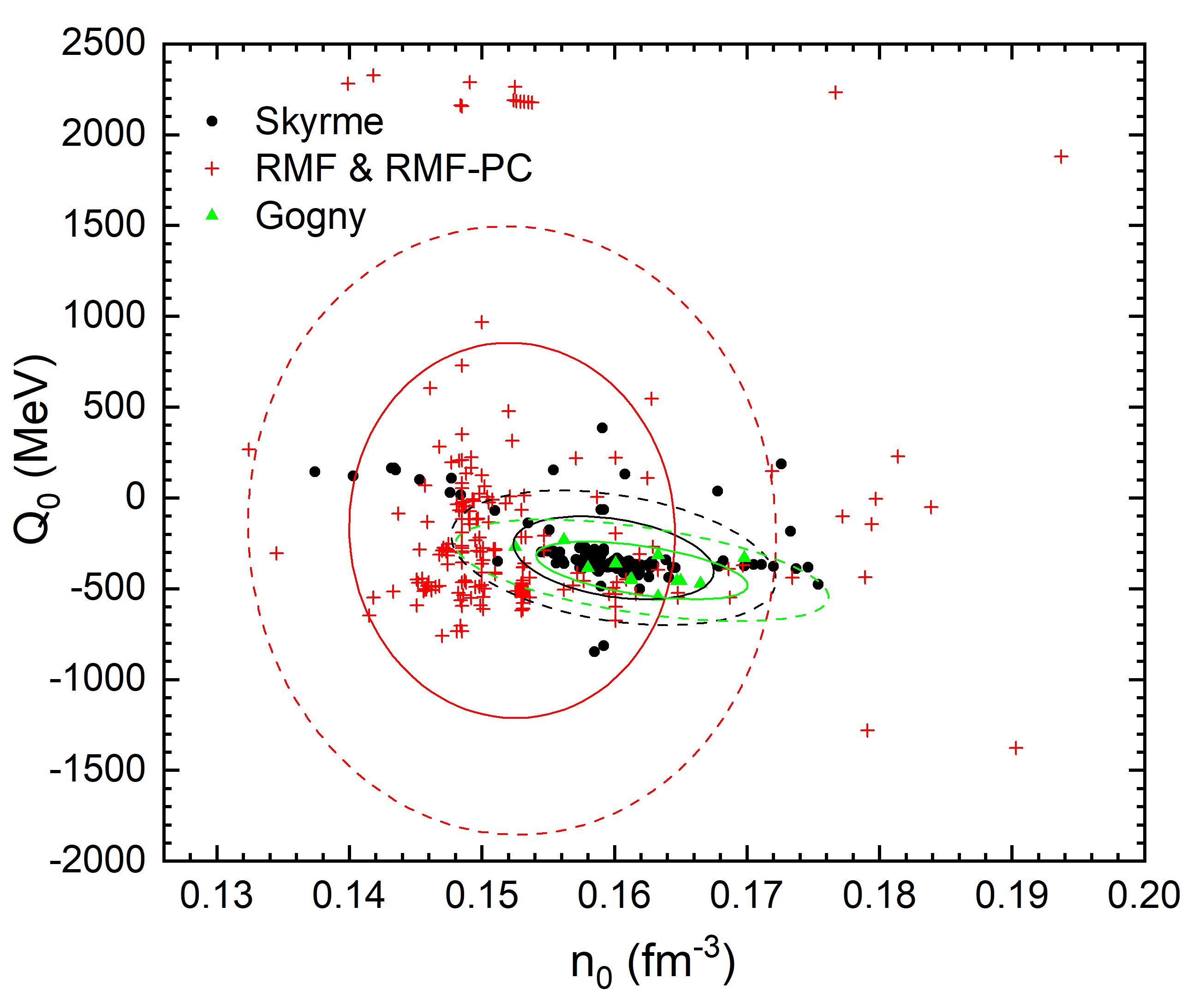}
     \includegraphics[width=8.5 cm,angle=0]{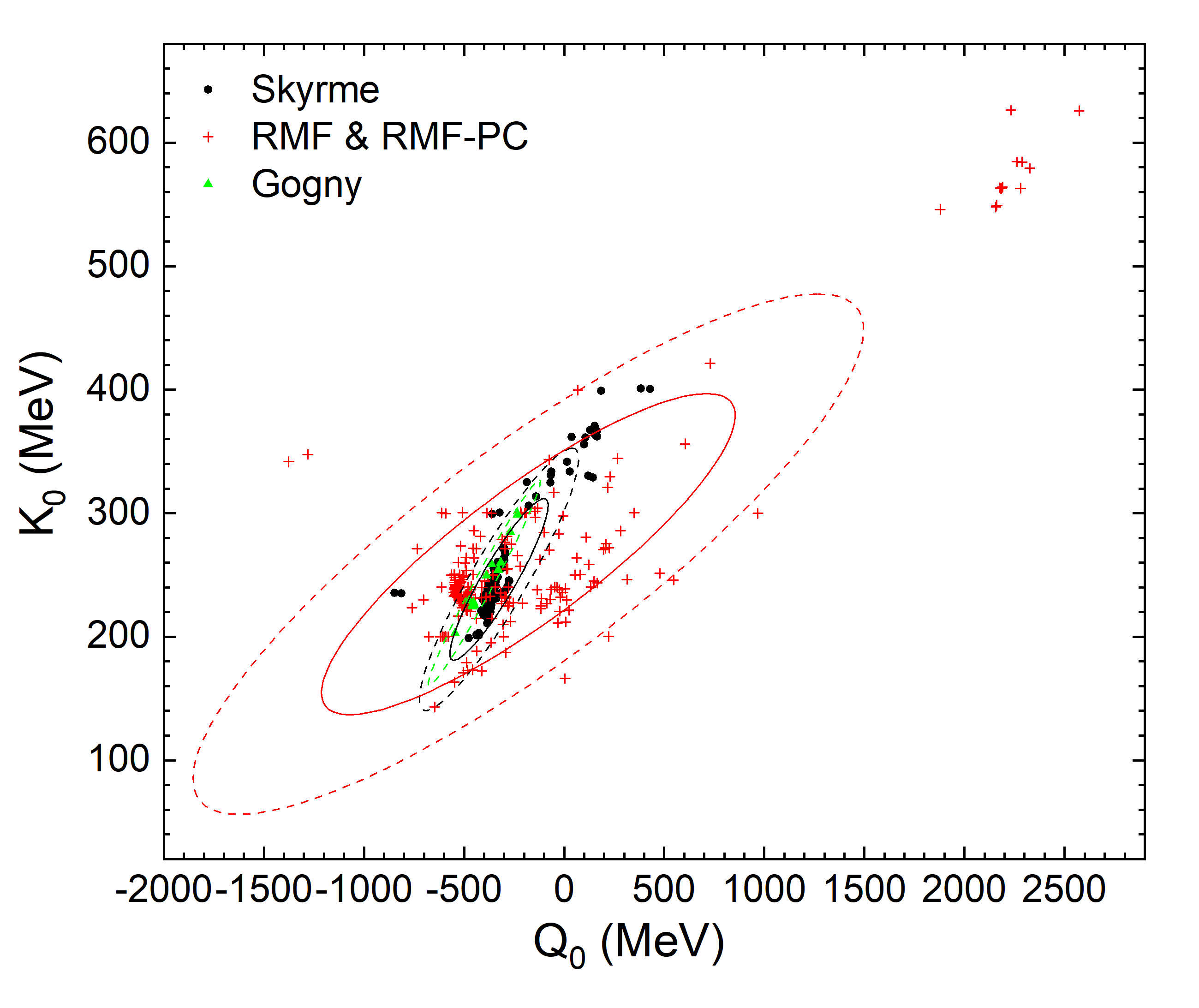}
\end{adjustwidth}
\vspace*{-0.2cm}\caption{Similar to Figure \ref{fig:Bn0}, but for the correlations between the symmetric matter  skewness $Q_0$  with $n_0$ (left panel) and $Q_0$ with $K_0$ (right panel).}\label{fig:Q0n0}
\end{figure}

%\subsection{Comparison of Saturation Properties Among Different Models}
Figure \ref{fig:Bn0} displays the symmetric matter densities $n_0$, energies $-E_0$ and incompressibilities $K_0$ at saturation for symmetric matter for the Skyrme, RMF and Gogny forces considered in this paper.   Similarly, the left panel of Figure \ref{fig:Q0n0} compares $n_0$, $K_0$ and skewnesses $Q_0$ at saturation.  
It is notable that $n_0, E_0$ and $K_0$ cluster in regions, often called the saturation window, which therefore represents predictions resulting from fitting these forces mainly to binding energies.  There are distinct differences in the saturation windows for Skyrme and RMF forces; RMF forces typically suggest a smaller saturation density, for example, as well as larger mean values of $-E_0$, $K_0$ and $Q_0$.  
The saturation window for Skyrme forces is defined by $n_0=0.160\pm0.006$ fm$^{-3}$, $E_0=-15.96\pm0.55$ MeV and $K_0=246.4\pm43.0$ MeV, and for RMF forces $n_0=0.152\pm0.008$ fm$^{-3}$, $E_0=-16.11\pm0.59$ MeV and $K_0=266.9\pm485.2$ MeV.  For reference, for Skyrme forces $Q_0=-326\pm161$ MeV and, for RMF forces, $Q_0=-179\pm678$ MeV. The parameters $n_0$, $E_0$, and $K_0$ are largely uncorrelated.  $Q_0$ is not as well localized as the other parameters, but their values are highly correlated with $K_0$ (and with $n_0$ for Skyrme forces), as seen in Fig. \ref{fig:Q0n0}. 

\begin{figure}[H]
\centering
\begin{adjustwidth}{0cm}{0cm}
     \includegraphics[width=8.5 cm,angle=0]{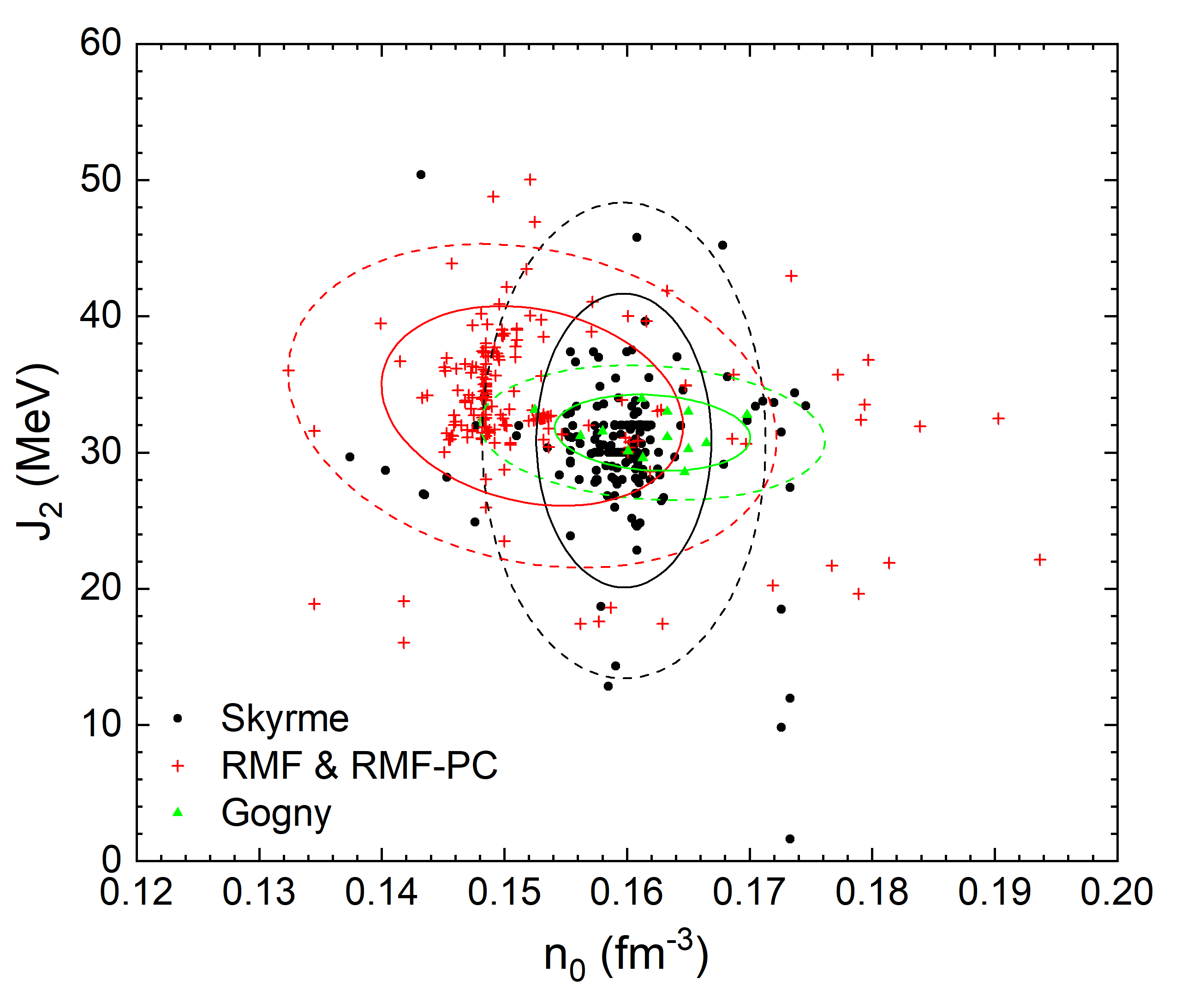}
     \includegraphics[width=8.5 cm,angle=0]{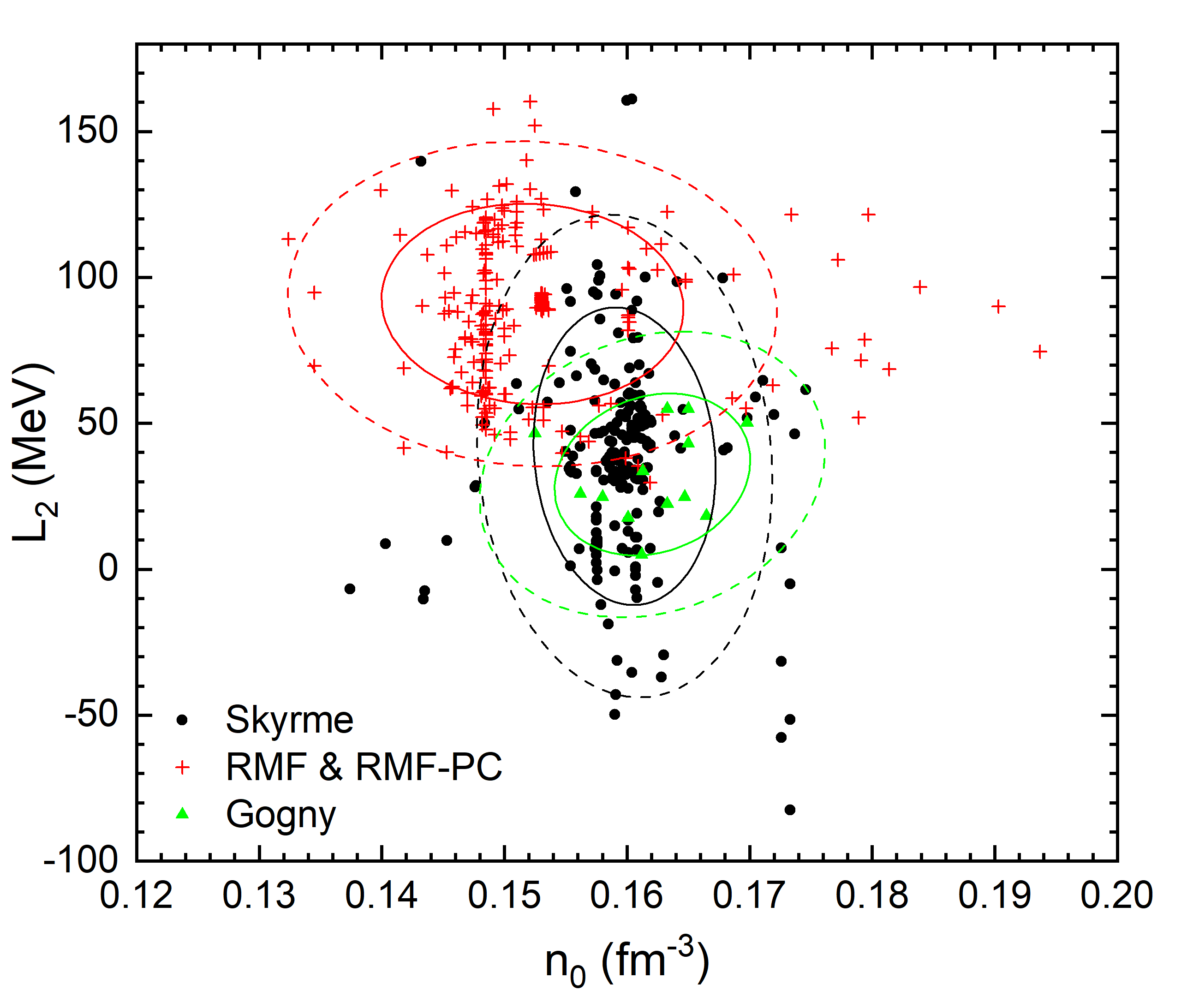}
\end{adjustwidth}
\vspace*{0cm}\caption{Similar to Figure \ref{fig:Bn0}, but for correlations between the parameters $J_2$ and $n_0$ (left panel) and $L_2$ and $n_0$ (right panel).}\label{fig:jln0}
\end{figure}

\begin{figure}[H]
\centering
\vspace*{-0.6cm}
\hspace{-.5cm}
    \includegraphics[width=12cm,angle=0]{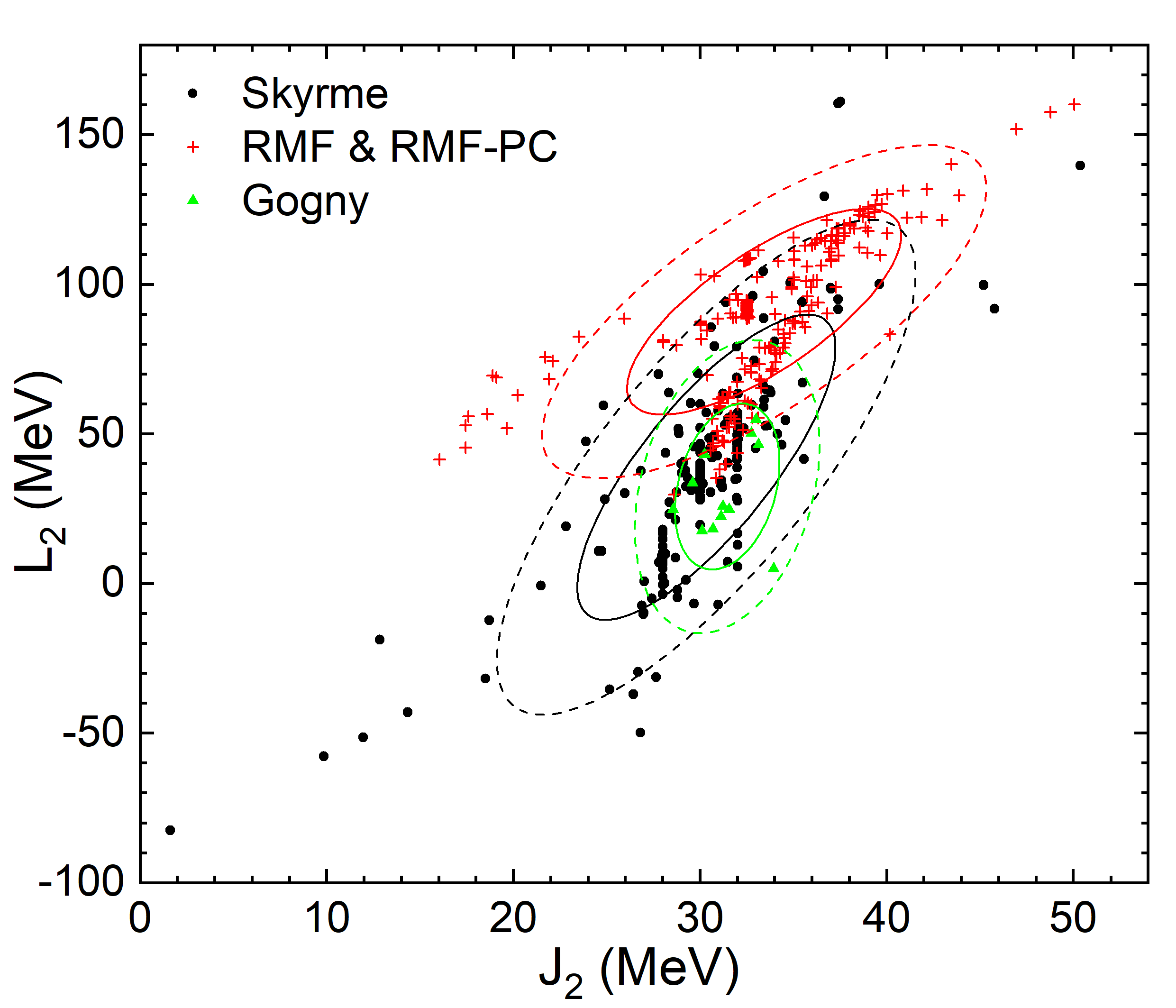}
  \vspace*{-0.5cm}  \caption{Similar to Figure \ref{fig:Bn0} but showing the correlation between $L_2$ and $J_2$.}
  \label{fig:jl}
\end{figure}

It is also useful to compare symmetry properties at the saturation density.  Figure \ref{fig:jln0} displays the saturation densities and the symmetry parameters $J_2$ and $L_2$.  It is seen that RMF forces have larger average values of $J_2$ and $L_2$ than do Skyrme and Gogny forces, and correlations between $n_0$ and both $J_2$ and $L_2$ are very weak.  On the other hand, as shown in Figure \ref{fig:jl}, $J_2$ and $L_2$ are strongly correlated for all kinds of forces. This correlation is one of the most powerful predictions of nuclear mass-fitting, and, interestingly, has a similar slope and centroid to that determined from chiral effective field theory expansions of nuclear matter \cite{Lattimer_2023}. RMF forces show a smaller correlation slope than for Skyrme forces.


\begin{figure}[H]
\centering
\begin{adjustwidth}{0cm}{0cm}
\includegraphics[width=8.5 cm,angle=0]{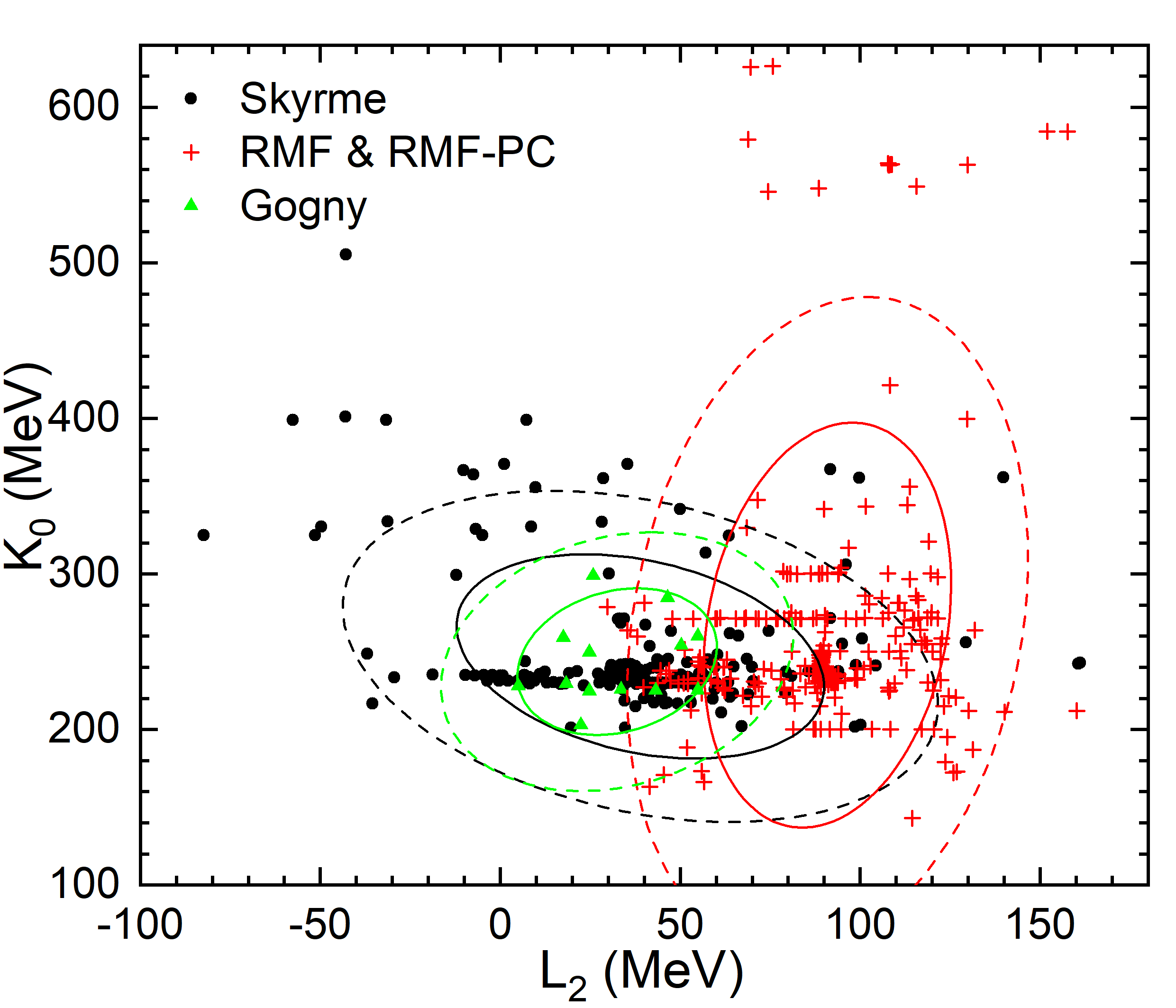}
\includegraphics[width=8.5 cm,angle=0]{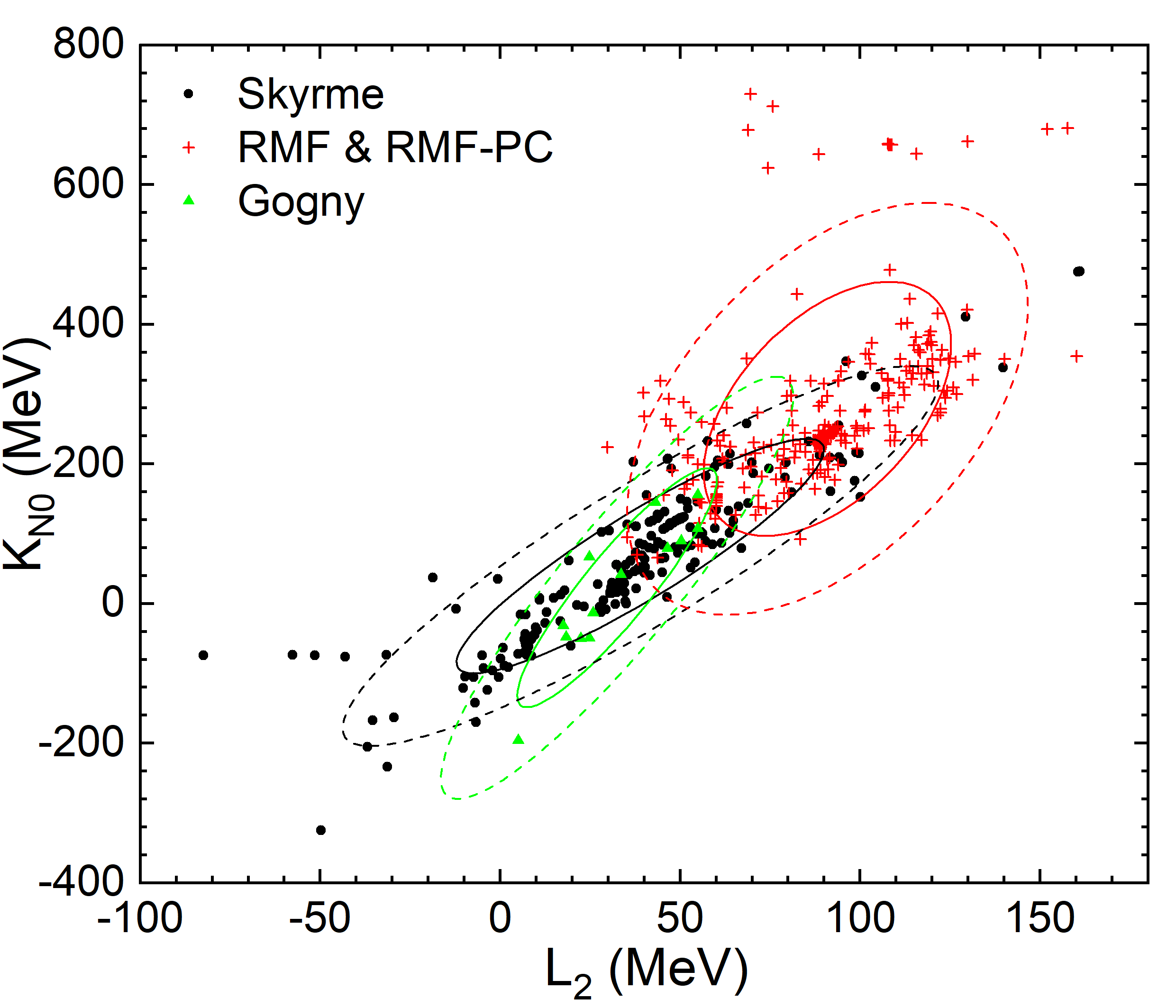}
\end{adjustwidth}
\vspace*{0cm}\caption{Similar to Figure \ref{fig:Bn0}, but for correlations between $K_0$ and $L_2$ (left panel) and $K_{N0}$ and $L_2$ (right panel).}\label{fig:kl}
\end{figure}

\begin{figure}[H]
\centering
\begin{adjustwidth}{0cm}{0cm}
\includegraphics[width=8.5 cm,angle=0]{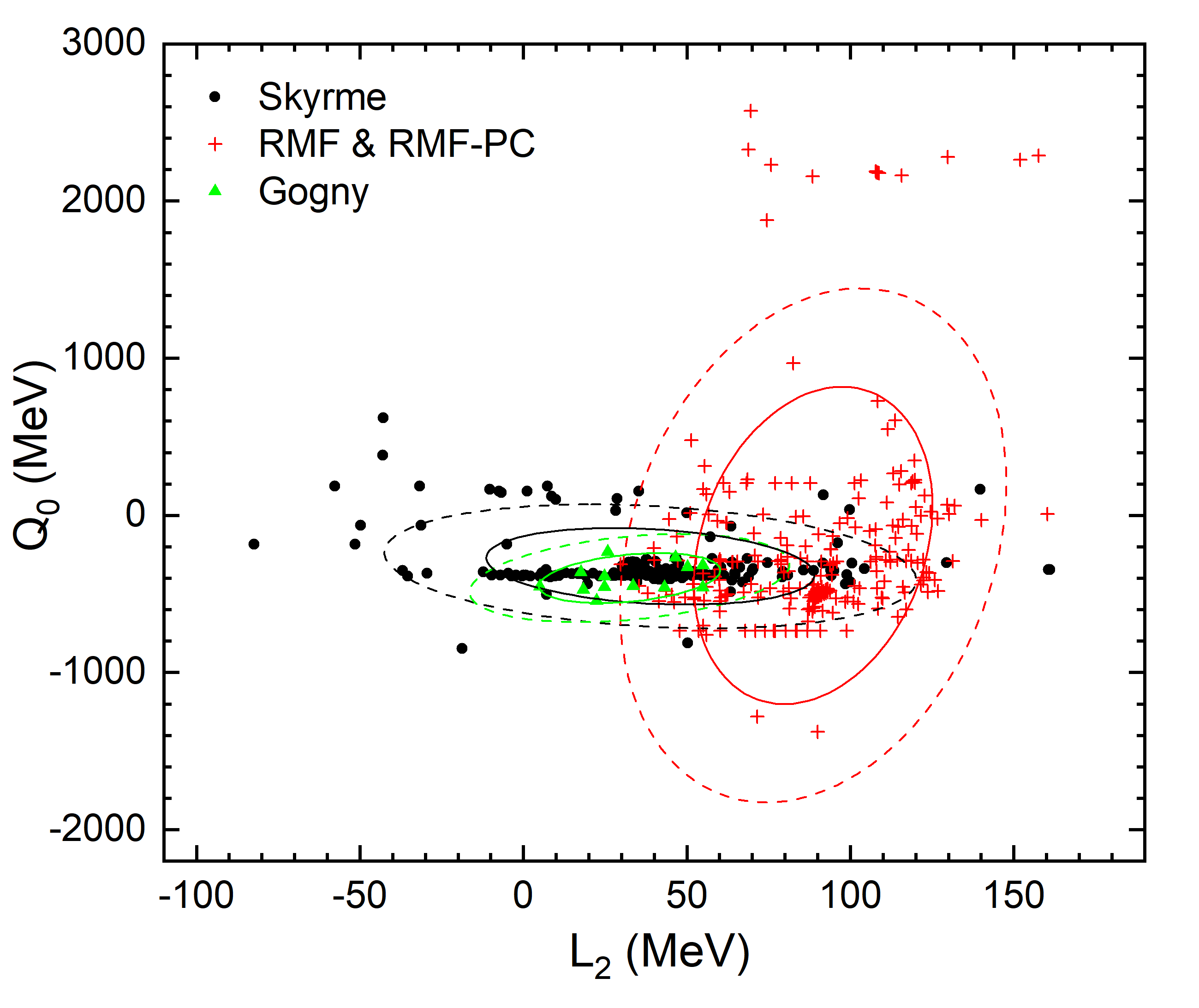}
\includegraphics[width=8.5 cm,angle=0]{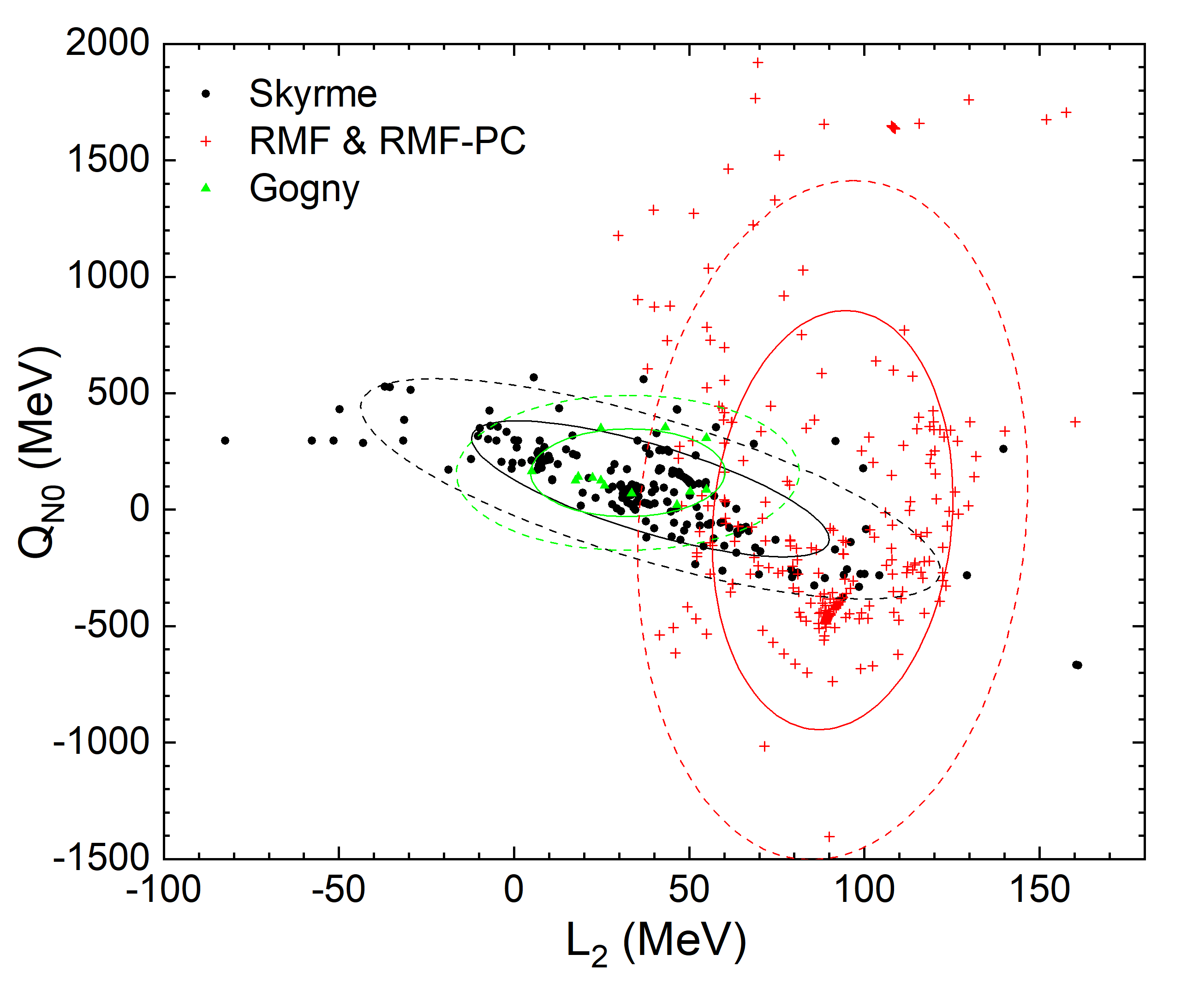}
\end{adjustwidth}
\vspace*{0cm}\caption{Similar to Figure \ref{fig:Bn0}, but for correlations between $Q_0$ and $L_2$ (left panel) and $Q_{N0}$ and $L_2$ (right panel).}\label{fig:ql}
\end{figure}

Figure \ref{fig:kl} compares the incompressibilities of both symmetric and pure neutron matter with the symmetry energy slope $L_2$, at the saturation density.  Similarly, Figure \ref{fig:ql} shows the skewnesses versus $L_2$.
All types of forces show stronger correlations between $K_{N0}$ and $L_2$ than between $K_0$ and $L_2$.  Globally, both skewness parameters are largely uncorrelated with $L_2$, except that $Q_{N0}$ for Skyrme forces does correlate with $L_2$.  The average values of these parameters for neutron matter are all larger than for those of symmetric matter.  These differences are especially significant for both $L$ and $K$.

\begin{figure}[H]
\centering
\begin{adjustwidth}{0cm}{0cm}
\hspace*{0cm}\includegraphics[width=8.5 cm,angle=0]{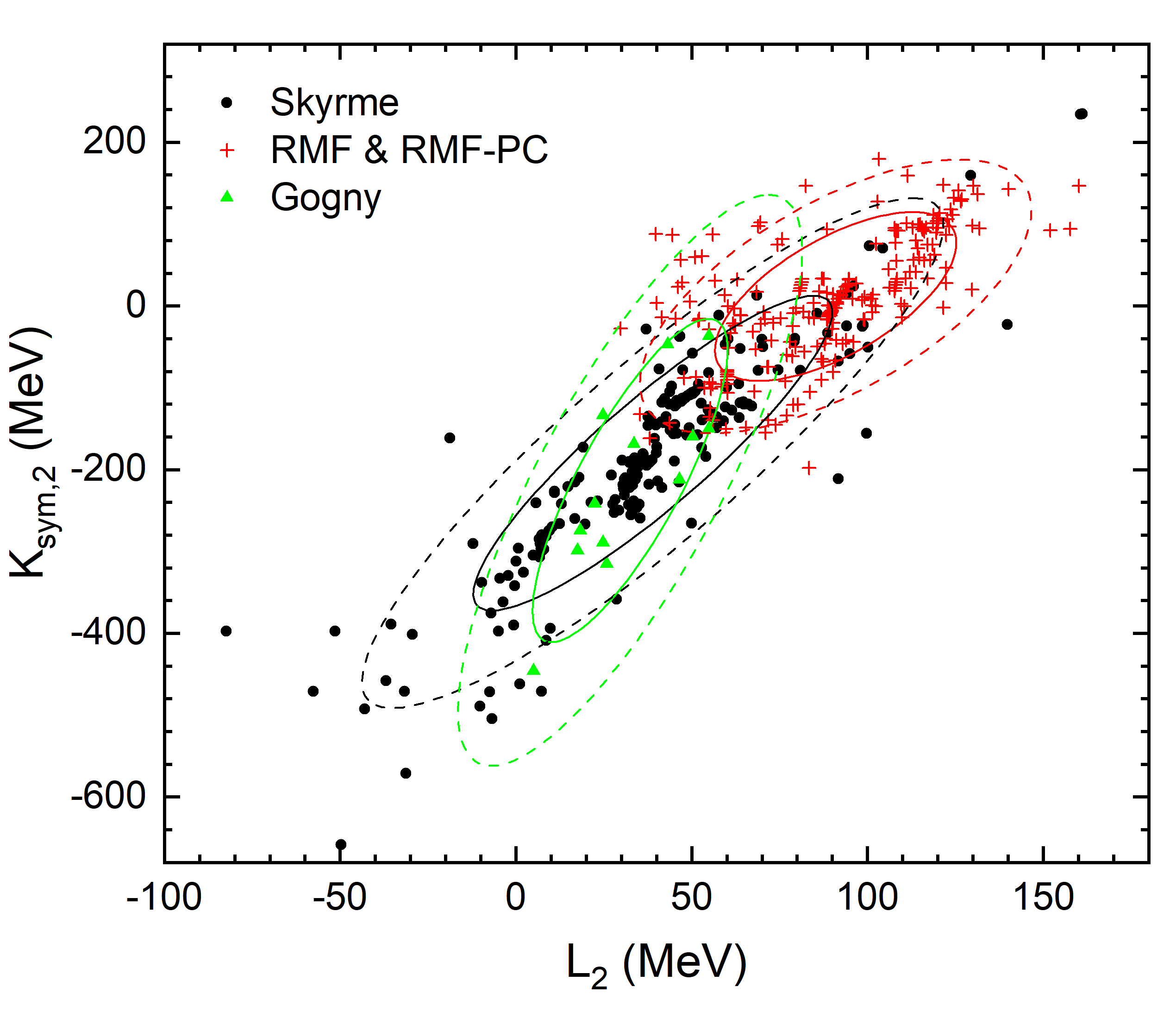}
\hspace*{0cm}\includegraphics[width=8.5 cm,angle=0]{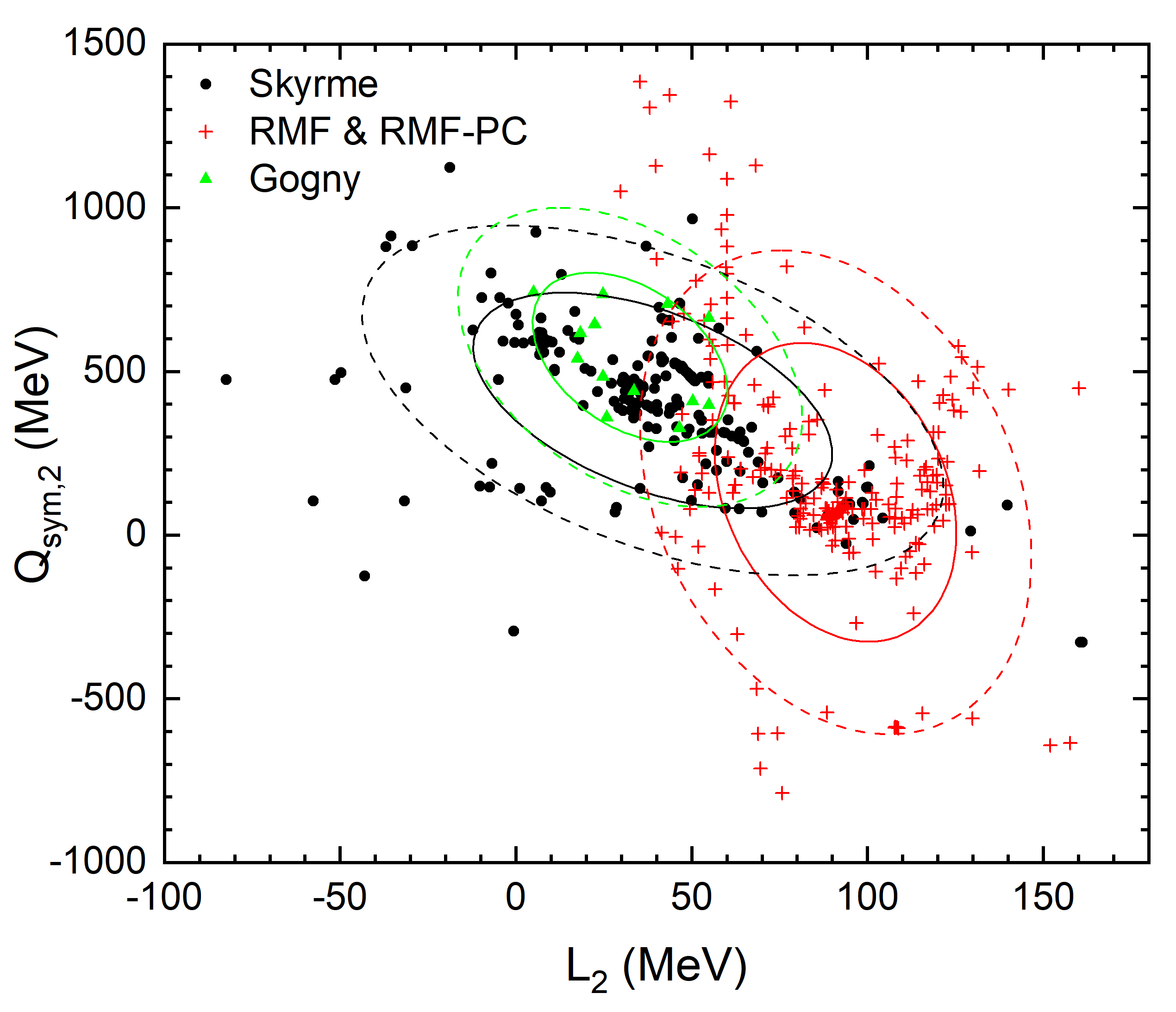}
\end{adjustwidth}
\vspace*{0cm}\caption{Similar to Figure \ref{fig:Bn0}, but for correlations between $K_{sym}$ and $L_2$ (left panel) $Q_{sym}$ and $L_2$ (right panel).}\label{fig:KsymQsymL}
\end{figure}

The moderate degrees of correlation of $K_{N0}$ and $Q_{N0}$ with $L_2$ suggest a similar degree of correlation between the corresponding symmetry energy parameters with $L_2$.  This is confirmed in Figure \ref{fig:KsymQsymL}.  The slopes of these correlations differ depending on the type of force considered, especially for $Q_{sym2}$.

\begin{figure}[htbp]
\centering
\vspace*{0cm}
\subfigure[\quad Symmetry energy  Make x axis $L_2$. ]{
\includegraphics[width=8.5cm]{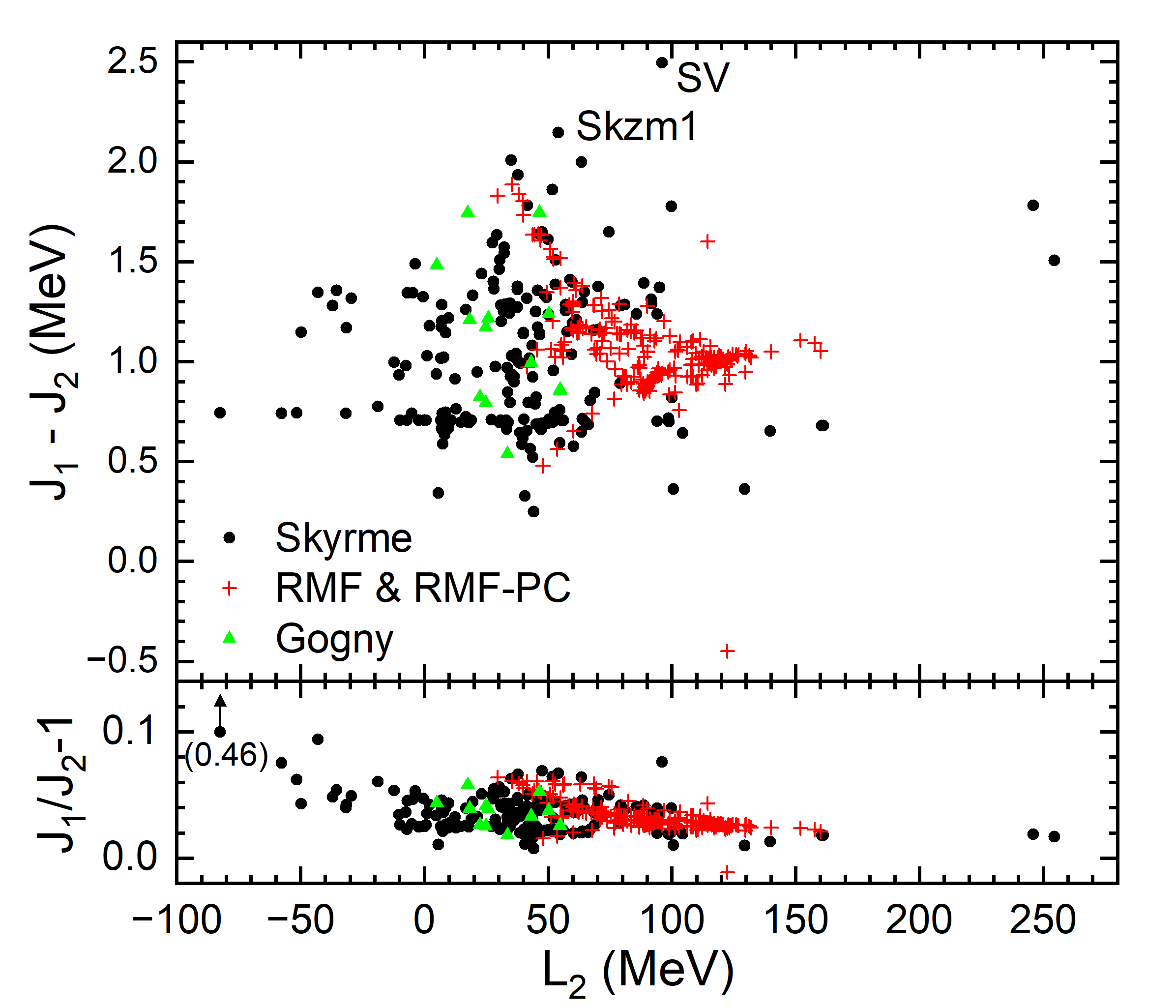}
}
\subfigure[\quad The slope of symmetric energy]{
\includegraphics[width=8.5cm]{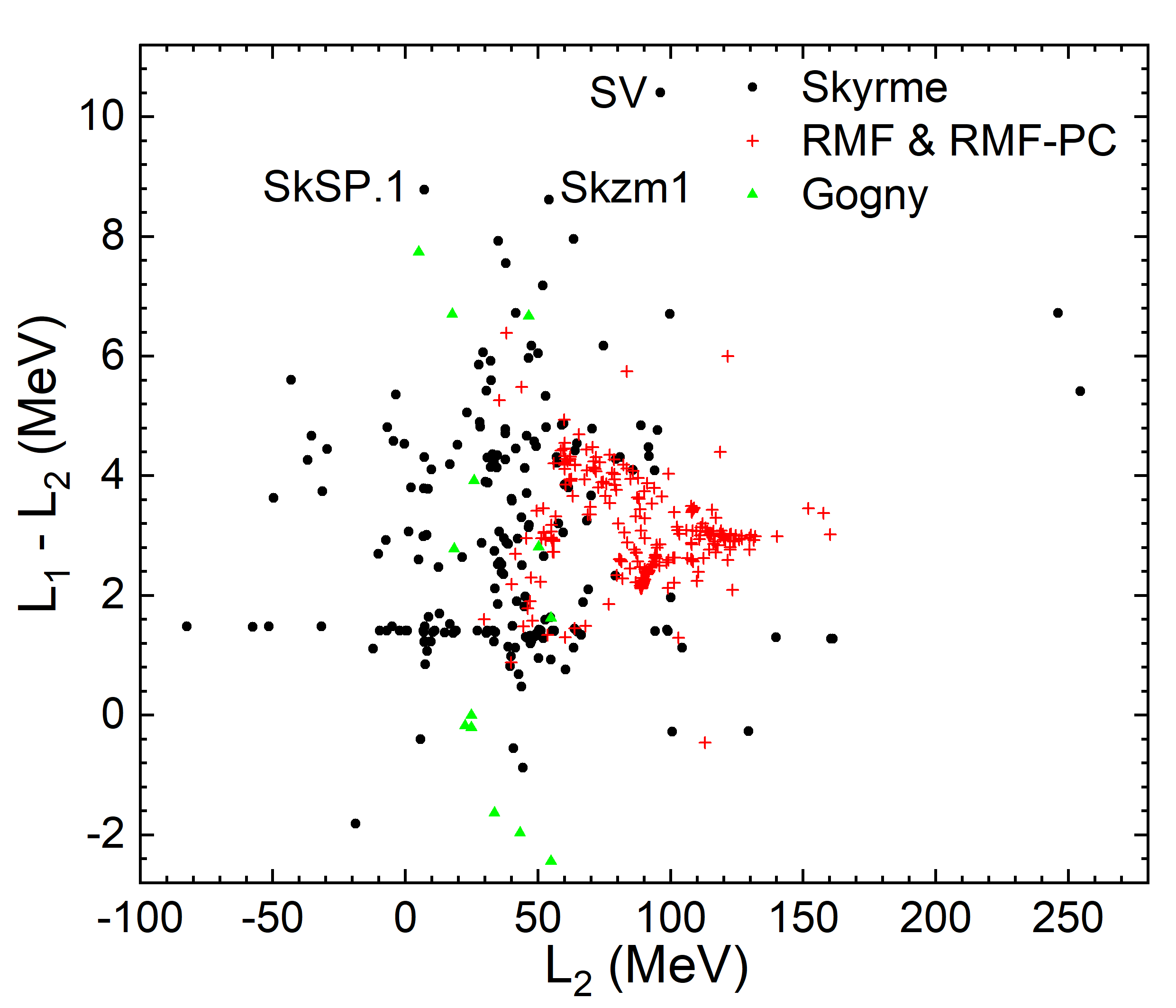}
}
\subfigure[\quad Volume incompressibility]{
\includegraphics[width=8.5cm]{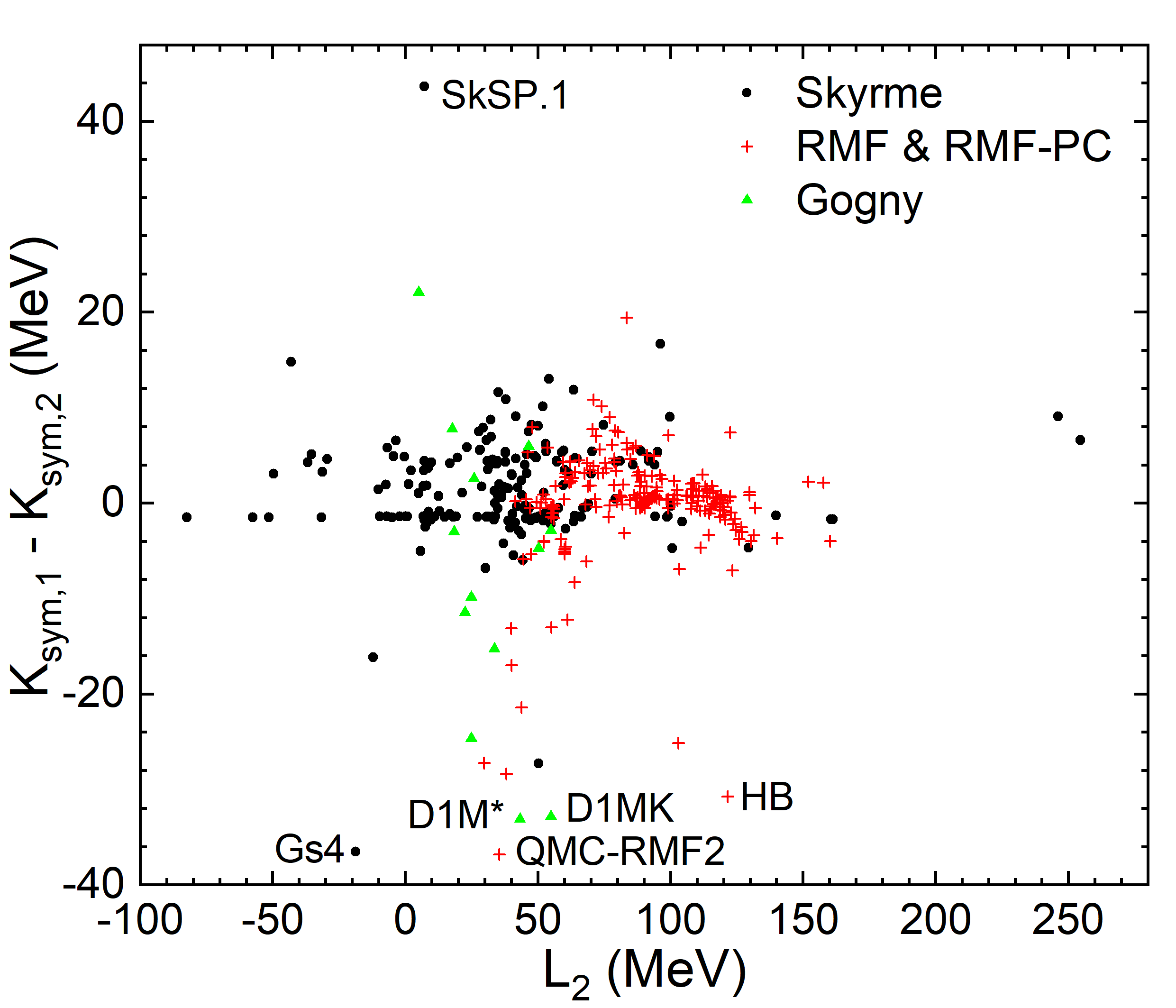}
}
\subfigure[\quad Skewness]{
\includegraphics[width=8.5cm]{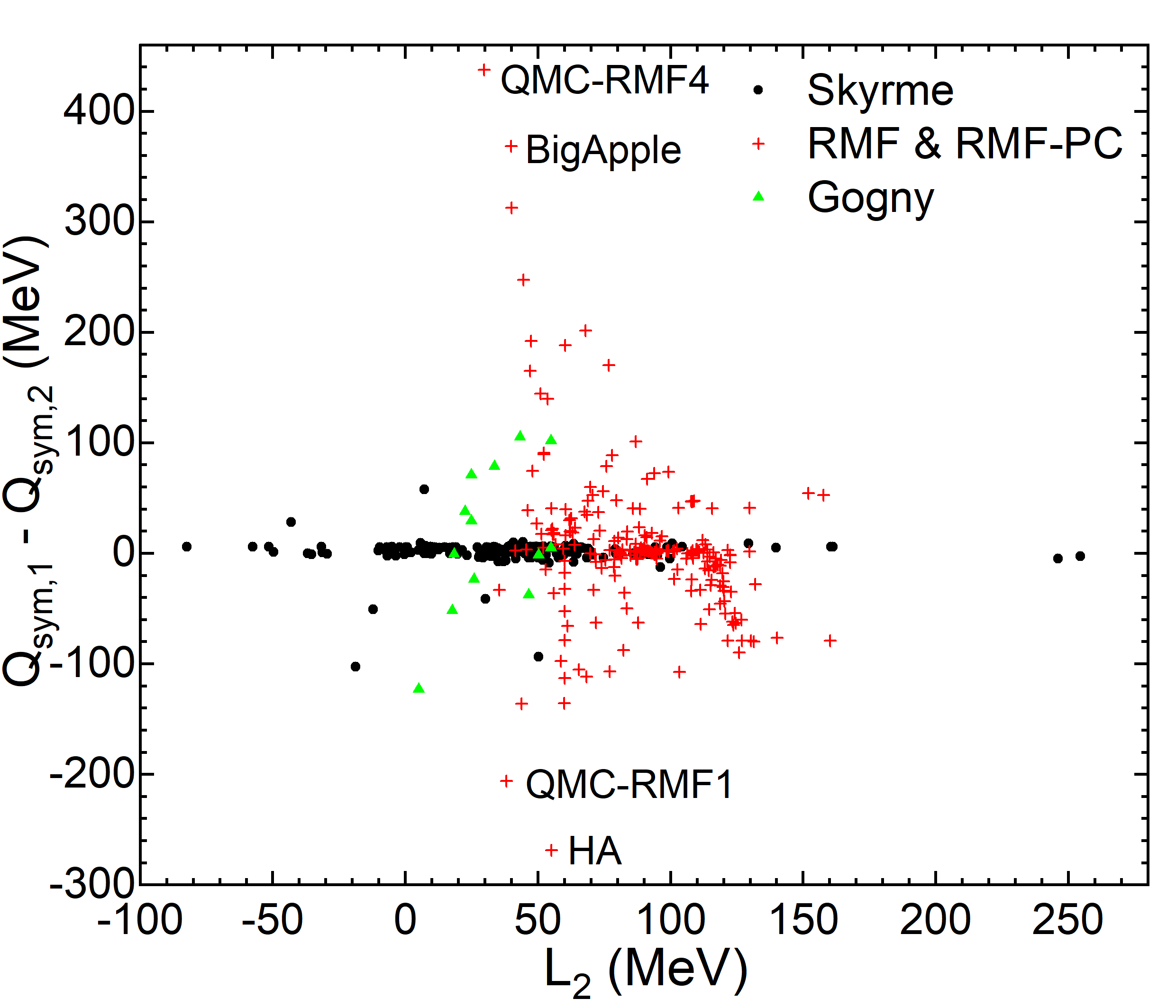}
}
\caption{Differences between the two methods for determining symmetry energy parameters.}\label{fig:comp}
\end{figure}

There are two methods of computing the symmetry energy parameters at the saturation density.  Figures \ref{fig:jln0} - \ref{fig:KsymQsymL} employed the second method, using the 2nd order term of a Taylor expansion of the symmetry energy in the neutron excess.  Figure \ref{fig:comp} shows the extent to which the two methods differ.  The mean difference $<J_1-J_2>=1.00\pm0.30$ MeV, while $<L_1-L_2>=2.58\pm8.31$ MeV, $<K_{sym,1}-K_{sym,2}>=1.40\pm4.44$ MeV and $<Q_{sym,1}-Q_{sym,2}>=2.79\pm38.6$ MeV.  The mean differences are quite small and nearly identical for Skyrme and RMF forces, but the standard deviations for Skyrme forces are about double those of RMF forces, except for the Skyrme skewness standard deviation which is 4.5 times smaller than that for RMF.  The primary contribution to the differences in the symmetry parameters arises from their kinetic energy contributions.  For Skyrme forces, this leads to
\begin{eqnarray}
 J_{1,kin}-J_{2,kin}&=&{\hbar^2\over M}\left({3\pi^2n_0\over2}\right)^{2/3}\left[{3\over10}\left(2^{2/3}-1\right)-{1\over6}\right]\approx0.69{\rm~MeV},\cr
 L_{1,kin}-L_{2,kin}&=&2\left(J_{1,kin}-J_{2,kin}\right)\approx1.39{\rm~MeV},\cr
 K_{sym,1,kin}-K_{2,sym,kin}&=&-2\left(J_{1,kin}-J_{2,kin}\right)\approx-1.39{~\rm MeV},\cr
 Q_{sym,1,kin}-Q_{sym,2,kin}&=&8\left(J_{1,kin}-J_{2,kin}\right)\approx5.56{\rm~MeV}.
\end{eqnarray}
RMF forces give a similar result.  The kinetic terms are important, but not the sole, contributions to these average differences, which are, nevertheless, relatively small, being of order 1\% or less for all symmetry parameters.
To the extent that the differences between the two methods are insignificant, which is not always the case as seen in Fig. \ref{fig:comp}, truncation of the Taylor expansion at quadratic order is justified.  The few forces with the largest deviations from the quadratic approximation are identified in each panel of Figure \ref{fig:comp}.  It should be emphasized that Fig. \ref{fig:comp} displays the deviations at saturation density; these deviations become larger at higher densities.

\newpage
\section{Nuclear Structure Properties}
\label{sec:surface}
The previous sections developed expressions for the energy density and pressure of uniform matter.  It is useful to also study the properties of nonuniform matter as exists within nuclei, so that deductions about nuclear structure can be made.  Within nuclei, in general, the neutron/proton ratio also varies with position.  In consequence, contributions to the total energy of a nucleus from the existence of a surface, as well as global properties such as the neutron skin thickness and dipole polarizability, can be evaluated.  In this paper, we will evaluate such quantities in a Thomas-Fermi approach and using the DM for nuclear structure \cite{myers1980droplet}.  The hydrodynamical model \cite{steiner2005isospin} gives the same results.  Where available, we will compare with the neutron skin thicknesses computed through the more accurate Hartree-Fock approximation.

The general expression for the total energy density of nucleonic matter within a nucleus, ignoring Coulomb, spin-orbit, shell and pairing effects, in Skyrme-like interactions can be expressed as 
\begin{eqnarray}
{\cal E} = {\cal E}_B + \frac{1}{2}\left[Q_{nn}(\vec{\nabla} n_n)^2+2Q_{np}\vec{\nabla} n_n \cdot\vec{\nabla} n_p+Q_{pp}(\vec\nabla n_p)^2\right], \label{eq:basicham}
\end{eqnarray}
where ${\cal E}_B(n,x)$ describes spatially homogeneous infinite bulk matter and the last factor expresses the contributions due to density gradients (as exist, for example, within nuclei). 
The coefficients $Q_{ij}$ are given in Eq. (\ref{eq:qnn}).
%\begin{eqnarray}
%Q_{nn}=Q_{pp} &=& {3\over16} \left[ t_1 \left(1 - x_1 \right) - t_2\left( 1 + x_2 \right) +t_4n^\delta\left(1-x_4+{4\delta\over3}\left[1+{x_4\over2}-x_t\left({1\over2}+x_4\right)\right]\right)-t_5n^\gamma(1+x_5)\right]\,,\nonumber\\
%Q_{np}=Q_{pn} &=& {1\over8}\left[ 3 t_1 \left(1 + \frac{x_1}{2} \right) - t_2 \left( 1 + \frac{x_2}{2} \right)+{3t_4n^\delta\over2}(2+x_4+\delta)-t_5n^\gamma\left(1+{x_5\over2}\right) \right]\,.
%\end{eqnarray}
These coefficients are spatially constant unless $t_4$ or $t_5$ are non-zero.

For RMF models, the expression for the total energy density is not conveniently expressed in terms of neutron and proton density gradients, but rather in terms of the gradient contributions from the meson fields  as given in Eq. (\ref{eq:eps0}),
%\begin{equation}
%\mathcal{E}=\mathcal{E}_B+{1\over2\hbar c}\left[\left({d\sigma\over dr}\right)^2-\left({d\omega\over dr}\right)^2-\left({d\rho\over dr}\right)^2+\left(\frac{d\delta}{dr}\right)^2\right].
%\label{eq:eps00}\end{equation} 
which is valid in the case of both constant and density-dependent coupling models.
The surface properties of point-coupling and Gogny interactions, which have different gradient contributions, will not be considered further in this paper.  

We will be particularly interested in the case of symmetric matter, which is sufficient in the DM to determine the surface symmetry energy, the neutron skin thickness, and the dipole polarizability. For Skyrme interactions, in this case,
\begin{equation}
{\cal E}={\cal E}_B+{Q_{nn}+Q_{np}\over4}\left({dn\over dr}\right)^2\equiv{\cal E}_B+{{\cal Q}\over2}\left({dn\over dr}\right)^2,
\label{eq:ebs}\end{equation}
which defines ${\cal Q}$.
For RMF interactions, we will assume the Thomas-Fermi approximation, which holds that the field gradients are small enough that the fields at any point within the nucleus are given by their uniform matter expressions, i.e., those given by Euler-Lagrange minimization Eq. (\ref{eq:field}).  In the case of symmetric matter, comparison of Eqs. (\ref{eq:ebs}) and (\ref{eq:eps0}) allow the identification of the corresponding expression for ${\cal Q}$,
\begin{equation}
{\cal Q}={\sigma^{\prime2}-\omega^{\prime2}-\rho^{\prime2}+\delta^{\prime2}\over\hbar c} ={\sigma^{\prime2}-\omega^{\prime2}\over\hbar c}.
\label{eq:epsq}\end{equation}
Note that in RMF interactions, ${\cal Q}$ always varies with density.  It would also be possible, by algebraic manipulation, to evaluate corresponding expressions for $Q_{nn}$ and $Q_{np}$ individually, but this will not be necessary in this paper. 

\subsection{Semi-Infinite Interface and the Surface Energy}
In general, phase equilibrium exists in zero temperature but arbitrary proton fraction matter, as long as the average density is below the nuclear saturation density for that proton fraction (where the pressure of a single phase of uniform matter would be negative).  In order to minimize the total energy, matter separates into two phases with different densities and proton fractions. Neglecting the Coulomb interaction, in the limit that the surface region is negligible compared to the overall volume, the system may be considered as consisting of two semi-infinite regions (the dense region corresponding to the matter within a nucleus, and the vacuous region corresponding to the region outside the nucleus) separated by a plane-parallel surface.  The Euler-Lagrange minimization of the total energy per unit area of this system, for fixed numbers of neutrons and protons per unit area, leads to an expression, which, after integrating by parts, is
\begin{equation}
 \mathcal{E}-\mu_{n0}n_n-\mu_{p0}n_p= {1\over2}\sum_{ij}Q_{ij}{dn_i\over dr}{dn_j\over dr},
\label{eq:eps3}\end{equation} 
where the Lagrange parameters for fixed neutrons and protons per unit area are the chemical potentials $\mu_{n0}$ and $\mu_{p0}$, which are constants.  This equation turns out to be the same whether or not the coefficients $Q_{ij}$ are constant or spatially varying, or on whether or not the interaction is a non-relativistic Skyrme-type or an RMF-type force.  It states that the bulk and gradient contributions to the total energy density are equal.  This expression provides explicit representations of the density gradients as functions of the densities.  In the case of symmetric matter, this expression can be straightforwardly integrated to yield the density profile $n(r)$ throughout the interface.

The left-hand side of Eq. (\ref{eq:eps3}) is simply the thermodynamic potential density.  For a semi-infinite geometry, the net surface thermodynamic potential per unit area, often referred to as the surface tension $\sigma$, is 
\begin{equation}
\sigma(\delta_L)=\int_{-\infty}^{+\infty}\left[\mathcal{E}-\mu_{n0}n_n-\mu_{p0}n_p\right]dz,
\label{eq:sigmadelta}
\end{equation}
where $\delta_L$ is the neutron excess in the dense phase far from the interface and $z$ is the coordinate perpendicular to the surface.  This is recognizable as the difference of the total energy per unit area between the configuration in which the density profile is optimized through the interface and the configuration in which the densities are taken to be spatially uniform with a sharp discontinuity at the interface where the density changes abruptly from its value ($n_0$ in the case of $\delta_L=0$) to 0, at least in the case when $\delta_L\alt0.2$, i.e., less than the neutron-drip value where the density in the external phase becomes finite. 
It is precisely the surface tension, $\sigma$, that is maximized when performing the Euler-Lagrange minimization of the total energy density per unit area of the semi-infinite system.

In the case of small, nonzero, asymmetry $\delta_L$, Eq, (\ref{eq:sigmadelta}) can be expanded as a Taylor series in $\delta_L^2$,  
\begin{equation}
    \sigma(\delta_L)=\sigma_o-\sigma_\delta\delta_L^2+\cdots,  
\label{eq:sasym}\end{equation}
where $\sigma_o$ is the surface tension for symmetric matter $\delta_L=0$ and $\sigma_\delta$ is the surface tension symmetry parameter. In the case of symmetric matter, $\mu_{n0}=\mu_{p0}=M+E_0$ (in the case of non-relativistic interactions, $M$ is not included in the chemical potentials).  Also, since ${\cal E}_B={\cal E}_{1/2}$ is the energy density for uniform symmetric matter, one finds
\begin{equation}
\sigma_o\equiv\sigma(\delta_L=0)=2\int_{-\infty}^{+\infty}\left(E_{1/2}(n)-E_0\right)ndz =\int_{-\infty}^{+\infty}{\cal Q}\left({dn\over dz}\right)^2dz.
\end{equation}
The gradient $dn/dz$ can be eliminated by substituting Eq. (\ref{eq:eps3}), noting that for a semi-infinite interface $r\rightarrow z$, which removes the spatial dependence to form the quadrature
\begin{equation}
\sigma_o=\int_0^{n_0}\sqrt{2{\cal Q}n(E_{1/2}(n)-E_0)}dn.
\label{eq:sigma0}
\end{equation}
The integrand of this equation vanishes at both boundaries.

\begin{figure}[H]
\centering
\begin{adjustwidth}{0cm}{0cm}
    \includegraphics[width=8.5cm,angle=0]{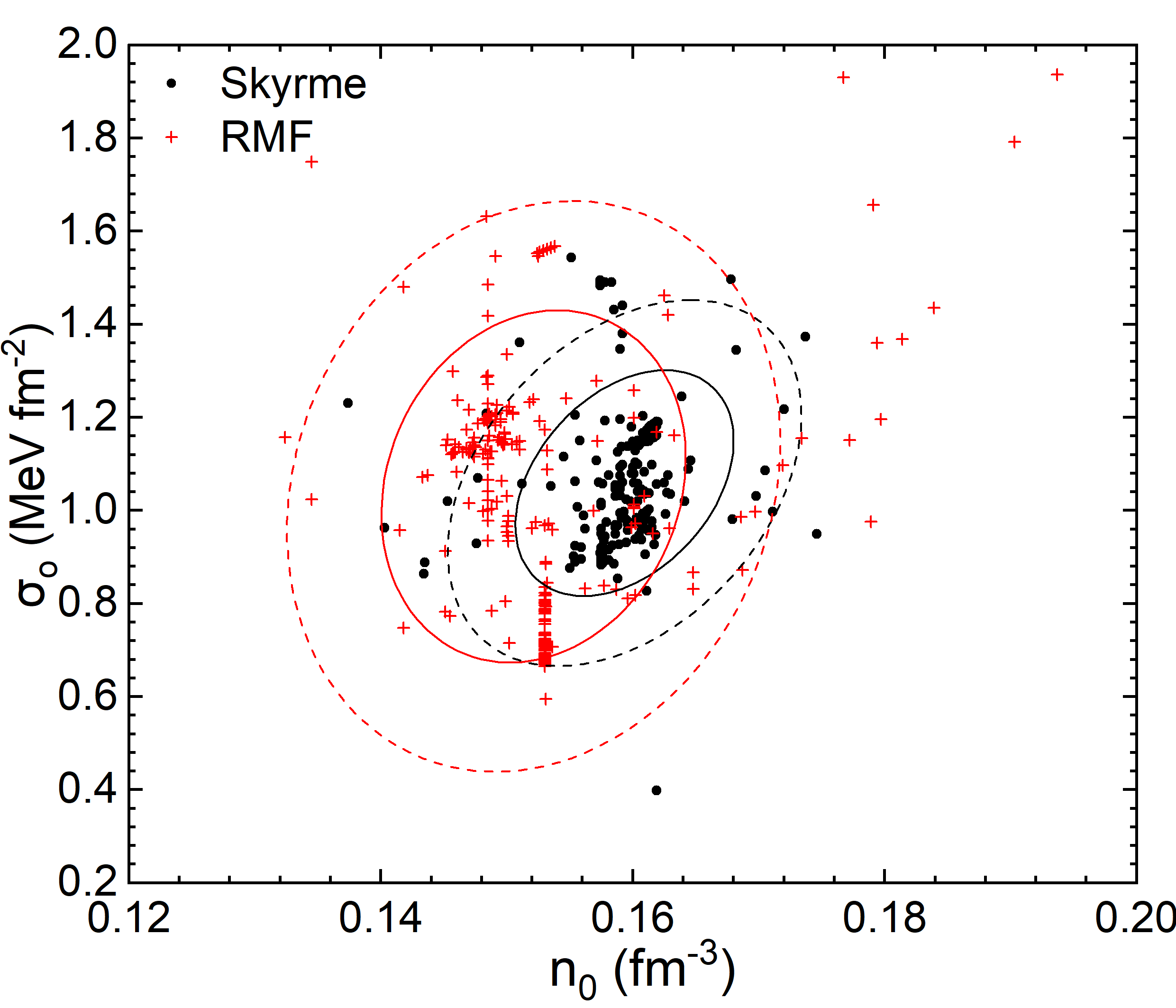}
    \includegraphics[width=8.5cm,angle=0]{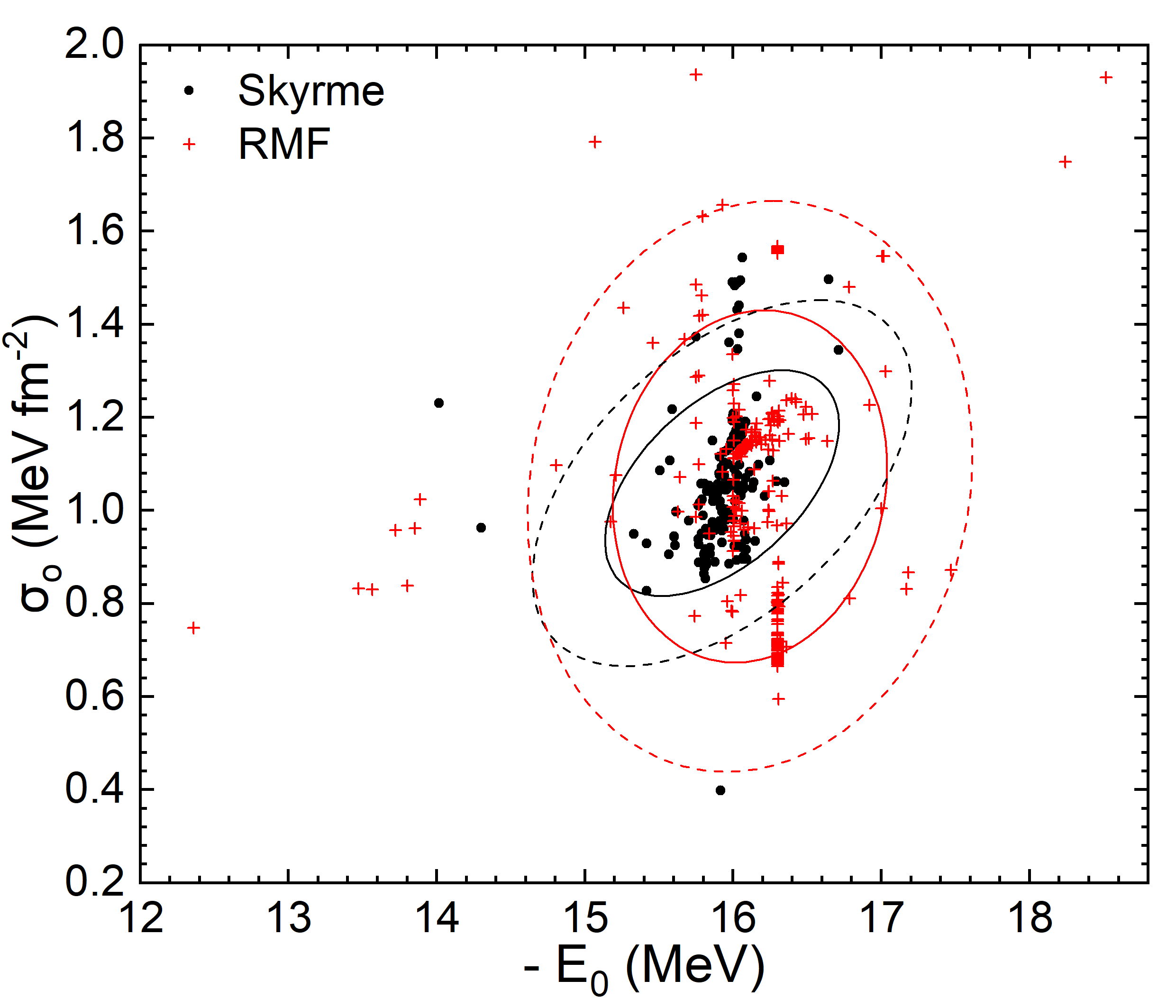}
\end{adjustwidth}
\vspace*{0cm}\caption{Correlations between the surface tension $\sigma_0$ and the saturation density $n_0$ (left panel) and the binding energy $-E_0$ at saturation (right panel) in Skyrme (black) and RMF (red) models.  $1\sigma$ and $2\sigma$ confidence ellipses are indicated by solid and dashed lines for each force type. \label{fig:correl3}}
\end{figure}

\begin{figure}[H]
\centering
\begin{adjustwidth}{0cm}{0cm}
 \includegraphics[width=8.5 cm,angle=0]{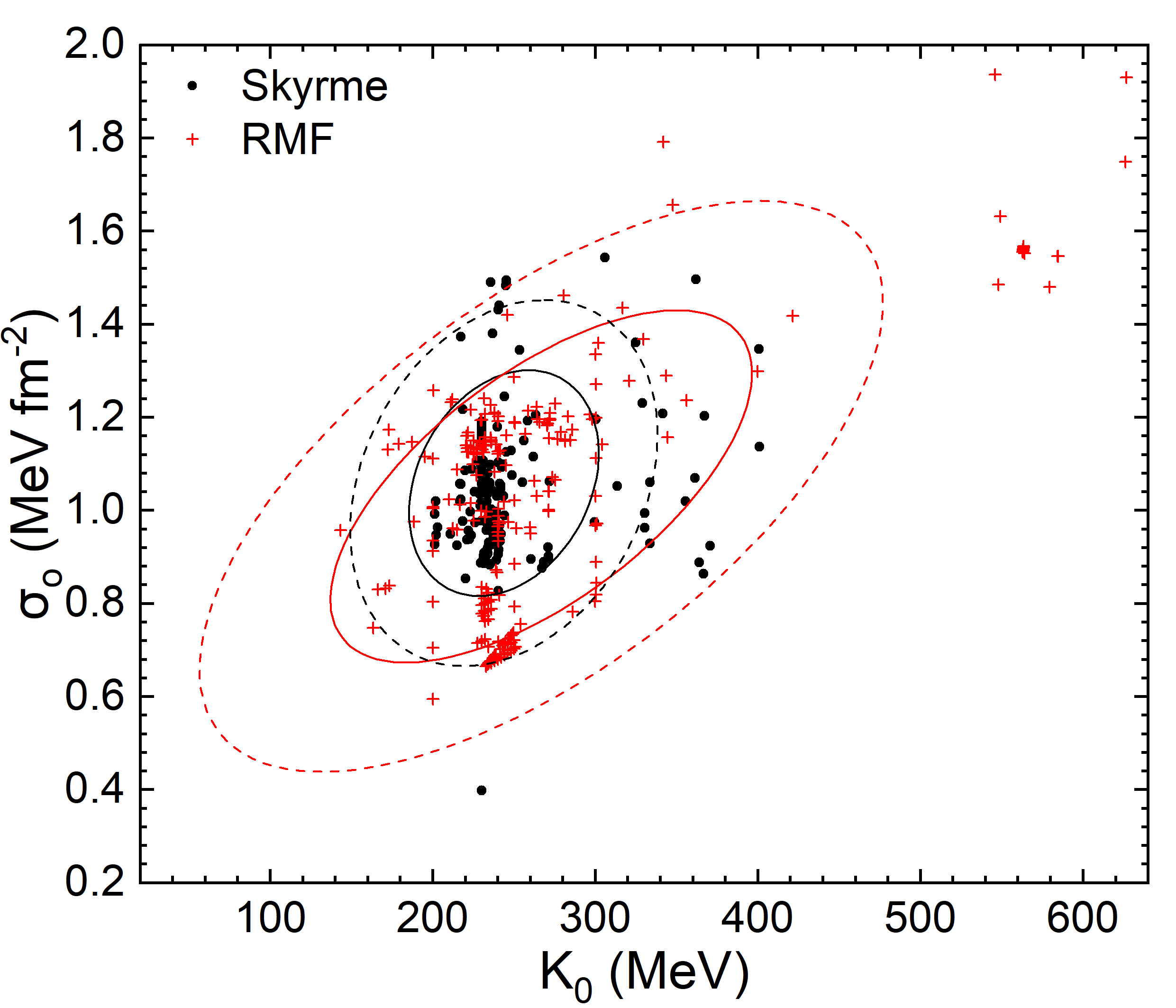}
     \includegraphics[width=8.5cm,angle=0]{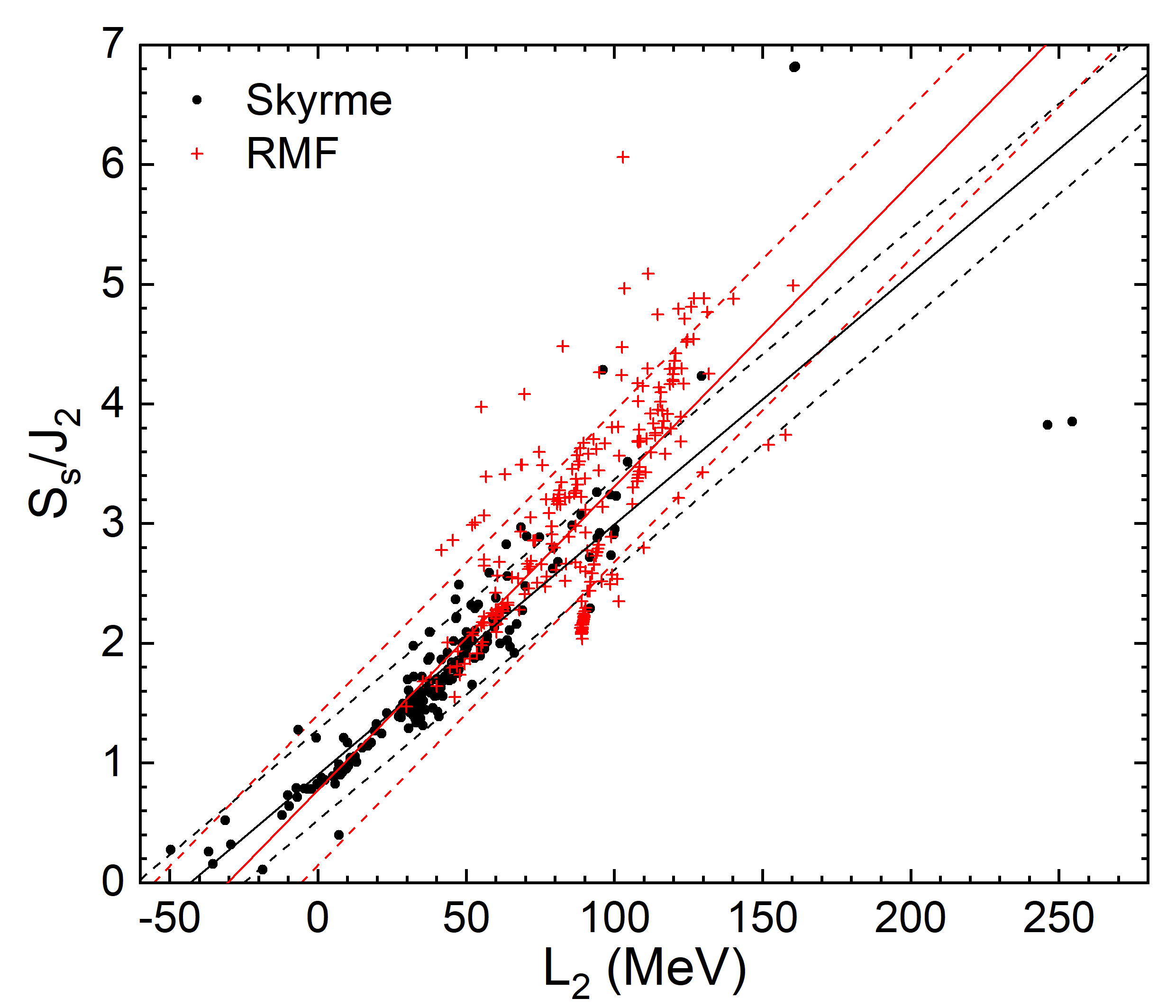}
\end{adjustwidth}
\vspace*{0cm}  \caption{The left panel is the same as Fig. \ref{fig:correl3} except for the surface tension $\sigma_o$ and the incompressibility at saturation $K_0$. The right panel shows the nearly linear correlations, with standard deviations, between $S_s/J_2$ and the symmetry energy slope $L_2$. }\label{fig:correl4}
\end{figure}

Figure \ref{fig:correl3} shows that the parameter $\sigma_0$ is only weakly correlated with either $n_0$ or $E_0$ for both types of forces.  However, it is more strongly correlated with $K_0$, especially for RMF forces, as shown in the left panel of Figure \ref{fig:correl4}.  This can be understood from Eq. (\ref{eq:sigma0}) if one approximates $E_{1/2}(n)=E_0+(K_0/18)(n-n_0)^2+\cdots$.  Then one finds that $\sigma_o\propto\sqrt{K_0}$.

The surface tension symmetry parameter $\sigma_\delta$ in Eq. (\ref{eq:sasym}) can be found by expanding Eq. (\ref{eq:sigmadelta}), after substituting Eq. (\ref{eq:eps1}), in powers of $\delta_L^2$.  This leads to
\begin{equation}
\sigma(\delta_L)= \int_0^{ n_0}\sqrt{{\cal Q}n\over2}\left[E_{1/2}(n)-E_0- \delta_L^2\left(\frac{J_2}{{\cal S}_2( n)}-1\right)\right]^{1/2}d n,
\label{eq:sigmadelta1}\end{equation}
and further expanding the integrand of Eq. (\ref{eq:sigmadelta1}), one finds 
\begin{equation}
\sigma_\delta = \frac{J_2}{\sqrt{2}}\int_0^{ n_0} \sqrt{{\cal Q}n\over E_{1/2}(n)-E_0}\left[\frac{J_2}{{\cal S}_2( n)}-1\right]d n,
\label{eq:sigmadelta2}\end{equation}
which is also, therefore, another quadrature.   The integral Eq. (\ref{eq:sigmadelta2}) is well-behaved:  in the limit where $n\rightarrow0$, ${\cal Q}(n=0)/(E_{1/2}(n=0)-E_0)$ is a constant and ${\cal S}_2(n)\rightarrow n^{2/3}$, so the integrand diverges as $n^{-1/6}$.  But its contribution to the integral at small density is proportional to $n^{5/6}$ and therefore vanishes.  In the opposite limit where $n\rightarrow n_0$, one has $E_{1/2}(n)-E_0\rightarrow K_{1/2}(1-u)^2/18$ and $J_2/{\cal S}_2(n)-1\rightarrow L_2(u-1)/(3J_2)$, where $u=n/n_0$, so the integrand approaches the constant $L_2\sqrt{{\cal Q}(n=n_0)n_0/K_{1/2}}$.

We note that in the DM, the surface energy $E_S$ and the surface symmetry energy $S_S$ parameters are respectively given as
\begin{align}
    E_S &= 4\pi r_o^2\sigma_o, \qquad     S_S = 4\pi r_o^2\sigma_\delta,
\label{eq:esss}\end{align}
where the nucleon radius $r_o$ satisfies $4\pi r_o^3n_0/3=1$. Values of $\sigma_o$, $\sigma_\delta$ and $S_S$ of Skyrme (RMF) parameterizations are included in Table III (IX).  Integration of Eq. (\ref{eq:sigmadelta2}) results in a strong correlation between  $S_S/J_2$ and $L_2$, which is, furthermore, linear to an excellent approximation as shown in the right panel of Fig. \ref{fig:correl4}.  For Skyrme and RMF forces, they are, respectively,
\begin{equation}
{S_S\over J_2}=(0.905\pm0.378) + 0.02091{L_2\over{\rm MeV}};\qquad {S_S\over J_2}=(0.777\pm0.439) +0.02535{L_2\over{\rm MeV}}.
\end{equation}

\subsection{Neutron Skin Thickness}
\begin{figure}[H]
\centering
\begin{adjustwidth}{0cm}{0cm}
    \includegraphics[width=8.5cm,angle=0]{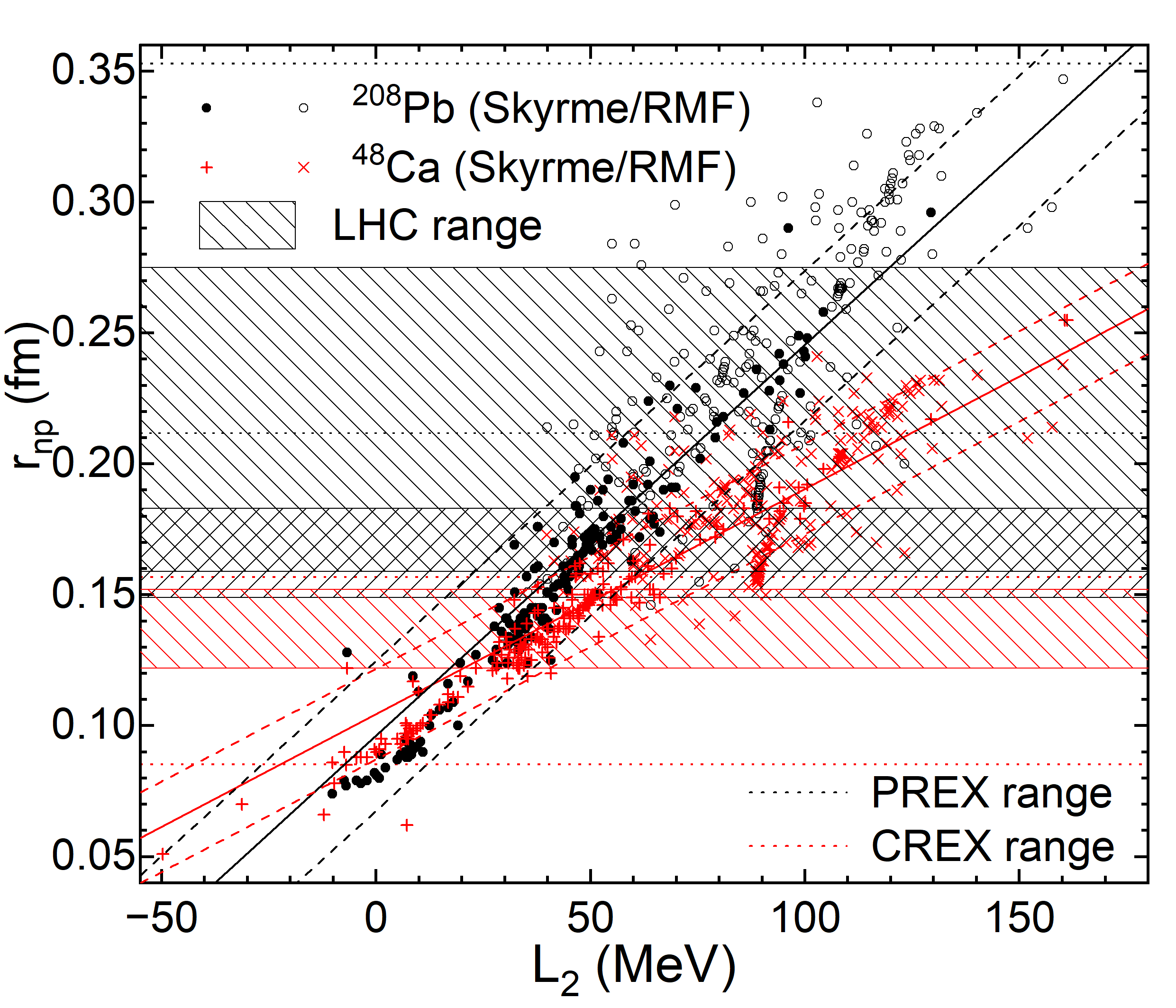}
    \includegraphics[width=8.5cm,angle=0]{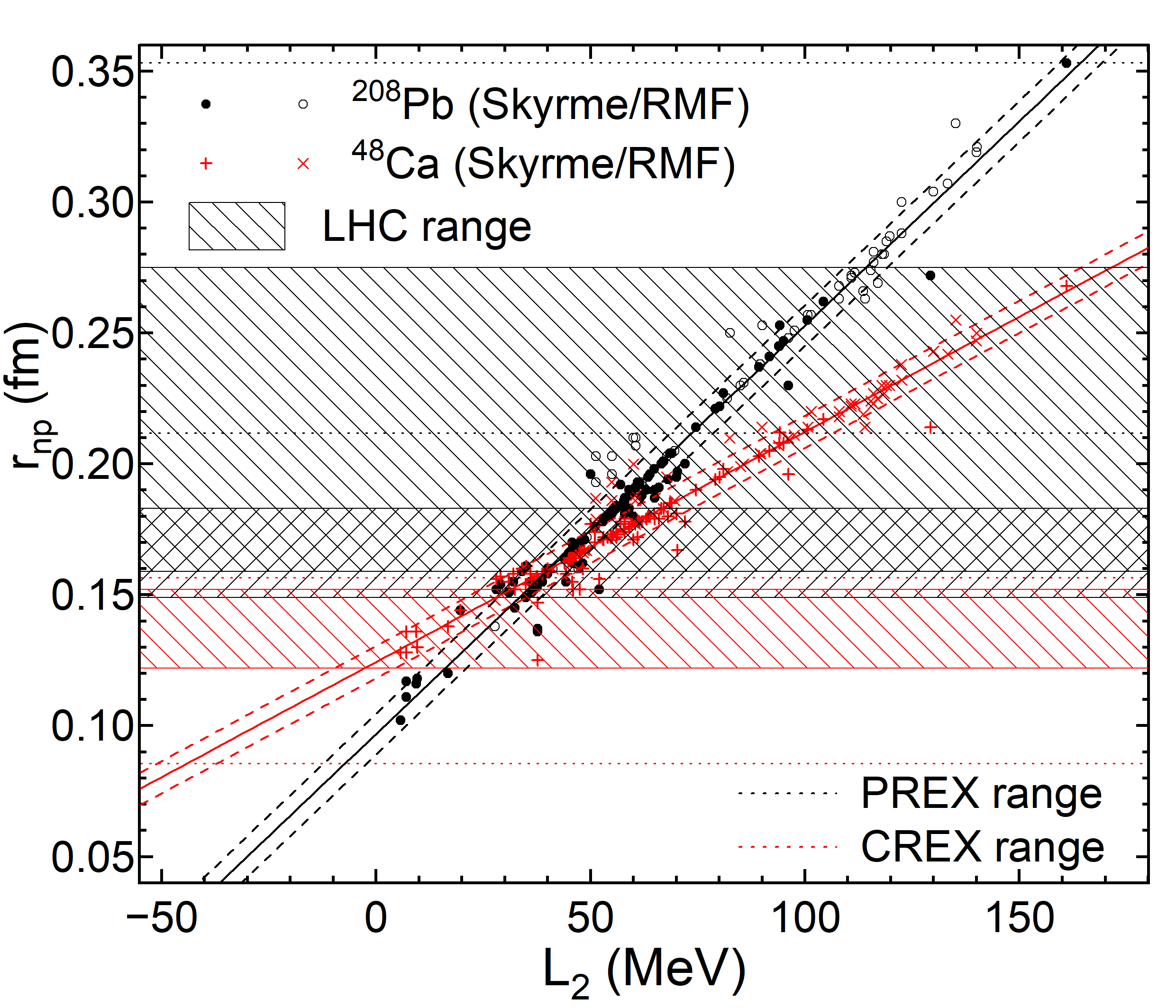}
\end{adjustwidth}
 \vspace*{0cm}  \caption{Skyrme and RMF predictions of the neutron skin thicknesses of $^{48}$Ca (red) and $^{208}$Pb (black) for the DM, including surface diffuseness corrections, (left panel) and Hartree-Fock calculations (right panel). Linear correlations and 1 standard deviation are shown as solid and dashed diagonal lines, respectively. The narrow hatched horizontal bands indicate the 1 standard deviation ranges of the previous averaged experimental results for these nuclei~\cite{Lattimer_2023}.  The dotted black (red) lines indicate the 1 standard deviation ranges of $r_{np}^{208}$ ($r_{np}^{48}$) from PREX I+II~\cite{adhikari2021accurate} (CREX~\cite{adhikari2022precision}).  The wide black hatched band shows the 1 standard deviation range of the LHC $^{208}$Pb neutron skin measurement \cite{Giacalone2023}.}
    \label{fig:skin}
\end{figure}

Other quantities of interest are the neutron skin thickness and the dipole polarizability.  Both the DM and hydrodynamical (h) models predict that the difference between the mean neutron and proton radii for a nucleus with mass $A$ and charge $Z$ is~\cite{Myersswiatecki1974,steiner2005isospin,lattimer2016equation}
\begin{equation}
R_n-R_p={2r_o\over3}\left(1+{S_S\over J_2A^{1/3}}\right)^{-1}\left[I{S_S\over J_2}-{3Ze^2\over140r_oJ_2}\left(1+{10S_S\over3J_2A^{1/3}}\right)\right],
\label{eq:rnrp}\end{equation}
where the nuclear isospin parameter is $I=(N-Z)/A$ [not to be confused with the RMF potential energy density $I(\delta)$].
The neutron skin thickness is defined as the difference of the mean square neutron and proton radii, or $r_{np}\approx\sqrt{3/5}(R_n-R_p)$. The nominal nuclear radius is  $R=r_oA^{1/3}$. The last term in the square brackets of these formulae represents the effects of Coulomb polarization.   In addition, one can include a correction
\begin{equation}
    \Delta r_{np}^{\rm surf}=\sqrt{3\over5}{5\left(b_n^2-b_p^2\right)\over2R} \label{eq: surcrn}
\end{equation}
due to the difference of the surface widths $b_n$ and $b_p$ of the neutron and proton density distributions \cite{myers1980droplet, PhysRevC.80.024316}, respectively.  For $^{208}$Pb, this correction is about 0.05 fm and for $^{48}$Ca it is about 0.04 fm \cite{PhysRevC.80.024316}, amounts which are not negligible.  We included this correction to the DM values of $r_{np}^{48}$ and $r_{np}^{208}$ for Skyrme (RMF) parameterizations in Table III (IX).  Equation (\ref{eq:rnrp}) indicates that $r_{np}$ should be highly correlated with $S_S/J_2$, and therefore, with $L_2$.  In spite of the various non-linear factors in Eq. (\ref{eq:rnrp}),
they largely cancel, resulting in a nearly linear correlation between $r_{np}$ and $L_2$, as indicated in Fig. \ref{fig:skin}.  This is true for both nuclei, and also for both Skyrme and RMF force models.  The slopes of these linear correlations, to the lowest order, are proportional to $I$ and therefore steeper for Pb relative to Ca.

For predicting neutron skin thicknesses, the DM leaves a lot to be desired.  Coulomb corrections and density variations within the nucleus are only treated approximately, and the DM truncates the leptodermous expansion so that curvature and higher-order corrections are ignored.  Figure \ref{fig:skin} compares the correlations between $r_{np}$ and $L_2$ for the DM with those of more accurate Hartree-Fock (HF) computations from Ref. \cite{tagami2022neutron}.  It is therefore somewhat surprising that the mean trends, including the slopes, averaged over all forces are nearly the same for both models.  The greater accuracy of HF computations results in linear correlations between $r_{np}$ and $L_2$ that have significantly smaller standard deviations for both nuclei by a factor $\approx2-2.5$ for $^{208}$Pb and $\approx3-4$ for $^{48}$Ca.

 The overall differences in the correlations between force types are small; for example, for the HF computations,  the slope of the Skyrme (RMF) correlation for $^{208}$Pb, is 0.00149 (0.00153) fm MeV$^{-1}$ and for $^{48}$Ca, the slope of the Skyrme (RMF) correlation is 0.000840 (0.000836) fm MeV$^{-1}$.  Similarly, small differences are found for DM computations as well.
 
\begin{figure}[H]
\centering
\hspace*{-1.1cm}\includegraphics[width=8.5 cm,angle=0]{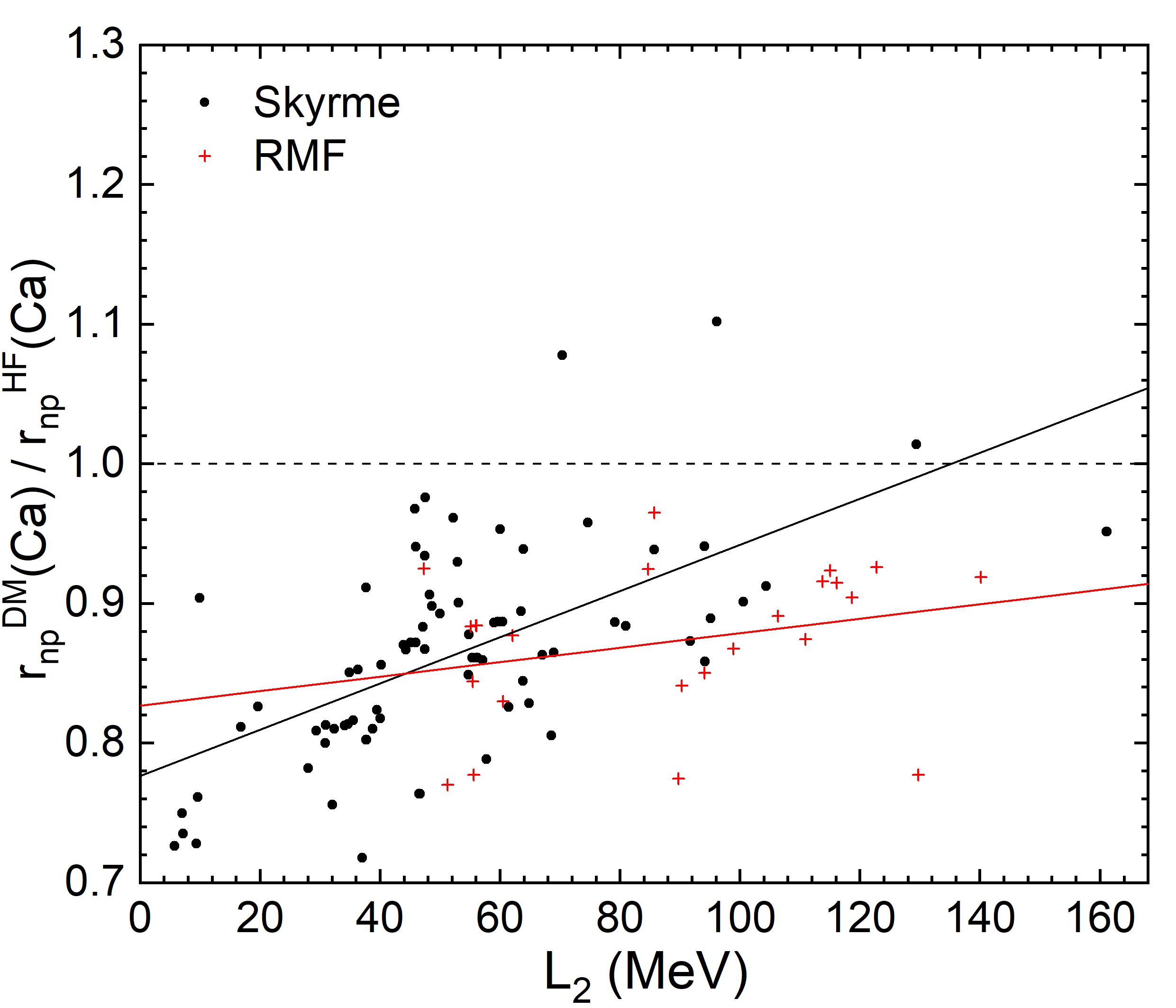} \qquad\qquad
\hspace*{-1.1cm}\includegraphics[width=8.5 cm,angle=0]{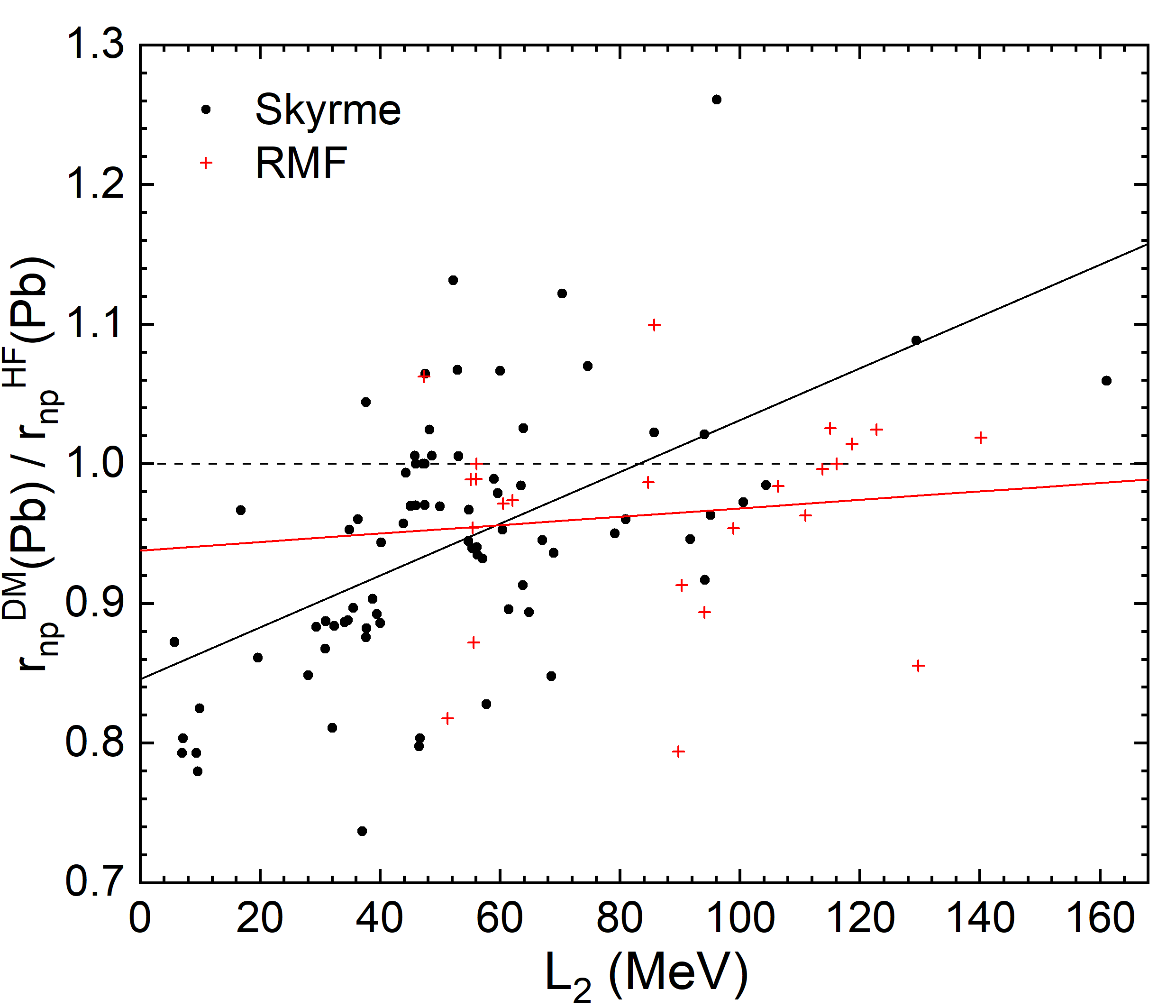}
\vspace*{0cm}\caption{Comparison of the neutron skin thicknesses of $^{48}$Ca and $^{208}$Pb calculated from the DM including surface diffuseness corrections with those from the HF method, for selected Skyrme (left panel) and RMF (right panel) models.\label{fig:r1dr2Ca}}
\end{figure}

To explore the differences between DM and HF predictions for neutron skin thicknesses in more detail, in Figure \ref{fig:r1dr2Ca} we display the ratio of the DM and HF values of $r_{np}^{48}$ (left panel) and $r_{np}^{208}$ (right panel).  In both cases, the average DM skin thickness is smaller than the HF values, especially for forces with small $L_2$ values.   The DM predictions are, however, closer to the HF predictions for Pb than for Ca.  The contribution of the Coulomb term in Eq. (\ref{eq:rnrp}) compared to the leading-order asymmetry ($IS_S/J_2$) term scales like $Z/(IS_S)$. It is reasonable that DM predictions would increase in accuracy for larger $A$ where the Coulomb corrections and the leptodermous truncation are more accurately represented.  The trend that DM predictions become more accurate as $L_2$ is increased for a given nuclei is due to the decreasing importance of Coulomb corrections with increasing $S_S$, and therefore, $L_2$.

\subsection{Dipole Polarizability\label{sec:dipole}}
\begin{figure}[H]
\centering
\begin{adjustwidth}{0cm}{0cm}
    \includegraphics[width=8.5cm,angle=0]{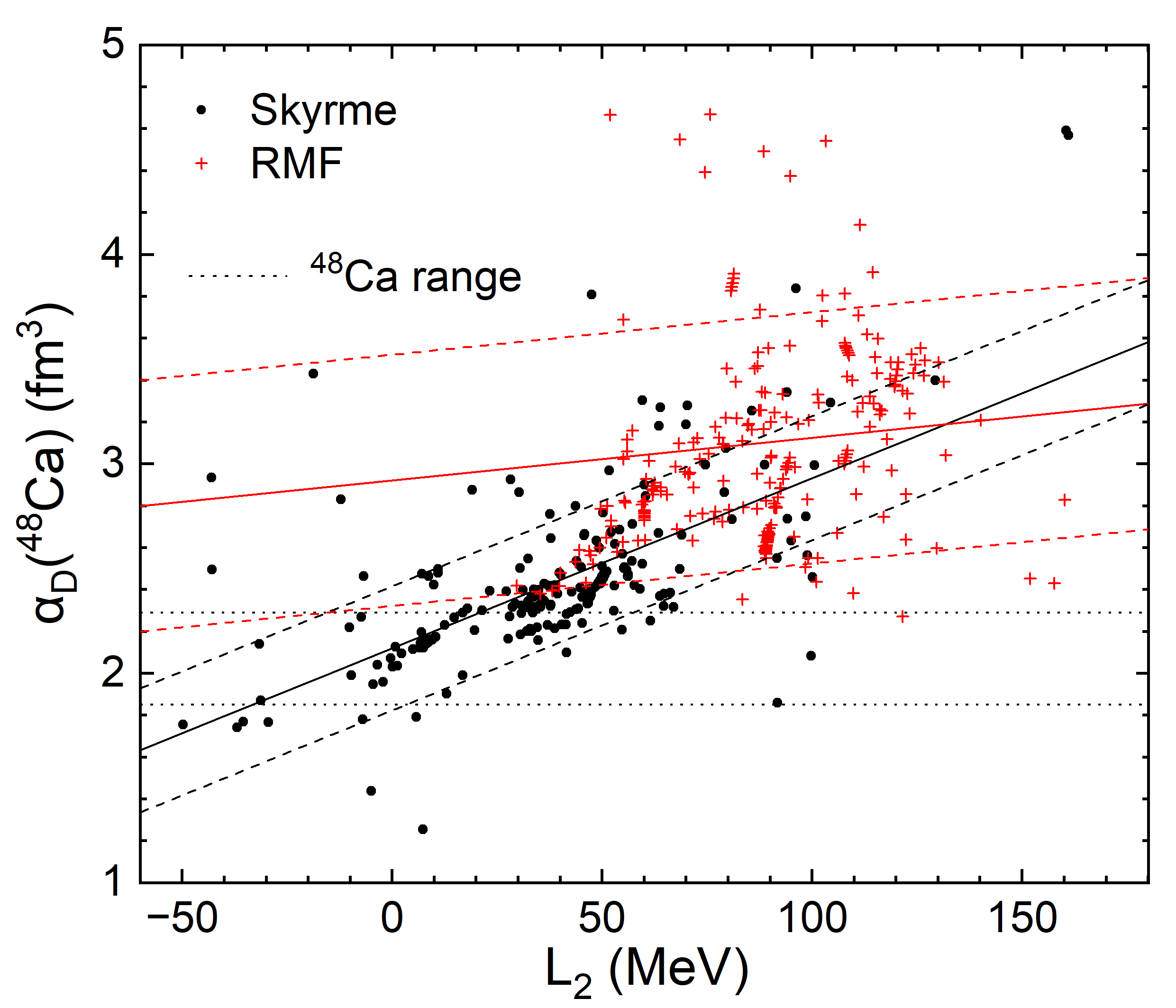}
    \includegraphics[width=8.5cm,angle=0]{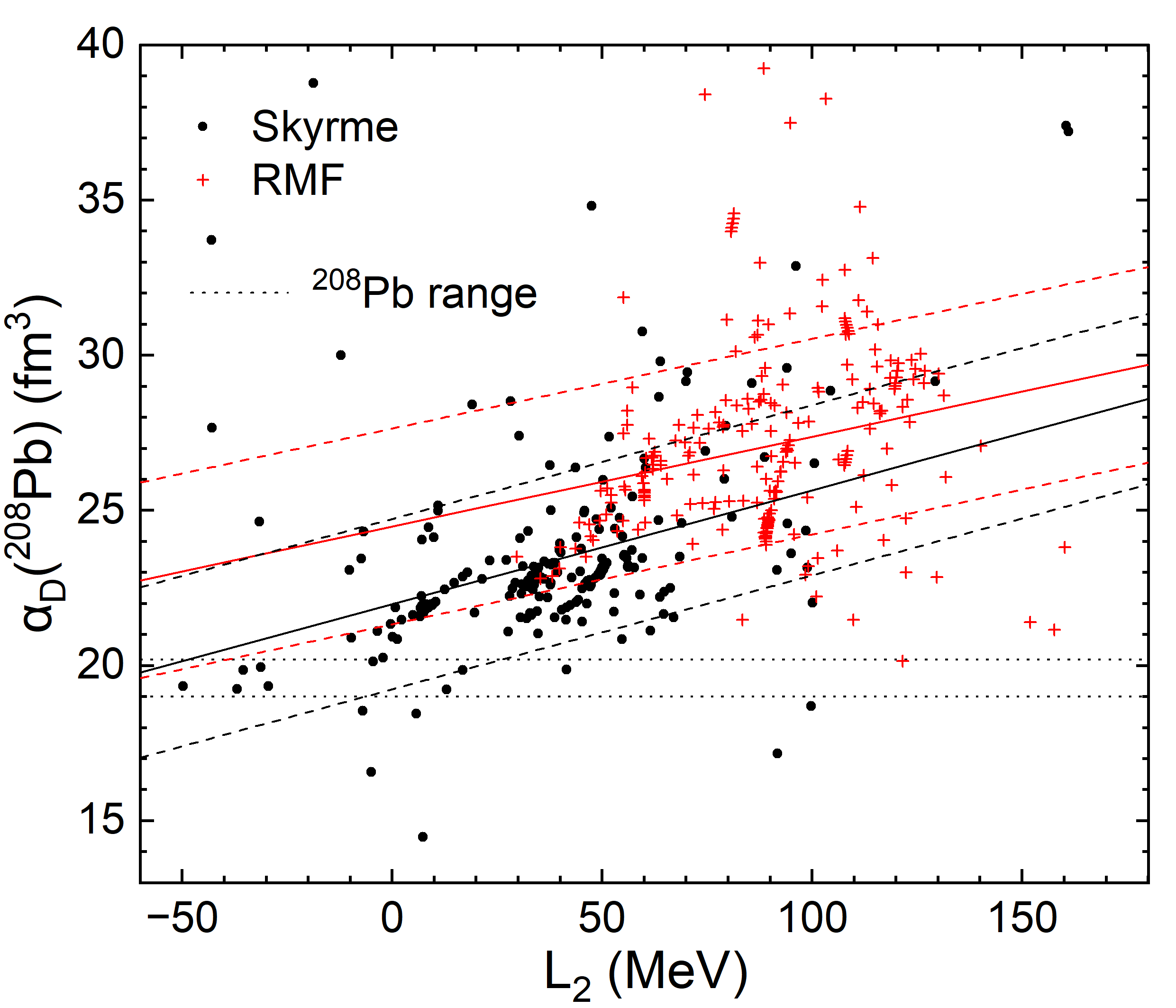}
\end{adjustwidth}
 \vspace*{0cm}  \caption{Skyrme and RMF predictions of the dipole polarizability of $^{48}$Ca (left) and $^{208}$Pb (right) calculated with hydrodynamical model. The dotted black lines in both figures indicate the ranges of the experimental results for these nuclei, Refs. \cite{Birkhan17} and \cite{Tamii11}, respectively.} \label{fig:dipole}
\end{figure}

The liquid droplet \cite{Myersswiatecki1974} and the hydrodynamical \cite{steiner2005isospin} models predict slightly different dipole polarizabilities
\begin{eqnarray}
\alpha_D^{\rm h}&=&{1\over20}{AR^2\over J_2}\left(1+{5S_S\over3J_2A^{1/3}}\right),\cr
\alpha_D^{\rm DM}&=&{\pi e^2\over90}{AR^2\over J_2}\left(1+{5S_S\over3J_2A^{1/3}}\right).
\label{eq:dipole}\end{eqnarray}
The factor $\pi e^2/90\approx1/19.894$ in the DM model is almost identical to the factor 1/20 from the hydrodynamical model, so the two predictions differ by only 0.5\%. Hydrodynamical model predictions of $\alpha_D$ for $^{48}$Ca and $^{208}$Pb for Skyrme (RMF) parameterizations are included in Table III (IX), and are shown in Fig. (\ref{fig:dipole}).  This figure shows that, in comparison to the experimentally measured values (from Ref. \cite{Birkhan17} and \cite{Tamii11}, respectively) the hydrodynamical (or DM) model largely overpredicts their values.  The theoretical results indicate a positive correlation with $L_2$, and reconciling them with experimental data would suggest rather small values of $L_2\approx0$ MeV.  No Hartree-Fock estimates are readily available in tabular form for the forces considered here.

\section{Selected Neutron Star Properties}
We are also interested in the dependence of neutron star properties on the equation of state (EOS).  However, it is unreasonable to naively extrapolate EOSs determined from fitting nuclear structure properties, for which the maximum density is $n_0$, to $(5-10)n_0$, the central density of the most massive neutron stars.  Therefore, we largely confine our attention to low-mass neutron stars, those with $M\alt1.6M_\odot$, for which the central densities are just a few $n_0$.  An exception is made to include the calculation of $M_{max}$, the maximum neutron star mass predicted by an EOS.  Many of the EOSs included here cannot satisfy the minimal constraint that $M_{max}\agt2M_\odot$, the mass of the most massive measured pulsar, but in practice one could replace a given EOS at densities in excess of $n_0$ with another model that could satisfy this condition.   Since this replacement is arbitrary, given the lack of experimental information for densities in excess of $n_0$, we do not pursue such replacements in this paper.

In principle, each EOS predicts a specific core-crust boundary, and a crust EOS at lower densities that contains finite nuclei in addition to nucleons.  Ideally, one would compute this crustal EOS for each force using a scheme such as that of Ref. \cite{LATTIMER1991331}.  Instead, for simplicity and computational expediency, we assume the crust EOS is a piecewise-polytropic EOS fitted to the Skyrme EOS SLy4 \cite{Zhao_2022}. The crust EOS is assumed for $n\alt0.04 $ fm$^{-3}$. For Skyrme and Gogny forces, we use the given EOS as PNM at densities in excess of 0.12 fm$^{-3}$ and smoothly interpolate in between with a smooth scheme that matches energies, pressures and incompressibilities at both boundaries.  For RMF forces, we use the same crust EOS below 0.04 fm$^{-3}$, and the EOS in the in-between range is obtained through an interpolation based on Neville's algorithm.
We checked that small variations in the EOS resulting from  these interpolations do not significantly affect the quantities we seek.  The quantity most sensitive to the way this region is treated is the tidal deformability, for which a lack of smoothness in the transition region has large effects.  But varying the interpolation scheme, as long as it is smooth, has only very small effect.   Although neutron-star matter has a small proton fraction, determined by beta equilibrium, and a slightly smaller pressure than PNM, we ignore this complication.  Primarily this omission will affect predictions of $M_{max}$, but given the unreliability of these interactions at the relevant densities, these differences have little consequence.

\subsection{Radius}
The basic relations determining the mass-radius relation are the traditional Tolman-Oppenheimer-Volkoff (TOV) differential equations
\begin{align}
\frac{dP}{dr}&=-\frac{G(\mathcal{E}+P)(mc^2+4\pi r^3P)}{rc^2(rc^2-2Gm)},\qquad{dm\over dr}=4\pi r^2\frac{\mathcal{E}}{c^2},
\end{align}
where $m(r)$ is the gravitational mass interior to the radius $r$.  Boundary conditions for these equations are $m(r=0)=0$ and $(dP/dr)_{r=0}=0$.  The integrations are terminated when $P=0$, which defines the surface $r=R$.  A specified value of central pressure $P_c=P(r=0)$ determines the total mass $M=m(r=R)$.  Note that the only EOS relation needed is the pressure-energy density relation $P(\mathcal{E})$, which in the neutron star core is assumed to be PNM.  Tables IV, X and XIII give the radii $R_{1.2}, R_{1.4}$ and $R_{1.6}$ corresponding to the masses $1.2M_\odot$, $1.4M_\odot$, and $1.6M_\odot$, for Skyrme, RMF and Gogny forces, respectively. Radii for larger masses are not considered, due to the large central densities of such structures.  For informational purposes, the maximum masses $M_{max}$ are also tabulated.
Besides the radii of low-mass neutron stars, we also consider other potentially observable quantities, including moments of inertia, tidal deformabilities and binding energies.

\begin{figure}[htbp]
\centering
\subfigure[\quad $R_{1.4}$ vs. $L_2$.  Correlation is given by Eq. (\ref{eq:r14}).]{
\includegraphics[width=8.5cm]{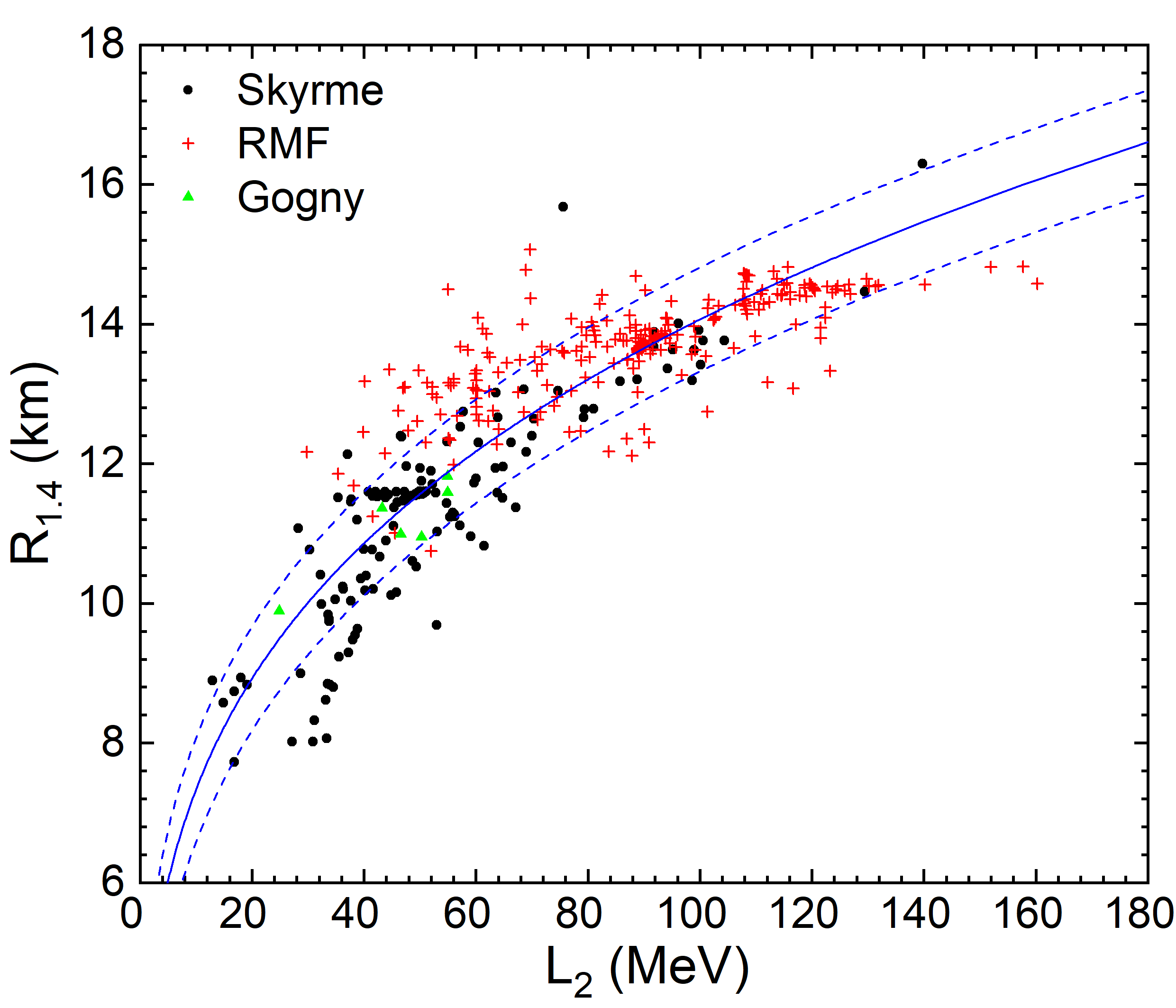}
}
\subfigure[\quad $\Lambda_{1.4}$ vs. $L_2$.  Correlation is given by Eq. (\ref{eq:l14}).]{
\includegraphics[width=8.5cm]{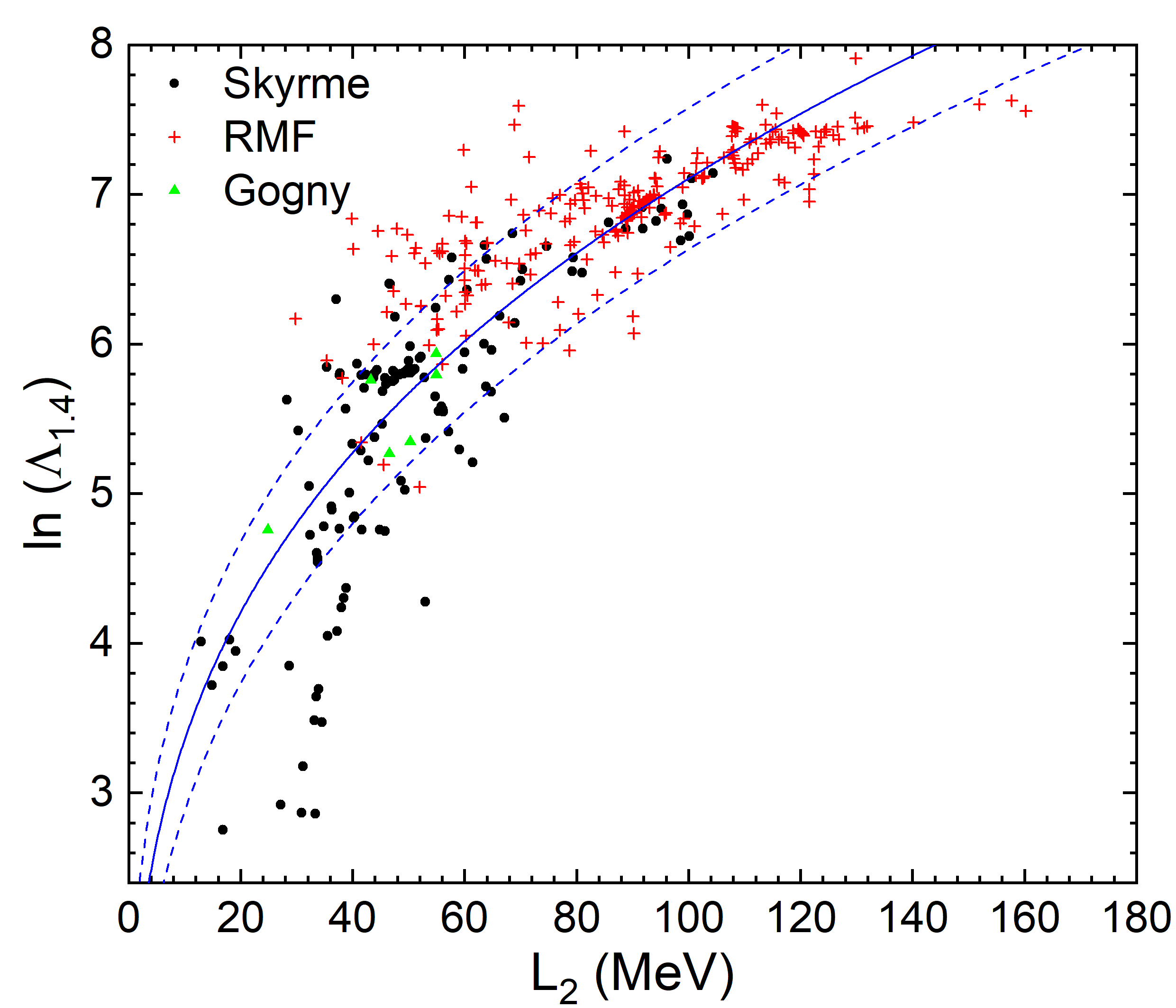}
}
\subfigure[\quad $I_{1.4}$ vs. $L_2$.  Correlation is given by Eq. (\ref{eq:i14}). ]{
\includegraphics[width=8.5cm]{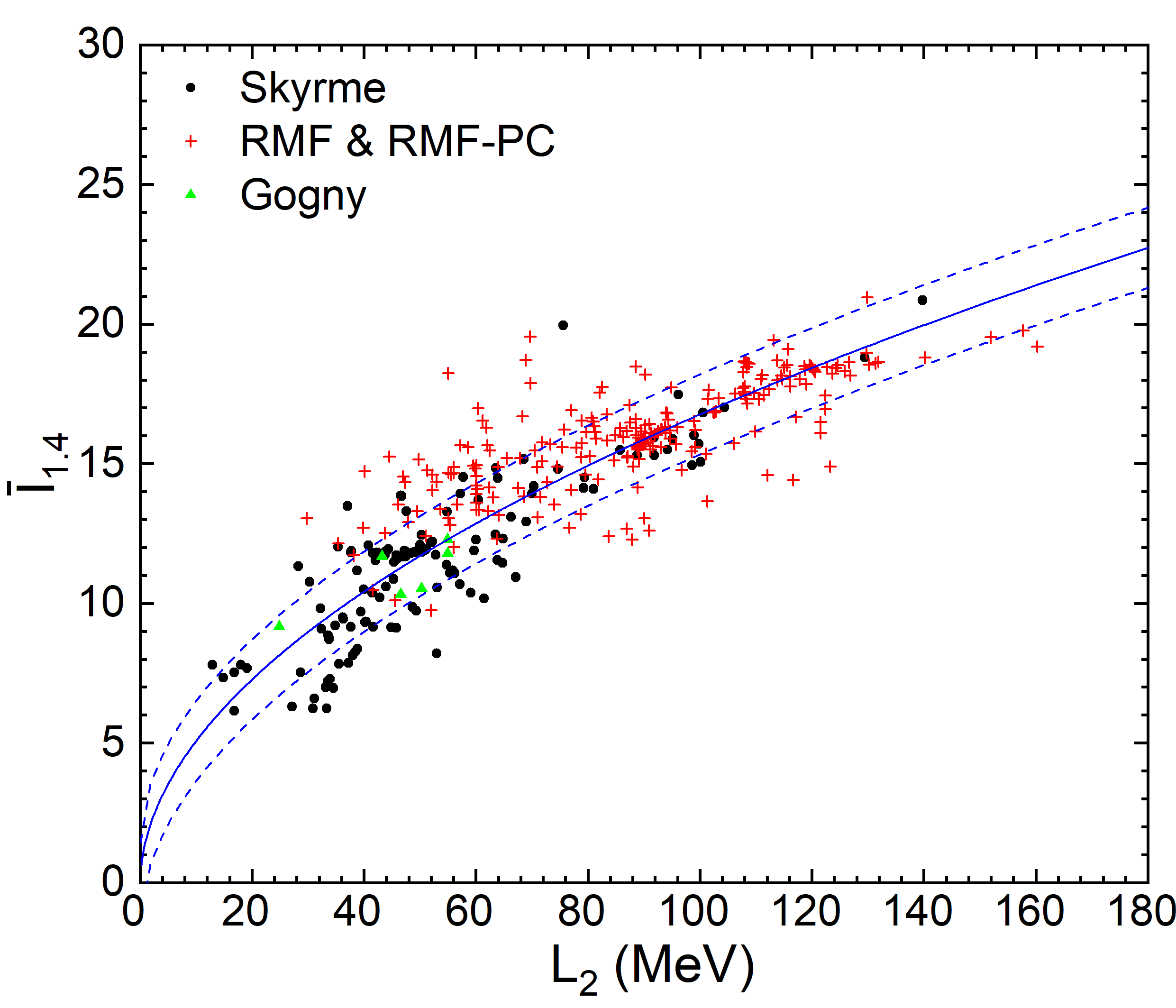}
}
\subfigure[\quad (BE/$M$)$_{1.4}$ vs. $L_2$.  Correlation is given by Eq. (\ref{eq:be14}).]{
\includegraphics[width=8.5cm]{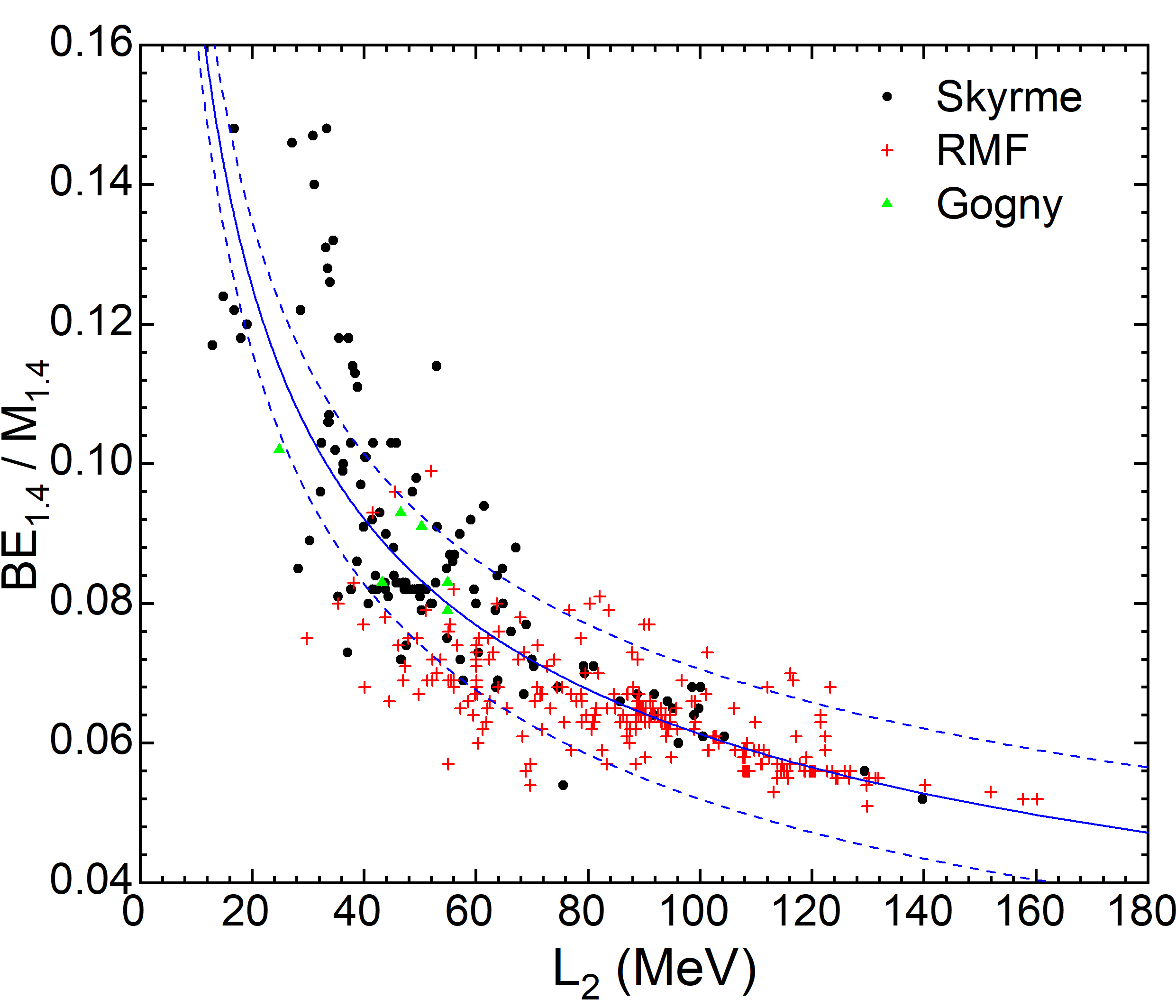}
}
\caption{Relationship between neutron star parameters and slope of symmetric energy}\label{fig:rlib}
\end{figure}

It has long been known that there is a high degree of correlation between the typical (i.e., $1.4M_\odot$) neutron star radius $R_{1.4}$ and the pressure of neutron-star matter in the range $n_0-2n_0$ \cite{Lattimer_2001}.   We update the relations found by Ref. \cite{Lattimer_2001} at $n_s$ and $1.5n_s$, assuming $M_{max}\ge2M_\odot$:
\begin{equation}
    R_{1.4}=(9.46\pm0.38)\left({P(n_s)\over{\rm MeV~fm^{-3}}}\right)^{1/4}{\rm km}=(6.99\pm0.26)\left({P(1.5n_s)\over{\rm MeV~fm^{-3}}}\right)^{1/4}{\rm km}.
\end{equation}
This implies a similar correlation exists between $R_{1.4}$ and $L_2$, which is shown in Fig. \ref{fig:rlib}.  The fitted curve satisfies
\begin{equation}
    R_{1.4}=(3.84\pm0.74)(L_2/{\rm MeV})^{0.282} {\rm~km};
\label{eq:r14}
\end{equation}
note the exponent is close to 1/4, but the correlation is less significant than with $P(n_s)$, partly due to variations of $n_s$, that $L_2\ne L_1=3P_N(n_s)/n_s$, and that the former correlation assumes $M_{max}\ge2M_\odot$.

\subsection{Moment of inertia}
The moment of inertia of a neutron star, the ratio of its rotational angular momentum to its angular velocity, is given for a slowly uniformly rotating star by \cite{1994ApJ...424..846R}
\begin{equation}
I = \frac{c^2}{G}\frac{R^3 u_R}{6+2u_R},
\end{equation}
where $u_R=u(R)$ is the surface value of a function $u(r)$ that obeys the first-order differential equation
\label{fig:r14}\begin{equation}
\frac{du}{dr} = \frac{4\pi Gr^2}{c^2}\frac{(\mathcal{E}+P)(4+u)}{c^2r-2Gm}-\frac{u}{r}(3+u),
\end{equation}
with the boundary condition $u(0)=0$ that is solved together with the TOV equations.  Note that $u$ is not to be confused with the earlier usage as $u=n/n_0$.
It is useful to define the dimensionless moment of inertia as
\begin{equation}
\bar{I} = \frac{c^4I}{G^2M^3}={u_R\over\beta^3(6+2u_R)},
\end{equation}
where $\beta=GM/(Rc^2)$ is the dimensionless compactness parameter. Like the $M-R$ relation, the only EOS relation required is $P(\mathcal{E})$. Tables IV, X and XIII give the quantities $\bar I_{1.2}, \bar I_{1.4}$ and $\bar I_{1.6}$  for Skyrme, RMF and Gogny forces, respectively.

\begin{figure}[H]
\centering
\begin{adjustwidth}{0cm}{0cm}
    \includegraphics[width=8.5cm,angle=0]{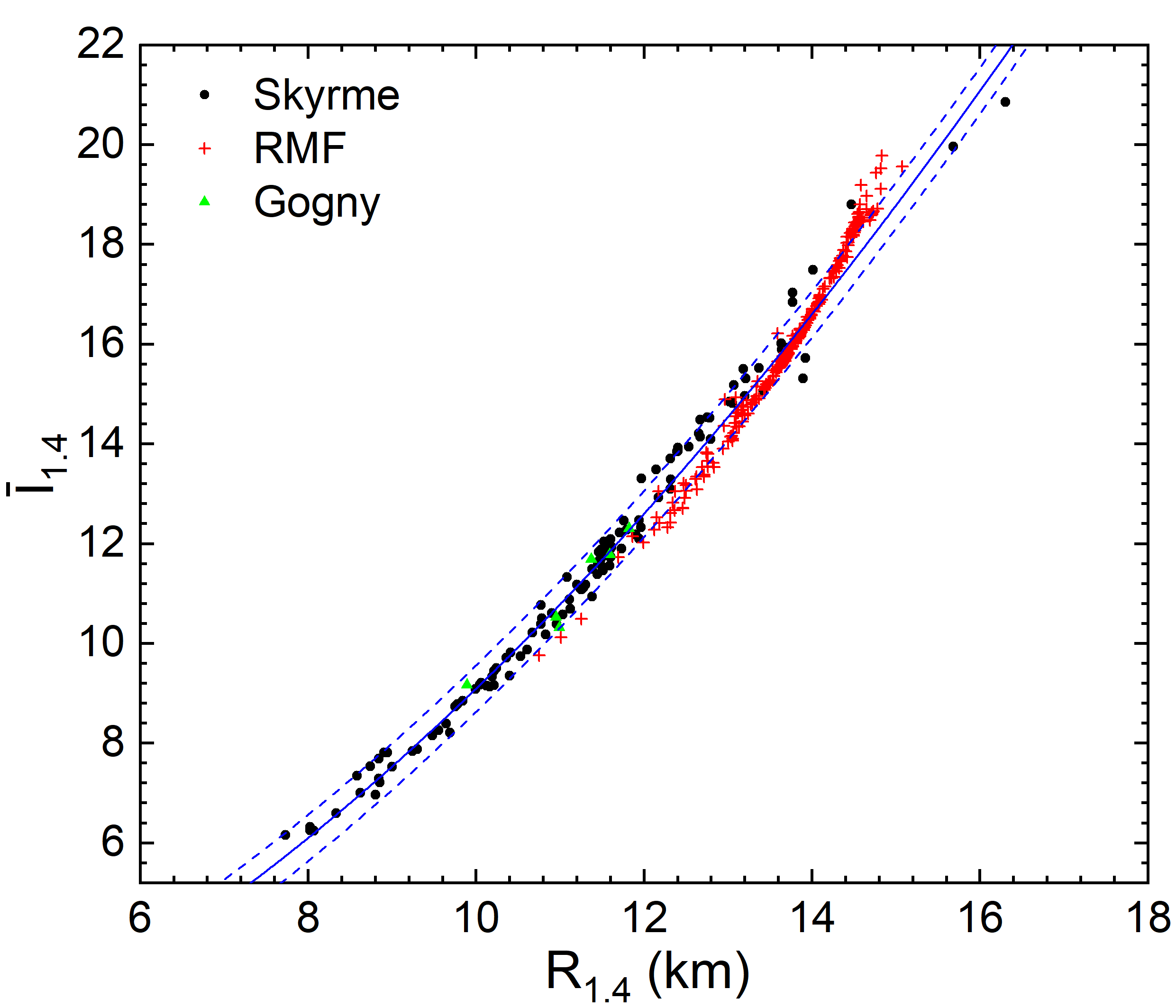}
    \includegraphics[width=8.5cm,angle=0]{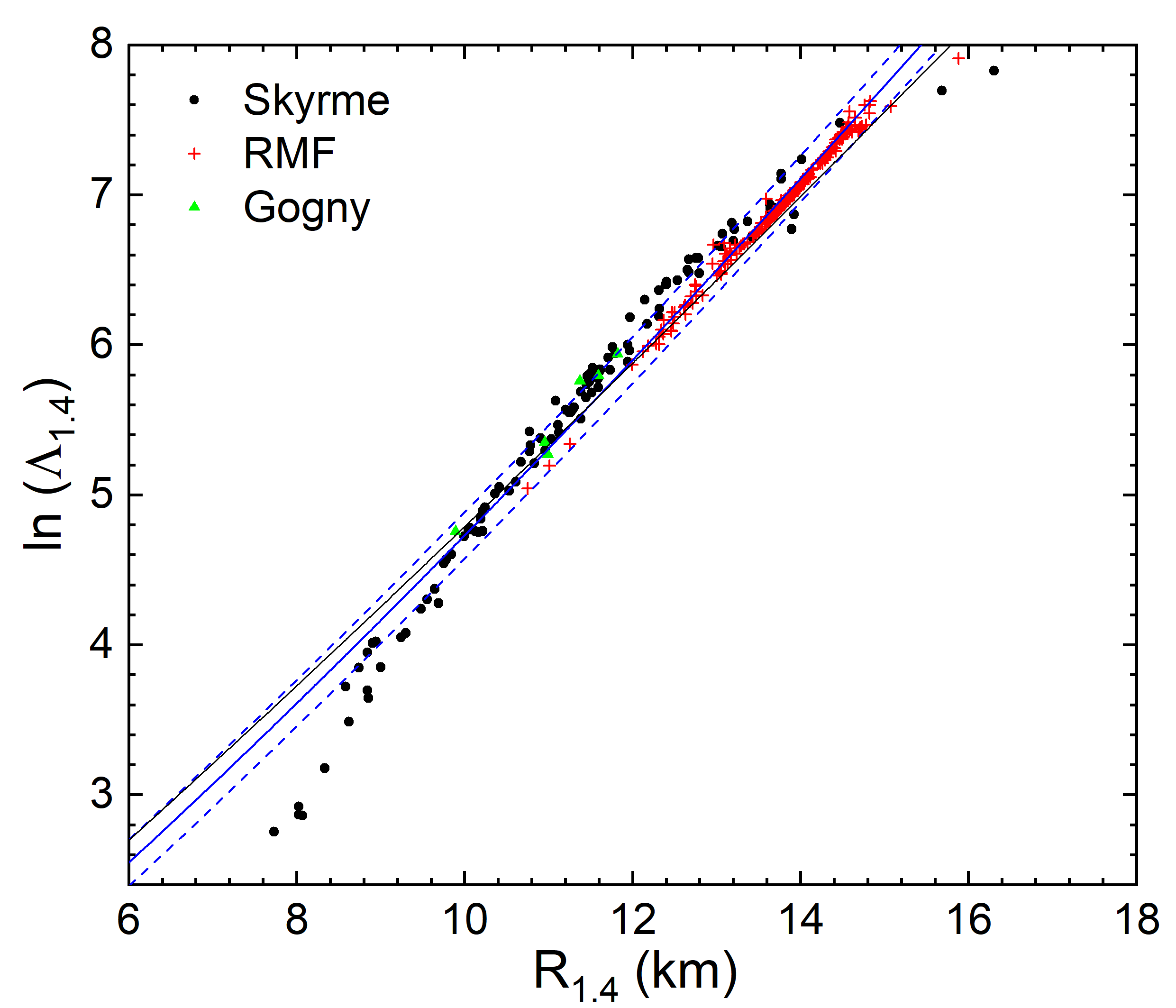}
\end{adjustwidth}
    \vspace*{0cm} \caption{Left: $I_{1.4}$ vs. $R_{1.4}$.  The correlation is given in Eq. (\ref{eq:IR}). Right: $\ln\Lambda_{1.4}$ vs. $R_{1.4}$.  The correlation is given in Eq. (\ref{eq:l14}).}
  \label{fig:IR}
\end{figure}

Since one expects that $I\propto MR^2$ in the Newtonian case, a correlation between $\bar I_{1.4}$ and $L$ is anticipated.  This is shown in Fig. \ref{fig:rlib}, where the fitted relation satisfies
\begin{equation}
    \bar I_{1.4}=0.542(L_2/{\rm MeV})^{0.183}\pm1.435.
\label{eq:i14}
\end{equation}
The correlation between $\bar I_{1.4}$ and $R_{1.4}$ itself is shown in Fig. \ref{fig:IR}, where the fitted curve is given by
\begin{equation}
    \bar I_{1.4} = 0.1478(R_{1.4}/{\rm km})^{1.789}\pm0.4649.
\label{eq:IR}\end{equation}
Note that this relation is less steep than $\bar I\propto R^2$, due to general relativistic corrections that become more important for small radii.  Indeed, the average correlation for RMF forces is noticeably steeper than for Skyrme forces, due to the fact that RMF forces typically have larger values of $R_{1.4}$.   Also note that the moment of inertia correlation with radius is much more accurate than with $L_2$.
%I-R: a=0.1478, b=1.789, SD=0.4649

\subsection{Tidal Deformability}
The tidal deformability $\lambda$ quantifies how easily a star is deformed when subjected to an external tidal field.  A larger tidal deformability signals a larger, less compact star that is easily deformable.  Specifically, the tidal deformability is defined as the ratio of the induced quadrupole moment to the external perturbing tidal field.    In the Newtonian limit, the perturbing tidal field is defined as the second spatial derivative of the external field.  Formally, the tidal deformability is defined as $\lambda=(2/5)k_2R^5$, where $k_2$ is the gravitational Love number \cite{Postnikov_2010}
\begin{equation}
 k_2 = \frac{8\beta^5}{5\mathcal{R}}(1-2\beta)^2\left[2-y_R+2\beta(y_R-1)\right],
\end{equation}
the parameter $\mathcal{R}$ being
\begin{eqnarray}
\mathcal{R} &=& 6\beta\left[2-y_R+\beta(5y_R-8)\right]+3(1-\beta^2)\left[ 2-y_R+2\beta(y_R-1)\right]\ln(1-2\beta) \nonumber\\
&\,& +4\beta^3\left[ 13-11y_R+\beta(3y_R-2)+2\beta^2(1+y_R)\right].
\end{eqnarray}
$y_R=y(R)$ is the surface value of the dimensionless function $y(r)$, which is the solution of the differential equation
\begin{equation}
\frac{dy(r)}{dr}=-\frac{y(r)^2}{r}-\frac{y(r)F(r)}{r}-rQ(r)
\end{equation}
that is solved together with the TOV equations.  The central boundary condition is $y(0)=2$.  The functions $F(r)$ and $Q(r)$ are expressed as
\begin{eqnarray}
F(r)&=&\left( 1-\frac{4\pi G}{c^4} r^2\left[\mathcal{E}(r)-P(r)\right] \right) \left[ 1-\frac{2Gm(r)}{rc^2}\right] ^{-1}, \nonumber\\
Q(r)&=&\frac{4\pi G}{c^4}\left[ 5\mathcal{E}(r)+9P(r)+\frac{\mathcal{E}(r)+P(r)}{dP(r)/d\mathcal{E}(r)}-\frac{6}{4\pi r^2}\right]\left[ 1-\frac{2Gm(r)}{rc^2}\right] ^{-1} \nonumber\\
&\,&-\left(\frac{2Gm(r)}{r^2c^2}\right)^2\left[ 1+\frac{4\pi r^3P(r)}{m(r)c^2}\right] ^2\left[ 1-\frac{2Gm(r)}{rc^2}\right] ^{-2}.
\label{eq:fq}\end{eqnarray}
Note that $dP/d\mathcal{E}$ is the square of the sound speed in units of the speed of light, and, once again, only $P(\mathcal{E})$ is the only EOS relation required. The appearance of the sound speed in a denominator in a term of $Q(r)$ means that the evaluation of $y_R$ is sensitive to how the core-crust interface is treated.  Physically unrealistic discontinuities in thermodynamic quantities near the interface will introduce anomalous contributions to $y_R$, so smoothing the $P(\mathcal{E})$ relation is important. 

As for the moment of inertia, it is convenient to introduce the dimensionless tidal deformability
\begin{equation}
\Lambda = \frac{2k_2}{3\beta^5}.
\label{eq:lam}
\end{equation} 
Tables IV, X and XIII give the quantities $\Lambda_{1.2}, \Lambda_{1.4}$ and $\Lambda_{1.6}$  for Skyrme, RMF and Gogny forces, respectively.

$\Lambda$ is known to scale approximately as $(R/M)^6$ \cite{Zhao2018}, rather than as $(R/M)^5$ as suggested by Eq. (\ref{eq:lam}), a result we validate:
\begin{equation}
    \Lambda_{1.4}= 0.0001363R_{1.4}^{6.021}\pm 102.7.
\label{eq:l14p}\end{equation}
Thus correlations among $\Lambda_{1.4}, R_{1.4}$ and $L_2$ are expected. These are shown in Figs. \ref{fig:rlib} and Fig. \ref{fig:IR}, where the relations
\begin{equation}
    \ln\Lambda_{1.4}=1.591(L_2/{\rm MeV})^{0.3251}\pm0.4728
 ~=0.3605(R_{1.4}/{\rm km})^{1.123}\pm0.2127\label{eq:l14}
\end{equation}
are drawn.  These relations can be made even more accurate if the restriction $M_{max}>2M_\odot$, which also implies $R_{1.4}>11$ km for the Skyrme and RMF forces used here, is applied, as the right-hand panel of Fig. \ref{fig:IR} suggests.

\begin{figure}[H]
\centering
\vspace*{0cm}
\hspace{0cm}
    \includegraphics[width=12cm,angle=0]{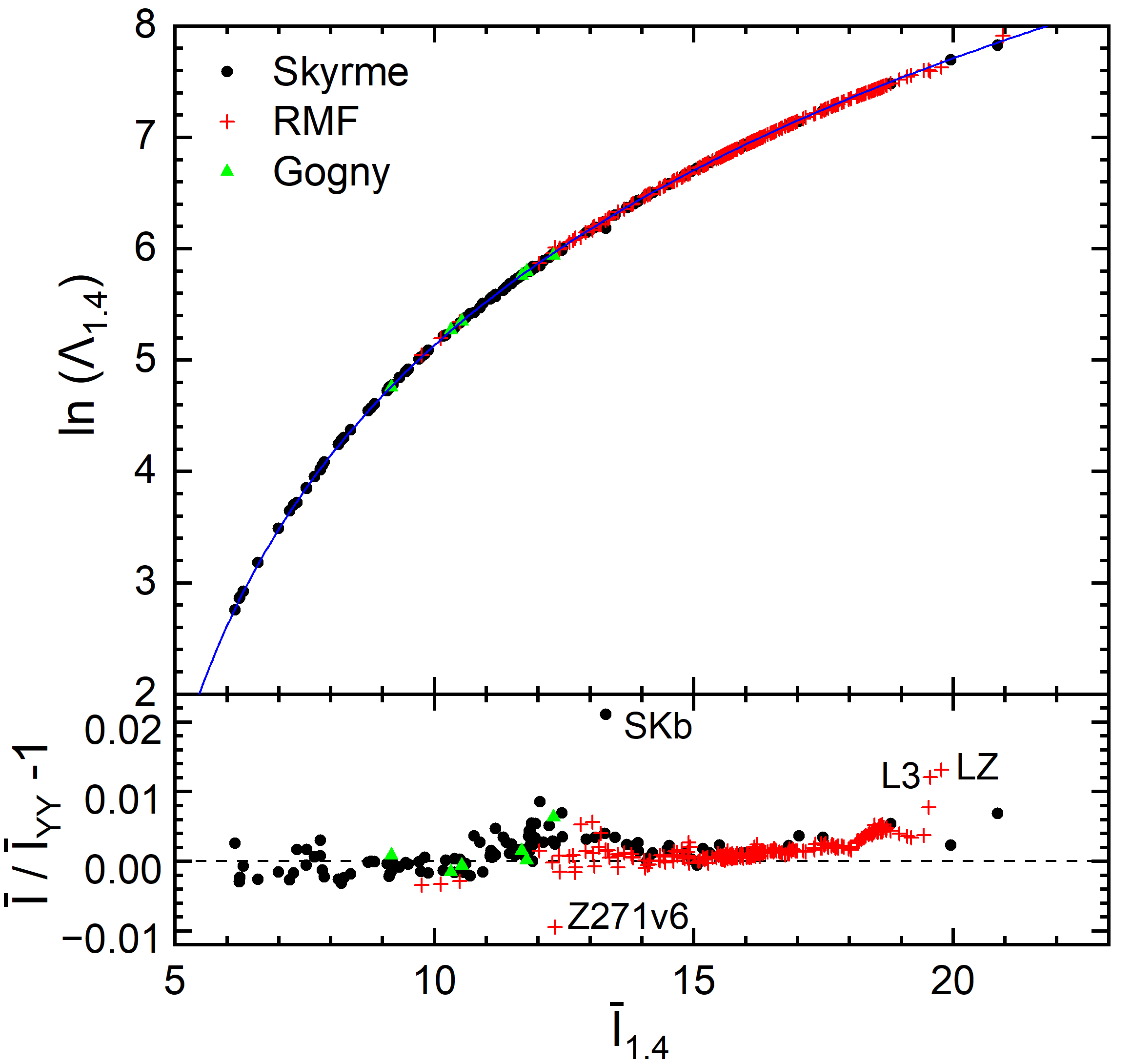}
  \vspace*{0cm}  \caption{The relation between tidal deformability and moment of inertia for neutron stars with 1.4 solar mass is shown in the upper panel.  The line in the upper panel is the Yagi-Yunes (YY) universal relation \cite{YAGIYUNES}, and the lower panel shows the deviation from it. Equations of state with the largest deviations, which in no case exceed 2\%, are indicated.}
  \label{fig:LI}
\end{figure}

Ref. \cite{YAGIYUNES} discovered a powerful correlation connecting the dimensionless tidal deformability and moment of inertia.  This correlation does not depend on the mass.  For the case of $1.4M_\odot$ stars only, in order to limit the number of data points to a legible value, we confirm their result in Fig. \ref{fig:LI}.

\subsection{Binding Energy}
In addition to the gravitational mass $M$ of a neutron star, one can consider the total baryon mass $N_Bm_B$ where $N_B$ is the number of baryons in the star and $m_B$ is the mass of a single baryon.  $N_B$ is defined as
 \cite{bagchi2011role}
\begin{equation}
    N_B = \int_0^R 4\pi r^2\left[1-\frac{2Gm(r)}{rc^2}\right]^{-1/2}n(r)dr,
\end{equation}
where $n(r)$ is the baryon number density at $r$. The binding energy of a neutron star is the energy released during its formation through gravitational collapse, which is
\begin{equation}
    {\rm BE} = (M_B - M)c^2.
\end{equation}
Calculation of the binding energy thus requires specification of the baryon density $n(r)$, which is found from the differential equation
\begin{equation}
    {d\mathcal{E}\over dn}={\mathcal{E}+P\over n},
\end{equation}
with the boundary condition $\mu_0=[(\mathcal{E}+P)/n]_{n\rightarrow0}\approx-9$ MeV, the binding energy, per nucleon, of iron.
 Tables IV, X and XIII give the quantities (BE$/Mc^2)_{1.2}$, (BE$/Mc^2)_{1.4}$ and (BE$/Mc^2)_{1.6}$  for Skyrme, RMF and Gogny forces, respectively.

The binding energy per unit mass is correlated with $L$ as shown in Fig. \ref{fig:rlib}, where the relation
\begin{equation}
({\rm BE}/Mc^2)_{1.4}=0.474(L_2/{\rm MeV})^{-0.4444}\pm0.00931
\label{eq:be14}\end{equation}
is shown. Since in Newtonian gravity, the binding energy of  a uniform sphere is $(3/5)GM^2/R$, and Eq. (\ref{eq:r14}) indicates $R_{1.4}\propto L_2^{0.28}$, one expects an inverse relationship between BE/$M$ and $L_2$.  Correlations between (BE$/Mc^2)_{1.4}$ and $R_{1.4}$ or $\beta_{1.4}$ have even more significance.

\section{Discussion and Conclusions}
In this paper we have compiled a large number of interaction parameters for  Skyrme, RMF and Gogny forces, largely culled from the publications of Dutra et al. \cite{dutra2012skyrme, dutra2014relativistic}.  Importantly, we created a complete database of interaction parameters for these forces including Table I (Skyrme), Table V, VI, VII (different types of RMF) and Table XI (Gogny), which are not contained in Refs. \cite{dutra2012skyrme,dutra2014relativistic}.  This should greatly ease future studies using these force collections.  We have given explicit formulae for the calculation of properties for both cold symmetric and pure neutron uniform matter.  We have also given prescriptions for calculating symmetry energy parameters, where the symmetry energy is alternately defined as the difference between the symmetric and pure neutron uniform matter energies or as the quadratic term in a Taylor expansion of the uniform matter energy in the neutron excess away from symmetric matter. 

In addition, selected properties related to nuclear structure, such as the surface energy, neutron skin thickness,  and dipole polarizability were computed for most forces.  We compared the approximate liquid droplet model determinations of the neutron skin thicknesses for both $^{48}$Ca and $^{208}$Pb with those from more accurate Hartree-Fock computations.  We found that the high degree of correlation between the neutron skin thickness and $L$ seen in HF computations is remarkably well reproduced by the DM.  We also computed the properties of relatively low-mass neutron stars (including radii, moments of inertia, tidal deformabilities and binding energies), which involve computations at densities for which the EOSs included here remain plausibly relevant.  The predicted neutron star maximum mass was found for all forces, although we argued that the specific values should be treated with caution since they necessarily involve  the properties of matter at densities far in excess of those found in nuclei, which renders their $P(\mathcal{E})$ relations unreliable.

\begin{figure}[H]
\centering
\vspace*{-0.6cm}
    \includegraphics[width=17cm,angle=0]{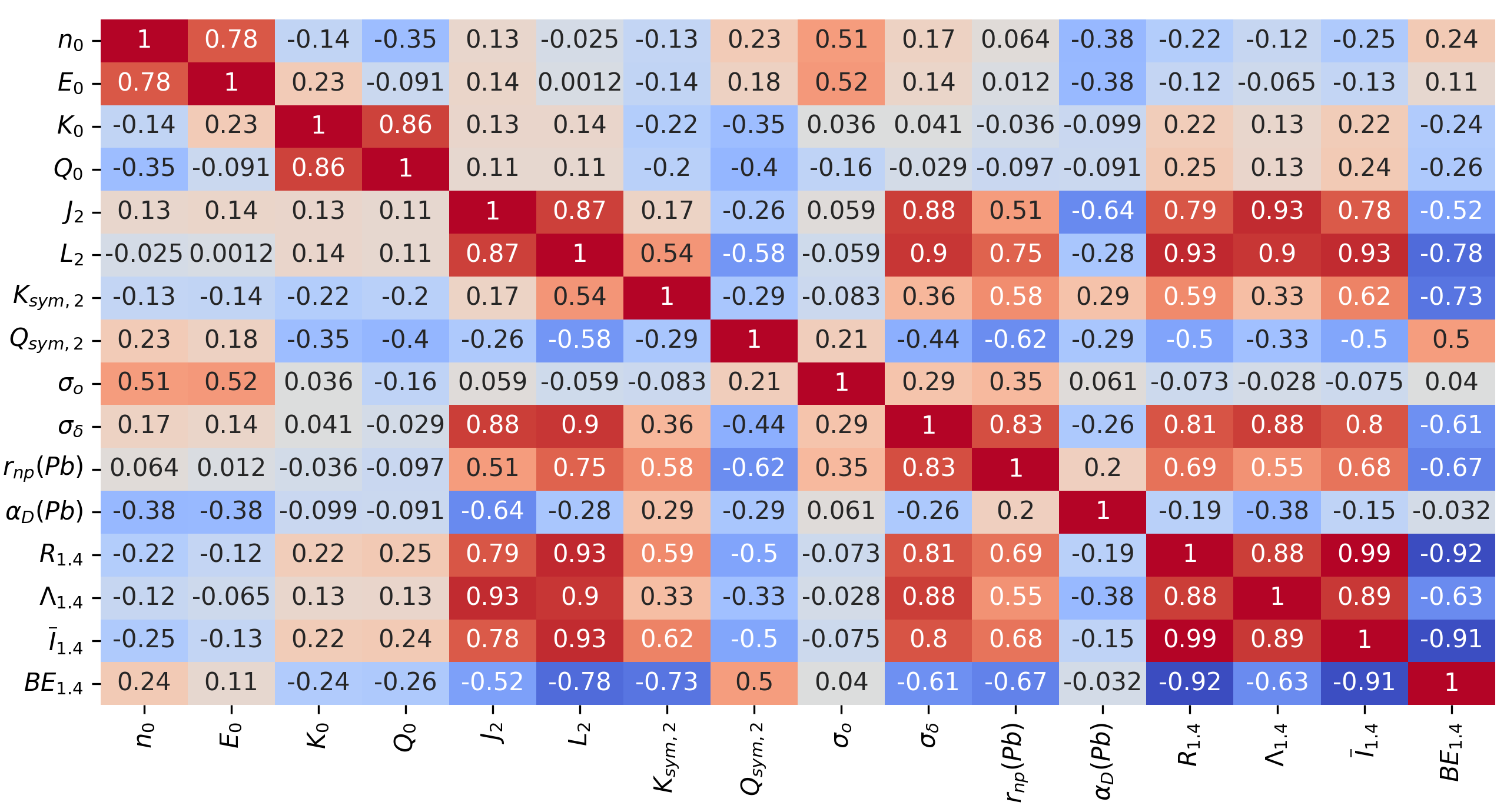}
  \vspace*{-0.2cm} \caption{Correlation matrix for Skyrme forces. }
\label{fig:correls}
\end{figure}

\begin{figure}[H]
\centering
\vspace*{-0.6cm}
    \includegraphics[width=17cm,angle=0]{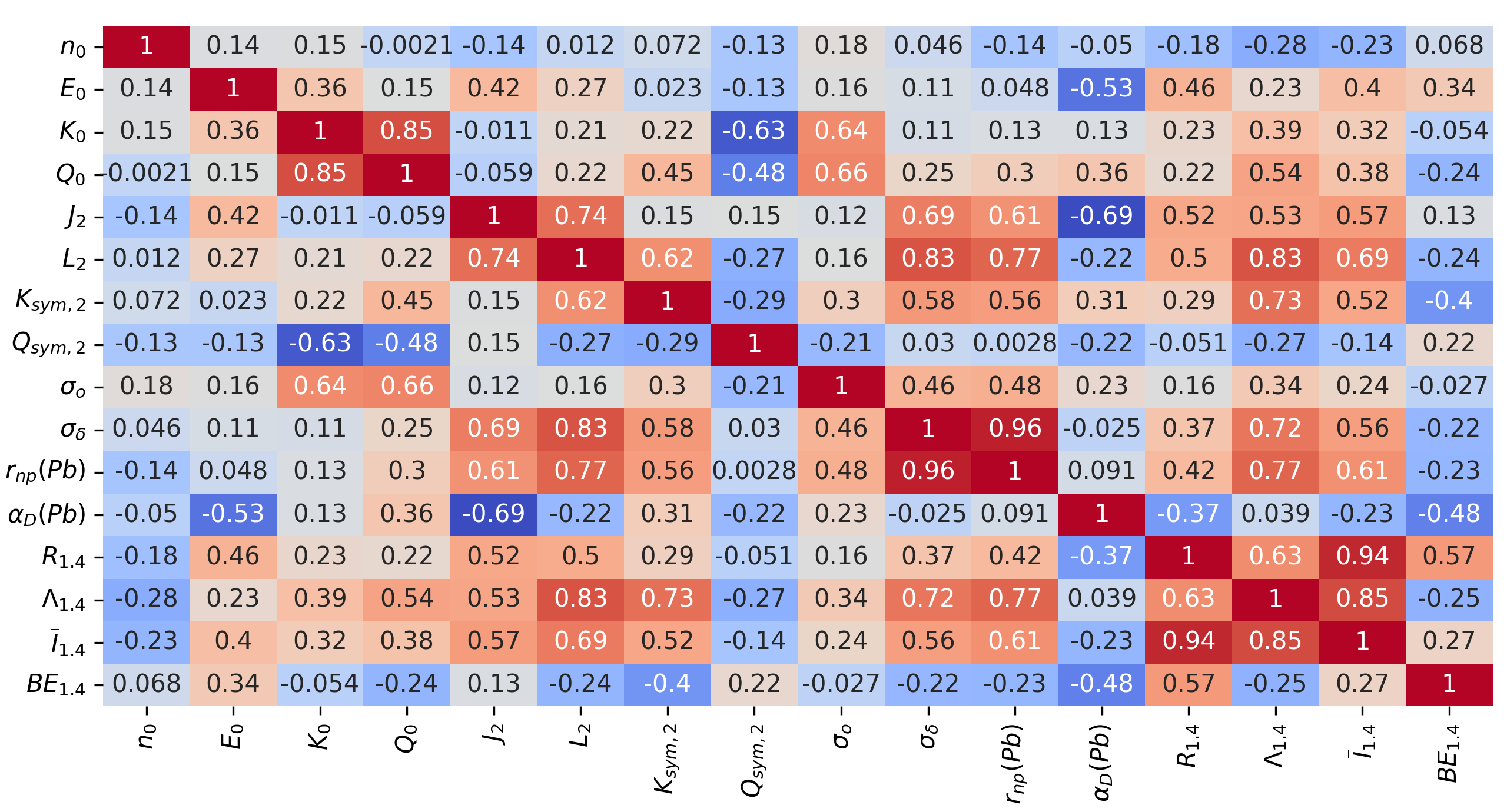}
  \vspace*{-0.2cm} \caption{Correlation matrix for RMF forces.}
\label{fig:correlr}
\end{figure}

Many of the calculated properties are highly correlated, which can lead to semi-universal relations among them.  Those relations involving nuclear structure are very useful in interpreting experiments, while those involving neutron star properties are important in  interpreting astronomical observations of neutron stars.  Those correlations that cross over between nuclear physics and astrophysics are especially interesting.  Figures \ref{fig:correls} and \ref{fig:correlr} show correlation matrices involving many of these parameters for Skyrme and RMF forces, respectively.
\subsection{Correlation Matrices}

The highest degrees of correlation exist among $J_2, L_2, r_{np}(\rm{Pb}), \sigma_\delta, R_{1.4}, \Lambda_{1.4}, \bar I_{1.4}$, and BE$_{1.4}$.  In most cases, the correlations are stronger for Skyrme forces than for RMF forces.  In particular, this is true for computations of BE$_{1.4}$, which exhibits only mild correlations with $R_{1.4}$ and none with the other parameters for RMF models.  The parameters $K_0$ and $Q_0$ are highly correlated, and $K_{sym,2}$ and $L_2$ are mildly correlated, for both types of forces.   For RMF forces, the parameter $\sigma_o$ is mildly correlated with $K_0$ and $Q_0$ but this is not the case for Skyrme forces.  For Skyrme forces, the parameters $n_0$ and $E_0$ are moderately correlated, but this is not the case for RMF forces.

In many cases where relatively strong correlations exist, we have determined in \S VI and \S VII simple fits with standard deviations which can be regarded as semi-universal relations.  We caution that since we do not explicitly consider interactions leading to strong phase transitions in dense matter, those semi-universal relations involving neutron stars in such cases could be violated.

In a future work, there are a number of improvements that could be made, including
\begin{itemize}
    \item computation of  Hartree-Fock values for surface energies, dipole polarizabilities, neutron skin thicknesses and giant resonance frequencies,
\item inclusion of additional forces, including those in more recent publications and those from the extensive compilation of Ref. \cite{tagami2022neutron}, who computed HF neutron skin thicknesses for many additional forces but did not tabulate their model parameters,
\item more detailed explorations of correlations among low-mass neutron star properties and bulk matter properties, not just at the saturation density, but at other densities such as $0.5n_0, 1.5n_0$ and $2n_0$.
\end{itemize}

\section*{Acknowledgements}

We acknowledge useful discussions with Yeunhwan Lim and Tianqi Zhao, and thank M. Dutra for providing valuable information, including some interaction parameters from models in Refs. \cite{dutra2012skyrme,dutra2014relativistic}.  We also thank the anonymous referee for valuable corrections and suggestions, and the CompOSE team as well as Hristijan Kochankovski for proposing the inclusion of additional interactions. Funding for this research was provided by the US Department of Energy under Grant DE-FG02-87ER40317.

\appendix
\begin{widetext}

\section{Parameters and Properties of Skyrme Models}
%    \subsection{Parameters}
\vspace*{-0.7cm}
\LTcapwidth=\textwidth
\begingroup
\centering
\squeezetable
\begin{longtable*}{@{\extracolsep{\fill}}lccccccccccrrr}
\caption{\small Parameters of Skyrme models. The unit for $t_0$ is $\rm{MeV\;fm^3}$. $t_1$ and $t_2$ are both in $\rm{MeV\;fm^5}$, $t_{3i}$ is in $\rm{MeV\;fm^{3+3\alpha_i}}$, and $t_4$ and $t_5$ are in $\rm{MeV\;fm^{5+3\delta}}$ and $\rm{MeV\;fm^{5+3\gamma}}$, respectively. Other parameters are  dimensionless.}\label{Tab:10}\\
\hline\hline
\endfirsthead
\caption* {\textbf{TABLE \ref{Tab:10}. (\rm{\textit{Continued.}}) }}\\
\hline\hline
 Model&$t_0$&$t_1$&$t_2$&$t_{31}$&$t_{32}$&$t_{33}$&$t_4$&$t_5$&$\sigma_1$&$\sigma_2$&$\sigma_3$&$\delta$&$\gamma$\\
    &$x_0$&$x_1$&$x_2$&$x_{31}$&$x_{32}$&$x_{33}$&$x_4$&$x_5$ \\\hline
\endhead
\hline
\endfoot
\hline\hline
\endlastfoot
 Model&$t_0$&$t_1$&$t_2$&$t_{31}$&$t_{32}$&$t_{33}$&$t_4$&$t_5$&$\sigma_1$&$\sigma_2$&$\sigma_3$&$\delta$&$\gamma$\\
      &$x_0$&$x_1$&$x_2$&$x_{31}$&$x_{32}$&$x_{33}$&$x_4$&$x_5$\\\hline
 BSk1\cite{samyn2002hartree} & -1830.45 & 262.97 & -296.446 & 13444.7 & - & - & - & - & 1/3 & - & - & - & -\\
      & 0.599988 & -0.5 & -0.5 & 0.823074 & - & - & - & - \\
 BSk2\cite{goriely2002hartree} & -1790.62 & 260.996 & -147.167 & 13215.1 & - & - & - & - & 0.343295 & - & - & - & -\\
      & 0.498986 & -0.08975 & 0.224411 & 0.515675 & - & - & - & -\\
    
 BSk2'\cite{goriely2002hartree} &  -1792.71 & 259.053 & -146.768 & 13267.9 & - & - & - & - & 0.343295 & - & - & - & -\\
      & 0.498612 & -0.08976 & 0.242854 & 0.509818 & - & - & - & -\\
\end{longtable*}
\endgroup
%\subsection{Saturation and Symmetry Properties}
\LTcapwidth=\textwidth
\begingroup
\centering
\squeezetable
\begin{longtable*}{@{\extracolsep{\fill}}lccccccccrrrr}
\caption {\small Properties of Skyrme models at their saturation densities. Entries are in $\rm{MeV}$ except for the saturation density $n_0$, which has units of $\rm{fm^{-3}}$.} \label{Tab:1}\\
\hline\hline
\endfirsthead
\caption* {\textbf{TABLE \ref{Tab:1}. (\rm{\textit{Continued.}}) }}\\
\hline\hline
 Model&$ n_0$&$E_0$&$K_0$&$Q_0$&$J_1$&$J_2$&$L_1$&$L_2$&\multicolumn{1}{c}{$K_{sym,1}$}&\multicolumn{1}{c}{$K_{sym,2}$}&\multicolumn{1}{c}{$Q_{sym,1}$}&\multicolumn{1}{c}{$Q_{sym,2}$}\\ \hline
\endhead
\hline
\endfoot
\hline\hline
\endlastfoot
 Model&$ n_0$&$E_0$&$K_0$&$Q_0$&$J_1$&$J_2$&$L_1$&$L_2$&\multicolumn{1}{c}{$K_{sym,1}$}&\multicolumn{1}{c}{$K_{sym,2}$}&\multicolumn{1}{c}{$Q_{sym,1}$}&\multicolumn{1}{c}{$Q_{sym,2}$}\\ \hline
 BSk1    &  0.1574&  -15.82&  231.45& -385.93&   28.46&   27.80&    8.37&    7.14& -283.67& -281.95&  612.56&  606.67\\
 BSk2    &  0.1575&  -15.81&  233.80& -380.40&   29.01&   27.99&   10.94&    7.93& -295.26& -297.10&  560.42&  558.09\\
 BSk2'   &  0.1575&  -15.81&  233.47& -381.19&   29.01&   27.99&   10.74&    7.74& -296.33& -298.15&  561.17&  558.81\\
 BSk3    &  0.1575&  -15.82&  234.95& -381.17&   28.94&   27.93&    9.72&    6.73& -305.24& -307.04&  552.91&  550.53\\
 BSk4    &  0.1575&  -15.79&  236.99& -367.49&   28.91&   27.99&   14.97&   12.50& -265.27& -266.04&  561.99&  558.58\\
 BSk5    &  0.1575&  -15.82&  237.34& -368.19&   29.64&   28.69&   24.02&   21.38& -239.28& -240.39&  503.14&  500.08\\
 BSk6    &  0.1575&  -15.76&  229.29& -370.96&   28.72&   27.99&   18.33&   16.81& -216.39& -215.26&  609.03&  603.72\\
\end{longtable*}
\endgroup
%\subsection{Surface Properties}
%\newpage
\begingroup
\centering
\squeezetable
\begin{longtable*}{@{\extracolsep{\fill}}lccccccc}
\caption{\small Nuclear structural properties of Skyrme models. The surface parameters $\sigma_o, \sigma_\delta$ and $S_S$ are calculated using a semi-infinite interface from Eqs. (\ref{eq:sigma0}),  (\ref{eq:sigmadelta2}), and (\ref{eq:esss}), respectively.  The neutron skin thicknesses $r_{np}$ and dipole polarizabilities $\alpha_D$ are calculated using the droplet model and the hydrodynamical model from Eqs. (\ref{eq:rnrp}) and (\ref{eq:dipole}), respectively. Also, the surface correction for the neutron skin thicknesses from Eq. (\ref{eq: surcrn}) has been included.}\label{Tab:SurSkyrme}\\
\hline\hline
\endfirsthead
\caption* {\textbf{TABLE \ref{Tab:SurSkyrme}. (\rm{\textit{Continued.}}) }}\\
\hline\hline
Model&$\sigma_o$&$\sigma_\delta$&$S_S$&$r_{np}(\ce{^{208}Pb})$&$r_{np}(\ce{^{48}Ca})$&$\alpha_{D}(\ce{^{208}Pb})$&$\alpha_{D}(\ce{^{48}Ca})$\\ 
& $\rm{MeV\;fm^{-2}}$ & $\rm{MeV\;fm^{-2}}$ & $\rm{MeV}$ & $\rm {fm}$ & $\rm {fm}$ & $\rm{fm^3}$ & $\rm{fm^3}$\\ \hline
\endhead
\hline
\endfoot
\hline\hline
\endlastfoot
Model&$\sigma_o$&$\sigma_\delta$&$S_S$&$r_{np}(\ce{^{208}Pb})$&$r_{np}(\ce{^{48}Ca})$&$\alpha_{D}(\ce{^{208}Pb})$&$\alpha_{D}(\ce{^{48}Ca})$\\
& $\rm{MeV\;fm^{-2}}$ & $\rm{MeV\;fm^{-2}}$ & $\rm{MeV}$ & $\rm {fm}$ & $\rm {fm}$ & $\rm{fm^3}$ & $\rm{fm^3}$\\ \hline
 BSk1    &   0.909&   1.539&   25.53&   0.089&   0.096&  21.936&   2.151\\
 BSk2    &   0.908&   1.624&   26.92&   0.093&   0.099&  21.984&   2.165\\
 BSk2'   &   0.906&   1.618&   26.82&   0.093&   0.099&  21.964&   2.162\\
 BSk3    &   0.883&   1.558&   25.83&   0.090&   0.097&  21.853&   2.144\\
 BSk4    &   0.951&   1.786&   29.60&   0.100&   0.104&  22.449&   2.231\\
 BSk5    &   0.961&   2.155&   35.72&   0.117&   0.115&  22.790&   2.302\\
\end{longtable*}
\endgroup
%\subsection{Neutron Star Properties}
\LTcapwidth=\textwidth

\begingroup
\centering
\squeezetable
\begin{longtable*}{@{\extracolsep{\fill}}lcccccccccccrr}
\caption{\small Neutron star properties of Skyrme models. Subscripts $n$ indicate values for stars of mass $nM_\odot$. 
$-1$ indicates where $M>M_{max}$. Radii and masses have units of km and $M_\odot$, respectively. Other quantities are dimensionless.}
%The last column $M_{max}$ is the maximum value of neutron star mass in our calculation in $M_\odot$. The tidal deformability and moment of inertia are dimensionless.}
\label{Tab:NSSk}\\
\hline\hline
\endfirsthead
\caption* {\textbf{TABLE \ref{Tab:NSSk}. (\rm{\textit{Continued.}}) }}\\
\hline\hline
\rule{0pt}{8pt}
Model&$R_{1.2}$&$\Lambda_{1.2}$&$\bar{I}_{1.2}$&${\rm BE}/M_{1.2}$&$R_{1.4}$&$\Lambda_{1.4}$&$\bar{I}_{1.4}$&${\rm BE}/M_{1.4}$&$R_{1.6}$&$\Lambda_{1.6}$&$\bar{I}_{1.6}$&${\rm BE}/M_{1.6}$&$M_{max}$\\ \hline
\endhead
\hline
\endfoot
\hline\hline
\endlastfoot
\rule{0pt}{8pt}
Model&$R_{1.2}$&$\Lambda_{1.2}$&$\bar{I}_{1.2}$&${\rm BE}/M_{1.2}$&$R_{1.4}$&$\Lambda_{1.4}$&$\bar{I}_{1.4}$&${\rm BE}/M_{1.4}$&$R_{1.6}$&$\Lambda_{1.6}$&$\bar{I}_{1.6}$&${\rm BE}/M_{1.6}$&$M_{max}$\\ \hline
 BSk1     &     -1 &     -1  &     -1   &     -1   &     -1 &     -1  &     -1   &     -1   &     -1 &     -1  &     -1   &     -1   &    0.09\\
 BSk2     &     -1 &     -1  &     -1   &     -1   &     -1 &     -1  &     -1   &     -1   &     -1 &     -1  &     -1   &     -1   &    0.09\\
 BSk2'    &     -1 &     -1  &     -1   &     -1   &     -1 &     -1  &     -1   &     -1   &     -1 &     -1  &     -1   &     -1   &    0.09\\
 BSk3     &     -1 &     -1  &     -1   &     -1   &     -1 &     -1  &     -1   &     -1   &     -1 &     -1  &     -1   &     -1   &    0.09\\
 BSk4     &     -1 &     -1  &     -1   &     -1   &     -1 &     -1  &     -1   &     -1   &     -1 &     -1  &     -1   &     -1   &    0.95\\
 BSk5     &    6.62&     11.5&      5.83&     0.159&     -1 &     -1  &     -1   &     -1   &     -1 &     -1  &     -1   &     -1   &    1.20\\
 BSk6     &    8.89&    160.7&      9.89&     0.095&    8.74&     46.9&      7.54&     0.122&    8.27&      9.4&      5.68&     0.162&    1.64\\
\end{longtable*}
\endgroup

\newpage
\section{Parameters and Properties for RMF and RMF-PC models}
%\subsection{Parameters}
%\subsection{RMF Models}
\begingroup
\centering
\squeezetable
\begin{longtable*}{@{\extracolsep{\fill}}lccccccccr}
\caption{\small Parameters of RMF models.  Masses, $A$, $\alpha_1$ and $\alpha_2$ are in MeV.  Other quantities %$g$'s, $B$, $C$, $\alpha_1^\prime$, $\alpha_2^\prime$ and $\alpha_3^\prime$ 
are dimensionless.  
}\label{Tab:20}\\
\hline\hline
\endfirsthead
\caption* {\textbf{TABLE \ref{Tab:20}. (\rm{\textit{Continued.}}) }}\\
\hline\hline
 Model&$M_n$&$m_\sigma$&$m_\omega$&$m_\rho$&$m_\delta$&$g_\sigma$&$g_\omega$&$g_\rho$&$g_\delta$\\
      &$A$&$B$&$C$&$\alpha_1$&$\alpha_1^\prime$&$\alpha_2$&$\alpha_2^\prime$&$\alpha_3^\prime$ \\\hline
\endhead
\hline
\endfoot
\hline\hline
\endlastfoot
 Model&$M_n$&$m_\sigma$&$m_\omega$&$m_\rho$&$m_\delta$&$g_\sigma$&$g_\omega$&$g_\rho$&$g_\delta$\\
      &$A$&$B$&$C$&$\alpha_1$&$\alpha_1^\prime$&$\alpha_2$&$\alpha_2^\prime$&$\alpha_3^\prime$ \\\hline
\multicolumn{10}{c}{linear finite range models (type 1)} \\
\hline
H1\cite{haddad1999density} & 939 & 550 & 783 & 770 & - & 11.079 & 13.806 & 5.258 & - \\
   & - & - & - & - & - & - & - & - \\
L1\cite{reinhard1989relativistic} & 938 & 550 & 783 & - & - & 10.3 & 12.6 & - & -\\
& - & - & - & - & - & - & - & - \\
L2\cite{reinhard1989relativistic} & 938 & 546.94 & 780 & - & - & 11.397 & 14.248 & - & -\\
\end{longtable*}
\endgroup
%\subsection{Density-dependent Models}

\begingroup
\centering
\squeezetable
\begin{longtable*}{@{\extracolsep{\fill}}lcccccccrr}
\caption{\small Parameters of density-dependent RMF (type 5) models. Masses are in $\rm MeV$ and $n_0$ has units of fm$^{-3}$. Other entries are dimensionless.}\label{Tab:PDD}\\
\hline\hline
\endfirsthead
\caption* {\textbf{TABLE \ref{Tab:PDD}. (\rm{\textit{Continued.}}) }}\\
\hline\hline
 Model&$M_n$&$m_\sigma$&$m_\omega$&$m_\rho$&$\Gamma_\sigma(n_0)$&$a_\sigma$&$b_\sigma$&$c_\sigma$&$d_\sigma$\\
      &$\Gamma_\omega(n_0)$&$a_\omega$&$b_\omega$&$c_\omega$&$d_\omega$&$\Gamma_\rho(n_0)$&$a_\rho$&$n_0$ \\\hline
\endhead
\hline
\endfoot
\hline\hline
\endlastfoot
 Model&$M_n$&$m_\sigma$&$m_\omega$&$m_\rho$&$\Gamma_\sigma(n_0)$&$a_\sigma$&$b_\sigma$&$c_\sigma$&$d_\sigma$\\
      &$\Gamma_\omega(n_0)$&$a_\omega$&$b_\omega$&$c_\omega$&$d_\omega$&$\Gamma_\rho(n_0)$&$a_\rho$&$n_0$ \\\hline
DD\cite{typel2005relativistic}  &   939.0&   547.2&   783.0&   763.0&  10.685&  1.3715&  0.6441&  1.0346&  0.5676 \\
    &13.312&  1.3856&  0.5217&  0.8700&  0.6190&   7.278&  0.4987&  0.1487\\
DDF\cite{klahn2006constraints}   &   939.0&   555.0&   783.0&   763.0&  11.024&  1.4867&  0.1956&  0.4282&  0.8823 \\
      &13.575&  1.5449&  0.1838&  0.4397&  0.8707&   7.290&  0.4479&  0.1469\\
DD-ME1\cite{nikvsic2002relativistic}   &   939.0&   549.5&   783.0&   763.0&  10.443&  1.3854&  0.9781&  1.5342&  0.4661 \\
         &12.894&  1.3879&  0.8525&  1.3566&  0.4957&   7.611&  0.5008&  0.1520\\
\end{longtable*}
\endgroup
%\subsection{RMF-PC Models}

\newpage
\begingroup
\centering
\squeezetable
\begin{longtable*}{@{\extracolsep{\fill}}lcccccc}
\caption{\small Parameters of RMF-PC (type 6) models.}\label{Tab:PPC}\\
\hline\hline
\endfirsthead
\caption* {\textbf{TABLE \ref{Tab:PPC}. (\rm{\textit{Continued.}}) }}\\
\hline\hline
 Model&$M_n\,\left({\rm MeV}\right)$&$\alpha_{\mbox{\tiny S}}\,\left({\rm MeV^{-2}}\right)$&$\beta_{\mbox{\tiny S}}\,\left({\rm MeV^{-5}}\right)$&$\gamma_{\mbox{\tiny S}}\,\left({\rm MeV^{-8}}\right)$&$\alpha_{\mbox{\tiny V}}\,\left({\rm MeV^{-2}}\right)$&$\gamma_{\mbox{\tiny V}}\,\left({\rm MeV^{-8}}\right)$\\
      &$\alpha_{\mbox{\tiny TV}}\,\left({\rm MeV^{-2}}\right)$&$\gamma_{\mbox{\tiny TV}}\,\left({\rm MeV^{-8}}\right)$&$\alpha_{\mbox{\tiny TS}}\,\left({\rm MeV^{-2}}\right)$&$\eta_1\,\left({\rm MeV^{-5}}\right)$&$\eta_2\,\left({\rm MeV^{-8}}\right)$&$\eta_3\,\left({\rm MeV^{-5}}\right)$\\\hline
\endhead
\hline
\endfoot
\hline\hline
\endlastfoot
 Model&$M_n\,\left({\rm MeV}\right)$&$\alpha_{\mbox{\tiny S}}\,\left({\rm MeV^{-2}}\right)$&$\beta_{\mbox{\tiny S}}\,\left({\rm MeV^{-5}}\right)$&$\gamma_{\mbox{\tiny S}}\,\left({\rm MeV^{-8}}\right)$&$\alpha_{\mbox{\tiny V}}\,\left({\rm MeV^{-2}}\right)$&$\gamma_{\mbox{\tiny V}}\,\left({\rm MeV^{-8}}\right)$\\
      &$\alpha_{\mbox{\tiny TV}}\,\left({\rm MeV^{-2}}\right)$&$\gamma_{\mbox{\tiny TV}}\,\left({\rm MeV^{-8}}\right)$&$\alpha_{\mbox{\tiny TS}}\,\left({\rm MeV^{-2}}\right)$&$\eta_1\,\left({\rm MeV^{-5}}\right)$&$\eta_2\,\left({\rm MeV^{-8}}\right)$&$\eta_3\,\left({\rm MeV^{-5}}\right)$\\\hline
 FZ0\cite{rusnak1997relativistic} & 939 & $-4.371\cdot10^{-4}$ & - & - & $3.373\cdot10^{-4}$ & - \\
     & $3.842\cdot10^{-5}$ & - & - & - & - & -\\
     
 PC-F1\cite{burvenich2002nuclear} & 939 & $-3.836\cdot10^{-4}$ & $7.686\cdot10^{-11}$ & $-2.904\cdot10^{-17}$ & $2.593\cdot10^{-4}$ & $-3.879\cdot10^{-18}$ \\
       & $3.468\cdot10^{-5}$ & - & - & - & - & -\\
       
 PC-F2\cite{burvenich2002nuclear} & 939 & $-3.836\cdot10^{-4}$ & $7.684\cdot10^{-11}$ & $-2.911\cdot10^{-17}$ & $2.594\cdot10^{-4}$ & $-3.823\cdot10^{-18}$ \\
       & $3.241\cdot10^{-5}$ & - & $2.34\cdot10^{-6}$ & - & - & -\\
\end{longtable*}
\endgroup
%\subsection{Saturation Properties}
\LTcapwidth=\textwidth
\begingroup
\centering
\squeezetable
\begin{longtable*}{@{\extracolsep{\fill}}lccccccccrrrr}
\caption{\small The same as Table \ref{Tab:1}, except for RMF and RMF-PC models.}\label{Tab:5}\\
\hline\hline
\endfirsthead
\caption* {\textbf{TABLE \ref{Tab:5}. (\rm{\textit{Continued.}}) }}\\
\hline\hline
Model&$n_0$&$E_0$&$K_0$&$Q_0$&$J_1$&$J_2$&$L_1$&$L_2$&\multicolumn{1}{c}{$K_{sym,1}$}&\multicolumn{1}{c}{$K_{sym,2}$}&\multicolumn{1}{c}{$Q_{sym,1}$}&\multicolumn{1}{c}{$Q_{sym,2}$}\\ \hline
\endhead
\hline
\endfoot
\hline\hline
\endlastfoot
Model&$n_0$&$E_0$&$K_0$&$Q_0$&$J_1$&$J_2$&$L_1$&$L_2$&\multicolumn{1}{c}{$K_{sym,1}$}&\multicolumn{1}{c}{$K_{sym,2}$}&\multicolumn{1}{c}{$Q_{sym,1}$}&\multicolumn{1}{c}{$Q_{sym,2}$}\\ \hline
\multicolumn{13}{c}{linear finite range models (type 1)} \\
\hline
 H1      &  0.1485&  -15.75&  547.98& 2156.48&   27.03&   25.95&   91.90&   88.46&   95.36&   93.56& -500.97& -541.41\\
 L1      &  0.1767&  -18.52&  626.43& 2231.43&   22.91&   21.69&   79.56&   75.69&   85.46&   81.83& -709.62& -788.55\\
 L2      &  0.1418&  -16.78&  579.31& 2325.86&   20.14&   19.08&   72.21&   68.85&   99.22&   97.43& -559.89& -607.50\\
 L3      &  0.1345&  -18.24&  625.85& 2573.09&   19.91&   18.87&   72.94&   69.59&  104.01&  102.15& -653.15& -713.38\\
 LBF     &  0.1525&  -17.01&  584.53& 2262.38&   48.04&   46.93&  155.39&  151.94&   94.82&   92.58& -587.49& -641.77\\
 LHS     &  0.1484&  -15.79&  549.20& 2161.57&   36.09&   35.01&  119.12&  115.68&   95.43&   93.62& -503.55& -544.35\\
\end{longtable*}
 \endgroup
%\subsection{Surface Properties}
\begingroup
\centering
\squeezetable
\begin{longtable}{@{\extracolsep{\fill}}lccccccc}
\caption{\small The same as Table \ref{Tab:SurSkyrme} except for RMF models.
%The surface tension $\sigma_0$ and $\sigma_\delta$ are in $\rm{MeV\;fm^{-2}}$ and the surface energy parameter $S_S$ is in $\rm{MeV}$; the neutron skin thickness is in $\rm {fm}$ and the dipole polarizability is in $\rm{fm^2\;MeV^{-1}}$.
}\label{Tab:SurRMF}\\
\hline\hline
\endfirsthead
\caption* {\textbf{TABLE \ref{Tab:SurRMF}. (\rm{\textit{Continued.}}) }}\\
\hline\hline
Model&$\sigma_o$&$\sigma_\delta$&$S_S$&$r_{np}(\ce{^{208}Pb})$&$r_{np}(\ce{^{48}Ca})$&$\alpha_{D}(\ce{^{208}Pb})$&$\alpha_{D}(\ce{^{48}Ca})$\\
& $\rm{MeV\;fm^{-2}}$ & $\rm{MeV\;fm^{-2}}$ & $\rm{MeV}$ & $\rm {fm}$ & $\rm {fm}$ & $\rm{fm^3}$ & $\rm{fm^3}$\\ \hline
\endhead
\hline
\endfoot
\hline\hline
\endlastfoot
Model&$\sigma_o$&$\sigma_\delta$&$S_S$&$r_{np}(\ce{^{208}Pb})$&$r_{np}(\ce{^{48}Ca})$&$\alpha_{D}(\ce{^{208}Pb})$&$\alpha_{D}(\ce{^{48}Ca})$\\
& $\rm{MeV\;fm^{-2}}$ & $\rm{MeV\;fm^{-2}}$ & $\rm{MeV}$ & $\rm {fm}$ & $\rm {fm}$ & $\rm{fm^3}$ & $\rm{fm^3}$\\ \hline
\multicolumn{8}{c}{linear finite range models (type 1)} \\
\hline
 H1       &   1.485&   5.462&   94.20&   0.247&   0.199&  39.243&   4.492\\
 L1       &   1.930&   4.927&   75.67&   0.206&   0.180&  40.976&   4.668\\
 L2       &   1.480&   3.746&   66.62&   0.211&   0.189&  53.985&   6.151\\
 L3       &   1.749&   4.185&   77.10&   0.239&   0.205&  61.277&   7.113\\
 LBF      &   1.546&  10.133&  171.68&   0.290&   0.210&  21.397&   2.452\\
\end{longtable}
\endgroup
%\subsection{Neutron Star Propertiess}
\begingroup
\centering
\squeezetable
\begin{longtable*}{@{\extracolsep{\fill}}lccccccccccccr}
\caption{\small The same as Table \ref{Tab:NSSk} except for RMF and RMF-PC models.
%The substitutes of physical quantities $n$ symbol n times of solar mass. For instance, $R_{1.2}$ refers to the radius of neutron star with mass equal to $1.2{\rm M_\odot}$ in the unit of ${\rm km}$. The last column $M_{max}$ is the maximum value of neutron star mass in our calculation in the unit of solar mass $\rm M_\odot$. The tidal deformability and moment of inertia are dimensionless.
}\label{Tab:NSRMF}\\
\hline\hline
\endfirsthead
\caption* {\textbf{TABLE \ref{Tab:NSRMF}. (\rm{\textit{Continued.}}) }}\\
\hline\hline
\rule{0pt}{8pt}
Model&$R_{1.2}$&$\Lambda_{1.2}$&$\bar{I}_{1.2}$&${\rm BE}/M_{1.2}$&$R_{1.4}$&$\Lambda_{1.4}$&$\bar{I}_{1.4}$&${\rm BE}/M_{1.4}$&$R_{1.6}$&$\Lambda_{1.6}$&$\bar{I}_{1.6}$&${\rm BE}/M_{1.6}$&$M_{max}$\\ \hline
\endhead
\hline
\endfoot
\hline\hline
\endlastfoot
\rule{0pt}{8pt}
Model&$R_{1.2}$&$\Lambda_{1.2}$&$\bar{I}_{1.2}$&${\rm BE}/M_{1.2}$&$R_{1.4}$&$\Lambda_{1.4}$&$\bar{I}_{1.4}$&${\rm BE}/M_{1.4}$&$R_{1.6}$&$\Lambda_{1.6}$&$\bar{I}_{1.6}$&${\rm BE}/M_{1.6}$&$M_{max}$\\ \hline
\multicolumn{14}{c}{linear finite range models (type 1)} \\
\hline
 H1      &     14.38&   3377.0&     23.07&     0.047&     14.69&   1675.2&     18.49&     0.057&     14.95&    855.1&     15.24&     0.066&    3.05\\
 L1      &     13.30&   2280.0&     20.23&     0.053&     13.59&   1070.3&     16.22&     0.063&     13.82&    531.1&     13.39&     0.074&    2.79\\
 L2      &     14.45&   3465.5&     23.27&     0.047&     14.78&   1750.4&     18.72&     0.056&     15.06&    903.7&     15.48&     0.065&    3.13\\
 L3      &     14.71&   4179.8&     24.31&     0.045&     15.07&   1984.1&     19.56&     0.054&     15.38&   1056.4&     16.20&     0.063&    3.25\\
 LBF     &     14.44&   4217.2&     24.35&     0.045&     14.82&   2003.0&     19.53&     0.054&     15.14&   1039.3&     16.11&     0.063&    3.11\\
 LHS     &     14.49&   3855.5&     23.86&     0.046&     14.82&   1889.5&     19.12&     0.055&     15.11&    960.4&     15.74&     0.065&    3.08\\
\end{longtable*}
\endgroup
\newpage
\section{Parameters and Properties of Gogny Models}
%\subsection{Parameters}
%Parameters of Gogny models used in this work are listed in Table IV.
\begingroup
\centering
\squeezetable
\begin{table*}[h]
\caption {\small Parameters of Gogny models. For each interaction, the first row corresponds quantities with subscript $i = 1$ and the second row to $i = 2$. $W_i$, $B_i$, $H_i$ and $M_i$ are in $\rm{MeV}$, $\mu_i$ in $\rm{fm}$, $t_{3i}$ in $\rm{MeV\;fm^{3+3\alpha_i}}$, while $x_{3i}$ and $\alpha_i$ are dimensionless.   We do not discuss more complex Gogny models having parameters with $i=3$.}
\begin{ruledtabular}
\begin{tabular}{ccccccccccccc}
Model&$W_i$&$B_i$&$\mathbf{H}_i$&$M_i$&$\mu_i$&$t_{3i}$&$x_{3i}$&$\alpha_i$\\ \hline
 D1\cite{blaizot1995microscopic} & -402.400 & -100.0000 & -496.200 & -23.5600 & 0.7 & 1350.000 & 1 & 1/3\\
    & -21.3000 & -11.770 & 37.270 & -68.8100 & 1.2 & - & - & - \\
    
 D1M\cite{goriely2009first} & -12797.57 & 14048.85 & -15144.43 & 11963.81 & 0.5 & 1562.220 & 1 & 1/3\\
    & 490.9500 & -752.270 & 675.120 & -693.570 & 1.0 & - & - & - \\
    
 D1M*\cite{gonzalez2018new} & -17242.014 & 19604.4056 & -20699.986 & 16408.3344 & 0.5 & 1561.220 & 1 & 1/3\\
    & 675.386 & -982.815 & 905.665 & -878.006 & 1.0 & - & - & - \\
    
 D1MK\cite{tagami2022neutron} & -17242.014 & 19604.4056 & -20699.986 & 16408.6002 & 0.5 & 1561.7167 & 1 & 1/3\\
    & 642.600 & -941.150 & 865.572 & -845.3008 & 1.0 & 0 & -1 & 1 \\   
    
 D1N\cite{chappert2008towards} & -2047.610 & 1700.0000 & -2414.930 & 1519.3500 & 0.8 & 1609.460 & 1 & 1/3\\
    & 293.020 & -300.780 & 414.590 & -316.8400 & 1.2 & - & - & - \\   
    
 D1P\cite{farine1999towards} & -372.890 & 62.6900 & -464.510 & -31.4900 & 0.9 & 1025.900 & 1.16 & 1/3\\
    & 34.6200 & -14.080 & 70.950 & -20.9600 & 1.44 & 256.020 & -2.007 & 0.92 \\ 
    
 D1PK\cite{tagami2022neutron} & -465.028 & 155.1345 & -506.775 & 117.7499 & 0.9 & 981.0654 & 1 & 1/3\\
    & 34.6200 & -14.080 & 70.950 & -41.3518 & 1.44 & 534.1557 & -1 & 1 \\ 
    
 D1S\cite{blaizot1995microscopic} & -1720.300 & 1300.00 & -1813.530 & 1397.600 & 0.7 & 1390.600 & 1 & 1/3\\
    & 103.640 & -163.480 & 162.810 & -223.930 & 1.2 & - & - & - \\   
    
 D250\cite{blaizot1995microscopic} & -1045.960 & 900.00 & -1127.639 & 1009.162 & 0.7 & 1350.000 & 1 & 2/3\\
    & 45.731 & -121.465 & 102.549 & -184.271 & 1.2 & - & - & - \\   
    
 D260\cite{blaizot1995microscopic} & 1396.200 & -2000.00 & 1232.414 & -1972.945 & 0.7 & 1300.500 & 1 & 1/3\\
    & -171.827 & 174.285 & -105.179 & 121.308 & 1.2 & - & - & - \\   
    
 D280\cite{blaizot1995microscopic} & 1689.093 & -2000.00 & 1097.591 & -2400.662 & 0.7 & 1301.000 & 1 & 1/3\\
    & -194.420 & 190.813 & -89.561 & 176.047 & 1.2 & - & - & - \\   
    
 D300\cite{blaizot1995microscopic} & 673.794 & -1000.00 & 446.057 & -1036.897 & 0.7 & 1450.000 & 1 & 2/3\\
    & -115.173 & 103.365 & -43.235 & 55.678 & 1.2 & - & - & - \\ 
    
 GT2\cite{otsuka2006mean} & 2311.000 & -3480.00 & 2962.00 & -2800.00 & 0.7 & 1400.000 & 1 & 1/3\\
    & -339.000 & 388.000 & -370.000 & 260.000 & 1.2 & - & - & - \\  
\end{tabular}
\end{ruledtabular}
\end{table*}
\endgroup
%\subsection{Saturation Properties}
\begingroup
\centering
\squeezetable
\begin{table*}[h]
\caption {\small The same as Table \ref{Tab:1} except for Gogny models.}
\begin{ruledtabular}
\begin{tabular}{ccccccccccccc}
 Model&$ n_0$&$E_0$&$K_0$&$Q_0$&$J_1$&$J_2$&$L_1$&$L_2$&$K_{sym1}$&$K_{sym2}$&$Q_{sym1}$&$Q_{sym2}$\\ \hline
 D1 & 0.1665 & -16.32 & 229.35 & -473.47 & 31.90 & 30.69 & 21.11 & 18.33 & -277.61 & -274.62 & 615.98 & 616.79\\
 D1M & 0.1647 & -16.04 & 224.75 & -456.81 & 29.72 & 28.54 & 24.61 & 24.82 & -157.87 & -133.21 & 806.61 & 735.63\\
 D1M* & 0.1650 & -16.07 & 225.20 & -458.33 & 31.24 & 30.24 & 41.19 & 43.16 & -80.22 & -47.09 & 811.72 & 706.21\\
 D1MK & 0.1650 & -16.08 & 225.24 & -458.44 & 33.85 & 32.99 & 52.41 & 54.86 & -69.71 & -36.83 & 766.16 & 664.14\\
 D1N & 0.1613 & -15.98 & 225.76 & -449.23 & 30.13 & 29.59 & 31.92 & 33.55 & -183.85 & -168.56 & 518.82 & 440.17\\
 D1P & 0.1698 & -15.27 & 254.10 & -329.36 & 33.98 & 32.75 & 53.08 & 50.26 & -164.08 & -159.35 & 406.62 & 408.31\\
 D1PK & 0.1633 & -16.02 & 260.01 & -316.42 & 33.85 & 32.99 & 56.53 & 54.91 & -152.69 & -149.82 & 402.40 & 397.27\\
 D1S & 0.1633 & -16.02 & 202.82 & -544.42 & 31.94 & 31.12 & 22.23 & 22.41 & -252.95 & -241.50 & 682.29 & 644.22\\
 D250 & 0.1580 & -15.86 & 249.71 & -387.56 & 32.35 & 31.56 & 24.81 & 24.81 & -299.17 & -289.33 & 513.59 & 484.31\\
 D260 & 0.1601 & -16.27 & 259.10 & -362.01 & 31.84 & 30.10 & 24.28 & 17.58 & -290.81 & -298.60 & 487.57 & 539.16\\
 D280 & 0.1525 & -16.35 & 284.80 & -268.80 & 34.87 & 33.13 & 53.19 & 46.52 & -205.92 & -211.83 & 288.67 & 326.06\\
 D300 & 0.1562 & -16.23 & 298.98 & -232.31 & 32.43 & 31.22 & 29.75 & 25.83 & -312.50 & -315.05 & 336.25 & 359.50\\
 GT2 & 0.1612 & -16.03 & 227.88 & -453.45 & 35.42 & 33.93 & 12.75 & 5.013 & -423.72 & -445.80 & 617.85 & 740.97\\
\end{tabular}
\end{ruledtabular}
\end{table*}
\endgroup
%\subsection{Neutron Star Properties}
\begingroup
\centering
\squeezetable
%\end{widetext}
%\bibliography{Database}% Produces the bibliography via BibTeX.
%\end{document}
\begin{table*}[h]
\caption{\small The same as Table \ref{Tab:NSSk} except for Gogny models. 
%The substitutes of physical quantities $n$ symbol n times of solar mass. For instance, $R_{1.2}$ refers to the radius of neutron star with mass equal to $1.2M_\odot$ in unit $km$. The last column $M_{max}$ is the maximum value of neutron star mass in our calculation in $M_\odot$. The tidal deformability and moment of inertia are dimensionless.
}
%\section*{Bibliography}
\begin{ruledtabular}
\begin{tabular}{cccccccccccccc}
Model&$R_{1.2}$&$\Lambda_{1.2}$&$\bar{I}_{1.2}$&${\rm BE}/M_{1.2}$&$R_{1.4}$&$\Lambda_{1.4}$&$\bar{I}_{1.4}$&${\rm BE}/M_{1.4}$&$R_{1.6}$&$\Lambda_{1.6}$&$\bar{I}_{1.6}$&${\rm BE}/M_{1.6}$&$M_{max}$\\ \hline
 D1       &     -1 &     -1  &     -1   &     -1   &     -1 &     -1  &     -1   &     -1   &     -1 &     -1  &     -1   &     -1   &    0.0081\\
 D1M      &    9.97&    347.4&     12.00&     0.081&    9.88&    116.4&      9.17&     0.102&    9.61&     33.9&      7.07&     0.129&    1.74\\
 D1M*     &   11.32&    816.7&     15.07&     0.067&   11.36&    316.9&     11.69&     0.083&   11.32&    123.4&      9.30&     0.101&    2.04\\
 D1MK     &   11.79&   1007.2&     15.93&     0.064&   11.81&    379.2&     12.30&     0.079&   11.75&    152.1&      9.77&     0.097&    2.08\\
 D1N      &    8.14&     59.3&      7.86&     0.119&     -1 &     -1  &     -1   &     -1   &     -1 &     -1  &     -1   &     -1   &    1.23\\
 D1P      &   11.09&    606.1&     13.92&     0.072&   10.95&    210.2&     10.53&     0.091&   10.69&     69.1&      8.16&     0.113&    1.89\\
 D1PK     &   11.63&    888.3&     15.39&     0.066&   11.59&    328.3&     11.78&     0.083&   11.46&    123.0&      9.28&     0.101&    2.07\\
 D1S      &     -1 &     -1  &     -1   &     -1   &     -1 &     -1  &     -1   &     -1   &     -1 &     -1  &     -1   &     -1   &    0.016\\
 D250     &     -1 &     -1  &     -1   &     -1   &     -1 &     -1  &     -1   &     -1   &     -1 &     -1  &     -1   &     -1   &    0.0001\\
 D260     &     -1 &     -1  &     -1   &     -1   &     -1 &     -1  &     -1   &     -1   &     -1 &     -1  &     -1   &     -1   &    0.014\\
 D280     &   11.33&    655.1&     14.16&     0.071&   10.98&    194.2&     10.32&     0.093&   10.22&     41.4&      7.32&     0.127&    1.65\\
 D300     &     -1 &     -1  &     -1   &     -1   &     -1 &     -1  &     -1   &     -1   &     -1 &     -1  &     -1   &     -1   &    0.72\\
 GT2      &     -1 &     -1  &     -1   &     -1   &     -1 &     -1  &     -1   &     -1   &     -1 &     -1  &     -1   &     -1   &    0.0007\\
\end{tabular}
\end{ruledtabular}
\end{table*}
\endgroup
\nocite{*}
\small
%\section*{Bibliography}
\end{widetext}
%\fi
\newpage
\bibliography{Database}% Produces the bibliography via BibTeX.

\begin{thebibliography}{164}
\expandafter\ifx\csname natexlab\endcsname\relax\def\natexlab#1{#1}\fi
\expandafter\ifx\csname bibnamefont\endcsname\relax
  \def\bibnamefont#1{#1}\fi
\expandafter\ifx\csname bibfnamefont\endcsname\relax
  \def\bibfnamefont#1{#1}\fi
\expandafter\ifx\csname citenamefont\endcsname\relax
  \def\citenamefont#1{#1}\fi
\expandafter\ifx\csname url\endcsname\relax
  \def\url#1{\texttt{#1}}\fi
\expandafter\ifx\csname urlprefix\endcsname\relax\def\urlprefix{URL }\fi
\providecommand{\bibinfo}[2]{#2}
\providecommand{\eprint}[2][]{\url{#2}}

\bibitem[{\citenamefont{Dutra et~al.}(2012)\citenamefont{Dutra, Louren\ifmmode~\mbox{\c{c}}\else \c{c}\fi{}o, S\'a~Martins, Delfino, Stone, and Stevenson}}]{dutra2012skyrme}
\bibinfo{author}{\bibfnamefont{M.}~\bibnamefont{Dutra}}, \bibinfo{author}{\bibfnamefont{O.}~\bibnamefont{Louren\ifmmode~\mbox{\c{c}}\else \c{c}\fi{}o}}, \bibinfo{author}{\bibfnamefont{J.~S.} \bibnamefont{S\'a~Martins}}, \bibinfo{author}{\bibfnamefont{A.}~\bibnamefont{Delfino}}, \bibinfo{author}{\bibfnamefont{J.~R.} \bibnamefont{Stone}}, \bibnamefont{and} \bibinfo{author}{\bibfnamefont{P.~D.} \bibnamefont{Stevenson}}, \bibinfo{journal}{Phys. Rev. C} \textbf{\bibinfo{volume}{85}}, \bibinfo{pages}{035201} (\bibinfo{year}{2012}), \urlprefix\url{https://link.aps.org/doi/10.1103/PhysRevC.85.035201}.

\bibitem[{\citenamefont{Dutra et~al.}(2014)\citenamefont{Dutra, Louren{\c{c}}o, Avancini, Carlson, Delfino, Menezes, Provid{\^e}ncia, Typel, and Stone}}]{dutra2014relativistic}
\bibinfo{author}{\bibfnamefont{M.}~\bibnamefont{Dutra}}, \bibinfo{author}{\bibfnamefont{O.}~\bibnamefont{Louren{\c{c}}o}}, \bibinfo{author}{\bibfnamefont{S.}~\bibnamefont{Avancini}}, \bibinfo{author}{\bibfnamefont{B.~V.} \bibnamefont{Carlson}}, \bibinfo{author}{\bibfnamefont{A.}~\bibnamefont{Delfino}}, \bibinfo{author}{\bibfnamefont{D.~P.} \bibnamefont{Menezes}}, \bibinfo{author}{\bibfnamefont{C.}~\bibnamefont{Provid{\^e}ncia}}, \bibinfo{author}{\bibfnamefont{S.}~\bibnamefont{Typel}}, \bibnamefont{and} \bibinfo{author}{\bibfnamefont{J.~R.} \bibnamefont{Stone}}, \bibinfo{journal}{Phys. Rev. C} \textbf{\bibinfo{volume}{90}}, \bibinfo{pages}{055203} (\bibinfo{year}{2014}).

\bibitem[{\citenamefont{Carlson et~al.}(2023)\citenamefont{Carlson, Dutra, Louren\ifmmode~\mbox{\c{c}}\else \c{c}\fi{}o, and Margueron}}]{PhysRevC.107.035805}
\bibinfo{author}{\bibfnamefont{B.~V.} \bibnamefont{Carlson}}, \bibinfo{author}{\bibfnamefont{M.}~\bibnamefont{Dutra}}, \bibinfo{author}{\bibfnamefont{O.}~\bibnamefont{Louren\ifmmode~\mbox{\c{c}}\else \c{c}\fi{}o}}, \bibnamefont{and} \bibinfo{author}{\bibfnamefont{J.}~\bibnamefont{Margueron}}, \bibinfo{journal}{Phys. Rev. C} \textbf{\bibinfo{volume}{107}}, \bibinfo{pages}{035805} (\bibinfo{year}{2023}), \urlprefix\url{https://link.aps.org/doi/10.1103/PhysRevC.107.035805}.

\bibitem[{\citenamefont{Typel}(2018)}]{typel2018relativistic}
\bibinfo{author}{\bibfnamefont{S.}~\bibnamefont{Typel}}, \bibinfo{journal}{Particles} \textbf{\bibinfo{volume}{1}}, \bibinfo{pages}{3} (\bibinfo{year}{2018}).

\bibitem[{\citenamefont{Vautherin and Brink}(1972)}]{vautherin1972hartree}
\bibinfo{author}{\bibfnamefont{D.}~\bibnamefont{Vautherin}} \bibnamefont{and} \bibinfo{author}{\bibfnamefont{D.~M.} \bibnamefont{Brink}}, \bibinfo{journal}{Phys. Rev. C} \textbf{\bibinfo{volume}{5}}, \bibinfo{pages}{626} (\bibinfo{year}{1972}).

\bibitem[{\citenamefont{Agrawal et~al.}(2006)\citenamefont{Agrawal, Dhiman, and Kumar}}]{agrawal2006exploring}
\bibinfo{author}{\bibfnamefont{B.~K.} \bibnamefont{Agrawal}}, \bibinfo{author}{\bibfnamefont{S.~K.} \bibnamefont{Dhiman}}, \bibnamefont{and} \bibinfo{author}{\bibfnamefont{R.}~\bibnamefont{Kumar}}, \bibinfo{journal}{Phys. Rev. C} \textbf{\bibinfo{volume}{73}}, \bibinfo{pages}{034319} (\bibinfo{year}{2006}).

\bibitem[{\citenamefont{Chamel et~al.}(2008)\citenamefont{Chamel, Goriely, and Pearson}}]{chamel2008further}
\bibinfo{author}{\bibfnamefont{N.}~\bibnamefont{Chamel}}, \bibinfo{author}{\bibfnamefont{S.}~\bibnamefont{Goriely}}, \bibnamefont{and} \bibinfo{author}{\bibfnamefont{J.~M.} \bibnamefont{Pearson}}, \bibinfo{journal}{Nucl. Phys. A} \textbf{\bibinfo{volume}{812}}, \bibinfo{pages}{72} (\bibinfo{year}{2008}).

\bibitem[{\citenamefont{Chamel et~al.}(2009)\citenamefont{Chamel, Goriely, and Pearson}}]{chamel2009further}
\bibinfo{author}{\bibfnamefont{N.}~\bibnamefont{Chamel}}, \bibinfo{author}{\bibfnamefont{S.}~\bibnamefont{Goriely}}, \bibnamefont{and} \bibinfo{author}{\bibfnamefont{J.~M.} \bibnamefont{Pearson}}, \bibinfo{journal}{Phys. Rev. C} \textbf{\bibinfo{volume}{80}}, \bibinfo{pages}{065804} (\bibinfo{year}{2009}).

\bibitem[{\citenamefont{Steiner et~al.}(2005)\citenamefont{Steiner, Prakash, Lattimer, and Ellis}}]{steiner2005isospin}
\bibinfo{author}{\bibfnamefont{A.~W.} \bibnamefont{Steiner}}, \bibinfo{author}{\bibfnamefont{M.}~\bibnamefont{Prakash}}, \bibinfo{author}{\bibfnamefont{J.~M.} \bibnamefont{Lattimer}}, \bibnamefont{and} \bibinfo{author}{\bibfnamefont{P.~J.} \bibnamefont{Ellis}}, \bibinfo{journal}{Phys. Rep.} \textbf{\bibinfo{volume}{411}}, \bibinfo{pages}{325} (\bibinfo{year}{2005}).

\bibitem[{\citenamefont{Nikolaus et~al.}(1992)\citenamefont{Nikolaus, Hoch, and Madland}}]{nikolaus1992nuclear}
\bibinfo{author}{\bibfnamefont{B.~A.} \bibnamefont{Nikolaus}}, \bibinfo{author}{\bibfnamefont{T.}~\bibnamefont{Hoch}}, \bibnamefont{and} \bibinfo{author}{\bibfnamefont{D.~G.} \bibnamefont{Madland}}, \bibinfo{journal}{Phys. Rev. C} \textbf{\bibinfo{volume}{46}}, \bibinfo{pages}{1757} (\bibinfo{year}{1992}).

\bibitem[{\citenamefont{Gonzalez-Boquera et~al.}(2017)\citenamefont{Gonzalez-Boquera, Centelles, Vi{\~n}as, and Rios}}]{gonzalez2017higher}
\bibinfo{author}{\bibfnamefont{C.}~\bibnamefont{Gonzalez-Boquera}}, \bibinfo{author}{\bibfnamefont{M.}~\bibnamefont{Centelles}}, \bibinfo{author}{\bibfnamefont{X.}~\bibnamefont{Vi{\~n}as}}, \bibnamefont{and} \bibinfo{author}{\bibfnamefont{A.}~\bibnamefont{Rios}}, \bibinfo{journal}{Phys. Rev. C} \textbf{\bibinfo{volume}{96}}, \bibinfo{pages}{065806} (\bibinfo{year}{2017}).

\bibitem[{\citenamefont{Lattimer}(2023)}]{Lattimer_2023}
\bibinfo{author}{\bibfnamefont{J.~M.} \bibnamefont{Lattimer}}, \bibinfo{journal}{Particles} \textbf{\bibinfo{volume}{6}}, \bibinfo{pages}{30} (\bibinfo{year}{2023}), \urlprefix\url{https://doi.org/10.3390%2Fparticles6010003}.

\bibitem[{\citenamefont{Myers and Swiatecki}(1980)}]{myers1980droplet}
\bibinfo{author}{\bibfnamefont{W.~D.} \bibnamefont{Myers}} \bibnamefont{and} \bibinfo{author}{\bibfnamefont{W.~J.} \bibnamefont{Swiatecki}}, \bibinfo{journal}{Nucl. Phys. A} \textbf{\bibinfo{volume}{336}}, \bibinfo{pages}{267} (\bibinfo{year}{1980}).

\bibitem[{\citenamefont{Myers and Swiatecki}(1974)}]{Myersswiatecki1974}
\bibinfo{author}{\bibfnamefont{W.~D.} \bibnamefont{Myers}} \bibnamefont{and} \bibinfo{author}{\bibfnamefont{W.~J.} \bibnamefont{Swiatecki}}, \bibinfo{journal}{Ann. Phys. (NY)} \textbf{\bibinfo{volume}{84}}, \bibinfo{pages}{186} (\bibinfo{year}{1974}).

\bibitem[{\citenamefont{Lattimer and Prakash}(2016)}]{lattimer2016equation}
\bibinfo{author}{\bibfnamefont{J.~M.} \bibnamefont{Lattimer}} \bibnamefont{and} \bibinfo{author}{\bibfnamefont{M.}~\bibnamefont{Prakash}}, \bibinfo{journal}{Phys. Rep.} \textbf{\bibinfo{volume}{621}}, \bibinfo{pages}{127} (\bibinfo{year}{2016}).

\bibitem[{\citenamefont{Warda et~al.}(2009)\citenamefont{Warda, Vi\~nas, Roca-Maza, and Centelles}}]{PhysRevC.80.024316}
\bibinfo{author}{\bibfnamefont{M.}~\bibnamefont{Warda}}, \bibinfo{author}{\bibfnamefont{X.}~\bibnamefont{Vi\~nas}}, \bibinfo{author}{\bibfnamefont{X.}~\bibnamefont{Roca-Maza}}, \bibnamefont{and} \bibinfo{author}{\bibfnamefont{M.}~\bibnamefont{Centelles}}, \bibinfo{journal}{Phys. Rev. C} \textbf{\bibinfo{volume}{80}}, \bibinfo{pages}{024316} (\bibinfo{year}{2009}), \urlprefix\url{https://link.aps.org/doi/10.1103/PhysRevC.80.024316}.

\bibitem[{\citenamefont{Adhikari et~al.}(2021)\citenamefont{Adhikari, Albataineh, Androic, Aniol, Armstrong, Averett, Gayoso, Barcus, Bellini, Beminiwattha et~al.}}]{adhikari2021accurate}
\bibinfo{author}{\bibfnamefont{D.}~\bibnamefont{Adhikari}}, \bibinfo{author}{\bibfnamefont{H.}~\bibnamefont{Albataineh}}, \bibinfo{author}{\bibfnamefont{D.}~\bibnamefont{Androic}}, \bibinfo{author}{\bibfnamefont{K.}~\bibnamefont{Aniol}}, \bibinfo{author}{\bibfnamefont{D.~S.} \bibnamefont{Armstrong}}, \bibinfo{author}{\bibfnamefont{T.}~\bibnamefont{Averett}}, \bibinfo{author}{\bibfnamefont{C.~A.} \bibnamefont{Gayoso}}, \bibinfo{author}{\bibfnamefont{S.}~\bibnamefont{Barcus}}, \bibinfo{author}{\bibfnamefont{V.}~\bibnamefont{Bellini}}, \bibinfo{author}{\bibfnamefont{R.~S.} \bibnamefont{Beminiwattha}}, \bibnamefont{et~al.}, \bibinfo{journal}{Phys. Rev. Lett.} \textbf{\bibinfo{volume}{126}}, \bibinfo{pages}{172502} (\bibinfo{year}{2021}).

\bibitem[{\citenamefont{Adhikari et~al.}(2022)\citenamefont{Adhikari, Albataineh, Androic, Aniol, Armstrong, Averett, Gayoso, Barcus, Bellini, Beminiwattha et~al.}}]{adhikari2022precision}
\bibinfo{author}{\bibfnamefont{D.}~\bibnamefont{Adhikari}}, \bibinfo{author}{\bibfnamefont{H.}~\bibnamefont{Albataineh}}, \bibinfo{author}{\bibfnamefont{D.}~\bibnamefont{Androic}}, \bibinfo{author}{\bibfnamefont{K.~A.} \bibnamefont{Aniol}}, \bibinfo{author}{\bibfnamefont{D.~S.} \bibnamefont{Armstrong}}, \bibinfo{author}{\bibfnamefont{T.}~\bibnamefont{Averett}}, \bibinfo{author}{\bibfnamefont{C.~A.} \bibnamefont{Gayoso}}, \bibinfo{author}{\bibfnamefont{S.~K.} \bibnamefont{Barcus}}, \bibinfo{author}{\bibfnamefont{V.}~\bibnamefont{Bellini}}, \bibinfo{author}{\bibfnamefont{R.~S.} \bibnamefont{Beminiwattha}}, \bibnamefont{et~al.}, \bibinfo{journal}{Phys. Rev. Lett.} \textbf{\bibinfo{volume}{129}}, \bibinfo{pages}{042501} (\bibinfo{year}{2022}).

\bibitem[{\citenamefont{Giacalone et~al.}(2023)\citenamefont{Giacalone, Nijs, and van~der Scheee}}]{Giacalone2023}
\bibinfo{author}{\bibfnamefont{G.}~\bibnamefont{Giacalone}}, \bibinfo{author}{\bibfnamefont{G.}~\bibnamefont{Nijs}}, \bibnamefont{and} \bibinfo{author}{\bibfnamefont{W.}~\bibnamefont{van~der Scheee}}, \bibinfo{journal}{Phys. Rev. Lett.} \textbf{\bibinfo{volume}{131}}, \bibinfo{pages}{202302} (\bibinfo{year}{2023}).

\bibitem[{\citenamefont{Tagami et~al.}(2022)\citenamefont{Tagami, Wakasa, Takechi, Matsui, and Yahiro}}]{tagami2022neutron}
\bibinfo{author}{\bibfnamefont{S.}~\bibnamefont{Tagami}}, \bibinfo{author}{\bibfnamefont{T.}~\bibnamefont{Wakasa}}, \bibinfo{author}{\bibfnamefont{M.}~\bibnamefont{Takechi}}, \bibinfo{author}{\bibfnamefont{J.}~\bibnamefont{Matsui}}, \bibnamefont{and} \bibinfo{author}{\bibfnamefont{M.}~\bibnamefont{Yahiro}}, \bibinfo{journal}{Results in Physics} \textbf{\bibinfo{volume}{33}}, \bibinfo{pages}{105155} (\bibinfo{year}{2022}), ISSN \bibinfo{issn}{2211-3797}.

\bibitem[{\citenamefont{Birkhan et~al.}(2011)\citenamefont{Birkhan, Miorelli, Bacca, Bassauer, Bertulani, Hagen, Matsubara, von Neumann-Cosel, Papenbrock, Pietralla et~al.}}]{Birkhan17}
\bibinfo{author}{\bibfnamefont{J.}~\bibnamefont{Birkhan}}, \bibinfo{author}{\bibfnamefont{M.}~\bibnamefont{Miorelli}}, \bibinfo{author}{\bibfnamefont{S.}~\bibnamefont{Bacca}}, \bibinfo{author}{\bibfnamefont{S.}~\bibnamefont{Bassauer}}, \bibinfo{author}{\bibfnamefont{C.}~\bibnamefont{Bertulani}}, \bibinfo{author}{\bibfnamefont{G.}~\bibnamefont{Hagen}}, \bibinfo{author}{\bibfnamefont{H.}~\bibnamefont{Matsubara}}, \bibinfo{author}{\bibfnamefont{P.}~\bibnamefont{von Neumann-Cosel}}, \bibinfo{author}{\bibfnamefont{T.}~\bibnamefont{Papenbrock}}, \bibinfo{author}{\bibfnamefont{N.}~\bibnamefont{Pietralla}}, \bibnamefont{et~al.}, \bibinfo{journal}{Phys. Rev. Lett.} \textbf{\bibinfo{volume}{118}}, \bibinfo{pages}{252501} (\bibinfo{year}{2011}).

\bibitem[{\citenamefont{Tamii}(2011)}]{Tamii11}
\bibinfo{author}{\bibfnamefont{A.}~\bibnamefont{Tamii}}, \bibinfo{journal}{Phys. Rev. Lett.} \textbf{\bibinfo{volume}{107}}, \bibinfo{pages}{062502} (\bibinfo{year}{2011}).

\bibitem[{\citenamefont{Lattimer and {D. Swesty}}(1991)}]{LATTIMER1991331}
\bibinfo{author}{\bibfnamefont{J.~M.} \bibnamefont{Lattimer}} \bibnamefont{and} \bibinfo{author}{\bibfnamefont{F.}~\bibnamefont{{D. Swesty}}}, \bibinfo{journal}{Nucl. Phys. A} \textbf{\bibinfo{volume}{535}}, \bibinfo{pages}{331} (\bibinfo{year}{1991}), ISSN \bibinfo{issn}{0375-9474}, \urlprefix\url{https://www.sciencedirect.com/science/article/pii/037594749190452C}.

\bibitem[{\citenamefont{Zhao and Lattimer}(2022)}]{Zhao_2022}
\bibinfo{author}{\bibfnamefont{T.}~\bibnamefont{Zhao}} \bibnamefont{and} \bibinfo{author}{\bibfnamefont{J.~M.} \bibnamefont{Lattimer}}, \bibinfo{journal}{Phys. Rev. D} \textbf{\bibinfo{volume}{106}}, \bibinfo{pages}{123002} (\bibinfo{year}{2022}), \urlprefix\url{https://link.aps.org/doi/10.1103/PhysRevD.106.123002}.

\bibitem[{\citenamefont{Lattimer and Prakash}(2001)}]{Lattimer_2001}
\bibinfo{author}{\bibfnamefont{J.~M.} \bibnamefont{Lattimer}} \bibnamefont{and} \bibinfo{author}{\bibfnamefont{M.}~\bibnamefont{Prakash}}, \bibinfo{journal}{Astrophys. J.} \textbf{\bibinfo{volume}{550}}, \bibinfo{pages}{426} (\bibinfo{year}{2001}), \urlprefix\url{https://dx.doi.org/10.1086/319702}.

\bibitem[{\citenamefont{{Ravenhall} and {Pethick}}(1994)}]{1994ApJ...424..846R}
\bibinfo{author}{\bibfnamefont{D.~G.} \bibnamefont{{Ravenhall}}} \bibnamefont{and} \bibinfo{author}{\bibfnamefont{C.~J.} \bibnamefont{{Pethick}}}, \bibinfo{journal}{\apj} \textbf{\bibinfo{volume}{424}}, \bibinfo{pages}{846} (\bibinfo{year}{1994}).

\bibitem[{\citenamefont{Postnikov et~al.}(2010)\citenamefont{Postnikov, Prakash, and Lattimer}}]{Postnikov_2010}
\bibinfo{author}{\bibfnamefont{S.}~\bibnamefont{Postnikov}}, \bibinfo{author}{\bibfnamefont{M.}~\bibnamefont{Prakash}}, \bibnamefont{and} \bibinfo{author}{\bibfnamefont{J.~M.} \bibnamefont{Lattimer}}, \bibinfo{journal}{Phys. Rev. D} \textbf{\bibinfo{volume}{82}}, \bibinfo{pages}{024016} (\bibinfo{year}{2010}), \urlprefix\url{https://link.aps.org/doi/10.1103/PhysRevD.82.024016}.

\bibitem[{\citenamefont{Zhao and Lattimer}(2018)}]{Zhao2018}
\bibinfo{author}{\bibfnamefont{T.}~\bibnamefont{Zhao}} \bibnamefont{and} \bibinfo{author}{\bibfnamefont{J.~M.} \bibnamefont{Lattimer}}, \bibinfo{journal}{Phys. Rev. D} \textbf{\bibinfo{volume}{98}}, \bibinfo{pages}{063020} (\bibinfo{year}{2018}).

\bibitem[{\citenamefont{Yagi and Yunes}(2017)}]{YAGIYUNES}
\bibinfo{author}{\bibfnamefont{K.}~\bibnamefont{Yagi}} \bibnamefont{and} \bibinfo{author}{\bibfnamefont{N.}~\bibnamefont{Yunes}}, \bibinfo{journal}{Phys. Rep.} \textbf{\bibinfo{volume}{681}}, \bibinfo{pages}{1} (\bibinfo{year}{2017}), ISSN \bibinfo{issn}{0370-1573}, \urlprefix\url{https://www.sciencedirect.com/science/article/pii/S0370157317300492}.

\bibitem[{\citenamefont{Bagchi}(2011)}]{bagchi2011role}
\bibinfo{author}{\bibfnamefont{M.}~\bibnamefont{Bagchi}}, \bibinfo{journal}{MNRAS Lett.} \textbf{\bibinfo{volume}{413}}, \bibinfo{pages}{L47} (\bibinfo{year}{2011}).

\bibitem[{\citenamefont{Samyn et~al.}(2002)\citenamefont{Samyn, Goriely, Heenen, Pearson, and Tondeur}}]{samyn2002hartree}
\bibinfo{author}{\bibfnamefont{M.}~\bibnamefont{Samyn}}, \bibinfo{author}{\bibfnamefont{S.}~\bibnamefont{Goriely}}, \bibinfo{author}{\bibfnamefont{P.-H.} \bibnamefont{Heenen}}, \bibinfo{author}{\bibfnamefont{J.~M.} \bibnamefont{Pearson}}, \bibnamefont{and} \bibinfo{author}{\bibfnamefont{F.}~\bibnamefont{Tondeur}}, \bibinfo{journal}{Nucl. Phys. A} \textbf{\bibinfo{volume}{700}}, \bibinfo{pages}{142} (\bibinfo{year}{2002}).

\bibitem[{\citenamefont{Goriely et~al.}(2002)\citenamefont{Goriely, Samyn, Heenen, Pearson, and Tondeur}}]{goriely2002hartree}
\bibinfo{author}{\bibfnamefont{S.}~\bibnamefont{Goriely}}, \bibinfo{author}{\bibfnamefont{M.}~\bibnamefont{Samyn}}, \bibinfo{author}{\bibfnamefont{P.-H.} \bibnamefont{Heenen}}, \bibinfo{author}{\bibfnamefont{J.~M.} \bibnamefont{Pearson}}, \bibnamefont{and} \bibinfo{author}{\bibfnamefont{F.}~\bibnamefont{Tondeur}}, \bibinfo{journal}{Phys. Rev. C} \textbf{\bibinfo{volume}{66}}, \bibinfo{pages}{024326} (\bibinfo{year}{2002}).

\bibitem[{\citenamefont{Samyn et~al.}(2003)\citenamefont{Samyn, Goriely, and Pearson}}]{samyn2003further}
\bibinfo{author}{\bibfnamefont{M.}~\bibnamefont{Samyn}}, \bibinfo{author}{\bibfnamefont{S.}~\bibnamefont{Goriely}}, \bibnamefont{and} \bibinfo{author}{\bibfnamefont{J.~M.} \bibnamefont{Pearson}}, \bibinfo{journal}{Nucl. Phys. A} \textbf{\bibinfo{volume}{725}}, \bibinfo{pages}{69} (\bibinfo{year}{2003}).

\bibitem[{\citenamefont{Goriely et~al.}(2003)\citenamefont{Goriely, Samyn, Bender, and Pearson}}]{goriely2003further}
\bibinfo{author}{\bibfnamefont{S.}~\bibnamefont{Goriely}}, \bibinfo{author}{\bibfnamefont{M.}~\bibnamefont{Samyn}}, \bibinfo{author}{\bibfnamefont{M.}~\bibnamefont{Bender}}, \bibnamefont{and} \bibinfo{author}{\bibfnamefont{J.~M.} \bibnamefont{Pearson}}, \bibinfo{journal}{Phys. Rev. C} \textbf{\bibinfo{volume}{68}}, \bibinfo{pages}{054325} (\bibinfo{year}{2003}).

\bibitem[{\citenamefont{Samyn et~al.}(2004)\citenamefont{Samyn, Goriely, Bender, and Pearson}}]{samyn2004further}
\bibinfo{author}{\bibfnamefont{M.}~\bibnamefont{Samyn}}, \bibinfo{author}{\bibfnamefont{S.}~\bibnamefont{Goriely}}, \bibinfo{author}{\bibfnamefont{M.}~\bibnamefont{Bender}}, \bibnamefont{and} \bibinfo{author}{\bibfnamefont{J.~M.} \bibnamefont{Pearson}}, \bibinfo{journal}{Phys. Rev. C} \textbf{\bibinfo{volume}{70}}, \bibinfo{pages}{044309} (\bibinfo{year}{2004}).

\bibitem[{\citenamefont{Goriely et~al.}(2005)\citenamefont{Goriely, Samyn, Pearson, and Onsi}}]{goriely2005further}
\bibinfo{author}{\bibfnamefont{S.}~\bibnamefont{Goriely}}, \bibinfo{author}{\bibfnamefont{M.}~\bibnamefont{Samyn}}, \bibinfo{author}{\bibfnamefont{J.~M.} \bibnamefont{Pearson}}, \bibnamefont{and} \bibinfo{author}{\bibfnamefont{M.}~\bibnamefont{Onsi}}, \bibinfo{journal}{Nucl. Phys. A} \textbf{\bibinfo{volume}{750}}, \bibinfo{pages}{425} (\bibinfo{year}{2005}).

\bibitem[{\citenamefont{Goriely et~al.}(2006)\citenamefont{Goriely, Samyn, and Pearson}}]{goriely2006further}
\bibinfo{author}{\bibfnamefont{S.}~\bibnamefont{Goriely}}, \bibinfo{author}{\bibfnamefont{M.}~\bibnamefont{Samyn}}, \bibnamefont{and} \bibinfo{author}{\bibfnamefont{J.~M.} \bibnamefont{Pearson}}, \bibinfo{journal}{Nucl. Phys. A} \textbf{\bibinfo{volume}{773}}, \bibinfo{pages}{279} (\bibinfo{year}{2006}).

\bibitem[{\citenamefont{Goriely et~al.}(2007)\citenamefont{Goriely, Samyn, and Pearson}}]{goriely2007further}
\bibinfo{author}{\bibfnamefont{S.}~\bibnamefont{Goriely}}, \bibinfo{author}{\bibfnamefont{M.}~\bibnamefont{Samyn}}, \bibnamefont{and} \bibinfo{author}{\bibfnamefont{J.~M.} \bibnamefont{Pearson}}, \bibinfo{journal}{Phys. Rev. C} \textbf{\bibinfo{volume}{75}}, \bibinfo{pages}{064312} (\bibinfo{year}{2007}).

\bibitem[{\citenamefont{Goriely and Pearson}(2008)}]{goriely2008further}
\bibinfo{author}{\bibfnamefont{S.}~\bibnamefont{Goriely}} \bibnamefont{and} \bibinfo{author}{\bibfnamefont{J.~M.} \bibnamefont{Pearson}}, \bibinfo{journal}{Phys. Rev. C} \textbf{\bibinfo{volume}{77}}, \bibinfo{pages}{031301(R)} (\bibinfo{year}{2008}).

\bibitem[{\citenamefont{Goriely et~al.}(2009{\natexlab{a}})\citenamefont{Goriely, Chamel, and Pearson}}]{goriely2009skyrme}
\bibinfo{author}{\bibfnamefont{S.}~\bibnamefont{Goriely}}, \bibinfo{author}{\bibfnamefont{N.}~\bibnamefont{Chamel}}, \bibnamefont{and} \bibinfo{author}{\bibfnamefont{J.~M.} \bibnamefont{Pearson}}, \bibinfo{journal}{Phys. Rev. Lett.} \textbf{\bibinfo{volume}{102}}, \bibinfo{pages}{152503} (\bibinfo{year}{2009}{\natexlab{a}}).

\bibitem[{\citenamefont{Goriely et~al.}(2010)\citenamefont{Goriely, Chamel, and Pearson}}]{goriely2010further}
\bibinfo{author}{\bibfnamefont{S.}~\bibnamefont{Goriely}}, \bibinfo{author}{\bibfnamefont{N.}~\bibnamefont{Chamel}}, \bibnamefont{and} \bibinfo{author}{\bibfnamefont{J.~M.} \bibnamefont{Pearson}}, \bibinfo{journal}{Phys. Rev. C} \textbf{\bibinfo{volume}{82}}, \bibinfo{pages}{035804} (\bibinfo{year}{2010}).

\bibitem[{\citenamefont{Goriely et~al.}(2013)\citenamefont{Goriely, Chamel, and Pearson}}]{goriely2013further}
\bibinfo{author}{\bibfnamefont{S.}~\bibnamefont{Goriely}}, \bibinfo{author}{\bibfnamefont{N.}~\bibnamefont{Chamel}}, \bibnamefont{and} \bibinfo{author}{\bibfnamefont{J.~M.} \bibnamefont{Pearson}}, \bibinfo{journal}{Phys. Rev. C} \textbf{\bibinfo{volume}{88}}, \bibinfo{pages}{024308} (\bibinfo{year}{2013}).

\bibitem[{\citenamefont{Friedrich and Reinhard}(1986)}]{friedrich1986skyrme}
\bibinfo{author}{\bibfnamefont{J.}~\bibnamefont{Friedrich}} \bibnamefont{and} \bibinfo{author}{\bibfnamefont{P.-G.} \bibnamefont{Reinhard}}, \bibinfo{journal}{Phys. Rev. C} \textbf{\bibinfo{volume}{33}}, \bibinfo{pages}{335} (\bibinfo{year}{1986}).

\bibitem[{\citenamefont{Lesinski et~al.}(2006)\citenamefont{Lesinski, Bennaceur, Duguet, and Meyer}}]{lesinski2006isovector}
\bibinfo{author}{\bibfnamefont{T.}~\bibnamefont{Lesinski}}, \bibinfo{author}{\bibfnamefont{K.}~\bibnamefont{Bennaceur}}, \bibinfo{author}{\bibfnamefont{T.}~\bibnamefont{Duguet}}, \bibnamefont{and} \bibinfo{author}{\bibfnamefont{J.}~\bibnamefont{Meyer}}, \bibinfo{journal}{Phys. Rev. C} \textbf{\bibinfo{volume}{74}}, \bibinfo{pages}{044315} (\bibinfo{year}{2006}).

\bibitem[{\citenamefont{Dutra}()}]{dutra}
\bibinfo{author}{\bibfnamefont{M.}~\bibnamefont{Dutra}}, \bibinfo{howpublished}{private communication}.

\bibitem[{\citenamefont{Guo-Qiang}(1991)}]{guo1991systematic}
\bibinfo{author}{\bibfnamefont{L.}~\bibnamefont{Guo-Qiang}}, \bibinfo{journal}{J. Phys. G: Nucl. Part. Phys.} \textbf{\bibinfo{volume}{17}}, \bibinfo{pages}{1} (\bibinfo{year}{1991}).

\bibitem[{\citenamefont{Agrawal et~al.}(2005)\citenamefont{Agrawal, Shlomo, and Au}}]{agrawal2005determination}
\bibinfo{author}{\bibfnamefont{B.~K.} \bibnamefont{Agrawal}}, \bibinfo{author}{\bibfnamefont{S.}~\bibnamefont{Shlomo}}, \bibnamefont{and} \bibinfo{author}{\bibfnamefont{V.~K.} \bibnamefont{Au}}, \bibinfo{journal}{Phys. Rev. C} \textbf{\bibinfo{volume}{72}}, \bibinfo{pages}{014310} (\bibinfo{year}{2005}).

\bibitem[{\citenamefont{Cao et~al.}(2006)\citenamefont{Cao, Lombardo, Shen, and Giai}}]{cao2006brueckner}
\bibinfo{author}{\bibfnamefont{L.~G.} \bibnamefont{Cao}}, \bibinfo{author}{\bibfnamefont{U.}~\bibnamefont{Lombardo}}, \bibinfo{author}{\bibfnamefont{C.~W.} \bibnamefont{Shen}}, \bibnamefont{and} \bibinfo{author}{\bibfnamefont{N.~V.} \bibnamefont{Giai}}, \bibinfo{journal}{Phys. Rev. C} \textbf{\bibinfo{volume}{73}}, \bibinfo{pages}{014313} (\bibinfo{year}{2006}).

\bibitem[{\citenamefont{Tondeur et~al.}(2000)\citenamefont{Tondeur, Goriely, Pearson, and Onsi}}]{tondeur2000towards}
\bibinfo{author}{\bibfnamefont{F.}~\bibnamefont{Tondeur}}, \bibinfo{author}{\bibfnamefont{S.}~\bibnamefont{Goriely}}, \bibinfo{author}{\bibfnamefont{J.~M.} \bibnamefont{Pearson}}, \bibnamefont{and} \bibinfo{author}{\bibfnamefont{M.}~\bibnamefont{Onsi}}, \bibinfo{journal}{Phys. Rev. C} \textbf{\bibinfo{volume}{62}}, \bibinfo{pages}{024308} (\bibinfo{year}{2000}).

\bibitem[{\citenamefont{Farine et~al.}(2001)\citenamefont{Farine, Pearson, and Tondeur}}]{farine2001skyrme}
\bibinfo{author}{\bibfnamefont{M.}~\bibnamefont{Farine}}, \bibinfo{author}{\bibfnamefont{J.~M.} \bibnamefont{Pearson}}, \bibnamefont{and} \bibinfo{author}{\bibfnamefont{F.}~\bibnamefont{Tondeur}}, \bibinfo{journal}{Nucl. Phys. A} \textbf{\bibinfo{volume}{696}}, \bibinfo{pages}{396} (\bibinfo{year}{2001}).

\bibitem[{\citenamefont{Goriely et~al.}(2001)\citenamefont{Goriely, Tondeur, and Pearson}}]{goriely2001hartree}
\bibinfo{author}{\bibfnamefont{S.}~\bibnamefont{Goriely}}, \bibinfo{author}{\bibfnamefont{F.}~\bibnamefont{Tondeur}}, \bibnamefont{and} \bibinfo{author}{\bibfnamefont{J.~M.} \bibnamefont{Pearson}}, \bibinfo{journal}{Atom. Data and Nucl. Data Tables} \textbf{\bibinfo{volume}{77}}, \bibinfo{pages}{311} (\bibinfo{year}{2001}).

\bibitem[{\citenamefont{Sharma et~al.}(1995)\citenamefont{Sharma, Lalazissis, K{\"o}nig, and Ring}}]{sharma1995isospin}
\bibinfo{author}{\bibfnamefont{M.~M.} \bibnamefont{Sharma}}, \bibinfo{author}{\bibfnamefont{G.~A.} \bibnamefont{Lalazissis}}, \bibinfo{author}{\bibfnamefont{J.}~\bibnamefont{K{\"o}nig}}, \bibnamefont{and} \bibinfo{author}{\bibfnamefont{P.}~\bibnamefont{Ring}}, \bibinfo{journal}{Phys. Rev. Lett.} \textbf{\bibinfo{volume}{74}}, \bibinfo{pages}{3744} (\bibinfo{year}{1995}).

\bibitem[{\citenamefont{Chen et~al.}(2010)\citenamefont{Chen, Ko, Li, and Xu}}]{chen2010density}
\bibinfo{author}{\bibfnamefont{L.-W.} \bibnamefont{Chen}}, \bibinfo{author}{\bibfnamefont{C.~M.} \bibnamefont{Ko}}, \bibinfo{author}{\bibfnamefont{B.-A.} \bibnamefont{Li}}, \bibnamefont{and} \bibinfo{author}{\bibfnamefont{J.}~\bibnamefont{Xu}}, \bibinfo{journal}{Phys. Rev. C} \textbf{\bibinfo{volume}{82}}, \bibinfo{pages}{024321} (\bibinfo{year}{2010}).

\bibitem[{\citenamefont{Lee and Mekjian}(2001)}]{lee2001liquid}
\bibinfo{author}{\bibfnamefont{S.~J.} \bibnamefont{Lee}} \bibnamefont{and} \bibinfo{author}{\bibfnamefont{A.~Z.} \bibnamefont{Mekjian}}, \bibinfo{journal}{Phys. Rev. C} \textbf{\bibinfo{volume}{63}}, \bibinfo{pages}{044605} (\bibinfo{year}{2001}).

\bibitem[{\citenamefont{Rayet et~al.}(1982)\citenamefont{Rayet, Arnould, Paulus, and Tondeur}}]{rayet1982nuclear}
\bibinfo{author}{\bibfnamefont{M.}~\bibnamefont{Rayet}}, \bibinfo{author}{\bibfnamefont{M.}~\bibnamefont{Arnould}}, \bibinfo{author}{\bibfnamefont{G.}~\bibnamefont{Paulus}}, \bibnamefont{and} \bibinfo{author}{\bibfnamefont{F.}~\bibnamefont{Tondeur}}, \bibinfo{journal}{Astron. Astrophys.} \textbf{\bibinfo{volume}{116}}, \bibinfo{pages}{183} (\bibinfo{year}{1982}).

\bibitem[{\citenamefont{Shen et~al.}(2019)\citenamefont{Shen, Col\`o, and Roca-Maza}}]{PhysRevC.99.034322}
\bibinfo{author}{\bibfnamefont{S.}~\bibnamefont{Shen}}, \bibinfo{author}{\bibfnamefont{G.}~\bibnamefont{Col\`o}}, \bibnamefont{and} \bibinfo{author}{\bibfnamefont{X.}~\bibnamefont{Roca-Maza}}, \bibinfo{journal}{Phys. Rev. C} \textbf{\bibinfo{volume}{99}}, \bibinfo{pages}{034322} (\bibinfo{year}{2019}), \urlprefix\url{https://link.aps.org/doi/10.1103/PhysRevC.99.034322}.

\bibitem[{\citenamefont{Van~Giai and Sagawa}(1981)}]{van1981spin}
\bibinfo{author}{\bibfnamefont{N.}~\bibnamefont{Van~Giai}} \bibnamefont{and} \bibinfo{author}{\bibfnamefont{H.}~\bibnamefont{Sagawa}}, \bibinfo{journal}{Phys. Lett. B} \textbf{\bibinfo{volume}{106}}, \bibinfo{pages}{379} (\bibinfo{year}{1981}).

\bibitem[{\citenamefont{Shen et~al.}(2009)\citenamefont{Shen, Han, Guo et~al.}}]{shen2009isospin}
\bibinfo{author}{\bibfnamefont{Q.-b.} \bibnamefont{Shen}}, \bibinfo{author}{\bibfnamefont{Y.-l.} \bibnamefont{Han}}, \bibinfo{author}{\bibfnamefont{H.-r.} \bibnamefont{Guo}}, \bibnamefont{et~al.}, \bibinfo{journal}{Phys. Rev. C} \textbf{\bibinfo{volume}{80}}, \bibinfo{pages}{024604} (\bibinfo{year}{2009}).

\bibitem[{\citenamefont{Lattimer and Ravenhall}(1978)}]{lattimer1978}
\bibinfo{author}{\bibfnamefont{J.~M.} \bibnamefont{Lattimer}} \bibnamefont{and} \bibinfo{author}{\bibfnamefont{D.~G.} \bibnamefont{Ravenhall}}, \bibinfo{journal}{Astrophys. J.} \textbf{\bibinfo{volume}{223}}, \bibinfo{pages}{314} (\bibinfo{year}{1978}).

\bibitem[{\citenamefont{Beiner et~al.}(1975)\citenamefont{Beiner, Flocard, Van~Giai, and Quentin}}]{beiner1975nuclear}
\bibinfo{author}{\bibfnamefont{M.}~\bibnamefont{Beiner}}, \bibinfo{author}{\bibfnamefont{H.}~\bibnamefont{Flocard}}, \bibinfo{author}{\bibfnamefont{N.}~\bibnamefont{Van~Giai}}, \bibnamefont{and} \bibinfo{author}{\bibfnamefont{P.}~\bibnamefont{Quentin}}, \bibinfo{journal}{Nucl. Phys. A} \textbf{\bibinfo{volume}{238}}, \bibinfo{pages}{29} (\bibinfo{year}{1975}).

\bibitem[{\citenamefont{Giannoni and Quentin}(1980)}]{giannoni1980mass}
\bibinfo{author}{\bibfnamefont{M.~J.} \bibnamefont{Giannoni}} \bibnamefont{and} \bibinfo{author}{\bibfnamefont{P.}~\bibnamefont{Quentin}}, \bibinfo{journal}{Phys. Rev. C} \textbf{\bibinfo{volume}{21}}, \bibinfo{pages}{2076} (\bibinfo{year}{1980}).

\bibitem[{\citenamefont{Agrawal et~al.}(2003)\citenamefont{Agrawal, Shlomo, and Kim~Au}}]{agrawal2003nuclear}
\bibinfo{author}{\bibfnamefont{B.~K.} \bibnamefont{Agrawal}}, \bibinfo{author}{\bibfnamefont{S.}~\bibnamefont{Shlomo}}, \bibnamefont{and} \bibinfo{author}{\bibfnamefont{V.}~\bibnamefont{Kim~Au}}, \bibinfo{journal}{Phys. Rev. C} \textbf{\bibinfo{volume}{68}}, \bibinfo{pages}{031304(R)} (\bibinfo{year}{2003}), \urlprefix\url{https://link.aps.org/doi/10.1103/PhysRevC.68.031304}.

\bibitem[{\citenamefont{K{\"o}hler}(1976)}]{kohler1976skyrme}
\bibinfo{author}{\bibfnamefont{H.~S.} \bibnamefont{K{\"o}hler}}, \bibinfo{journal}{Nucl. Phys. A} \textbf{\bibinfo{volume}{258}}, \bibinfo{pages}{301} (\bibinfo{year}{1976}).

\bibitem[{\citenamefont{Reinhard and Flocard}(1995)}]{reinhard1995nuclear}
\bibinfo{author}{\bibfnamefont{P.-G.} \bibnamefont{Reinhard}} \bibnamefont{and} \bibinfo{author}{\bibfnamefont{H.}~\bibnamefont{Flocard}}, \bibinfo{journal}{Nucl. Phys. A} \textbf{\bibinfo{volume}{584}}, \bibinfo{pages}{467} (\bibinfo{year}{1995}).

\bibitem[{\citenamefont{Nazarewicz et~al.}(1996)\citenamefont{Nazarewicz, Dobaczewski, Werner, Maruhn, Reinhard, Rutz, Chinn, Umar, and Strayer}}]{nazarewicz1996structure}
\bibinfo{author}{\bibfnamefont{W.}~\bibnamefont{Nazarewicz}}, \bibinfo{author}{\bibfnamefont{J.}~\bibnamefont{Dobaczewski}}, \bibinfo{author}{\bibfnamefont{T.~R.} \bibnamefont{Werner}}, \bibinfo{author}{\bibfnamefont{J.~A.} \bibnamefont{Maruhn}}, \bibinfo{author}{\bibfnamefont{P.-G.} \bibnamefont{Reinhard}}, \bibinfo{author}{\bibfnamefont{K.}~\bibnamefont{Rutz}}, \bibinfo{author}{\bibfnamefont{C.~R.} \bibnamefont{Chinn}}, \bibinfo{author}{\bibfnamefont{A.~S.} \bibnamefont{Umar}}, \bibnamefont{and} \bibinfo{author}{\bibfnamefont{M.~R.} \bibnamefont{Strayer}}, \bibinfo{journal}{Phys. Rev. C} \textbf{\bibinfo{volume}{53}}, \bibinfo{pages}{740} (\bibinfo{year}{1996}).

\bibitem[{\citenamefont{Krivine et~al.}(1980)\citenamefont{Krivine, Treiner, and Bohigas}}]{krivine1980derivation}
\bibinfo{author}{\bibfnamefont{H.}~\bibnamefont{Krivine}}, \bibinfo{author}{\bibfnamefont{J.}~\bibnamefont{Treiner}}, \bibnamefont{and} \bibinfo{author}{\bibfnamefont{O.}~\bibnamefont{Bohigas}}, \bibinfo{journal}{Nucl. Phys. A} \textbf{\bibinfo{volume}{336}}, \bibinfo{pages}{155} (\bibinfo{year}{1980}).

\bibitem[{\citenamefont{Li and Heenen}(1996)}]{li1996self}
\bibinfo{author}{\bibfnamefont{X.}~\bibnamefont{Li}} \bibnamefont{and} \bibinfo{author}{\bibfnamefont{P.-H.} \bibnamefont{Heenen}}, \bibinfo{journal}{Phys. Rev. C} \textbf{\bibinfo{volume}{54}}, \bibinfo{pages}{1617} (\bibinfo{year}{1996}).

\bibitem[{\citenamefont{Bartel et~al.}(1982)\citenamefont{Bartel, Quentin, Brack, Guet, and H{\aa}kansson}}]{bartel1982towards}
\bibinfo{author}{\bibfnamefont{J.}~\bibnamefont{Bartel}}, \bibinfo{author}{\bibfnamefont{P.}~\bibnamefont{Quentin}}, \bibinfo{author}{\bibfnamefont{M.}~\bibnamefont{Brack}}, \bibinfo{author}{\bibfnamefont{C.}~\bibnamefont{Guet}}, \bibnamefont{and} \bibinfo{author}{\bibfnamefont{H.-B.} \bibnamefont{H{\aa}kansson}}, \bibinfo{journal}{Nucl. Phys. A} \textbf{\bibinfo{volume}{386}}, \bibinfo{pages}{79} (\bibinfo{year}{1982}).

\bibitem[{\citenamefont{Bennour et~al.}(1989)\citenamefont{Bennour, Heenen, Bonche, Dobaczewski, and Flocard}}]{bennour1989charge}
\bibinfo{author}{\bibfnamefont{L.}~\bibnamefont{Bennour}}, \bibinfo{author}{\bibfnamefont{P.-H.} \bibnamefont{Heenen}}, \bibinfo{author}{\bibfnamefont{P.}~\bibnamefont{Bonche}}, \bibinfo{author}{\bibfnamefont{J.}~\bibnamefont{Dobaczewski}}, \bibnamefont{and} \bibinfo{author}{\bibfnamefont{H.}~\bibnamefont{Flocard}}, \bibinfo{journal}{Phys. Rev. C} \textbf{\bibinfo{volume}{40}}, \bibinfo{pages}{2834} (\bibinfo{year}{1989}).

\bibitem[{\citenamefont{Reinhard et~al.}(1999)\citenamefont{Reinhard, Dean, Nazarewicz, Dobaczewski, Maruhn, and Strayer}}]{reinhard1999shape}
\bibinfo{author}{\bibfnamefont{P.-G.} \bibnamefont{Reinhard}}, \bibinfo{author}{\bibfnamefont{D.~J.} \bibnamefont{Dean}}, \bibinfo{author}{\bibfnamefont{W.}~\bibnamefont{Nazarewicz}}, \bibinfo{author}{\bibfnamefont{J.}~\bibnamefont{Dobaczewski}}, \bibinfo{author}{\bibfnamefont{J.~A.} \bibnamefont{Maruhn}}, \bibnamefont{and} \bibinfo{author}{\bibfnamefont{M.~R.} \bibnamefont{Strayer}}, \bibinfo{journal}{Phys. Rev. C} \textbf{\bibinfo{volume}{60}}, \bibinfo{pages}{014316} (\bibinfo{year}{1999}).

\bibitem[{\citenamefont{Dobaczewski et~al.}(1984)\citenamefont{Dobaczewski, Flocard, and Treiner}}]{dobaczewski1984hartree}
\bibinfo{author}{\bibfnamefont{J.}~\bibnamefont{Dobaczewski}}, \bibinfo{author}{\bibfnamefont{H.}~\bibnamefont{Flocard}}, \bibnamefont{and} \bibinfo{author}{\bibfnamefont{J.}~\bibnamefont{Treiner}}, \bibinfo{journal}{Nucl. Phys. A} \textbf{\bibinfo{volume}{422}}, \bibinfo{pages}{103} (\bibinfo{year}{1984}).

\bibitem[{\citenamefont{Rashdan}(2000)}]{rashdan2000skyrme}
\bibinfo{author}{\bibfnamefont{M.}~\bibnamefont{Rashdan}}, \bibinfo{journal}{Modern Phys. Lett. A} \textbf{\bibinfo{volume}{15}}, \bibinfo{pages}{1287} (\bibinfo{year}{2000}).

\bibitem[{\citenamefont{G{\'o}mez et~al.}(1992)\citenamefont{G{\'o}mez, Prieto, and Navarro}}]{gomez1992improved}
\bibinfo{author}{\bibfnamefont{J.}~\bibnamefont{G{\'o}mez}}, \bibinfo{author}{\bibfnamefont{C.}~\bibnamefont{Prieto}}, \bibnamefont{and} \bibinfo{author}{\bibfnamefont{J.}~\bibnamefont{Navarro}}, \bibinfo{journal}{Nucl. Phys. A} \textbf{\bibinfo{volume}{549}}, \bibinfo{pages}{125} (\bibinfo{year}{1992}).

\bibitem[{\citenamefont{Pearson et~al.}(1991)\citenamefont{Pearson, Aboussir, Dutta, Nayak, Farine, and Tondeur}}]{pearson1991thomas}
\bibinfo{author}{\bibfnamefont{J.~M.} \bibnamefont{Pearson}}, \bibinfo{author}{\bibfnamefont{Y.}~\bibnamefont{Aboussir}}, \bibinfo{author}{\bibfnamefont{A.}~\bibnamefont{Dutta}}, \bibinfo{author}{\bibfnamefont{R.}~\bibnamefont{Nayak}}, \bibinfo{author}{\bibfnamefont{M.}~\bibnamefont{Farine}}, \bibnamefont{and} \bibinfo{author}{\bibfnamefont{F.}~\bibnamefont{Tondeur}}, \bibinfo{journal}{Nucl. Phys. A} \textbf{\bibinfo{volume}{528}}, \bibinfo{pages}{1} (\bibinfo{year}{1991}).

\bibitem[{\citenamefont{Aboussir et~al.}(1992)\citenamefont{Aboussir, Pearson, Dutta, and Tondeur}}]{aboussir1992thomas}
\bibinfo{author}{\bibfnamefont{Y.}~\bibnamefont{Aboussir}}, \bibinfo{author}{\bibfnamefont{J.~M.} \bibnamefont{Pearson}}, \bibinfo{author}{\bibfnamefont{A.}~\bibnamefont{Dutta}}, \bibnamefont{and} \bibinfo{author}{\bibfnamefont{F.}~\bibnamefont{Tondeur}}, \bibinfo{journal}{Nucl. Phys. A} \textbf{\bibinfo{volume}{549}}, \bibinfo{pages}{155} (\bibinfo{year}{1992}).

\bibitem[{\citenamefont{Pearson and Nayak}(2000)}]{pearson2000nuclear}
\bibinfo{author}{\bibfnamefont{J.~M.} \bibnamefont{Pearson}} \bibnamefont{and} \bibinfo{author}{\bibfnamefont{R.~C.} \bibnamefont{Nayak}}, \bibinfo{journal}{Nucl. Phys. A} \textbf{\bibinfo{volume}{668}}, \bibinfo{pages}{163} (\bibinfo{year}{2000}).

\bibitem[{\citenamefont{Onsi et~al.}(1994)\citenamefont{Onsi, Przysiezniak, and Pearson}}]{onsi1994equation}
\bibinfo{author}{\bibfnamefont{M.}~\bibnamefont{Onsi}}, \bibinfo{author}{\bibfnamefont{H.}~\bibnamefont{Przysiezniak}}, \bibnamefont{and} \bibinfo{author}{\bibfnamefont{J.~M.} \bibnamefont{Pearson}}, \bibinfo{journal}{Phys. Rev. C} \textbf{\bibinfo{volume}{50}}, \bibinfo{pages}{460} (\bibinfo{year}{1994}).

\bibitem[{\citenamefont{Pethick et~al.}(1995)\citenamefont{Pethick, Ravenhall, and Lorenz}}]{pethick1995inner}
\bibinfo{author}{\bibfnamefont{C.~J.} \bibnamefont{Pethick}}, \bibinfo{author}{\bibfnamefont{D.~G.} \bibnamefont{Ravenhall}}, \bibnamefont{and} \bibinfo{author}{\bibfnamefont{C.~P.} \bibnamefont{Lorenz}}, \bibinfo{journal}{Nucl. Phys. A} \textbf{\bibinfo{volume}{584}}, \bibinfo{pages}{675} (\bibinfo{year}{1995}).

\bibitem[{\citenamefont{Ko et~al.}(1974)\citenamefont{Ko, Pauli, Brack, and Brown}}]{ko1974microscopic}
\bibinfo{author}{\bibfnamefont{C.~M.} \bibnamefont{Ko}}, \bibinfo{author}{\bibfnamefont{H.~C.} \bibnamefont{Pauli}}, \bibinfo{author}{\bibfnamefont{M.}~\bibnamefont{Brack}}, \bibnamefont{and} \bibinfo{author}{\bibfnamefont{G.~E.} \bibnamefont{Brown}}, \bibinfo{journal}{Nucl. Phys. A} \textbf{\bibinfo{volume}{236}}, \bibinfo{pages}{269} (\bibinfo{year}{1974}).

\bibitem[{\citenamefont{Treiner and Krivine}(1976)}]{treiner1976simple}
\bibinfo{author}{\bibfnamefont{J.}~\bibnamefont{Treiner}} \bibnamefont{and} \bibinfo{author}{\bibfnamefont{H.}~\bibnamefont{Krivine}}, \bibinfo{journal}{J. Phys. G: Nucl. Phys.} \textbf{\bibinfo{volume}{2}}, \bibinfo{pages}{285} (\bibinfo{year}{1976}).

\bibitem[{\citenamefont{Brown}(1998)}]{brown1998new}
\bibinfo{author}{\bibfnamefont{B.~A.} \bibnamefont{Brown}}, \bibinfo{journal}{Phys. Rev. C} \textbf{\bibinfo{volume}{58}}, \bibinfo{pages}{220} (\bibinfo{year}{1998}).

\bibitem[{\citenamefont{Lim and Holt}(2017)}]{PhysRevC.95.065805}
\bibinfo{author}{\bibfnamefont{Y.}~\bibnamefont{Lim}} \bibnamefont{and} \bibinfo{author}{\bibfnamefont{J.~W.} \bibnamefont{Holt}}, \bibinfo{journal}{Phys. Rev. C} \textbf{\bibinfo{volume}{95}}, \bibinfo{pages}{065805} (\bibinfo{year}{2017}), \urlprefix\url{https://link.aps.org/doi/10.1103/PhysRevC.95.065805}.

\bibitem[{\citenamefont{Obaid and Alzubadi}(2023)}]{Obaid2023}
\bibinfo{author}{\bibfnamefont{R.}~\bibnamefont{Obaid}} \bibnamefont{and} \bibinfo{author}{\bibfnamefont{A.}~\bibnamefont{Alzubadi}}, \bibinfo{journal}{East European Journal of Physics} pp. \bibinfo{pages}{76--84} (\bibinfo{year}{2023}).

\bibitem[{\citenamefont{Brown et~al.}(2007)\citenamefont{Brown, Shen, Hillhouse, Meng, and Trzci{\'n}ska}}]{brown2007neutron}
\bibinfo{author}{\bibfnamefont{B.~A.} \bibnamefont{Brown}}, \bibinfo{author}{\bibfnamefont{G.}~\bibnamefont{Shen}}, \bibinfo{author}{\bibfnamefont{G.~C.} \bibnamefont{Hillhouse}}, \bibinfo{author}{\bibfnamefont{J.}~\bibnamefont{Meng}}, \bibnamefont{and} \bibinfo{author}{\bibfnamefont{A.}~\bibnamefont{Trzci{\'n}ska}}, \bibinfo{journal}{Phys. Rev. C} \textbf{\bibinfo{volume}{76}}, \bibinfo{pages}{034305} (\bibinfo{year}{2007}).

\bibitem[{\citenamefont{Margueron et~al.}(2002)\citenamefont{Margueron, Navarro, and Van~Giai}}]{margueron2002instabilities}
\bibinfo{author}{\bibfnamefont{J.}~\bibnamefont{Margueron}}, \bibinfo{author}{\bibfnamefont{J.}~\bibnamefont{Navarro}}, \bibnamefont{and} \bibinfo{author}{\bibfnamefont{N.}~\bibnamefont{Van~Giai}}, \bibinfo{journal}{Phys. Rev. C} \textbf{\bibinfo{volume}{66}}, \bibinfo{pages}{014303} (\bibinfo{year}{2002}), \urlprefix\url{https://link.aps.org/doi/10.1103/PhysRevC.66.014303}.

\bibitem[{\citenamefont{Chabanat}(1995)}]{chabanat1995effective}
\bibinfo{author}{\bibfnamefont{E.}~\bibnamefont{Chabanat}}, \bibinfo{type}{Tech. Rep.}, \bibinfo{institution}{Lyon-1 Univ.} (\bibinfo{year}{1995}).

\bibitem[{\citenamefont{Chabanat et~al.}(1997)\citenamefont{Chabanat, Bonche, Haensel, Meyer, and Schaeffer}}]{chabanat1997skyrme}
\bibinfo{author}{\bibfnamefont{E.}~\bibnamefont{Chabanat}}, \bibinfo{author}{\bibfnamefont{P.}~\bibnamefont{Bonche}}, \bibinfo{author}{\bibfnamefont{P.}~\bibnamefont{Haensel}}, \bibinfo{author}{\bibfnamefont{J.}~\bibnamefont{Meyer}}, \bibnamefont{and} \bibinfo{author}{\bibfnamefont{R.}~\bibnamefont{Schaeffer}}, \bibinfo{journal}{Nucl. Phys. A} \textbf{\bibinfo{volume}{627}}, \bibinfo{pages}{710} (\bibinfo{year}{1997}).

\bibitem[{\citenamefont{Chabanat et~al.}(1998)\citenamefont{Chabanat, Bonche, Haensel, Meyer, and Schaeffer}}]{chabanat1998skyrme}
\bibinfo{author}{\bibfnamefont{E.}~\bibnamefont{Chabanat}}, \bibinfo{author}{\bibfnamefont{P.}~\bibnamefont{Bonche}}, \bibinfo{author}{\bibfnamefont{P.}~\bibnamefont{Haensel}}, \bibinfo{author}{\bibfnamefont{J.}~\bibnamefont{Meyer}}, \bibnamefont{and} \bibinfo{author}{\bibfnamefont{R.}~\bibnamefont{Schaeffer}}, \bibinfo{journal}{Nucl. Phys. A} \textbf{\bibinfo{volume}{635}}, \bibinfo{pages}{231} (\bibinfo{year}{1998}).

\bibitem[{\citenamefont{Guichon and Thomas}(2004)}]{guichon2004quark}
\bibinfo{author}{\bibfnamefont{P.~A.~M.} \bibnamefont{Guichon}} \bibnamefont{and} \bibinfo{author}{\bibfnamefont{A.~W.} \bibnamefont{Thomas}}, \bibinfo{journal}{Phys. Rev. Lett.} \textbf{\bibinfo{volume}{93}}, \bibinfo{pages}{132502} (\bibinfo{year}{2004}).

\bibitem[{\citenamefont{Guichon et~al.}(2006)\citenamefont{Guichon, Matevosyan, Sandulescu, and Thomas}}]{guichon2006physical}
\bibinfo{author}{\bibfnamefont{P.~A.~M.} \bibnamefont{Guichon}}, \bibinfo{author}{\bibfnamefont{H.~H.} \bibnamefont{Matevosyan}}, \bibinfo{author}{\bibfnamefont{N.}~\bibnamefont{Sandulescu}}, \bibnamefont{and} \bibinfo{author}{\bibfnamefont{A.~W.} \bibnamefont{Thomas}}, \bibinfo{journal}{Nucl. Phys. A} \textbf{\bibinfo{volume}{772}}, \bibinfo{pages}{1} (\bibinfo{year}{2006}).

\bibitem[{\citenamefont{Kl{\"u}pfel et~al.}(2009)\citenamefont{Kl{\"u}pfel, Reinhard, B{\"u}rvenich, and Maruhn}}]{klupfel2009variations}
\bibinfo{author}{\bibfnamefont{P.}~\bibnamefont{Kl{\"u}pfel}}, \bibinfo{author}{\bibfnamefont{P.-G.} \bibnamefont{Reinhard}}, \bibinfo{author}{\bibfnamefont{T.~J.} \bibnamefont{B{\"u}rvenich}}, \bibnamefont{and} \bibinfo{author}{\bibfnamefont{J.~A.} \bibnamefont{Maruhn}}, \bibinfo{journal}{Phys. Rev. C} \textbf{\bibinfo{volume}{79}}, \bibinfo{pages}{034310} (\bibinfo{year}{2009}).

\bibitem[{\citenamefont{Tondeur et~al.}(1984)\citenamefont{Tondeur, Brack, Farine, and Pearson}}]{tondeur1984static}
\bibinfo{author}{\bibfnamefont{F.}~\bibnamefont{Tondeur}}, \bibinfo{author}{\bibfnamefont{M.}~\bibnamefont{Brack}}, \bibinfo{author}{\bibfnamefont{M.}~\bibnamefont{Farine}}, \bibnamefont{and} \bibinfo{author}{\bibfnamefont{J.~M.} \bibnamefont{Pearson}}, \bibinfo{journal}{Nucl. Phys. A} \textbf{\bibinfo{volume}{420}}, \bibinfo{pages}{297} (\bibinfo{year}{1984}).

\bibitem[{\citenamefont{Lesinski et~al.}(2007)\citenamefont{Lesinski, Bender, Bennaceur, Duguet, and Meyer}}]{lesinski2007tensor}
\bibinfo{author}{\bibfnamefont{T.}~\bibnamefont{Lesinski}}, \bibinfo{author}{\bibfnamefont{M.}~\bibnamefont{Bender}}, \bibinfo{author}{\bibfnamefont{K.}~\bibnamefont{Bennaceur}}, \bibinfo{author}{\bibfnamefont{T.}~\bibnamefont{Duguet}}, \bibnamefont{and} \bibinfo{author}{\bibfnamefont{J.}~\bibnamefont{Meyer}}, \bibinfo{journal}{Phys. Rev. C} \textbf{\bibinfo{volume}{76}}, \bibinfo{pages}{014312} (\bibinfo{year}{2007}).

\bibitem[{\citenamefont{Kortelainen et~al.}(2012)\citenamefont{Kortelainen, McDonnell, Nazarewicz, Reinhard, Sarich, Schunck, Stoitsov, and Wild}}]{kortelainen2012nuclear}
\bibinfo{author}{\bibfnamefont{M.}~\bibnamefont{Kortelainen}}, \bibinfo{author}{\bibfnamefont{J.}~\bibnamefont{McDonnell}}, \bibinfo{author}{\bibfnamefont{W.}~\bibnamefont{Nazarewicz}}, \bibinfo{author}{\bibfnamefont{P.-G.} \bibnamefont{Reinhard}}, \bibinfo{author}{\bibfnamefont{J.}~\bibnamefont{Sarich}}, \bibinfo{author}{\bibfnamefont{N.}~\bibnamefont{Schunck}}, \bibinfo{author}{\bibfnamefont{M.~V.} \bibnamefont{Stoitsov}}, \bibnamefont{and} \bibinfo{author}{\bibfnamefont{S.~M.} \bibnamefont{Wild}}, \bibinfo{journal}{Phys. Rev. C} \textbf{\bibinfo{volume}{85}}, \bibinfo{pages}{024304} (\bibinfo{year}{2012}).

\bibitem[{\citenamefont{Pearson and Goriely}(2001)}]{pearson2001isovector}
\bibinfo{author}{\bibfnamefont{J.~M.} \bibnamefont{Pearson}} \bibnamefont{and} \bibinfo{author}{\bibfnamefont{S.}~\bibnamefont{Goriely}}, \bibinfo{journal}{Phys. Rev. C} \textbf{\bibinfo{volume}{64}}, \bibinfo{pages}{027301} (\bibinfo{year}{2001}).

\bibitem[{\citenamefont{Haddad}(1999)}]{haddad1999density}
\bibinfo{author}{\bibfnamefont{S.}~\bibnamefont{Haddad}}, \bibinfo{journal}{EuroPhys. Lett.} \textbf{\bibinfo{volume}{48}}, \bibinfo{pages}{505} (\bibinfo{year}{1999}).

\bibitem[{\citenamefont{Reinhard}(1989)}]{reinhard1989relativistic}
\bibinfo{author}{\bibfnamefont{P.-G.} \bibnamefont{Reinhard}}, \bibinfo{journal}{Rep. Prog. Phys.} \textbf{\bibinfo{volume}{52}}, \bibinfo{pages}{439} (\bibinfo{year}{1989}).

\bibitem[{\citenamefont{ben Ali~Dadi}(2010)}]{ben2010parametrization}
\bibinfo{author}{\bibfnamefont{A.}~\bibnamefont{ben Ali~Dadi}}, \bibinfo{journal}{Phys. Rev. C} \textbf{\bibinfo{volume}{82}}, \bibinfo{pages}{025203} (\bibinfo{year}{2010}).

\bibitem[{\citenamefont{Rufa et~al.}(1988)\citenamefont{Rufa, Reinhard, Maruhn, Greiner, and Strayer}}]{rufa1988optimal}
\bibinfo{author}{\bibfnamefont{M.}~\bibnamefont{Rufa}}, \bibinfo{author}{\bibfnamefont{P.-G.} \bibnamefont{Reinhard}}, \bibinfo{author}{\bibfnamefont{J.~A.} \bibnamefont{Maruhn}}, \bibinfo{author}{\bibfnamefont{W.}~\bibnamefont{Greiner}}, \bibnamefont{and} \bibinfo{author}{\bibfnamefont{M.~R.} \bibnamefont{Strayer}}, \bibinfo{journal}{Phys. Rev. C} \textbf{\bibinfo{volume}{38}}, \bibinfo{pages}{390} (\bibinfo{year}{1988}).

\bibitem[{\citenamefont{Piekarewicz}(2002)}]{piekarewicz2002correlating}
\bibinfo{author}{\bibfnamefont{J.}~\bibnamefont{Piekarewicz}}, \bibinfo{journal}{Phys. Rev. C} \textbf{\bibinfo{volume}{66}}, \bibinfo{pages}{034305} (\bibinfo{year}{2002}).

\bibitem[{\citenamefont{Glendenning}(2012)}]{glendenning2012compact}
\bibinfo{author}{\bibfnamefont{N.~K.} \bibnamefont{Glendenning}}, \emph{\bibinfo{title}{Compact stars: Nuclear Physics, Particle Physics and General Relativity}} (\bibinfo{publisher}{Springer Science \& Business Media}, \bibinfo{year}{2012}).

\bibitem[{\citenamefont{Glendenning}(1982)}]{glendenning1982hyperon}
\bibinfo{author}{\bibfnamefont{N.~K.} \bibnamefont{Glendenning}}, \bibinfo{journal}{Phys. Lett. B} \textbf{\bibinfo{volume}{114}}, \bibinfo{pages}{392} (\bibinfo{year}{1982}).

\bibitem[{\citenamefont{Glendenning and Moszkowski}(1991)}]{glendenning1991reconciliation}
\bibinfo{author}{\bibfnamefont{N.~K.} \bibnamefont{Glendenning}} \bibnamefont{and} \bibinfo{author}{\bibfnamefont{S.~A.} \bibnamefont{Moszkowski}}, \bibinfo{journal}{Phys. Rev. Lett.} \textbf{\bibinfo{volume}{67}}, \bibinfo{pages}{2414} (\bibinfo{year}{1991}).

\bibitem[{\citenamefont{Ghosh et~al.}(1995)\citenamefont{Ghosh, Phatak, and Sahu}}]{ghosh1995hybrid}
\bibinfo{author}{\bibfnamefont{S.~K.} \bibnamefont{Ghosh}}, \bibinfo{author}{\bibfnamefont{S.~C.} \bibnamefont{Phatak}}, \bibnamefont{and} \bibinfo{author}{\bibfnamefont{P.~K.} \bibnamefont{Sahu}}, \bibinfo{journal}{Zeitschrift f{\"u}r Physik A Hadrons and Nuclei} \textbf{\bibinfo{volume}{352}}, \bibinfo{pages}{457} (\bibinfo{year}{1995}).

\bibitem[{\citenamefont{Piekarewicz and Centelles}(2009)}]{piekarewicz2009incompressibility}
\bibinfo{author}{\bibfnamefont{J.}~\bibnamefont{Piekarewicz}} \bibnamefont{and} \bibinfo{author}{\bibfnamefont{M.}~\bibnamefont{Centelles}}, \bibinfo{journal}{Phys. Rev. C} \textbf{\bibinfo{volume}{79}}, \bibinfo{pages}{054311} (\bibinfo{year}{2009}).

\bibitem[{\citenamefont{Mueller and Serot}(1996)}]{mueller1996relativistic}
\bibinfo{author}{\bibfnamefont{H.}~\bibnamefont{Mueller}} \bibnamefont{and} \bibinfo{author}{\bibfnamefont{B.~D.} \bibnamefont{Serot}}, \bibinfo{journal}{Nucl. Phys. A} \textbf{\bibinfo{volume}{606}}, \bibinfo{pages}{508} (\bibinfo{year}{1996}).

\bibitem[{\citenamefont{Xia et~al.}(2022)\citenamefont{Xia, Maruyama, Li, Sun, Long, and Zhang}}]{Xia_2022}
\bibinfo{author}{\bibfnamefont{C.-J.} \bibnamefont{Xia}}, \bibinfo{author}{\bibfnamefont{T.}~\bibnamefont{Maruyama}}, \bibinfo{author}{\bibfnamefont{A.}~\bibnamefont{Li}}, \bibinfo{author}{\bibfnamefont{B.~Y.} \bibnamefont{Sun}}, \bibinfo{author}{\bibfnamefont{W.-H.} \bibnamefont{Long}}, \bibnamefont{and} \bibinfo{author}{\bibfnamefont{Y.-X.} \bibnamefont{Zhang}}, \bibinfo{journal}{Communications in Theoretical Physics} \textbf{\bibinfo{volume}{74}}, \bibinfo{pages}{095303} (\bibinfo{year}{2022}), \urlprefix\url{https://dx.doi.org/10.1088/1572-9494/ac71fd}.

\bibitem[{\citenamefont{Bender et~al.}(1999)\citenamefont{Bender, Rutz, Reinhard, Maruhn, and Greiner}}]{bender1999shell}
\bibinfo{author}{\bibfnamefont{M.}~\bibnamefont{Bender}}, \bibinfo{author}{\bibfnamefont{K.}~\bibnamefont{Rutz}}, \bibinfo{author}{\bibfnamefont{P.-G.} \bibnamefont{Reinhard}}, \bibinfo{author}{\bibfnamefont{J.~A.} \bibnamefont{Maruhn}}, \bibnamefont{and} \bibinfo{author}{\bibfnamefont{W.}~\bibnamefont{Greiner}}, \bibinfo{journal}{Phys. Rev. C} \textbf{\bibinfo{volume}{60}}, \bibinfo{pages}{034304} (\bibinfo{year}{1999}).

\bibitem[{\citenamefont{Liu et~al.}(2002)\citenamefont{Liu, Greco, Baran, Colonna, and Di~Toro}}]{liu2002asymmetric}
\bibinfo{author}{\bibfnamefont{B.}~\bibnamefont{Liu}}, \bibinfo{author}{\bibfnamefont{V.}~\bibnamefont{Greco}}, \bibinfo{author}{\bibfnamefont{V.}~\bibnamefont{Baran}}, \bibinfo{author}{\bibfnamefont{M.}~\bibnamefont{Colonna}}, \bibnamefont{and} \bibinfo{author}{\bibfnamefont{M.}~\bibnamefont{Di~Toro}}, \bibinfo{journal}{Phys. Rev. C} \textbf{\bibinfo{volume}{65}}, \bibinfo{pages}{045201} (\bibinfo{year}{2002}).

\bibitem[{\citenamefont{Centelles et~al.}(1998)\citenamefont{Centelles, Del~Estal, and Vinas}}]{centelles1998semiclassical}
\bibinfo{author}{\bibfnamefont{M.}~\bibnamefont{Centelles}}, \bibinfo{author}{\bibfnamefont{M.}~\bibnamefont{Del~Estal}}, \bibnamefont{and} \bibinfo{author}{\bibfnamefont{X.}~\bibnamefont{Vinas}}, \bibinfo{journal}{Nucl. Phys. A} \textbf{\bibinfo{volume}{635}}, \bibinfo{pages}{193} (\bibinfo{year}{1998}).

\bibitem[{\citenamefont{Lalazissis et~al.}(1997)\citenamefont{Lalazissis, K{\"o}nig, and Ring}}]{lalazissis1997new}
\bibinfo{author}{\bibfnamefont{G.~A.} \bibnamefont{Lalazissis}}, \bibinfo{author}{\bibfnamefont{J.}~\bibnamefont{K{\"o}nig}}, \bibnamefont{and} \bibinfo{author}{\bibfnamefont{P.}~\bibnamefont{Ring}}, \bibinfo{journal}{Phys. Rev. C} \textbf{\bibinfo{volume}{55}}, \bibinfo{pages}{540} (\bibinfo{year}{1997}).

\bibitem[{\citenamefont{Lalazissis et~al.}(2009)\citenamefont{Lalazissis, Karatzikos, Fossion, Arteaga, Afanasjev, and Ring}}]{lalazissis2009effective}
\bibinfo{author}{\bibfnamefont{G.~A.} \bibnamefont{Lalazissis}}, \bibinfo{author}{\bibfnamefont{S.}~\bibnamefont{Karatzikos}}, \bibinfo{author}{\bibfnamefont{R.}~\bibnamefont{Fossion}}, \bibinfo{author}{\bibfnamefont{D.~P.} \bibnamefont{Arteaga}}, \bibinfo{author}{\bibfnamefont{A.~V.} \bibnamefont{Afanasjev}}, \bibnamefont{and} \bibinfo{author}{\bibfnamefont{P.}~\bibnamefont{Ring}}, \bibinfo{journal}{Phys. Lett. B} \textbf{\bibinfo{volume}{671}}, \bibinfo{pages}{36} (\bibinfo{year}{2009}).

\bibitem[{\citenamefont{Nerlo-Pomorska and Sykut}(2004)}]{nerlo2004new}
\bibinfo{author}{\bibfnamefont{B.}~\bibnamefont{Nerlo-Pomorska}} \bibnamefont{and} \bibinfo{author}{\bibfnamefont{J.}~\bibnamefont{Sykut}}, \bibinfo{journal}{Int. J. Mod. Phys. E} \textbf{\bibinfo{volume}{13}}, \bibinfo{pages}{75} (\bibinfo{year}{2004}).

\bibitem[{\citenamefont{Serot and Walecka}(1997)}]{serot1997recent}
\bibinfo{author}{\bibfnamefont{B.~D.} \bibnamefont{Serot}} \bibnamefont{and} \bibinfo{author}{\bibfnamefont{J.~D.} \bibnamefont{Walecka}}, \bibinfo{journal}{Int. J. Mod. Phys. E} \textbf{\bibinfo{volume}{6}}, \bibinfo{pages}{515} (\bibinfo{year}{1997}).

\bibitem[{\citenamefont{Serot}(1992)}]{serot1992quantum}
\bibinfo{author}{\bibfnamefont{B.~D.} \bibnamefont{Serot}}, \bibinfo{journal}{Rep. Prog. Phys.} \textbf{\bibinfo{volume}{55}}, \bibinfo{pages}{1855} (\bibinfo{year}{1992}).

\bibitem[{\citenamefont{Chung et~al.}(2000)\citenamefont{Chung, Wang, Santiago, and Zhang}}]{chung2000nuclear}
\bibinfo{author}{\bibfnamefont{K.}~\bibnamefont{Chung}}, \bibinfo{author}{\bibfnamefont{C.~S.} \bibnamefont{Wang}}, \bibinfo{author}{\bibfnamefont{A.~J.} \bibnamefont{Santiago}}, \bibnamefont{and} \bibinfo{author}{\bibfnamefont{J.~W.} \bibnamefont{Zhang}}, \bibinfo{journal}{Eur. Phys. J. A-Hadrons and Nuclei} \textbf{\bibinfo{volume}{9}}, \bibinfo{pages}{453} (\bibinfo{year}{2000}).

\bibitem[{\citenamefont{Rashdan}(2001)}]{rashdan2001structure}
\bibinfo{author}{\bibfnamefont{M.}~\bibnamefont{Rashdan}}, \bibinfo{journal}{Phys. Rev. C} \textbf{\bibinfo{volume}{63}}, \bibinfo{pages}{044303} (\bibinfo{year}{2001}).

\bibitem[{\citenamefont{Reinhard}(1988)}]{reinhard1988nonlinearity}
\bibinfo{author}{\bibfnamefont{P.~G.} \bibnamefont{Reinhard}}, \bibinfo{journal}{Zeitschrift f{\"u}r Physik A Atomic Nuclei} \textbf{\bibinfo{volume}{329}}, \bibinfo{pages}{257} (\bibinfo{year}{1988}).

\bibitem[{\citenamefont{Sulaksono et~al.}(2005)\citenamefont{Sulaksono, Mart, and Bahri}}]{sulaksono2005nilsson}
\bibinfo{author}{\bibfnamefont{A.}~\bibnamefont{Sulaksono}}, \bibinfo{author}{\bibfnamefont{T.}~\bibnamefont{Mart}}, \bibnamefont{and} \bibinfo{author}{\bibfnamefont{C.}~\bibnamefont{Bahri}}, \bibinfo{journal}{Phys. Rev. C} \textbf{\bibinfo{volume}{71}}, \bibinfo{pages}{034312} (\bibinfo{year}{2005}).

\bibitem[{\citenamefont{Furnstahl et~al.}(1997)\citenamefont{Furnstahl, Serot, and Tang}}]{furnstahl1997chiral}
\bibinfo{author}{\bibfnamefont{R.~J.} \bibnamefont{Furnstahl}}, \bibinfo{author}{\bibfnamefont{B.~D.} \bibnamefont{Serot}}, \bibnamefont{and} \bibinfo{author}{\bibfnamefont{H.-B.} \bibnamefont{Tang}}, \bibinfo{journal}{Nucl. Phys. A} \textbf{\bibinfo{volume}{615}}, \bibinfo{pages}{441} (\bibinfo{year}{1997}).

\bibitem[{\citenamefont{Horowitz and Piekarewicz}(2001)}]{horowitz2001neutron}
\bibinfo{author}{\bibfnamefont{C.~J.} \bibnamefont{Horowitz}} \bibnamefont{and} \bibinfo{author}{\bibfnamefont{J.}~\bibnamefont{Piekarewicz}}, \bibinfo{journal}{Phys. Rev. Lett.} \textbf{\bibinfo{volume}{86}}, \bibinfo{pages}{5647} (\bibinfo{year}{2001}).

\bibitem[{\citenamefont{Gmuca}(1992{\natexlab{a}})}]{gmuca1992relativistic}
\bibinfo{author}{\bibfnamefont{S.}~\bibnamefont{Gmuca}}, \bibinfo{journal}{Zeitschrift f{\"u}r Physik A Hadrons and Nuclei} \textbf{\bibinfo{volume}{342}}, \bibinfo{pages}{387} (\bibinfo{year}{1992}{\natexlab{a}}).

\bibitem[{\citenamefont{Centelles et~al.}(1992)\citenamefont{Centelles, Vinas, Barranco, Marcos, and Lombard}}]{centelles1992semiclassical}
\bibinfo{author}{\bibfnamefont{M.}~\bibnamefont{Centelles}}, \bibinfo{author}{\bibfnamefont{X.}~\bibnamefont{Vinas}}, \bibinfo{author}{\bibfnamefont{M.}~\bibnamefont{Barranco}}, \bibinfo{author}{\bibfnamefont{S.}~\bibnamefont{Marcos}}, \bibnamefont{and} \bibinfo{author}{\bibfnamefont{R.~J.} \bibnamefont{Lombard}}, \bibinfo{journal}{Nucl. Phys. A} \textbf{\bibinfo{volume}{537}}, \bibinfo{pages}{486} (\bibinfo{year}{1992}).

\bibitem[{\citenamefont{Gmuca}(1992{\natexlab{b}})}]{gmuca1992finite}
\bibinfo{author}{\bibfnamefont{S.}~\bibnamefont{Gmuca}}, \bibinfo{journal}{Nucl. Phys. A} \textbf{\bibinfo{volume}{547}}, \bibinfo{pages}{447} (\bibinfo{year}{1992}{\natexlab{b}}).

\bibitem[{\citenamefont{Bunta and Gmuca}(2003)}]{bunta2003asymmetric}
\bibinfo{author}{\bibfnamefont{J.~K.} \bibnamefont{Bunta}} \bibnamefont{and} \bibinfo{author}{\bibfnamefont{{\v{S}}.}~\bibnamefont{Gmuca}}, \bibinfo{journal}{Phys. Rev. C} \textbf{\bibinfo{volume}{68}}, \bibinfo{pages}{054318} (\bibinfo{year}{2003}).

\bibitem[{\citenamefont{M{\"u}ller and Serot}(1995)}]{muller1995phase}
\bibinfo{author}{\bibfnamefont{H.}~\bibnamefont{M{\"u}ller}} \bibnamefont{and} \bibinfo{author}{\bibfnamefont{B.~D.} \bibnamefont{Serot}}, \bibinfo{journal}{Phys. Rev. C} \textbf{\bibinfo{volume}{52}}, \bibinfo{pages}{2072} (\bibinfo{year}{1995}).

\bibitem[{\citenamefont{Fattoyev and Piekarewicz}(2010)}]{fattoyev2010relativistic}
\bibinfo{author}{\bibfnamefont{F.~J.} \bibnamefont{Fattoyev}} \bibnamefont{and} \bibinfo{author}{\bibfnamefont{J.}~\bibnamefont{Piekarewicz}}, \bibinfo{journal}{Phys. Rev. C} \textbf{\bibinfo{volume}{82}}, \bibinfo{pages}{025805} (\bibinfo{year}{2010}).

\bibitem[{\citenamefont{Sharma et~al.}(2000)\citenamefont{Sharma, Farhan, and Mythili}}]{sharma2000shell}
\bibinfo{author}{\bibfnamefont{M.~M.} \bibnamefont{Sharma}}, \bibinfo{author}{\bibfnamefont{A.~R.} \bibnamefont{Farhan}}, \bibnamefont{and} \bibinfo{author}{\bibfnamefont{S.}~\bibnamefont{Mythili}}, \bibinfo{journal}{Phys. Rev. C} \textbf{\bibinfo{volume}{61}}, \bibinfo{pages}{054306} (\bibinfo{year}{2000}).

\bibitem[{\citenamefont{Long et~al.}(2004)\citenamefont{Long, Meng, Giai, and Zhou}}]{long2004new}
\bibinfo{author}{\bibfnamefont{W.}~\bibnamefont{Long}}, \bibinfo{author}{\bibfnamefont{J.}~\bibnamefont{Meng}}, \bibinfo{author}{\bibfnamefont{N.~V.} \bibnamefont{Giai}}, \bibnamefont{and} \bibinfo{author}{\bibfnamefont{S.-G.} \bibnamefont{Zhou}}, \bibinfo{journal}{Phys. Rev. C} \textbf{\bibinfo{volume}{69}}, \bibinfo{pages}{034319} (\bibinfo{year}{2004}), \urlprefix\url{https://link.aps.org/doi/10.1103/PhysRevC.69.034319}.

\bibitem[{\citenamefont{Sugahara and Toki}(1994)}]{sugahara1994relativistic}
\bibinfo{author}{\bibfnamefont{Y.}~\bibnamefont{Sugahara}} \bibnamefont{and} \bibinfo{author}{\bibfnamefont{H.}~\bibnamefont{Toki}}, \bibinfo{journal}{Nucl. Phys. A} \textbf{\bibinfo{volume}{579}}, \bibinfo{pages}{557} (\bibinfo{year}{1994}).

\bibitem[{\citenamefont{Fattoyev et~al.}(2020)\citenamefont{Fattoyev, Horowitz, Piekarewicz, and Reed}}]{fattoyev2020gw190814}
\bibinfo{author}{\bibfnamefont{F.~J.} \bibnamefont{Fattoyev}}, \bibinfo{author}{\bibfnamefont{C.~J.} \bibnamefont{Horowitz}}, \bibinfo{author}{\bibfnamefont{J.}~\bibnamefont{Piekarewicz}}, \bibnamefont{and} \bibinfo{author}{\bibfnamefont{B.}~\bibnamefont{Reed}}, \bibinfo{journal}{Phys. Rev. C} \textbf{\bibinfo{volume}{102}}, \bibinfo{pages}{065805} (\bibinfo{year}{2020}).

\bibitem[{\citenamefont{Agrawal}(2010)}]{agrawal2010asymmetric}
\bibinfo{author}{\bibfnamefont{B.~K.} \bibnamefont{Agrawal}}, \bibinfo{journal}{Phys. Rev. C} \textbf{\bibinfo{volume}{81}}, \bibinfo{pages}{034323} (\bibinfo{year}{2010}).

\bibitem[{\citenamefont{Dhiman et~al.}(2007)\citenamefont{Dhiman, Kumar, and Agrawal}}]{dhiman2007nonrotating}
\bibinfo{author}{\bibfnamefont{S.~K.} \bibnamefont{Dhiman}}, \bibinfo{author}{\bibfnamefont{R.}~\bibnamefont{Kumar}}, \bibnamefont{and} \bibinfo{author}{\bibfnamefont{B.~K.} \bibnamefont{Agrawal}}, \bibinfo{journal}{Phys. Rev. C} \textbf{\bibinfo{volume}{76}}, \bibinfo{pages}{045801} (\bibinfo{year}{2007}).

\bibitem[{\citenamefont{Cai and Chen}(2012)}]{cai2012nuclear}
\bibinfo{author}{\bibfnamefont{B.-J.} \bibnamefont{Cai}} \bibnamefont{and} \bibinfo{author}{\bibfnamefont{L.-W.} \bibnamefont{Chen}}, \bibinfo{journal}{Phys. Rev. C} \textbf{\bibinfo{volume}{85}}, \bibinfo{pages}{024302} (\bibinfo{year}{2012}).

\bibitem[{\citenamefont{Tolos et~al.}(2016)\citenamefont{Tolos, Centelles, and Ramos}}]{Tolos_2017}
\bibinfo{author}{\bibfnamefont{L.}~\bibnamefont{Tolos}}, \bibinfo{author}{\bibfnamefont{M.}~\bibnamefont{Centelles}}, \bibnamefont{and} \bibinfo{author}{\bibfnamefont{A.}~\bibnamefont{Ramos}}, \bibinfo{journal}{The Astrophysical Journal} \textbf{\bibinfo{volume}{834}}, \bibinfo{pages}{3} (\bibinfo{year}{2016}), \urlprefix\url{https://dx.doi.org/10.3847/1538-4357/834/1/3}.

\bibitem[{\citenamefont{Miyatsu et~al.}(2023)\citenamefont{Miyatsu, Cheoun, Kim, and Saito}}]{miyatsu2023can}
\bibinfo{author}{\bibfnamefont{T.}~\bibnamefont{Miyatsu}}, \bibinfo{author}{\bibfnamefont{M.-K.} \bibnamefont{Cheoun}}, \bibinfo{author}{\bibfnamefont{K.}~\bibnamefont{Kim}}, \bibnamefont{and} \bibinfo{author}{\bibfnamefont{K.}~\bibnamefont{Saito}}, \bibinfo{journal}{Phys. Lett. B} p. \bibinfo{pages}{138013} (\bibinfo{year}{2023}).

\bibitem[{\citenamefont{Todd-Rutel and Piekarewicz}(2005)}]{todd2005neutron}
\bibinfo{author}{\bibfnamefont{B.~G.} \bibnamefont{Todd-Rutel}} \bibnamefont{and} \bibinfo{author}{\bibfnamefont{J.}~\bibnamefont{Piekarewicz}}, \bibinfo{journal}{Phys. Rev. Lett.} \textbf{\bibinfo{volume}{95}}, \bibinfo{pages}{122501} (\bibinfo{year}{2005}).

\bibitem[{\citenamefont{Piekarewicz and Weppner}(2006)}]{piekarewicz2006insensitivity}
\bibinfo{author}{\bibfnamefont{J.}~\bibnamefont{Piekarewicz}} \bibnamefont{and} \bibinfo{author}{\bibfnamefont{S.~P.} \bibnamefont{Weppner}}, \bibinfo{journal}{Nucl. Phys. A} \textbf{\bibinfo{volume}{778}}, \bibinfo{pages}{10} (\bibinfo{year}{2006}).

\bibitem[{\citenamefont{Kumar et~al.}(2006)\citenamefont{Kumar, Agrawal, and Dhiman}}]{kumar2006effects}
\bibinfo{author}{\bibfnamefont{R.}~\bibnamefont{Kumar}}, \bibinfo{author}{\bibfnamefont{B.~K.} \bibnamefont{Agrawal}}, \bibnamefont{and} \bibinfo{author}{\bibfnamefont{S.~K.} \bibnamefont{Dhiman}}, \bibinfo{journal}{Phys. Rev. C} \textbf{\bibinfo{volume}{74}}, \bibinfo{pages}{034323} (\bibinfo{year}{2006}).

\bibitem[{\citenamefont{Sulaksono and Mart}(2006)}]{sulaksono2006low}
\bibinfo{author}{\bibfnamefont{A.}~\bibnamefont{Sulaksono}} \bibnamefont{and} \bibinfo{author}{\bibfnamefont{T.}~\bibnamefont{Mart}}, \bibinfo{journal}{Phys. Rev. C} \textbf{\bibinfo{volume}{74}}, \bibinfo{pages}{045806} (\bibinfo{year}{2006}).

\bibitem[{\citenamefont{Horowitz and Piekarewicz}(2002)}]{horowitz2002constraining}
\bibinfo{author}{\bibfnamefont{C.~J.} \bibnamefont{Horowitz}} \bibnamefont{and} \bibinfo{author}{\bibfnamefont{J.}~\bibnamefont{Piekarewicz}}, \bibinfo{journal}{Phys. Rev. C} \textbf{\bibinfo{volume}{66}}, \bibinfo{pages}{055803} (\bibinfo{year}{2002}).

\bibitem[{\citenamefont{Pradhan et~al.}(2023)\citenamefont{Pradhan, Chatterjee, Gandhi, and Schaffner-Bielich}}]{PRADHAN2023122578}
\bibinfo{author}{\bibfnamefont{B.~K.} \bibnamefont{Pradhan}}, \bibinfo{author}{\bibfnamefont{D.}~\bibnamefont{Chatterjee}}, \bibinfo{author}{\bibfnamefont{R.}~\bibnamefont{Gandhi}}, \bibnamefont{and} \bibinfo{author}{\bibfnamefont{J.}~\bibnamefont{Schaffner-Bielich}}, \bibinfo{journal}{Nuclear Physics A} \textbf{\bibinfo{volume}{1030}}, \bibinfo{pages}{122578} (\bibinfo{year}{2023}), ISSN \bibinfo{issn}{0375-9474}, \urlprefix\url{https://www.sciencedirect.com/science/article/pii/S0375947422002020}.

\bibitem[{\citenamefont{Alford et~al.}(2022)\citenamefont{Alford, Brodie, Haber, and Tews}}]{PhysRevC.106.055804}
\bibinfo{author}{\bibfnamefont{M.~G.} \bibnamefont{Alford}}, \bibinfo{author}{\bibfnamefont{L.}~\bibnamefont{Brodie}}, \bibinfo{author}{\bibfnamefont{A.}~\bibnamefont{Haber}}, \bibnamefont{and} \bibinfo{author}{\bibfnamefont{I.}~\bibnamefont{Tews}}, \bibinfo{journal}{Phys. Rev. C} \textbf{\bibinfo{volume}{106}}, \bibinfo{pages}{055804} (\bibinfo{year}{2022}), \urlprefix\url{https://link.aps.org/doi/10.1103/PhysRevC.106.055804}.

\bibitem[{\citenamefont{Haidari and Sharma}(2008)}]{haidari2008sigma}
\bibinfo{author}{\bibfnamefont{M.~M.} \bibnamefont{Haidari}} \bibnamefont{and} \bibinfo{author}{\bibfnamefont{M.~M.} \bibnamefont{Sharma}}, \bibinfo{journal}{Nucl. Phys. A} \textbf{\bibinfo{volume}{803}}, \bibinfo{pages}{159} (\bibinfo{year}{2008}).

\bibitem[{\citenamefont{Sharma}(2008)}]{sharma2008scalar}
\bibinfo{author}{\bibfnamefont{M.~M.} \bibnamefont{Sharma}}, \bibinfo{journal}{Phys. Lett. B} \textbf{\bibinfo{volume}{666}}, \bibinfo{pages}{140} (\bibinfo{year}{2008}).

\bibitem[{\citenamefont{Del~Estal et~al.}(2001)\citenamefont{Del~Estal, Centelles, Vinas, and Patra}}]{del2001effects}
\bibinfo{author}{\bibfnamefont{M.}~\bibnamefont{Del~Estal}}, \bibinfo{author}{\bibfnamefont{M.}~\bibnamefont{Centelles}}, \bibinfo{author}{\bibfnamefont{X.}~\bibnamefont{Vinas}}, \bibnamefont{and} \bibinfo{author}{\bibfnamefont{S.~K.} \bibnamefont{Patra}}, \bibinfo{journal}{Phys. Rev. C} \textbf{\bibinfo{volume}{63}}, \bibinfo{pages}{024314} (\bibinfo{year}{2001}).

\bibitem[{\citenamefont{Shen et~al.}(2020)\citenamefont{Shen, Ji, Hu, and Sumiyoshi}}]{Shen_2020}
\bibinfo{author}{\bibfnamefont{H.}~\bibnamefont{Shen}}, \bibinfo{author}{\bibfnamefont{F.}~\bibnamefont{Ji}}, \bibinfo{author}{\bibfnamefont{J.}~\bibnamefont{Hu}}, \bibnamefont{and} \bibinfo{author}{\bibfnamefont{K.}~\bibnamefont{Sumiyoshi}}, \bibinfo{journal}{The Astrophysical Journal} \textbf{\bibinfo{volume}{891}}, \bibinfo{pages}{148} (\bibinfo{year}{2020}), \urlprefix\url{https://dx.doi.org/10.3847/1538-4357/ab72fd}.

\bibitem[{\citenamefont{Menezes and Provid{\^e}ncia}(2004)}]{menezes2004delta}
\bibinfo{author}{\bibfnamefont{D.~P.} \bibnamefont{Menezes}} \bibnamefont{and} \bibinfo{author}{\bibfnamefont{C.}~\bibnamefont{Provid{\^e}ncia}}, \bibinfo{journal}{Phys. Rev. C} \textbf{\bibinfo{volume}{70}}, \bibinfo{pages}{058801} (\bibinfo{year}{2004}).

\bibitem[{\citenamefont{Typel}(2005)}]{typel2005relativistic}
\bibinfo{author}{\bibfnamefont{S.}~\bibnamefont{Typel}}, \bibinfo{journal}{Phys. Rev. C} \textbf{\bibinfo{volume}{71}}, \bibinfo{pages}{064301} (\bibinfo{year}{2005}).

\bibitem[{\citenamefont{Kl\"ahn et~al.}(2006)\citenamefont{Kl\"ahn, Blaschke, Typel, van Dalen, Faessler, Fuchs, Gaitanos, Grigorian, Ho, Kolomeitsev et~al.}}]{klahn2006constraints}
\bibinfo{author}{\bibfnamefont{T.}~\bibnamefont{Kl\"ahn}}, \bibinfo{author}{\bibfnamefont{D.}~\bibnamefont{Blaschke}}, \bibinfo{author}{\bibfnamefont{S.}~\bibnamefont{Typel}}, \bibinfo{author}{\bibfnamefont{E.~N.~E.} \bibnamefont{van Dalen}}, \bibinfo{author}{\bibfnamefont{A.}~\bibnamefont{Faessler}}, \bibinfo{author}{\bibfnamefont{C.}~\bibnamefont{Fuchs}}, \bibinfo{author}{\bibfnamefont{T.}~\bibnamefont{Gaitanos}}, \bibinfo{author}{\bibfnamefont{H.}~\bibnamefont{Grigorian}}, \bibinfo{author}{\bibfnamefont{A.}~\bibnamefont{Ho}}, \bibinfo{author}{\bibfnamefont{E.~E.} \bibnamefont{Kolomeitsev}}, \bibnamefont{et~al.}, \bibinfo{journal}{Phys. Rev. C} \textbf{\bibinfo{volume}{74}}, \bibinfo{pages}{035802} (\bibinfo{year}{2006}), \urlprefix\url{https://link.aps.org/doi/10.1103/PhysRevC.74.035802}.

\bibitem[{\citenamefont{Nik{\v{s}}i{\'c} et~al.}(2002)\citenamefont{Nik{\v{s}}i{\'c}, Vretenar, Finelli, and Ring}}]{nikvsic2002relativistic}
\bibinfo{author}{\bibfnamefont{T.}~\bibnamefont{Nik{\v{s}}i{\'c}}}, \bibinfo{author}{\bibfnamefont{D.}~\bibnamefont{Vretenar}}, \bibinfo{author}{\bibfnamefont{P.}~\bibnamefont{Finelli}}, \bibnamefont{and} \bibinfo{author}{\bibfnamefont{P.}~\bibnamefont{Ring}}, \bibinfo{journal}{Phys. Rev. C} \textbf{\bibinfo{volume}{66}}, \bibinfo{pages}{024306} (\bibinfo{year}{2002}).

\bibitem[{\citenamefont{Lalazissis et~al.}(2005)\citenamefont{Lalazissis, Nik{\v{s}}i{\'c}, Vretenar, and Ring}}]{lalazissis2005new}
\bibinfo{author}{\bibfnamefont{G.~A.} \bibnamefont{Lalazissis}}, \bibinfo{author}{\bibfnamefont{T.}~\bibnamefont{Nik{\v{s}}i{\'c}}}, \bibinfo{author}{\bibfnamefont{D.}~\bibnamefont{Vretenar}}, \bibnamefont{and} \bibinfo{author}{\bibfnamefont{P.}~\bibnamefont{Ring}}, \bibinfo{journal}{Phys. Rev. C} \textbf{\bibinfo{volume}{71}}, \bibinfo{pages}{024312} (\bibinfo{year}{2005}).

\bibitem[{\citenamefont{Typel et~al.}(2010)\citenamefont{Typel, R{\"o}pke, Kl{\"a}hn, Blaschke, and Wolter}}]{typel2010composition}
\bibinfo{author}{\bibfnamefont{S.}~\bibnamefont{Typel}}, \bibinfo{author}{\bibfnamefont{G.}~\bibnamefont{R{\"o}pke}}, \bibinfo{author}{\bibfnamefont{T.}~\bibnamefont{Kl{\"a}hn}}, \bibinfo{author}{\bibfnamefont{D.}~\bibnamefont{Blaschke}}, \bibnamefont{and} \bibinfo{author}{\bibfnamefont{H.~H.} \bibnamefont{Wolter}}, \bibinfo{journal}{Phys. Rev. C} \textbf{\bibinfo{volume}{81}}, \bibinfo{pages}{015803} (\bibinfo{year}{2010}).

\bibitem[{\citenamefont{Li and Sedrakian}(2023)}]{Li_2023}
\bibinfo{author}{\bibfnamefont{J.-J.} \bibnamefont{Li}} \bibnamefont{and} \bibinfo{author}{\bibfnamefont{A.}~\bibnamefont{Sedrakian}}, \bibinfo{journal}{The Astrophysical Journal} \textbf{\bibinfo{volume}{957}}, \bibinfo{pages}{41} (\bibinfo{year}{2023}), \urlprefix\url{https://dx.doi.org/10.3847/1538-4357/acfa73}.

\bibitem[{\citenamefont{Typel and Wolter}(1999)}]{typel1999relativistic}
\bibinfo{author}{\bibfnamefont{S.}~\bibnamefont{Typel}} \bibnamefont{and} \bibinfo{author}{\bibfnamefont{H.~H.} \bibnamefont{Wolter}}, \bibinfo{journal}{Nucl. Phys. A} \textbf{\bibinfo{volume}{656}}, \bibinfo{pages}{331} (\bibinfo{year}{1999}).

\bibitem[{\citenamefont{Rusnak and Furnstahl}(1997)}]{rusnak1997relativistic}
\bibinfo{author}{\bibfnamefont{J.~J.} \bibnamefont{Rusnak}} \bibnamefont{and} \bibinfo{author}{\bibfnamefont{R.~J.} \bibnamefont{Furnstahl}}, \bibinfo{journal}{Nucl. Phys. A} \textbf{\bibinfo{volume}{627}}, \bibinfo{pages}{495} (\bibinfo{year}{1997}).

\bibitem[{\citenamefont{B{\"u}rvenich et~al.}(2002)\citenamefont{B{\"u}rvenich, Madland, Maruhn, and Reinhard}}]{burvenich2002nuclear}
\bibinfo{author}{\bibfnamefont{T.~J.} \bibnamefont{B{\"u}rvenich}}, \bibinfo{author}{\bibfnamefont{D.~G.} \bibnamefont{Madland}}, \bibinfo{author}{\bibfnamefont{J.~A.} \bibnamefont{Maruhn}}, \bibnamefont{and} \bibinfo{author}{\bibfnamefont{P.-G.} \bibnamefont{Reinhard}}, \bibinfo{journal}{Phys. Rev. C} \textbf{\bibinfo{volume}{65}}, \bibinfo{pages}{044308} (\bibinfo{year}{2002}).

\bibitem[{\citenamefont{Zhao et~al.}(2010)\citenamefont{Zhao, Li, Yao, Meng et~al.}}]{zhao2010new}
\bibinfo{author}{\bibfnamefont{P.-W.} \bibnamefont{Zhao}}, \bibinfo{author}{\bibfnamefont{Z.-P.} \bibnamefont{Li}}, \bibinfo{author}{\bibfnamefont{J.-M.} \bibnamefont{Yao}}, \bibinfo{author}{\bibfnamefont{J.}~\bibnamefont{Meng}}, \bibnamefont{et~al.}, \bibinfo{journal}{Phys. Rev. C} \textbf{\bibinfo{volume}{82}}, \bibinfo{pages}{054319} (\bibinfo{year}{2010}).

\bibitem[{\citenamefont{Blaizot et~al.}(1995)\citenamefont{Blaizot, Berger, Decharg{\'e}, and Girod}}]{blaizot1995microscopic}
\bibinfo{author}{\bibfnamefont{J.~P.} \bibnamefont{Blaizot}}, \bibinfo{author}{\bibfnamefont{J.}~\bibnamefont{Berger}}, \bibinfo{author}{\bibfnamefont{J.}~\bibnamefont{Decharg{\'e}}}, \bibnamefont{and} \bibinfo{author}{\bibfnamefont{M.}~\bibnamefont{Girod}}, \bibinfo{journal}{Nucl. Phys. A} \textbf{\bibinfo{volume}{591}}, \bibinfo{pages}{435} (\bibinfo{year}{1995}).

\bibitem[{\citenamefont{Goriely et~al.}(2009{\natexlab{b}})\citenamefont{Goriely, Hilaire, Girod, and P{\'e}ru}}]{goriely2009first}
\bibinfo{author}{\bibfnamefont{S.}~\bibnamefont{Goriely}}, \bibinfo{author}{\bibfnamefont{S.}~\bibnamefont{Hilaire}}, \bibinfo{author}{\bibfnamefont{M.}~\bibnamefont{Girod}}, \bibnamefont{and} \bibinfo{author}{\bibfnamefont{S.}~\bibnamefont{P{\'e}ru}}, \bibinfo{journal}{Phys. Rev. Lett.} \textbf{\bibinfo{volume}{102}}, \bibinfo{pages}{242501} (\bibinfo{year}{2009}{\natexlab{b}}).

\bibitem[{\citenamefont{Gonzalez-Boquera et~al.}(2018)\citenamefont{Gonzalez-Boquera, Centelles, Vi{\~n}as, and Robledo}}]{gonzalez2018new}
\bibinfo{author}{\bibfnamefont{C.}~\bibnamefont{Gonzalez-Boquera}}, \bibinfo{author}{\bibfnamefont{M.}~\bibnamefont{Centelles}}, \bibinfo{author}{\bibfnamefont{X.}~\bibnamefont{Vi{\~n}as}}, \bibnamefont{and} \bibinfo{author}{\bibfnamefont{L.~M.} \bibnamefont{Robledo}}, \bibinfo{journal}{Phys. Lett. B} \textbf{\bibinfo{volume}{779}}, \bibinfo{pages}{195} (\bibinfo{year}{2018}).

\bibitem[{\citenamefont{Chappert et~al.}(2008)\citenamefont{Chappert, Girod, and Hilaire}}]{chappert2008towards}
\bibinfo{author}{\bibfnamefont{F.}~\bibnamefont{Chappert}}, \bibinfo{author}{\bibfnamefont{M.}~\bibnamefont{Girod}}, \bibnamefont{and} \bibinfo{author}{\bibfnamefont{S.}~\bibnamefont{Hilaire}}, \bibinfo{journal}{Phys. Lett. B} \textbf{\bibinfo{volume}{668}}, \bibinfo{pages}{420} (\bibinfo{year}{2008}).

\bibitem[{\citenamefont{Farine et~al.}(1999)\citenamefont{Farine, Von-Eiff, Schuck, Berger, Decharg{\'e}, and Girod}}]{farine1999towards}
\bibinfo{author}{\bibfnamefont{M.}~\bibnamefont{Farine}}, \bibinfo{author}{\bibfnamefont{D.}~\bibnamefont{Von-Eiff}}, \bibinfo{author}{\bibfnamefont{P.}~\bibnamefont{Schuck}}, \bibinfo{author}{\bibfnamefont{J.}~\bibnamefont{Berger}}, \bibinfo{author}{\bibfnamefont{J.}~\bibnamefont{Decharg{\'e}}}, \bibnamefont{and} \bibinfo{author}{\bibfnamefont{M.}~\bibnamefont{Girod}}, \bibinfo{journal}{J. Phys. G: Nucl. Part. Phys.} \textbf{\bibinfo{volume}{25}}, \bibinfo{pages}{863} (\bibinfo{year}{1999}).

\bibitem[{\citenamefont{Otsuka et~al.}(2006)\citenamefont{Otsuka, Matsuo, and Abe}}]{otsuka2006mean}
\bibinfo{author}{\bibfnamefont{T.}~\bibnamefont{Otsuka}}, \bibinfo{author}{\bibfnamefont{T.}~\bibnamefont{Matsuo}}, \bibnamefont{and} \bibinfo{author}{\bibfnamefont{D.}~\bibnamefont{Abe}}, \bibinfo{journal}{Phys. Rev. Lett.} \textbf{\bibinfo{volume}{97}}, \bibinfo{pages}{162501} (\bibinfo{year}{2006}).

\end{thebibliography}
\end{document}